\documentclass[11pt]{article} 

\usepackage[showcomments]{ajtex}

\usepackage{caption}
\usepackage{subcaption}
\usepackage{ytableau}
\usepackage{xfp}

\title{Notes on symmetries in particle physics}

\author{Akash Jain}\email{ajainphysics@gmail.com}

\affiliation{Department of Physics \& Astronomy, University of Victoria, PO
  Box 1700 STN CSC, Victoria, BC, V8W 2Y2, Canada}

\abstract{These are introductory notes on symmetries in quantum field theory and
  how they apply to particle physics. The notes cover the fundamentals of group
  theory, their representations, Lie groups, and Lie algebras, along with an
  elaborate discussion of the representations of SU(N), Lorentz, and Poincar\'e
  groups and their respective algebras. We spend a lot of time on the
  realisation of these symmetry groups in quantum field theory, as both global
  and gauge symmetries, as well as their spontaneous breaking and the Higgs
  mechanism. In the end, we culminate all the lessons from the course to
  enumerate the symmetries and field content of the Standard Model of particle
  physics and write down the Standard Model Lagrangian. Special consideration is
  given to how the weak-force gauge bosons and the matter fields obtain their
  mass via the Higgs mechanism.

  These lecture notes were developed while teaching the graduate course PHYS 509
  - Standard Model Phenomenology at the University of Victoria in the spring of
  2021. They build upon and extend the hand-written notes of Prof. Adam Ritz who
  had developed and delivered the course in previous years. That prior work is
  acknowledged, and included here with permission.}

\usepackage[bbgreekl]{mathbbol}

\setlist[itemize]{itemsep=0pt,topsep=5pt}
\setlist[enumerate]{itemsep=0pt,topsep=5pt}

\usepackage{tikz}
\usetikzlibrary{decorations.markings,decorations.pathmorphing,arrows.meta}

\tikzset{wave/.style={decorate,decoration={snake,segment length=1.7mm,
      amplitude=0.5mm}}}
\tikzset{coil/.style={decorate,decoration={coil, aspect=0.7, amplitude = 0.5mm,
      segment length = 1mm}}}
\tikzset{weak/.style={decorate,decoration={zigzag,segment length=1.7mm,
      amplitude=0.5mm}}}

\tikzset{left/.style={arrows={Stealth[scale length=0.5, scale width=1.5]-}}}
\tikzset{right/.style={arrows={-[sep=-2pt]Stealth[scale length=0.5, scale
      width=1.5]}}}
\tikzset{leftright/.style={arrows={Stealth[scale length=0.5, scale
      width=1.5]-Stealth[scale length=0.5, scale width=1.5]}}}

\definecolor{lightblue}{RGB}{200,225,255}
\definecolor{lightblueborder}{RGB}{30,100,255}

\begin{document}

\maketitle

\newpage

\section{Symmetries in particle physics}

Symmetry is one of the foundational pillars on which our modern understanding of
physics is built. Symmetries underlying physical systems put strong constraints
on the mathematical models we build to describe them, often boiling the freedom
down to a handful of free parameters that can be fixed by experiments and
observations. By virtue of Noether's theorem~\cite{Noether1918, Noether:1918zz},
symmetries have also been associated with the conserved quantities that remain
constant throughout the evolution of a system, such as energy, momentum, angular
momentum, charge, and particle number. These play a vital role in characterising
the spectrum of physical states a system can exhibit, as well as its
time-evolution. On the other hand, in high-energy physics, symmetries have been
known to underlie the fundamental forces we see in nature, i.e. weak force,
strong force, electromagnetic force, and the gravitational force (albeit a
complete description of gravity is still lacking due to our inability to
reconcile it with the quantum nature of the universe). Our present understanding
of the former three of these fundamental forces is compiled into a unified
field-theoretic description, developed during the second-half of the 20th
century, known as the \emph{Standard Model of particle
  physics}~\cite{Oerter:2006iy}. The discovery of a part of the model, known as
the \emph{electroweak theory}, furnishing a unified description of weak and
electromagnetic forces, was awarded the 1979 Nobel Prize in
physics~\cite{nobel_physics_1979}.

Even the absence of symmetries has proved to be a powerful organisational tool
in physics. The physical phases of matter, including the traditional ones such
as solids, liquids, and gases, as well as the more exotic ones such as
superfluids, superconductors, and the Bose-Einstein condensate, can often be
classified based on symmetries alone (or the absence thereof). Careful
consideration of symmetries has also led to the prediction and discovery of
entirely new phases of matter previously unknown to physicists, and have even
been the subject of the 2016 Nobel Prize in
physics~\cite{nobel_physics_2016}. In particle physics, the (spontaneous)
breaking of the electroweak symmetry at low-energies is crucial to give mass to
the weak-force carriers W and Z bosons, making the weak-force short-ranged. This
so-called Higgs mechanism is also responsible for the masses of (almost) all the
quarks and leptons making up the entire observable matter in the universe. This
discovery led to the 2013 Nobel Prize in physics~\cite{nobel_physics_2013}. The
only exception to this general rule are neutrinos, which are left massless by
the Higgs mechanism, but have been observed to be massive in nature. This
discovery was awarded yet another Nobel Prize in
2015~\cite{nobel_physics_2015}. Therefore, it is no surprise that a deluge of
pre-prints continues to appear every day in theoretical physics, suggesting
ingenious ways of exploiting symmetries to learn more about the world around us.

The aim of this course is to set the groundwork for the discussion of symmetries
in (quantum) field theory, particularly aimed at the applications in particle
physics. As we expect from a theory of fundamental forces in nature, the
Standard Model of particle physics is invariant under spacetime symmetries --
arbitrary space and time translations, rotations, and the special relativistic
Lorentz boosts -- together known as the Poincar\'e transformations. The model
also features certain abstract ``internal'' $\SU(3)\times\SU(2)\times\rmU(1)$
``gauge'' symmetry comprised of $8+3+1$ independent symmetry transformations; we
will learn more about this nomenclature and notation as we navigate through the
course. Superficially speaking, the gauge symmetries reflect redundancies in the
description of the theory that arise in order to manifestly preserve locality
and Lorentz invariance. The 8 transformations making up the SU(3) part of the
internal symmetry correspond to the strong force, while the remaining 4
transformations making up the $\SU(2)\times\rmU(1)$ part correspond to the
electroweak force. Every particle or field occurring in nature furnishes a
representation of these symmetries, i.e. transforms in a well-defined manner
under these symmetry transformations, in a way that leaves the physics
invariant. At our day-to-day energy scales, the $\SU(2)\times\rmU(1)$ symmetry
is spontaneously broken down to a single $\rmU(1)$ sub-transformation, breaking
down the electroweak force into a short-ranged weak force and a long-ranged
electromagnetic force. The Standard Model is not invariant under the
spatial-parity (P) and/or time-reversal (T) operation, which are symmetries of
the underlying Minkowski spacetime, but is invariant under the combined CPT
symmetry including the charge conjugation (C) operation.

In these notes, we will not follow the historical flow of developments leading
up to the Standard Model; several excellent references already follow this
approach; see e.g.~\cite{Burgess:2006hbd}. Rather, we will take a
symmetry-oriented perspective where we will use symmetries, their
representations, and their breaking pattern to build the Standard Model
Lagrangian from the ground-up. The mathematical language for symmetry is
\emph{group theory}. The notes start with a formal discussion of the basics of
group theory in \cref{sec:group-theory}, specialising to the Abelian unitary
group U(1) in \cref{sec:U1}, the special unitary group SU(N) in \cref{sec:SUN}
and the Lorentz and Poincar\'e groups in \cref{sec:LorentzPoincare}, relevant
for the Standard Model. Building upon these formal developments, in
\cref{field-theory-symm}, we introduce how these symmetry groups can be realised
in (quantum) field theory, and eventually write down the Standard Model
Lagrangian in \cref{sec:standardmodel}. These latter two sections also cover the
spontaneous breaking of internal symmetries and the Higgs mechanism, which
generates masses for the matter fields and the weak-force gauge bosons. Finally,
in \cref{sec:outlook}, we close off with a outlook and further scope of the
course material.

Although we are not aware of any texts that closely follow the structure of this
course, there are a plethora of excellent references that individually cover
various of its aspects, albeit in much more detail than the scope of these
notes. A list of the suggested reading material appears on Prof.~Adam Ritz's
\href{https://www.uvic.ca/science/physics/vispa/people/faculty/_ritz/P509.php}{course
  website}. We will mention the suggested reading material for individual
sections as we make our way through the notes.

Throughout these notes we will use the mostly positive metric sign convention
$(-1,1,1,1)$. We will \emph{not} set any of the fundamental constants to one for
pedagogical reasons.

\newpage

\section{Basic group theory}
\label{sec:group-theory}

In this section, we will study the fundamentals of group theory. We will set the
stage with some illustrative examples of ``rotation'' and ``reflection''
symmetry groups to draw out the essential features of group theory. We will then
proceed to a formal discussion of group theory and representation theory. A main
feature of this section will be the discussion on Lie groups and Lie algebras
that are vital from the perspective of field theory. The discussion in this
section will be quite brief and focused to our needs; a more complete treatment
can be found in the standard texts geared towards the application of group
theory in theoretical physics such as~\cite{Georgi:1999wka, Jones:1990ti,
  Cornwell:1997ke}.

\subsection{Introductory examples}
\label{sec:illustration}

To facilitate the forthcoming abstract discussion of group theory, let us start
with some illustrative examples of symmetries and groups. In particular, we
shall be interested in the symmetries of a square and a circle, i.e. the
geometric operations that leave the respective objects invariant. 

\subsubsection{Symmetries of a square -- dihedral group}

Let us start with a square. It has a 4-fold rotational symmetry, i.e. rotations
by multiples of $\pi/2$ of the square about its center leave the square
invariant:
\begin{equation*}
  \begin{tikzpicture}
    \draw[thick] (-10mm,-10mm) -- (-10mm,10mm) -- (10mm,10mm) -- (10mm,-10mm) -- (-10mm,-10mm);
    \draw[right,thick] (-2.8mm,-2.8mm) arc (-135:135:4mm);
    \node at (0,0) {$0$};
    \node at (0,-15mm) {$e$};
  \end{tikzpicture} \qquad
  \begin{tikzpicture}
    \draw[thick] (-10mm,-10mm) -- (-10mm,10mm) -- (10mm,10mm) -- (10mm,-10mm) -- (-10mm,-10mm);
    \draw[right,thick] (-2.8mm,-2.8mm) arc (-135:135:4mm);
    \node at (0,0) {$\frac{\pi}{2}$};
    \node at (0,-15mm) {$R_1$};
  \end{tikzpicture} \qquad
  \begin{tikzpicture}
    \draw[thick] (-10mm,-10mm) -- (-10mm,10mm) -- (10mm,10mm) -- (10mm,-10mm) -- (-10mm,-10mm);
    \draw[right,thick] (-2.8mm,-2.8mm) arc (-135:135:4mm);
    \node at (0,0) {$\pi$};
    \node at (0,-15mm) {$R_2$};
  \end{tikzpicture} \qquad
  \begin{tikzpicture}
    \draw[thick] (-10mm,-10mm) -- (-10mm,10mm) -- (10mm,10mm) -- (10mm,-10mm) -- (-10mm,-10mm);
    \draw[right,thick] (-2.8mm,-2.8mm) arc (-135:135:4mm);
    \node at (0,0) {$\frac{3\pi}{2}$};
    \node at (0,-15mm) {$R_3$};
  \end{tikzpicture}
\end{equation*}
There are three non-trivial rotations by angles $\pi/2$, $\pi$, and $3\pi/2$,
along with a trivial rotation by angle $0$ or $2\pi$ that amounts to doing
nothing.  In addition, the square is also invariant under reflections (parity
transformations) about the horizontal and vertical axes, and about the two
diagonals:
\begin{equation*}
  \begin{tikzpicture}
    \draw[thick] (-10mm,-10mm) -- (-10mm,10mm) -- (10mm,10mm) -- (10mm,-10mm) -- (-10mm,-10mm);
    \draw[thick,dashed] (0,-12mm) -- (0,12mm);
    \node at (0,-15mm) {$P_0$};
  \end{tikzpicture} \qquad
  \begin{tikzpicture}
    \draw[thick] (-10mm,-10mm) -- (-10mm,10mm) -- (10mm,10mm) -- (10mm,-10mm) --
    (-10mm,-10mm);
    \draw[thick,dashed] (12mm,-12mm) -- (-12mm,12mm);
    \node at (0,-15mm) {$P_1$};
  \end{tikzpicture} \qquad
  \raisebox{0mm}{\begin{tikzpicture}
    \draw[thick] (-10mm,-10mm) -- (-10mm,10mm) -- (10mm,10mm) -- (10mm,-10mm) -- (-10mm,-10mm);
    \draw[thick,dashed] (-12mm,0) -- (12mm,0);
    \node at (0,-15mm) {$P_2$};
  \end{tikzpicture}} \qquad
  \begin{tikzpicture}
    \draw[thick] (-10mm,-10mm) -- (-10mm,10mm) -- (10mm,10mm) -- (10mm,-10mm) -- (-10mm,-10mm);
    \draw[thick,dashed] (-12mm,-12mm) -- (12mm,12mm);
    \node at (0,-15mm) {$P_3$};
  \end{tikzpicture}
\end{equation*}
These 8 transformations together form the \emph{dihedral group}
$\rmD_4 = \{e,R_1,R_2,R_3,P_0,P_1,P_2,P_3\}$. Note that composing any two
symmetry transformations together results in another element of the group. This
can be summarised via the multiplication table
\begin{equation*}
  \begin{tabular}[h]{c|ccc}
    & $g_1$ & $g_2$ & $\cdots$ \\
    \hline
    $g_1$ & $g_1 g_1$ & $g_2 g_1$ & $\cdots$ \\
    $g_2$ & $g_1 g_2$ & $g_2 g_2$ & $\cdots$ \\
    $\vdots$ & $\vdots$ & $\vdots$ & $\ddots$
  \end{tabular}
  \qquad\implies\qquad
  \begin{tabular}[h]{c|cccccccc}
    & $e$ & $R_1$ & $R_2$ & $R_3$ & $P_0$ & $P_1$ & $P_2$ & $P_3$ \\
    \hline
    $e$ & $e$ & $R_1$ & $R_2$ & $R_3$ & $P_0$ & $P_1$ & $P_2$ & $P_3$ \\
    $R_1$ & $R_1$ & $R_2$ & $R_3$ & $e$ & $P_3$ & $P_0$ & $P_1$ & $P_2$ \\ 
    $R_2$ & $R_2$ & $R_3$ & $e$ & $R_1$ & $P_2$ & $P_3$ & $P_0$ & $P_1$ \\
    $R_3$ & $R_3$ & $e$ & $R_1$ & $R_2$ & $P_1$ & $P_2$ & $P_3$ & $P_0$ \\
    $P_0$ & $P_0$ & $P_1$ & $P_2$ & $P_3$ & $e$ & $R_1$ & $R_2$ & $R_3$ \\
    $P_1$ & $P_1$ & $P_2$ & $P_3$ & $P_0$ & $R_3$ & $e$ & $R_1$ & $R_2$ \\
    $P_2$ & $P_2$ & $P_3$ & $P_0$ & $P_1$ & $R_2$ & $R_3$ & $e$ & $R_1$ \\
    $P_3$ & $P_3$ & $P_0$ & $P_1$ & $P_2$ & $R_1$ & $R_2$ & $R_3$ & $e$
  \end{tabular}
\end{equation*}
The composition $g_1g_2$ means the transformation $g_2$ followed by the
transformation $g_1$.  Note also that the composition does not, in general,
commute. For example, $P_1 P_2 \neq P_2 P_1$. However, the composition is
associative, i.e. $(g_1 g_2)g_3 = g_1(g_2g_3)$ for all the elements of the
group. Another property that the symmetry group has is that every transformation
has an inverse, which can be used to undo the said transformation. To wit
\begin{gather*}
  e^{-1} = e, \qquad
  R_1^{-1} = R_3, \qquad
  R_2^{-1} = R_2, \qquad
  R_3^{-1} = R_1, \\
  P_0^{-1} = P_0, \qquad
  P_1^{-1} = P_1, \qquad
  P_2^{-1} = P_2, \qquad
  P_3^{-1} = P_3.
\end{gather*}

The symmetry transformations in the dihedral group can be represented by
$2\times 2$ matrices representing their action on a point $(a,b)$ on the
square. We find
\begin{equation}
  \begin{gathered}
    e =
    \begin{pmatrix}
      1 & 0 \\ 0 & 1
    \end{pmatrix}, \qquad R_1 =
    \begin{pmatrix}
      0 & 1 \\ -1 & 0
    \end{pmatrix}, \qquad R_2 =
    \begin{pmatrix}
      -1 & 0 \\ 0 & -1
    \end{pmatrix}, \qquad R_3 =
    \begin{pmatrix}
      0 & -1 \\ 1 & 0
    \end{pmatrix}, \\
    P_0 =
    \begin{pmatrix}
      - 1 & 0 \\ 0 & 1
    \end{pmatrix}, \qquad P_1 =
    \begin{pmatrix}
      0 & -1 \\ -1 & 0
    \end{pmatrix}, \qquad P_2 =
    \begin{pmatrix}
      1 & 0 \\ 0 & -1
    \end{pmatrix}, \qquad P_3 =
    \begin{pmatrix}
      0 & 1 \\ 1 & 0
    \end{pmatrix}.
  \end{gathered}
  \label{eq:rep-D4}
\end{equation}
This is called a \emph{representation} of the group. It can be explicitly
checked that this matrix representation follows the group multiplication table
given above.

It should be noted that the entire group can be generated by multiplying
together two fundamental elements: the $\pi/2$ rotation $R_1$ and the reflection
about the vertical axis $P_0$. For instance, we have
\begin{equation}
  e = (R_1)^4, \quad R_2 = (R_1)^2, \quad R_3 = (R_1)^3, \qquad
  P_1 = R_1P_0, \quad
  P_2 = (R_1)^2P_0, \quad
  P_3 = (R_1)^3P_0.
\end{equation}
The elements $R_1$, $P_0$ can be understood as the generators of the dihedral
group.

There are various subsets, called \emph{subgroups}, of the dihedral group
$\rmD_4$ that are closed under composition and form a group of their own. Few
notable examples are the orientation preserving transformations of a square
called the ``cyclic group'' $\bbZ_4 = \{e,R_1,R_2,R_3\}$, dihedral group of a
line-segment (or of a rectangle) $\rmD_2 = \{e,R_2,P_0,P_2\}$, the cyclic group
$\bbZ_2 = \{e,R_2\}$, and the trivial subset $\{e\}$. The cyclic groups $\bbZ_4$
and $\bbZ_2$ can also be represented by the complex numbers
\begin{equation}
  e = 1, \qquad
  R_1 = \E{i\pi/2}, \qquad
  R_2 = \E{i\pi}, \qquad
  R_3 = \E{i3\pi/2}.
  \label{eq:rep-Z4}
\end{equation}
This furnishes a \emph{complex representation} of the said groups and
illustrates the idea that groups can have many distinct representations. We note
yet another representation of $\bbZ_4$ and $\bbZ_2$ as
\begin{equation}
  e =
  \begin{pmatrix}
    1 & 0 & 0 & 0 \\
    0 & 1 & 0 & 0 \\
    0 & 0 & 1 & 0 \\
    0 & 0 & 0 & 1
  \end{pmatrix}, \quad
  R_1 =
  \begin{pmatrix}
    0 & 0 & 0 & 1 \\
    1 & 0 & 0 & 0 \\
    0 & 1 & 0 & 0 \\
    0 & 0 & 1 & 0 
  \end{pmatrix}, \quad
  R_2 =
  \begin{pmatrix}
    0 & 0 & 1 & 0 \\
    0 & 0 & 0 & 1 \\
    1 & 0 & 0 & 0 \\
    0 & 1 & 0 & 0
  \end{pmatrix}, \quad
  R_3 =
  \begin{pmatrix}
    0 & 1 & 0 & 0 \\
    0 & 0 & 1 & 0 \\
    0 & 0 & 0 & 1 \\
    1 & 0 & 0 & 0
  \end{pmatrix}.
  \label{eq:reducibleRepZ4}
\end{equation}

A similar story can be repeated for other $n$-sided regular polygons, such a
equilateral triangles, regular pentagons etc., leading to other dihedral groups
$\rmD_n$.

\subsubsection{Symmetries of a circle -- orthogonal group}

A circle can be obtained as the $n\to\infty$ limit of a regular $n$-sided
polygon. Correspondingly, the dihedral group $\rmD_n$ in $n\to\infty$ limit
leads to the symmetry group of a circle, known as the 2-dimensional
\emph{orthogonal group} $\rmO(2)$. The transformations in this group consist of
arbitrary rotations about the center, and reflections about arbitrary diameters:
\begin{equation*}
  \begin{tikzpicture}
    \draw [thick] (0,0) circle (10mm);
    \draw[right,thick] (-2.8mm,-2.8mm) arc (-135:135:4mm);
    \node at (0,0) {$\theta$};
    \node at (0,-15mm) {$R(\theta) ~|~ \theta \in [0,2\pi)$};
  \end{tikzpicture} \qquad\qquad
  \begin{tikzpicture}
    \draw [thick] (0,0) circle (10mm);
    \draw [thick,dashed] (-7mm,10mm) -- (7mm,-10mm);
    \draw [thick,dotted] (0,0) -- (0,10mm);
    \node at (-1.5mm,6mm) {$\frac{\phi}{2}$};
    \node at (0,-15mm) {$P(\phi) ~|~ \phi \in [0,2\pi)$};
  \end{tikzpicture}
\end{equation*}
The identity element is given by $e=R(0)$ that does nothing. The respective
multiplication table can be written down as
\begin{equation*}
  \begin{tabular}[h]{c|cc} & $R(\theta_2)$ & $P(\phi_2)$ \\
    \hline
    $R(\theta_1)$ & $R(\theta_1+\theta_2)$ & $P(\phi_2-\theta_1)$ \\
    $P(\phi_1)$ & $P(\phi_1+\theta_2)$ & $R(\phi_1-\phi_2)$
  \end{tabular}
\end{equation*}
It is clearly seen that the group multiplication does not commute, because
$R(\theta)P(\phi) \neq P(\phi)R(\theta)$, however it is associative. The inverse
of the rotations $R(\theta)$ are $R(-\theta)$, while the inverse of the
reflections $P(\phi)$ are themselves. 

The orthogonal group $\rmO(2)$ can be represented by $2\times 2$ matrices acting
on an arbitrary point $(a,b)$ on the circle:
\begin{equation}
  R(\theta) =
  \begin{pmatrix}
    \cos\theta & \sin\theta \\ -\sin\theta & \cos\theta
  \end{pmatrix}, \qquad
  P(\phi) =
  \begin{pmatrix}
    -\cos\phi & -\sin\phi \\ -\sin\phi & \cos\phi
  \end{pmatrix}.
  \label{eq:rep-O2}
\end{equation}
This is the set of all $2\times 2$ orthogonal matrices satisfying
$M M^\rmT = \mathbb{1}$; hence the name. In can be checked that this \emph{group
  representation} respects the multiplication table mentioned above.

The generators of O(2) can be taken to be the infinitesimal rotation
$R_\epsilon = R(\epsilon)$ (for $\epsilon\to0$) and the y-axis reflection
$P_0 = P(0)$, leading to
\begin{equation}
  R(\theta) = \lim_{\epsilon\to0} (R_\epsilon)^{\theta/\epsilon}, \qquad
  P(\phi) = \lim_{\epsilon\to0} (R_\epsilon)^{\phi/\epsilon} P_0.
\end{equation}
For infinitesimal transformations, given a matrix representation, the generators
can be represented as an exponential $R_\epsilon = \exp(i\epsilon\sigma_2)$,
where $\sigma_2$ is the second Pauli matrix. The matrix $\sigma_2$ is called the
Lie algebra generator of rotations. In terms of this, we can represent the group
elements as $R(\theta) = \exp(i\theta\sigma_2)$ and
$P(\phi) = \exp(i\phi\sigma_2)P_0$.

All dihedral groups $\rmD_n$ and cyclic groups $\bbZ_n$ form a subgroup of the
orthogonal group $\rmO(2)$. These are given by
\begin{align}
  \rmD_n
  &= \big\{R_a = R(2a\pi/n), P_{a} = P(2a\pi/n)
    ~|~a=0,1,\ldots, n-1 \big\}, \nn\\
  \bbZ_n
  &= \big\{R_a = R(2a\pi/n)~|~a=0,1,\ldots, n-1\big\}.
\end{align}
Another interesting subgroup of $\rmO(2)$ comprises of the orientation
preserving transformations of a circle $\SO(2) = \{R(\theta)\}$, known as the
special orthogonal group of dimension 2. In the matrix representation above,
$\SO(2)$ is the set of all $2\times 2$ orthogonal matrices with unit determinant
$\det M = 1$. There is another representation of $\SO(2)$ given by complex
numbers
\begin{equation}
  R(\theta) = \E{i\theta},
  \label{eq:rep-SO2}
\end{equation}
formally known as the unitary group $\rmU(1)$. More generally, the unitary group
$\rmU(N)$ is the set of all $N\times N$ unitary matrices satisfying
$M M^\dagger = \mathbb{1}$.

Similarly, the symmetry group of an $N$-dimensional sphere is called $\rmO(N)$,
while its orientation preserving subgroup is called $\SO(N)$. The former can be
represented by $N\times N$ orthogonal matrices, while the latter by $N\times N$
orthogonal matrices with unit determinant.

\subsection{Basic definitions}

\paragraph*{Group:}

A \emph{group} $G$ is a set of elements with a product rule, such that
\begin{enumerate}
\item $G$ is \emph{closed} under group multiplication, i.e. $g_1g_2 \in G$ for
  all elements $g_1,g_2\in G$ --- combining two symmetry operations is also a
  symmetry.
\item The group product is \emph{associative}, i.e. $(g_1g_2)g_3 = g_1(g_2g_3)$
  for all elements $g_1,g_2,g_3\in G$ --- symmetry operations are associative.
\item There exists a unique identity element $e\in G$ such that $eg = ge = g$
  for any element $g\in G$ --- there exists a trivial symmetry operation where
  nothing is done.
\item For every element $g\in G$, there exists a unique inverse $g^{-1}\in G$
  such that $g^{-1}g = gg^{-1} = e$ --- symmetry operations can be inverted to
  return to the original state.
\end{enumerate}
If $g_1g_2=g_2g_1$ for all $g_1,g_2\in G$, the group $G$ is called an
\emph{Abelian group} with a commutative product. If $g_1g_2\neq g_2g_1$ for some
$g_1,g_2\in G$, the group $G$ is called a \emph{non-Abelian group} with a
non-commutative product.

\begin{itemize}
\item Example: from our illustrative examples in \cref{sec:illustration}, it can
  be explicitly checked that $\rmD_4$, $\bbZ_4$, $\rmO(2)$, and $\SO(2)$ are all
  groups. In particular, $\bbZ_4$ and $\SO(2)$ are Abelian groups, while
  $\rmD_4$ and $\rmO(2)$ are non-Abelian groups.
  
\item Example: symmetry transformations in a field theory form a group. If a set
  of invertible field transformations $G=\{a,b,c,\ldots\}$ on the fields
  $\varphi \to a\circ\varphi$ leave the action invariant, i.e.
  $S[a\circ\varphi] = S[\varphi]$ etc.: (1) successive symmetry transformations
  also leave the action invariant, (2) symmetry transformations are associative,
  (3) the identity element is the trivial transformation $\varphi \to \varphi$,
  and (4) symmetry transformations have well-defined inverses.
\end{itemize}

\paragraph*{Group isomorphism:} Two groups $G$ and $G'$ are said to be
isomorphic, denoted as $G\cong G'$, if there exists an invertible map
$\iota:G\to G'$ such that $\iota(e) = e'$,
$\iota(g_1g_2) = \iota(g_1)\iota(g_2)$, and $\iota(g^{-1}) = \iota(g)^{-1}$,
where $g,g_1,g_2\in G$ are arbitrary elements and $e\in G$ and $e'\in G'$ are
the identity elements.
\begin{itemize}
\item Example: from our illustrative examples in \cref{sec:illustration}, the
  groups U(1) and SO(2) are isomorphic, with the isomorphism
  \begin{equation}
    \iota\lb \E{i\theta} \in \rmU(1) \rb =
    \begin{pmatrix}
      \cos\theta & \sin\theta \\ -\sin\theta & \cos\theta
    \end{pmatrix} \in \SO(2).
    \label{eq:U1-SO2}
  \end{equation}
\end{itemize}

\paragraph*{Product groups:} Given two groups $G$ and $G'$, their product group
is given by the set of ordered pairs $G\times G' = \{(g,g')~|~g\in G,g'\in G'\}$
with the obvious product operation
\begin{equation}
  (g_1,g'_1)(g_2,g'_2) = (g_1g_2,g'_1g'_2).
\end{equation}
\begin{itemize}
\item Example: from our illustrative examples in \cref{sec:illustration}, the
  dihedral group $\rmD_2$ is isomorphic to the product group
  $\bbZ_2\times\bbZ_2$, with the isomorphism $\iota(e) = (e,e)$,
  $\iota(R_2) = (R_2,R_2)$, $\iota(P_0) = (e,R_2)$, and $\iota(P_2) =
  (R_2,e)$.

\item NB: A subgroup $G$ is always isomorphic to its product with the trivial
  group $G\times \{e\}$. We will not distinguish between these groups.
\end{itemize}

\paragraph*{Subgroup:} A subset $H \subset G$ is called a \emph{subgroup} of a
group $G$ if it forms a group within itself with the same product rule.

A subgroup $H\subset G$ is called a \emph{normal} or \emph{invariant} subgroup
of $G$ if $g^{-1}h g \in H$ for all $h\in H$ and $g \in G$.

\begin{itemize}
\item NB: given any group $G$, the trivial group element $\{e\}$ and the group
  $G$ itself are normal subgroups of $G$.
\item NB: every subgroup of an Abelian group is normal.
\item NB: given two groups $G$ and $G'$, the product group $G\times G'$ has
  normal subgroups $G\cong G\times \{e\}$ and $G' \cong \{e\}\times G'$.
\item Example: $\bbZ_4$, $\rmD_2$, $\bbZ_2$ are subgroups of $\rmD_4$, while
  $\rmD_n$, $\bbZ_n$, $\SO(2)$ are subgroups of $\rmO(2)$. In particular, the
  subgroups $\bbZ_4$ and $\SO(2)$ are normal subgroups of $\rmD_4$ and $\rmO(2)$
  respectively.
\end{itemize}

\paragraph*{Simple and semi-simple groups:} A group $G$ is said to be
\emph{simple} if its only normal subgroups are the trivial group $\{e\}$ and $G$
itself. A group is semi-simple if it is a product of simple groups.
\begin{itemize}
\item Example: the cyclic groups of prime order $\bbZ_p$ are simple.
\item Example: the group SU(N) of all $N\times N$ unitary matrices with unit
  determinant is simple.
\end{itemize}

\paragraph*{Centre:} The \emph{centre}
$Z(G) = \{z\in G ~|~ zg = gz ~\forall g \in G \}$ of a group $G$ is the maximal
subset of $G$ that commutes with all elements of $G$.

\begin{itemize}
\item NB: the centre of a group is a normal subgroup.
\item NB: the centre of an Abelian group is the entire group itself.
\item Example: the centre of all the groups $\rmD_4$, $\bbZ_4$, $\rmO(2)$,
  $\SO(2)$ is the normal subgroup $\bbZ_2 = \{e,R_2\}$; note that
  $R_2 = R(\pi)$.
\end{itemize}

\paragraph*{Cosets:} Let $G$ be a group and $H\subset G$ be a subgroup. Given an
element $g\in G$, a left coset of $H$ in $G$ are given as
$gH = \{gh ~|~ h\in H\}$. Similarly, a right coset of $H$ in $G$ are given as
$Hg = \{hg ~|~ h\in H\}$.
\begin{itemize}
\item NB: given a normal subgroup $H\subset G$, the left and right cosets of $H$
  in $G$ are the same, i.e. $gH = Hg$ for all $g\in G$, and are simply called
  cosets.
\end{itemize}

\paragraph*{Quotient group:} Given a group $G$ and a normal subgroup
$H\subset G$, the quotient group of $H$ in $G$ is defined to be the set of all
cosets $G/H = \{gH~|~g\in G\}$. The product rule in the quotient group is given
as $(g_1H)(g_2H) = (g_1g_2)H$ for all $g_1,g_2\in G$. Note that the identity
element in $G/H$ is simply $H$ and the inverse elements are
$(gH)^{-1} = g^{-1}H$.

\begin{itemize}
\item NB: given a product group $G\times G'$ and the normal subgroup
  $G\cong G\times\{e\}$, the quotient group $(G\times G')/G$ is isomorphic to
  $G'$. Similarly, $(G\times G')/G'$ is isomorphic to $G$. Note that the
  converse is not true: $H \times (G/H)$ is not generically isomorphic to $G$.

\item Example: $\bbZ_4$ is a normal subgroup of $\rmD_4$, and
  $\rmD_4/\bbZ_4 \cong \bbZ_2$. However,
  $\rmD_4\not\cong\bbZ_4\times\bbZ_2$. Similarly, $\SO(2)$ is a normal subgroup
  of $\rmO(2)$, and $\rmO(2)/\SO(2) \cong \bbZ_2$. However,
  $\rmO(2) \not\cong \SO(2)\times\bbZ_2$.

\end{itemize}

\paragraph*{Order:} the \emph{order} $|G|$ of a group $G$ is the number of
  elements in $G$.

\begin{itemize}
\item NB: continuous symmetry transformations, such as rotations of a circle,
  have order infinite and are known as infinite dimensional groups.
\item Example: $|\rmD_n| = 2n$, $|\bbZ_n| = n$, $|\rmO(2)| = \infty$, and
  $|\SO(2)| = \infty$.
\item NB: Given groups $G$ and $G'$, we have $|G\times G'| = |G|
  |G'|$. Similarly, given a group $G$ and normal subgroup $H\subset G$, we have
  $|G/H| = |G|/|H|$. Note that the latter is not well-defined when $H$ is an
  infinite dimensional group.
\end{itemize}

\subsection{Group representations}
\label{sec:group-reps}

An important aspect of group theory is that groups can have many
\emph{representations}. In addition to giving a tangible meaning to the abstract
group elements in terms of matrices and such, in field theory, it also implies
that the same symmetry can be realised by various different kinds of fields.

\paragraph*{Representation:} An $N$-dimensional \emph{matrix representation}
$D(G)$ of a group $G$ is a mapping of the elements of $G$ onto a set of
$N\times N$ invertible matrices $D:G\to \mathrm{GL(N})$, s.t.
\begin{enumerate}
\item $D(e) = \mathbb{1}$, where $\mathbb{1}\in\mathrm{GL(N})$ is the
  $N\times N$ identity matrix.
\item $D(g_1)D(g_2) = D(g_1 g_2)$ for all $g_1,g_2 \in G$.
\end{enumerate}
A representation is said to be \emph{faithful} if the mapping $D(G)$ is
one-to-one and invertible. In particular, for a faithful representation,
$D(g)=\mathbb{1} \implies g=e$. A representation $D(G)$ is said to be
\emph{unitary} if $D(g)$ is a unitary matrix for all $g\in G$,
i.e. $D(g)D(g)^\dagger = \mathbb1$.

The representation $D(g) = \mathbb{1}$ for all $g\in G$ is called the
\emph{trivial} or \emph{singlet} representation of the group $G$. Except for the
order 1 trivial group $G=\{e\}$, the trivial representation is never a faithful
representation. The trivial representation is a unitary representation.

For a complex matrix representation $D(G)$, the \emph{conjugate representation}
$D^*(G)$ is defined as $D^*(g) = D(g)^*$ for all $g\in G$.

\begin{itemize}
\item NB: a faithful representation of a group is isomorphic to the group.
\item Given a representation $D(G)$ of a group $G$ and a subgroup $H\subset G$,
  we can construct a representation $D(H)$ of $H$ by restricting to the elements
  of $H$ in the map.
\item Example: A 2-dimensional real matrix representation of $\rmD_4$ and
  $\bbZ_4$ is given in \cref{eq:rep-D4}, while a 1-dimensional complex matrix
  representation of $\bbZ_4$ is given in \cref{eq:rep-Z4}. Similarly, a
  2-dimensional real matrix representation of $\rmO(2)$ and $\SO(2)$ is given in
  \cref{eq:rep-O2}, while a 1-dimensional complex matrix representation of
  $\SO(2)$ is given in \cref{eq:rep-SO2}. All of these are faithful
  representations. All of these are unitary representations.
\item NB: in quantum field theory, we will particularly be interested in unitary
  representations. Given that a physical theory admits a symmetry group $G$, the
  physical states are expected to transform under a unitary representation of
  the group $\ket{\psi} \to D(G)\ket{\psi}$ for $g\in G$. This ensures that the
  inner product $\langle \psi|\psi\rangle$ remains invariant under the
  symmetry. Similarly, observable Hermitian operators also transform under a
  unitary representation of the group
  $\mathcal{O}\to D(G) \mathcal{O} D(G)^{-1}$, so that they remain Hermitian
  after the transformation.
\end{itemize}

\paragraph*{Equivalent representations:} Two representations $D(G)$, $D'(G)$ of
a group $G$ are said to be equivalent if there exists an invertible matrix $S$
such that $D'(g) = S^{-1}D(g)S$ for all $g\in G$. We will often not distinguish
between equivalent representations of a group because they are related by a mere
change of basis.

\begin{itemize}
\item NB: equivalent unitary representations $D(G)$, $D'(G)$ of a group $G$ are
  related via a unitary matrix $U$, i.e. $D'(g) = U^{\dagger}D(g)U$ for all
  $g\in G$.
\end{itemize}

\paragraph*{Direct sum representations:} Given two representations $D(G)$,
$D'(G)$ of a group $G$, we can construct a larger direct sum representation as
\begin{equation}
  (D\oplus D')(g) =
  \begin{pmatrix}
    D(g)  & 0 \\
    0 & D'(g)
  \end{pmatrix}
  \quad \forall g\in G
\end{equation}
We can iterate this procedure for any number of representations.

\paragraph*{Irreducible representation:} A representation $D(G)$ of a group $G$
is said to be \emph{reducible} if it is equivalent to a block upper-triangular
representation, i.e. there exists a matrix $S$ such that
\begin{equation}
  S^{-1} D(g) S =
  \begin{pmatrix}
    D_1(g) & D_{12}(g) \\
    0 & D_2(g)
  \end{pmatrix} \quad\forall g\in G,
\end{equation}
where the $\dim D_1(g) \neq \dim D(g)$. Note that, while $D_1(G)$, $D_2(G)$ are
smaller sub-representations of $G$, the map $D_{12}(G)$ is \emph{not} a
representation. A representation is said to be \emph{irreducible}, if it is not
reducible.

A representation is said to be \emph{completely reducible} if it is equivalent
to a direct sum of irreducible representations. 

\begin{itemize}
\item NB: an irreducible representation is automatically completely reducible.
\item NB: we state without proof an important result that every
  finite-dimensional unitary representation of a group is completely reducible
  into a direct sum of irreducible unitary representations. This implies that to
  study unitary representations of a group, it is sufficient to classify all of
  its irreducible unitary representations.
  
\item Example: \cref{eq:rep-D4} furnishes an irreducible representation of both
  $\rmD_4$ and $\bbZ_4$, and \cref{eq:rep-Z4} furnishes an irreducible
  representation of $\bbZ_4$.  While \cref{eq:rep-Z4}, being a one-dimensional
  representation, continues to furnish an irreducible representation of all the
  subgroups of $\bbZ_4$ as well, the representations of $\rmD_2$ and $\bbZ_2$
  furnished by \cref{eq:rep-D4} consist of diagonal matrices and hence are
  reducible. A similar story applies to the representations of the continuous
  groups $\rmO(2)$ and $\SO(2)$ given in \cref{eq:rep-O2,eq:rep-SO2}.

\item The representation of $\bbZ_4$ given in \cref{eq:reducibleRepZ4} is a
  reducible representation. Being a unitary representation, it is also a
  completely reducible representation. Using
  \begin{equation}
    S =
    \begin{pmatrix}
      1 & -1 & 1 & 0 \\
      1 & 1 & 0 & -1 \\
      1 & -1 & -1 & 0 \\
      1 & 1 & 0 & 1 \\
    \end{pmatrix},
  \end{equation}
  as the equivalence transformation, this representation turns into
  \begin{gather}
    e =
    \begin{pmatrix}
      1 & 0 & 0 & 0 \\
      0 & 1 & 0 & 0 \\
      0 & 0 & 1 & 0 \\
      0 & 0 & 0 & 1 \\
    \end{pmatrix}, \qquad
    R_1 =
    \begin{pmatrix}
      1 & 0 & 0 & 0 \\
      0 & -1 & 0 & 0 \\
      0 & 0 & 0 & 1 \\
      0 & 0 & -1 & 0 \\
    \end{pmatrix}, \nn\\
    R_3 =
    \begin{pmatrix}
      1 & 0 & 0 & 0 \\
      0 & 1 & 0 & 0 \\
      0 & 0 & -1 & 0 \\
      0 & 0 & 0 & -1 \\
    \end{pmatrix}, \qquad
    R_3 =
    \begin{pmatrix}
      1 & 0 & 0 & 0 \\
      0 & -1 & 0 & 0 \\
      0 & 0 & 0 & -1 \\
      0 & 0 & 1 & 0 \\
    \end{pmatrix}.
  \end{gather}
  This is a direct sum of the trivial representation, another 1-dimensional
  non-faithful representation $D(e)=D(R_2)=1$, $D(R_1)=D(R_3)=-1$, and the
  2-dimensional faithful representation given in \cref{eq:rep-D4}.
\end{itemize}

\subsection{Lie algebras}

Let us take a quick detour to study a related topic -- Lie algebras. These are
quintessential in the study of continuous groups and play a pivotal role in
high-energy physics. Their connection to group theory will be established later
in \cref{sec:lie-groups}.

\paragraph*{Lie algebra:} A \emph{Lie algebra} $\fg$ is a vector-space over some
field $F$ (real $\bbR$ or complex numbers $\bbC$) with a bilinear Lie bracket
operation $[\cdot,\cdot]: \fg\times\fg \to \fg$ satisfying
\begin{enumerate}
\item \emph{Alternativity}: $[X,X] = 0$ for all $X\in\fg$.
\item \emph{Anti-commutativity}: $[X,Y] = -[Y,X]$ for all $X,Y\in\fg$.
\item \emph{Bilinearity}: $[aX+bY,Z] = a[X,Z]+b[Y,Z]$ for all $X,Y,Z\in\fg$ and
  $a,b\in F$.
\item \emph{Jacobi identity}: $[X,[Y,Z]]+[Y,[Z,X]]+[Z,[X,Y]]=0$.
\end{enumerate}
The dimension of a Lie algebra $\dim\fg$ is defined to be the dimension of the
vector space. A Lie algebra $\fg$ is said to be \emph{Abelian} if $[X,Y]=0$ for
all $X,Y\in\fg$.

A Lie algebra is said to be Abelian if $[X,Y]=0$ for all $X,Y\in\fg$. A Lie
algebra is said to be non-Abelian if it is not Abelian.

\begin{itemize}
\item Example: examples of Lie algebras include the matrix algebras over
  $F=\bbR$:
  \begin{itemize}
  \item $\mathfrak{gl}(N,\bbR)$, $\mathfrak{gl}(N,\bbC)$: real and complex
    $N\times N$ matrices (dimension $N^2$, $2N^2$).
  \item $\mathfrak{sl}(N,\bbR)$, $\mathfrak{sl}(N,\bbC)$: real and complex
    $N\times N$ traceless matrices, $\tr M = 0$ (dimension $N^2-1$, $2N^2-2$).
  \item $\mathfrak{so}(N)$: imaginary $N\times N$ antisymmetric matrices,
    $M^\rmT = -M$ (dimension $N(N-1)/2$).
  \item $\mathfrak{u}(N)$: complex $N\times N$ Hermitian matrices,
    $M^\dagger = M$ (dimension $N^2$).
  \item $\mathfrak{su}(N)$: complex $N\times N$ traceless Hermitian matrices,
    $M^\dagger = M$ and $\tr M = 0$ (dimension $N^2-1$).
  \end{itemize}
  The respective Lie bracket operation is given by the matrix commutator
  $[X,Y] = XY - YX$. The Lie algebras $\mathfrak{gl}(N,\bbC)$ and
  $\mathfrak{sl}(N,\bbC)$ can also be defined over $F=\bbC$, but have dimensions
  $N^2$ and $N^2-1$ respectively.
\end{itemize}

\paragraph*{Generators and structure constants:} Given a Lie algebra $\fg$ over
a field $F$, we can choose a vector-space basis $T_a$ on $\fg$, known as the
\emph{generators} of $\fg$, so that an arbitrary element $X\in\fg$ can be
represented as $X = \sum_a\alpha_a T_a$ with $\alpha_a \in F$. In this basis,
the entire structure of the Lie algebra can be captured by the \emph{structure
  constants} $f_{abc}$, defined as
\begin{equation}
  [T_a,T_b] = i\hbar \sum_c f_{abc} T_c,
\end{equation}
where $\hbar$ is the reduced Planck's constant.\footnote{The factor of $i\hbar$
  is a purely physicist's convention and does not appear in pure maths
  literature. Physicists also typically set $\hbar$ to 1.}  We can choose a
particular set of generators orthonormalised such that
\begin{equation}
  \tr(T_a T_b) = C\hbar^2 \delta_{ab},
\end{equation}
for some convenient real number $C$. In such a basis, the structure constants
are totally antisymmetric; to wit
\begin{equation}
  i\hbar^3 C f_{abc}
  = \tr([T_a,T_b]T_c)
  = \tr([T_c,T_a]T_b)
  = \tr([T_b,T_c]T_a).
\end{equation}
\begin{itemize}
\item Example: the generator of $\mathfrak{so}(2)$ is given by the second Pauli
  matrix
  \begin{equation}
    J_1 =
    \hbar\begin{pmatrix}
      0 & -i \\ i & 0
    \end{pmatrix},
  \end{equation}
  with $f_{111} = 0$ and $C = 2$. Similarly, the generators of
  $\mathfrak{so}(3)$ are given by
  \begin{equation}\label{eq:so3-algebra}
    J_1 = \hbar
    \begin{pmatrix}
      0 & 0 & 0 \\ 0 & 0 & -i \\ 0 & i & 0
    \end{pmatrix}, \qquad
    J_2 = \hbar
    \begin{pmatrix}
      0 & 0 & i \\ 0 & 0 & 0 \\ -i & 0 & 0
    \end{pmatrix}, \qquad
    J_3 = \hbar
    \begin{pmatrix}
      0 & -i & 0 \\ i & 0 & 0 \\ 0 & 0 & 0
    \end{pmatrix},
  \end{equation}
  with $f_{abc} = \epsilon_{abc}$ being the Levi-Civita symbol, and $C=2$.

\item Example: the generators of $\mathfrak{su}(2)$ are given by the Pauli
  matrices (normalised by $1/2$)
  \begin{equation}\label{eq:su2-algebra}
    T_1 = \frac\hbar2
    \begin{pmatrix}
      0 & 1 \\ 1 & 0
    \end{pmatrix}, \qquad
    T_2 = \frac\hbar2
    \begin{pmatrix}
      0 & -i \\ i & 0
    \end{pmatrix}, \qquad
    T_3 = \frac\hbar2
    \begin{pmatrix}
      1 & 0 \\ 0 & -1
    \end{pmatrix},
  \end{equation}
  with again $f_{abc} = \epsilon_{abc}$ and $C = 1/2$. This should be
  familiar as the angular-momentum algebra in quantum mechanics. Similarly, the
  generators of $\mathfrak{su}(3)$ are given by the \emph{Gell-Mann matrices}
  (normalised by $\hbar/2$)
  \begin{gather}
    T_1 = \frac{\hbar}{2}
    \begin{pmatrix}
      0 & 1 & 0 \\ 1 & 0 & 0 \\ 0 & 0 & 0
    \end{pmatrix}, \qquad
    T_2 = \frac{\hbar}{2}
    \begin{pmatrix}
      0 & -i & 0 \\ i & 0 & 0 \\ 0 & 0 & 0
    \end{pmatrix}, \qquad
    T_3 = \frac{\hbar}{2}
    \begin{pmatrix}
      1 & 0 & 0 \\ 0 & -1 & 0 \\ 0 & 0 & 0
    \end{pmatrix}, \nn\\
    T_4 = \frac\hbar2
    \begin{pmatrix}
      0 & 0 & 1 \\ 0 & 0 & 0 \\ 1 & 0 & 0
    \end{pmatrix}, \qquad
    T_5 = \frac\hbar2
    \begin{pmatrix}
      0 & 0 & -i \\ 0 & 0 & 0 \\ i & 0 & 0
    \end{pmatrix}, \qquad
    T_6 = \frac\hbar2
    \begin{pmatrix}
      0 & 0 & 0 \\ 0 & 0 & 1 \\ 0 & 1 & 0
    \end{pmatrix}, \nn\\
    T_7 = \frac\hbar2
    \begin{pmatrix}
      0 & 0 & 0 \\ 0 & 0 & -i \\ 0 & i & 0
    \end{pmatrix}, \qquad
    T_8 = \frac{\hbar}{2\sqrt{3}}
    \begin{pmatrix}
      1 & 0 & 0 \\ 0 & 1 & 0 \\ 0 & 0 & -2
    \end{pmatrix},
    \label{eq:su3-algebra}
  \end{gather}
  with the structure constants
  \begin{equation}\label{eq:su3-structure-constants}
    f_{123} = 1, \qquad
    f_{147} = f_{165} = f_{246} = f_{257} = f_{345} = f_{376} = \frac12, \qquad
    f_{458} = f_{678} = \frac{\sqrt{3}}{2},
  \end{equation}
  while all others either 0 or determined by the total anti-symmetry of
  $f_{abc}$, and $C = 1/2$.
  
\item Example: The $\mathfrak{su}(2)$ and $\mathfrak{su}(3)$ algebras can be
  extended to $\fu(2)$ and $\fu(3)$ respectively by including an additional
  generator
  \begin{equation}
    T_0 = \frac\hbar2 \mathbb{1}_{2\times 2}
    \qquad\text{and}\qquad
    T_0 = \frac{\hbar}{\sqrt{6}} \mathbb{1}_{3\times 3},
  \end{equation}
  respectively. The additional structure constants $f_{0ab} = 0$, because $T_0$
  commutes with all the other generators in the respective algebras, while
  $C = 1/2$. The $T_0$ generator for generic $\fu(N)$ is given by $\hbar/\sqrt{2N}\,\mathbb{1}_{N\times N}$.

\item NB: Note that the Lie algebra $\mathfrak{su}(1)$ is trivial and only
  contains the element $0$, as there are no non-zero $1\times 1$ traceless
  Hermitian matrices. The Lie algebra $\fu(1)$, on the other hand, has generator
  $T_0 = \hbar/\sqrt{2}$, with $f_{000} = 0$ and $C = 1/2$. Note that $\fu(1)$
  is just the set of real numbers.
\end{itemize}

\paragraph*{Lie algebra isomorphism:} Two Lie algebras $\fg$ and $\fg'$ over a
field $F$ are said to be isomorphic, denoted as $\fg\cong\fg'$, if there exists
an invertible map $\iota:\fg\to\fg'$ such that
$\iota(aX+bY) = a\iota(X)+b\iota(Y)$ and $\iota([X,Y]) = [\iota(X),\iota(Y)]$,
where $X,Y\in \fg$ are arbitrary elements and $a,b\in F$.
\begin{itemize}
\item NB: given two isomorphic Lie algebras $\fg\cong\fg'$ and the generators
  $T_a$ of $\fg$, the set $\iota(T_a)$ serves as generators of $\fg'$. In this
  basis, the structure constants of the isomorphic Lie algebras $\fg$ and $\fg'$
  are the same.
  
\item Example: the Lie algebras $\mathfrak{so}(2)$ and $\mathfrak{u}(1)$ are
  isomorphic, with the isomorphism
  \begin{equation}
    \iota(\theta\in\fu(1)) = \theta\, J_1 \in \mathfrak{so}(2).
  \end{equation}
  The Lie algebras $\mathfrak{so}(3)$ and $\mathfrak{su}(2)$ are also
  isomorphic, with the isomorphism
  \begin{equation}
    \iota(\theta_aT_a\in \mathfrak{su}(2)) = \theta_a J_a \in \mathfrak{so}(3).
    \label{eq:so3-su2-map}
  \end{equation}
  However, unlike $\SO(2)\cong\rmU(1)$, the groups $\SO(3)$ and $\SU(2)$ are not
  isomorphic; more on this later in \cref{sec:lie-groups}.

\end{itemize}

\paragraph*{Direct sum of Lie algebras:} Given two Lie algebras $\fg$ and $\fg'$
over a field $F$, the direct sum Lie algebra is given by
$\fg\oplus\fg'= \{(X,X')~|~X\in\fg,X'\in\fg'\}$, such that
\begin{equation}
  a(X,X') + b(Y,Y') = (aX+bY, aX'+ bY') \quad
  \forall~a,b\in F,
\end{equation}
with the obvious commutator operation
\begin{equation}
  [(X,X'),(Y,Y')] = ([X,Y],[X',Y']).
\end{equation}
\begin{itemize}
\item Example: the Lie algebra $\mathfrak{u}(N)$ is isomorphic to the direct sum
  Lie algebra $\mathfrak{su}(N)\oplus\fu(1)$, with the isomorphism given by
  \begin{equation}
    \iota\Big( (\lambda_a T_a,\hbar\lambda_0) \in \mathfrak{su}(N)\oplus\fu(1) \Big)
    = \lambda_a T_a + \sqrt{2N}\,\lambda_0 T_0 \in \fu(N).
    \label{eq:un-sun-map}
  \end{equation}
  Here $T_a$ with $a=1,2,\ldots, N^2-1$ are the generators of
  $\mathfrak{su}(N)$, while the generator $T_0$ is proportional to the identity
  matrix. It should again be noted that the group $\rmU(N)$ is not isomorphic to
  the product group $\SU(N)\times\rmU(1)$; more on this later in
  \cref{sec:lie-groups}.
\end{itemize}

\paragraph*{Lie subalgebra:} Given a Lie algebra $\fg$, a subset $\fh\subset\fg$
is called a \emph{Lie subalgebra} of $\fg$ if it is closed under the Lie bracket
operation: $[X,Y]\in\fh$ for all $X,Y\in\fh$.

Given a Lie algebra $\fg$, a Lie subalgebra $\fh\subset\fg$ is said to be an
\emph{invariant subalgebra} or \emph{ideal} if $[X,Y]\in\fh$ for all $X\in\fg$
and $Y\in\fh$.

\begin{itemize}
\item NB: given a Lie algebra $\fg$, the trivial algebra $\{0\}$ and the Lie
  algebra $\fg$ itself are Lie subalgebras.

\item NB: given two Lie algebras $\fg$ and $\fg'$, the direct sum Lie algebra
  $\fg\oplus\fg'$ has invariant subalgebras $\fg\cong\fg\oplus\{0\}$ and
  $\fg'\cong\{0\}\oplus\fg'$.
  
\item Example: $\mathfrak{sl}(N,\bbR)\subset\mathfrak{gl}(N,\bbR)$,
  $\mathfrak{sl}(N,\bbC)\subset\mathfrak{gl}(N,\bbC)$,
  $\mathfrak{so}(N)\subset\mathfrak{sl}(N,\bbR)$,
  $\mathfrak{u}(N)\subset\mathfrak{gl}(N,\bbC)$,
  $\mathfrak{su}(N)\subset\mathfrak{sl}(N,\bbC)$,
  $\mathfrak{su}(N)\subset\mathfrak{u}(N)$ are examples of Lie
  subalgebras. Moreover, $\mathfrak{sl}(N,\bbR)\subset\mathfrak{gl}(N,\bbR)$,
  $\mathfrak{sl}(N,\bbC)\subset\mathfrak{gl}(N,\bbC)$, and
  $\mathfrak{su}(N)\subset\mathfrak{u}(N)$ are invariant subalgebras.
  
\item NB: given a Lie algebra $\fg$ over a field $F$ with generators $T_a$, it
  admits an infinite number of $\fu(1)$ subalgebras spanned by any linear
  combination of the generators:
  $\fh(\lambda_a) = \{\alpha \lambda_a T_a~|~\alpha\in\bbR\}$ for any
  $\lambda_a\in F$.
\item Example: The Lie algebra $\mathfrak{su}(3)$, with generators
  $T_{1,\ldots,8}$ given in \cref{eq:su3-algebra}, has an $\mathfrak{su}(2)$
  subalgebra spanned by the generators $T_{1,2,3}$.
\end{itemize}

\paragraph*{Simple and semi-simple Lie algebra:} A Lie algebra $\fg$ is said to
be \emph{simple} if it does not admit any invariant subalgebras except the
trivial algebra $\{0\}$ and $\fg$ itself.\footnote{Some authors restrict simple
  Lie algebras and Lie groups to be non-Abelian.\label{foot:simpledef}} A Lie algebra is said to be
\emph{semi-simple} if it is a direct sum of simple Lie algebras.
\begin{itemize}
\item Example: $\mathfrak{sl}(N,\bbC)$, $\mathfrak{sl}(N,\bbR)$ and
  $\mathfrak{su}(N)$ for $N\geq 1$ are simple Lie algebras.
\item Example: $\fu(N)\cong\mathfrak{su}(N)\oplus\fu(1)$ for $N\geq 1$ is a
  semi-simple Lie algebra.
\end{itemize}

\paragraph*{Centre:} The centre $Z(\fg) = \{X\in\fg~|~[X,Y]=0 ~\forall Y\in\fg\}$
of a Lie algebra $\fg$ is defined to the maximal subset of $\fg$ that commutes
with all the elements of $\fg$.

\begin{itemize}
\item NB: the centre of a Lie algebra is an invariant Lie subalgebra.
\item NB: the centre of an Abelian Lie algebra is itself.
\end{itemize}

\paragraph*{Quotient algebra:} Given a Lie algebra $\fg$ and an invariant
subalgebra $\fh$, the quotient algebra of $\fh$ in $\fg$ is defined as
$\fg/\fh = \{X+\fh~|~X\in\fg\}$, where $X+\fh = \{X+Y~|~Y\in\fh\}$ for all
$X\in\fg$. The commutator on the quotient algebra is defined as
$[X+\fh,Y+\fh] = [X,Y]+\fh$.
\begin{itemize}
\item NB: given a direct sum Lie algebra $\fg\oplus\fg'$ and the invariant
  subalgebra $\fg\cong\fg\oplus\{0\}$, we have that
  $(\fg\oplus\fg')/\fg \cong\fg'$. Similarly $(\fg\oplus\fg')/\fg' \cong\fg$.
\end{itemize}

\paragraph*{Lie algebra representation:} A \emph{matrix representation} $D(\fg)$
of a Lie algebra $\fg$ defined over a field $F$ is a linear map onto the set of
matrices $D:\fg \to \mathfrak{gl}(N)$ satisfying
\begin{enumerate}
\item $D(aX+bY) = a D(X) + b D(Y)$ for all $X,Y\in\fg$ and $a,b\in F$.
\item $[D(X),D(Y)] = D([X,Y])$ for all $X,Y\in\fg$.
\end{enumerate}

A Lie algebra representation is said to be \emph{faithful} if the mapping
$D(\fg)$ is one-to-one and invertible. In particular, for a faithful
representation, $D(X)=0\implies X=0$. A representation $D(\fg)$ of a Lie algebra
$\fg$ is said to be \emph{Hermitian} if $D(\fg)$ is a Hermitian matrix for all
$X\in\fg$, i.e. $D(X)^\dagger = D(X)$.

The definitions of equivalent representations, direct sum of representations,
and reducible, irreducible, and completely reducible representations for Lie
algebras are the same as for groups in \cref{sec:group-reps}.

\begin{itemize}
\item NB: a faithful representation of a Lie algebra is isomorphic to the Lie
  algebra.
\item NB: we state without proof that a Hermitian Lie algebra representation is
  completely reducible as a direct sum of irreducible Hermitian Lie algebra
  representations.
\end{itemize}

\paragraph*{Adjoint representation:} The \emph{adjoint representation} is a
$\dim{\fg}\times\dim{\fg}$ representation furnished by the structure constants,
with the matrix elements of the generators given by
\begin{equation}\label{eq:adjoint-rep}
  (T^{\text{adjoint}}_a)_{bc} = -i\hbar f_{abc}.
\end{equation}
Note that the Jacobi's identity implies
\begin{gather}
  \sum_d\lb f_{bcd}f_{ade} + f_{abd}f_{cde} + f_{cad}f_{bde} \rb = 0 \nn\\
  \implies
  \lB T^{\text{adjoint}}_{a},T^{\text{adjoint}}_{b}\rB_{de}
  = i\hbar \sum_c f_{abc}(T^{\text{adjoint}}_{c})_{de}.
\end{gather}
\begin{itemize}
  
\item Example: the $\mathfrak{so}(3)$ algebra given in \cref{eq:so3-algebra} is
  already in the adjoint representation. The adjoint representation for
  $\mathfrak{su}(2)$ is also given by \cref{eq:so3-algebra}. This supports
  $\mathfrak{so}(3)\cong\mathfrak{su}(2)$.

\item Example: the adjoint representation for $\mathfrak{su}(3)$ can be worked
  out using the structure constants in \cref{eq:su3-structure-constants}.

\item Example: The adjoint representation for $\fu(N)$ is not faithful, because
  $D(J_0)=0$. Similarly, the adjoint representations for $\mathfrak{so}(2)$ and
  $\fu(1)$ are not faithful.
\end{itemize}

\subsection{Lie groups}
\label{sec:lie-groups}

After our little excursion into Lie algebras, let us return to group theory. We
will now focus on continuous groups, called Lie groups. These play an important
role in quantum field theory as the groups of continuous symmetry operations
such as rotations. Lie groups will also formalise the relation between Lie
algebras and group theory.

\paragraph*{Lie group:} A \emph{Lie group} $G$ is an
infinite-dimensional group whose elements $g(\alpha)$ can be smoothly labelled
by finite set of continuous parameters $\alpha = \{\alpha_1,\alpha_2, \ldots\}$,
such that the group multiplication is a smooth operation
\begin{equation}
  g(\alpha)g(\beta) = g(\gamma(\alpha,\beta)), \qquad\qquad
  \gamma_i(\alpha,\beta) \text{ are smooth functions of $\alpha_i$ and $\beta_i$}.
\end{equation}

\begin{itemize}
\item Example: the orthogonal group $\rmO(2)$ and the special orthogonal group
  $\SO(2)$ from our examples in \cref{sec:illustration} are Lie groups, with the
  smooth mapping furnished by \cref{eq:rep-O2}.
\item Example: other examples of matrix Lie groups are
  \begin{itemize}
  \item $\mathrm{GL}(N,\bbR)$, $\mathrm{GL}(N,\bbC)$: real and complex
    $N\times N$ invertible matrices.
  \item $\mathrm{SL}(N,\bbR)$, $\mathrm{SL}(N,\bbC)$: real and complex
    $N\times N$ matrices with unit determinant, $\det M = 1$.
  \item $\rmO(N)$: real $N\times N$ orthogonal matrices, $MM^\rmT = \mathbb{1}$
    --- the symmetry group of an $N$-dimensional sphere.
  \item $\SO(N)$: real $N\times N$ orthogonal matrices with unit determinant,
    $MM^\rmT = \mathbb{1}$, $\det M = 1$ --- the symmetry group of rotations of
    an $N$-dimensional sphere.
  \item $\rmU(N)$: complex $N\times N$ unitary matrices,
    $M M^\dagger = \mathbb{1}$.
  \item $\SU(N)$: complex $N\times N$ unitary matrices with unit determinant,
    $M M^\dagger = \mathbb{1}$, $\det M = 1$.
  \end{itemize}
\end{itemize}

\paragraph*{Lie subgroup:} A subgroup $H\subset G$ is said to be a \emph{Lie
  subgroup} of $G$ if it forms a Lie group itself.

\begin{itemize}
\item Example: $\SL(N,\bbR)\subset\GL(N,\bbR)$, $\SL(N,\bbC)\subset\GL(N,\bbC)$,
  $\rmO(N)\subset\GL(N,\bbR)$, $\SO(N)\subset\SL(N,\bbR)$,
  $\rmU(N)\subset\GL(N,\bbC)$, $\SU(N)\subset\SL(N,\bbC)$,
  $\SO(N)\subset\rmO(N)$, and $\SU(N)\subset\rmU(N)$ are all examples of Lie
  subgroups. Moreover, $\SL(N,\bbR)\subset\GL(N,\bbR)$,
  $\SL(N,\bbC)\subset\GL(N,\bbC)$, $\SO(N)\subset\rmO(N)$, and
  $\SU(N)\subset\rmU(N)$ are normal/invariant subgroups.
\end{itemize}

\paragraph*{Connected Lie groups:} A Lie group $G$ is said to be
\emph{connected} if every group element can be continuously connected to the
identity element, i.e. there exists a parametrisation $g(\alpha)$ of $G$ so that
$g(\alpha=0)=e$ for all $g(\alpha)\in G$.

The \emph{connected part} of a Lie group $G$ is the Lie subgroup of $G$
comprising of all the elements of $G$ that are continuously connected to the
identity element. The connected part of a Lie group is a connected Lie group
itself.

\begin{itemize}
\item $\SO(N)$ is the connected part of $\rmO(N)$. This is because an $\rmO(N)$
  matrix has either determinant 1 or $-1$, while an $\SO(N)$ matrix has
  determinant 1. The identity matrix has determinant 1, which cannot be flipped
  to $-1$ using continuous deformations.
  
\item Unlike the orthogonal case, both $\SU(N)$ and $\rmU(N)$ are connected Lie
  groups. This is because a $\rmU(N)$ matrix has determinant $\exp(i\theta)$,
  and these can be continuously arrived at starting from the identity matrix by
  moving around the complex unit circle.
\end{itemize}


\paragraph*{Simple Lie groups:} A Lie group $G$ is said to be simple if is
connected and has no connected normal Lie subgroups except the trivial group
$\{e\}$ and $G$ itself; see \cref{foot:simpledef}.
\begin{itemize}
\item NB: a connected Lie group that is simple as a group is automatically a
  simple Lie group. However, a simple Lie group is not necessarily simple as a
  group. In particular, a simple Lie group is allowed to have discrete normal
  subgroups.
\item Example: U(1) is a simple Lie group. However, U(1) is not simple as group
  because it has discrete normal subgroups.
\end{itemize}

\paragraph*{Exponential parametrisation:} Given a faithful representation $D(G)$
of a Lie group $G$, a group element $g(\alpha)$ arbitrarily close to the
identity element $e$ can be represented as
\begin{equation}
  D(g(\alpha)) = \mathbb{1} + \frac{i}{\hbar} \sum_a \alpha_a D(T_a) + O(\alpha^2).
\end{equation}
Here $T_a$ are abstract objects known as the generators of $G$, while $D(T_a)$
is a faithful representation of the generators. The connected part of $G$ in the
representation $D(G)$ can be generated by multiplying together an infinite
number of infinitesimal elements $g(\alpha/k)$ with $k\to\infty$, i.e.
\begin{equation}
  D(g(\alpha)) = \lim_{k\to\infty} \lb \mathbb{1}
  + \frac{i}{\hbar k} \alpha_a D(T_a)  \rb^{k}
  = \exp\lb i/\hbar\, \alpha_a D(T_a) \rb.
\end{equation}
Summation over the index $a$ is understood. This is known as the
\emph{exponential parametrisation} of the connected part of the Lie group.

The generators $T_a$ span the Lie algebra
$\fg = \{\lambda_a T_a~|~\lambda_a\in\bbR\}$ associated with the Lie group $G$,
with the Lie bracket derived from the commutator operation on the representation
space: $[X,Y]=D^{-1}([D(X),D(Y)])$ for $X,Y\in\fg$. Note that $D(G)$ is a
faithful representation, so the map $D$ is invertible. Matrix commutators
identically satisfy all the properties of a Lie bracket. Closedness of the Lie
algebra under commutators follows from the group product
\begin{gather}
  \exp\lb \frac{i}{\hbar} \sum_a \alpha_a D(T_a) \rb
  \exp\lb \frac{i}{\hbar} \sum_a \beta_a D(T_a) \rb
  = \exp\lb \frac{i}{\hbar} \sum_a \gamma_a D(T_a) \rb
  \quad\text{for some }\gamma_a \nn\\
  \implies
  \gamma_a D(T_a)
  = \alpha_a D(T_a)
  + \beta_a D(T_a)
  + \frac{i}{2\hbar} \alpha_a\beta_b [D(T_a),D(T_b)]
  + \ldots \nn\\
  \implies
  [D(T_a),D(T_b)] = i\hbar f_{abc}D(T_c) \quad\text{for some } f_{abc} \nn\\
  \text{and}\qquad
  \gamma_a = \alpha_a + \beta_a + \frac{i}{2} f_{bca} \alpha_b\beta_c
  + \ldots.
\end{gather}
The structure constants $f_{abc}$ only depend on the structure of the group and
not on the representation employed.  The exponential parametrisation of a Lie
group $G$ and the relation to the associated Lie algebra $\fg$ can also be
established without making reference to a faithful representation, however we
will not concern ourselves with these formalities.

\begin{itemize}
\item Example: Lie algebras associated with the aforementioned matrix groups are
  \begin{itemize}
  \item $\mathrm{GL}(N,\bbR) \to \mathfrak{gl}(N,\bbR)$,
    $\mathrm{GL}(N,\bbC) \to \mathfrak{gl}(N,\bbC)$.
  \item $\mathrm{SL}(N,\bbR) \to \mathfrak{sl}(N,\bbR)$,
    $\mathrm{SL}(N,\bbC) \to \mathfrak{sl}(N,\bbC)$.
  \item $\rmO(N),\mathrm{SO}(N) \to \mathfrak{so}(N)$.
  \item $\rmU(N) \to \fu(N)$, $\mathrm{SU}(N) \to \mathfrak{su}(N)$.
  \end{itemize}
\item Example: note that the Lie algebra for both $\rmO(N)$ and $\SO(N)$ groups
  is $\mathfrak{so}(N)$. This is because the connected part of $\rmO(N)$ is its
  subgroup $\SO(N)$.

\item NB: let $G$, $G'$ be connected Lie groups with associated Lie algebras $\fg$,
  $\fg'$.  Exponentiating the direct sum Lie algebra $\fg\oplus\fg'$ leads to
  the direct product Lie group $G\times G'$ and vice-versa.

\item NB: given a Lie group $G$ with Lie algebra $\fg$, and a Lie subgroup
  $H\subset G$, the Lie algebra $\fh$ associated with $H$ is a Lie subalgebra of
  $\fg$.

\item NB: A simple Lie algebra generates a \emph{simple Lie group} via
  exponentiation.

\item NB: given a Lie group $G$ and the associated Lie algebra $\fg$, the
  concepts of equivalent representations, direct sum of representations, and
  reducible, irreducible, and completely reducible representations of invertibly
  translate into each other.

\item NB: given a Lie group $G$ and the associated Lie algebra $\fg$, a
  Hermitian representation of a Lie algebra generates a unitary representation
  of the Lie group and vice-versa.
  
\end{itemize}

\paragraph*{Isomorphic Lie algebras vs isomorphic Lie groups:} Let $G\cong G'$
be isomorphic connected Lie groups, then the associated Lie algebras are also
isomorphic $\fg\cong\fg'$. However, the converse is \emph{not} true: given two
isomorphic Lie algebras $\fg\cong\fg'$, the connected Lie groups $G$, $G'$
generated via exponentiation need not be isomorphic.
\begin{itemize}
\item Example: consider the isomorphic Lie algebras
  $\mathfrak{su}(2)\cong\mathfrak{so}(3)$ with the isomorphism
  $\iota:\mathfrak{su}(2)\to\mathfrak{so}(3)$ given in
  \cref{eq:so3-su2-map}. Upon exponentiation, this defines a map
  $\iota:\SU(2)\to\SO(3)$ as
  \begin{equation}
    \iota\big(\exp(i/\hbar\,\theta_a T_a)\in\SU(2) \big)
    = \exp(i/\hbar\,\iota(\theta_a T_a))
    = \exp(i/\hbar\,\theta_a J_a)
    \in \SO(3).
    \label{eq:so3-double-cover}
  \end{equation}
  However, this is not an invertible map because multiple entries in $\SU(2)$
  are mapped to the same entry in $\SO(3)$. Take, for example,
  $\theta_a=(0,0,0)$ and $\theta_a=(0,0,2\pi)$. This results in distinct
  elements $\exp(i/\hbar\, \theta_aT_a) = \mathbb1$ and
  $\exp(i/\hbar\,\theta_aT_a) = \exp(2\pi i/\hbar\, T_3) = -\mathbb1$ in
  $\SU(2)$, being mapped to the same element
  $\exp(i/\hbar\,\theta_aJ_a) = \mathbb1$ and
  $\exp(i/\hbar\,\theta_aJ_a) = \exp(2\pi i/\hbar\, J_3) = \mathbb1$ in
  $\SO(3)$. Therefore, the Lie groups $\SU(2)$ and $\SO(3)$ are not
  isomorphic. In fact, $\SU(2)$ is the double-cover of $\SO(3)$, where every
  element is covered twice.

  This has important implications in quantum mechanics. Fundamental particles in
  nature do not transform under unitary representations of the rotation group
  $\SO(3)$, but of its double-cover $\SU(2)$. This leads to the existence of
  fermions, which are particles that need to be completely rotated in space
  twice before returning to their original state. This is because a single
  $2\pi$ rotation in $\SO(3)$ corresponds to the inversion operator $-\mathbb 1$
  in $\SU(2)$, and a double $4\pi$ rotation is needed to return to the original
  state.

\item Example: consider the isomorphic Lie algebras
  $\mathfrak{su}(N)\oplus\fu(1)\cong\fu(N)$ with the isomorphism
  $\iota: \mathfrak{su}(N)\oplus\fu(1)\to\fu(N)$ given in
  \cref{eq:un-sun-map}. Upon exponentiation, this defines a map among the Lie
  groups $\iota:\SU(N)\times\rmU(1)\to\rmU(N)$ given as
  \begin{align}
    \iota\Big((\exp(i/\hbar\,\lambda_a T_a),
    \exp(i\lambda_0)) \in \SU(N)\times\rmU(1)
    \Big)
    &= \exp\lb i/\hbar\, \iota\big((\lambda_a T_a,\hbar\lambda_0)\big)\rb \nn\\
    &= \exp(i/\hbar\, \lambda_a T_a + i\lambda_0\mathbb{1}) \nn\\
    &= \E{i\lambda_0} \exp(i/\hbar\, \lambda_a T_a)
    \in \rmU(N).
  \end{align}
  Or more succinctly, given $(U,\alpha)\in \SU(N)\times\rmU(1)$, we have
  $\iota((U,\alpha)) = \alpha U \in \rmU(N)$. However, this is again not an
  invertible map. Let $\omega$ be one of the $N$ distinct $N$th roots of
  identity. Then $(\omega U,\alpha/\omega)\in \SU(N)\times\rmU(1)$ map to the
  same element $\alpha U\in\rmU(N)$ for every $\omega$. Therefore, the Lie
  groups $\SU(N)\times\rmU(1)$ and $\rmU(N)$ are not isomorphic for $N>1$. In
  fact, $\SU(N)\times\rmU(1)$ covers $\rmU(N)$ a total of $N$-times.
\end{itemize}

\newpage

\section{U(1) group}
\label{sec:U1}

U(1) is the simplest Lie group. It is the group of all $1\times 1$ unitary
matrices, which amounts to the set of all unit norm complex numbers. U(1) is an
Abelian group, because complex numbers commute among themselves. U(1) is also a
simple Lie group. An arbitrary element $\alpha\in\rmU(1)$ can be parametrised as
\begin{equation}
  \alpha = \exp(i\theta), \qquad
  \text{where}\quad \theta\in[0,2\pi).
\end{equation}
As mentioned in \cref{eq:U1-SO2}, U(1) is isomorphic to SO(2), the group of all
$2\times 2$ real orthogonal matrices. Therefore, U(1) can also be understood as
the group of all two-dimensional rotations.

The Lie algebra of U(1) is $\fu(1)$, which is the set of all $1\times1$
Hermitian matrices or, in other words, the set of all real numbers $\bbR$. The
generator of U(1) can be taken to be $T_0 = \hbar$. The structure constant
$f_{000} = 0$ because $[T_0,T_0]=0$. On the other hand, $\tr(T_0T_0) = \hbar^2$,
so $C=1$ in this basis.

The irreducible representations of U(1) can be characterised by a real number
$q$. To wit, the irreducible representation $D_q$ of U(1) is defined as
\begin{equation}
  D_q(\E{i\theta}) = \E{iq\theta} \qquad \forall \E{i\theta}\in\rmU(1)
  \quad\text{and}\quad \theta\in\bbR.
\end{equation}
Correspondingly, the action of $D_q$ on the Lie algebra is given as
$D_q(\theta) = q\theta$ for all $\theta\in\fu(1)$. In terms of the generator,
$D_q(T_0) = q\hbar$. Note that the conjugate representation to $D_q$ is
$D_{-q}$. A complex field $\phi$ is said to transform in the $D_q$
representation of U(1), or to have charge $q$ under U(1), if
$\E{i\theta}\in\rmU(1)$ acts as
\begin{equation}
  \phi \to \phi' = \E{iq\theta}\phi.
\end{equation}

The Abelian Lie group U(1) is different from its non-Abelian cousins, which we
will study in the next section, in that its irreducible representations are
parametrised by a continuous real parameter $q$. In other words, U(1) charges of
fields can be fractional or even irrational. This is a good news for quarks
which are observed to have fractional charges under the electromagnetic U(1)
symmetry. However, it does beg the question why the electromagnetic charges of
fundamental fields in our universe are quantised at all.

\newpage

\section{Special unitary group}
\label{sec:SUN}

In this section we specialise to the special unitary group $\SU(N)$, associated
with ``internal'' symmetries in the Standard Model of particle physics. This is
the group comprising of $N\times N$ unit determinant unitary matrices
$U\in\SU(N)$ satisfying $U U^\dagger = U^\dagger U = \mathbb1$ and $\det U =
1$. The unitarity condition can be represented in the index notation as
\begin{equation}
  U_i{}^k U^j{}_{k}  = U_k{}^j U^k{}_{i} = \delta_{i}^j,
\end{equation}
where $U_i{}^j$ denotes the components of $U$ and $U^{j}{}_i = (U_j{}^i)^*$
denotes the components of $(U^\dagger)^\rmT$. The repeated indices are
understood to be summed over $i,j,k,\ldots = 1,2,\ldots,N$. $\SU(N)$ is a Lie
group and the associated Lie algebra $\mathfrak{su}(N)$ is the set of all
$N\times N$ traceless Hermitian matrices. The dimension of $\mathfrak{su}(N)$ as
a vector-space is $N^2-1$ and the respective generators are denoted as $T_a$
with $a=1,2,\ldots, N^2-1$. We study the representations of $\SU(N)$ and
introduce the method of Young tableaux to classify all the irreducible
representations. Later in the section, we specialise to the specific examples of
the first few special unitary groups SU(2) and SU(3), which will be helpful in
the forthcoming discussion of the Standard Model. Majority of the discussion
here is derived from~\cite{Georgi:1999wka}.

\subsection{SU(N) representations}

\paragraph*{Fundamental representation:} The identity map $D(U)_i{}^j = U_i{}^j$
for all $U\in\SU(N)$ furnishes an obvious representation of the group $\SU(N)$,
known as the \emph{fundamental representation}. An $N$-component complex field
$\psi_i = (\psi_1,\ldots,\psi_N)$ is said to transform in the fundamental
representation of $\mathrm{SU}(N)$ if an arbitrary group transformation
$U\in\SU(N)$ acts as
\begin{equation}
  \psi_i \to \psi'_i = U_i{}^j\psi_j.
\end{equation}
The fundamental representation is often denoted by the dimension ``$\mathbf N$''
of the vector $\psi_i$ (for instance, ``$\mathbf 3$'' for the fundamental
representation of SU(3)). Suppressing the indices, the transformation rule for a
fundamental field can be represented as $\psi \to U\psi$.

\paragraph*{Anti-fundamental representation:} The \emph{anti-fundamental
  representation} is the conjugate representation to the fundamental
representation, given as $D(U)^i{}_j = (U_i{}^j)^*$. An $N$-component complex
field $\bar\psi^i$ is said to transform in the anti-fundamental representation
of $\SU(N)$ if an arbitrary group transformation $U\in\SU(N)$ acts as
\begin{equation}
  \bar\psi^i \to \bar\psi'^i = U^i{}_j \bar\psi^j.
\end{equation}
The anti-fundamental representation is often denoted with a bar over the
dimension ``$\bar{\mathbf N}$''. Suppressing the indices, the transformation
rule for an anti-fundamental field can be represented as
$\bar\psi \to \bar\psi U^\dagger$.
\begin{itemize}
\item NB: given a fundamental vector field $\psi_i$, the complex conjugate
  vector $\bar\psi^i = (\psi_i)^*$ transforms in the anti-fundamental
  representation. Similarly, given an anti-fundamental vector field
  $\bar\psi^i$, the complex conjugate vector $\psi_i = (\bar\psi^i)^*$
  transforms in the fundamental representation.
\end{itemize}

\paragraph*{Singlet representation:} The trivial map $D(U) = 1$ for all
$U\in\SU(N)$ furnishes the \emph{trivial} or \emph{singlet representation} of
the group $\SU(N)$. A complex field $\phi$ is said to transform in the
\emph{singlet representation} of $\SU(N)$ if it is invariant under a
$U\in\SU(N)$ transformation
\begin{equation}
  \phi \to \phi' = \phi.
\end{equation}
The trivial representation is denoted by ``$\mathbf 1$''.
\begin{itemize}
\item NB: given a fundamental field $\psi_i$, we can create a singlet field as
  $\psi_i\bar\psi^i$.
\end{itemize}

\paragraph*{Adjoint representation:} A complex field $\Psi_i{}^j$, with
$i,j=1,\ldots,N$ and the trace $\Psi^i{}_i = 0$, is said to transform in the
\emph{adjoint representation} of $\SU(N)$ if
\begin{equation}
  \Psi_i{}^j \to \Psi'_i{}^j = U_i{}^k U^j{}_l \Psi_k{}^l.
\end{equation}
Suppressing the indices, the transformation rule is given as
$\Psi\to U\Psi U^\dagger$. The adjoint representation is denoted by the
dimension ``$\mathbf{N^2-1}$'' of the field $\Psi_i{}^j$.
\begin{itemize}
\item NB: given a fundamental field $\psi_i$, we can create an adjoint field as
  $\psi_i \bar\psi^j - 1/N\,\delta_i^j \psi_k\bar\psi^k$.
\item NB: elements of the Lie algebra $\mathfrak{su}(N)$, i.e. $N\times N$
  traceless Hermitian matrices, transform in the adjoint representation of
  $\SU(N)$. Note that
  \begin{equation}
    X^\dagger = X \implies (UXU^{-1})^\dagger = UXU^{-1}, \qquad
    \tr X = 0 \implies \tr(UXU^{-1}) = 0.
  \end{equation}
\item NB: the relation to the adjoint representation \eqref{eq:adjoint-rep} of
  the Lie algebra $\mathfrak{su}(N)$ can be established as follows: given the
  $\mathfrak{su}(N)$ generators $T_a$ and the exponential representation
  $U=\exp(i/\hbar\,\theta^a T_a)$, the transformation rule of $T_a$ is given as
  \begin{equation}
    T_a \to U T_a U^{-1}
    = \exp(i/\hbar\,\theta^bT_b^{\text{adjoint}})^c{}_a T_c.
  \end{equation}
\end{itemize}

\paragraph*{Tensor representations:} An arbitrary tensor field
$\Psi^{j_1j_2\ldots}_{i_ii_2\ldots}$ that behaves under an $\SU(N)$
transformation according to
\begin{equation}
  \Psi_{i_1i_2\ldots}^{j_1j_2\ldots}
  \to \Psi'^{j_1j_2\ldots}_{i_1i_2\ldots}
  = (U_{i_1}{}^{k_1}U_{i_2}{}^{k_2}\ldots)
  (U^{j_1}{}_{l_1}U^{j_2}{}_{l_2}\ldots)
  \Psi_{k_1k_2\ldots}^{l_1l_2\ldots},
  \label{eq:SUN_tensor_trans}
\end{equation}
is said to transform in a \emph{tensor representation} of $\SU(N)$. Given two
tensor fields $\Psi_{i_1i_2\ldots}^{j_1j_2\ldots}$ and
$\Phi_{i_1i_2\ldots}^{j_1j_2\ldots}$, we can construct a higher-dimensional
\emph{product representation} as
\begin{equation}
  (\Psi\otimes\Phi)_{i_1i_2\ldots k_1k_2\ldots}^{j_1j_2\ldots l_1l_2\ldots} =
  \Psi_{i_1i_2\ldots}^{j_1j_2\ldots}\Phi_{k_1k_2\ldots}^{l_1l_2\ldots}.
\end{equation}

\begin{itemize}
\item NB: Kronecker delta symbol $\delta_i^j$ and Levi-Civita symbols
  $\epsilon_{i_1i_2\ldots i_N}$, $\epsilon^{i_1i_2\ldots i_N}$ are $\SU(N)$
  invariants
  \begin{gather}
    \delta_i^j \to U_i{}^k U^j{}_l \delta_k{}^l = \delta_i^j, \nn\\
    \epsilon_{i_1i_2\ldots i_N}
    \to (U_{i_1}{}^{j_1}U_{i_2}{}^{j_2}\ldots U_{i_N}{}^{j_N})
    \epsilon_{j_1j_2\ldots j_N}
    = (\det U) \epsilon_{i_1i_2\ldots i_N}, \nn\\
    \epsilon^{i_1i_2\ldots i_N}
    \to (U^{i_1}{}_{j_1}U^{i_2}{}_{j_2}\ldots U^{i_N}{}_{j_N})
    \epsilon^{j_1j_2\ldots j_N}
    = \frac{1}{\det U} \epsilon^{i_1i_2\ldots i_N},
  \end{gather}
  with $\det U = 1$. Note that the Kronecker delta symbol is also invariant
  under the group $\rmU(N)$, while the Levi-Civita symbols are not, because
  $\det U$ is not necessarily 1.
\end{itemize}

\paragraph*{Reducible and irreducible representations:} A tensor field
$\Psi_{i_1i_2\ldots}^{j_1j_2\ldots}$ is said to transform in a \emph{reducible
  representation} of $\SU(N)$, if it can be decomposed into a direct sum of
fields transforming under smaller representations. A tensor field that does not
transform under a reducible representation is said to transform in an
\emph{irreducible representation}.

The decomposition of a reducible tensor field
$\Psi_{i_1i_2\ldots}^{j_1j_2\ldots}$ into fields transforming under smaller
irreducible representations is known as its \emph{tensor decomposition}.
\begin{itemize}
\item NB: tensor product of irreducible representations is generically
  reducible.
  
\item Example: fundamental, anti-fundamental, singlet, and adjoint
  representations of $\SU(N)$ are all irreducible.
\item Example: A tensor field $\Psi_i{}^j$ (not necessarily traceless) is
  reducible into a singlet field $\Psi_i{}^i$ and an adjoint field
  $\Psi_i{}^j - 1/N\, \delta^j_i \Psi_k{}^k$. A tensor field $\Psi_{ij}$, on the
  other hand, is reducible into a symmetric tensor field
  $\Psi_{(ij)} = (\Psi_{ij}+\Psi_{ji})/2$ and an antisymmetric tensor field
  $\Psi_{[ij]} = (\Psi_{ij}-\Psi_{ji})/2$. Both of these transform independently
  under an $\SU(N)$ transformation.
\end{itemize}

\subsection{Young Tableaux}

\paragraph*{Young tableaux:} We can generate arbitrary irreducible
representations of $\SU(N)$ using the method of Young tableaux. Firstly, given a
tensor field $\Psi_{i_1i_2\ldots}^{j_1j_2\ldots}$, we can lower all the indices
using the $\epsilon_{i_1\ldots i_N}$ tensor to obtain the equivalent
representation
\begin{equation}
  \Psi_{i_1i_2\ldots k_1k_2\ldots l_1l_2\ldots}
  = \Psi_{i_1i_2\ldots}^{j_1j_2\ldots}
  \lb\epsilon_{j_1k_1l_1\ldots}
  \epsilon_{j_2k_2l_2\ldots}
  \ldots \rb.
\end{equation}
Given this, a typical Young tableaux representing an irreducible tensor field
$\Psi_{i_1j_1\ldots}$ has the form
\begin{equation*}
  \begin{ytableau}
    i_1 & j_1 & \cdots & \ & \ & \ \\
    i_2 & j_2 & \cdots & \ & \ \\
    \vdots & \vdots & \ddots & \ & \ \\
    \ & \ & \ \\
    \ 
  \end{ytableau}
\end{equation*}
with rules
\begin{enumerate}
\item The indices in a given column have to be anti-symmetrised.
\item The indices in a given row have to be symmetrised.
\item A row cannot contain more boxes than the row above.
\item A column cannot contain more than $N$ boxes because of anti-symmetry. A
  column with $N$ boxes corresponds to an uncontracted factor of
  $\epsilon_{i_1\ldots i_N}$ and can be removed.
\end{enumerate}
The dimensions of an irreducible representation can be computed as
\begin{equation}
  \dim = \frac{\prod_i (N+n_i)}{\prod_j d_j}, \qquad
  n_i: \quad
    \begin{ytableau}
      0 & 1 & 2 & 3 & 4 & 5 \\
      -1 & 0 & 1 & 2 & 3 \\
      -2 & -1 & 0 & 1 & 2 \\
      -3 & -2 & -1 \\
      -4
    \end{ytableau}, \qquad
  d_i: \quad
  \begin{ytableau}
      10 & 8 & 7 & 5 & 4 & 1 \\
      8 & 6 & 5 & 3 & 2 \\
      7 & 5 & 4 & 2 & 1 \\
      4 & 2 & 1 \\
      1
  \end{ytableau}
\end{equation}
$n_i$ for the top-left box is 0; moving right we increase by 1, while moving
down we decrease by 1. On the other hand, $d_i$ for a box is 1 plus the number
of boxes below and to the right of the box.
\begin{itemize}
\item Examples:
  \begin{gather}
    \text{Fundamental ``$\mathbf{N}$'' } (\psi_i) : \quad
    \raisebox{-1mm}{\begin{ytableau} i \end{ytableau}}, \quad
    \text{Anti-fundamental ``$\bar{\mathbf{N}}$'' } (\psi_{[i_1\ldots i_{N-1}]}
    = \epsilon_{ki_1\ldots i_{N-1}} \psi^k): \quad
    \begin{ytableau}
      i_1 \\ i_2 \\ \vdots \\ i_{N-1}
    \end{ytableau}
    , \nn\\
    \text{Singlet ``$\mathbf{1}$'' } (\psi): \quad
    \begin{ytableau}
      i_1 \\ i_2 \\ \vdots \\ i_{N}
    \end{ytableau} \quad{\text{or null}}, \quad
    \text{Adjoint ``$\mathbf{N^2-1}$'' } (\psi_{j[i_1\ldots i_{N-1}]}
    = \epsilon_{ki_1\ldots i_{N-1}} \psi_j{}^k): \quad
    \begin{ytableau}
      i_1 & j \\ i_2 \\ \vdots \\ i_{N-1}
    \end{ytableau}, \nn\\
    \text{Symmetric 2-tensor ``$\mathbf{N(N+1)/2}$'' } (\psi_{(ij)}): \quad
    \raisebox{-1mm}{\begin{ytableau} i & j \end{ytableau}}, \nn\\
    \text{Anti-symmetric 2-tensor ``$\mathbf{N(N-1)/2}$'' } (\psi_{[ij]}): \quad
    \begin{ytableau}
      i \\ j
    \end{ytableau}.
  \end{gather}
\end{itemize}

\paragraph*{Tensor decomposition:} Young tableaux can also be used to decompose
the tensor product of irreducible representations into the direct sum of
irreducible representations. To compute the tensor product of two
representations $A$ and $B$, we use the algorithm:
\begin{enumerate}
\item Label all the boxes in the first row of $B$ as ``$a_1$'', in the second
  row as ``$a_2$'', in the third row as ``$a_3$'', and so on.
\item Attach the ``$a_1$'' boxes from $B$ to $A$ in all possible ways to
  produce legitimate tableaux. Repeat for ``$a_2$'' boxes, and so on. Discard
  any tableaux with repeated labels in a column.
\item For a given tableaux, create a sequence by enlisting all the labels from
  right to left, from top row to the bottom. Reading this sequence from left to
  right, discard any tableaux where the number of $a_i$'s is less than the
  number of $a_{i+1}$'s at any point in the sequence.
\item Tableaux with same structure are inequivalent only if the respective
  sequences are different.
\end{enumerate}
In the final answer, the product of the dimensions of $A$ and $B$ must equal the
sum of the dimensions of the reduced tableaux.

\subsection{SU(2)}
\label{sec:SU2}

SU(2) is the simplest special unitary group. It is the group of all $2\times 2$
unit determinant unitary matrices. The generators of the associated Lie algebra
$\mathfrak{su}(2)$, i.e. the set of all $2\times 2$ traceless Hermitian
matrices, are given by $T_i = 1/2\,\sigma_i$, where $\sigma_i$ are the three
Pauli matrices; see \cref{eq:su2-algebra}. It is convenient to work in the basis
\begin{equation}
  T_\pm = T_1 \pm iT_2, \qquad T_3,
\end{equation}
with commutation relations
\begin{equation}
  [T_+,T_-] = 2\hbar\, T_3, \qquad
  [T_3,T_\pm] = \pm \hbar\,  T_\pm.
\end{equation}
Note that $T_\pm^\dagger = T_\mp$. There is a Casimir operator in the theory
$T^2 = T_1^2 + T_2^2 + T_3^2 = \half\{T_+, T_-\} + T_3^2$ that commutes with all
the operators in the algebra, i.e $[T^2,T_a]=0$. Note that the anti-commutator
of the generators is given by $\{T_i,T_j\} = \hbar^2/2\,\delta_{ij}\mathbb{1}$.

The fundamental and anti-fundamental representations of SU(2) are
equivalent. This trivially follows using the rules of Young tableaux, as both of
these can be related by a contraction with $\epsilon_{ij}$ and are represented
by the same tableaux
\begin{equation}
  \text{Fundamental ``$\mathbf{2}$'' } (\psi_i) : \quad
  \raisebox{-1mm}{\begin{ytableau} i \end{ytableau}}, \qquad
  \text{Anti-fundamental ``$\bar{\mathbf{2}}$'' }
  (\epsilon_{ij} \psi^j): \quad
  \raisebox{-1mm}{\begin{ytableau} i \end{ytableau}}.
\end{equation}
In fact, all inequivalent irreducible representations of SU(2) are given by
\begin{gather}
  \text{``$\mathbf{1}$'' } (j=0): \quad\text{null}, \qquad
  \text{``$\mathbf{2}$'' } (j=1/2): \quad
  \raisebox{-1mm}{\begin{ytableau} \ \end{ytableau}}, \qquad
  \text{``$\mathbf{3}$'' } (j=1): \quad
  \raisebox{-1mm}{\begin{ytableau} \ & \ \end{ytableau}}, \nn\\
  \text{``$\mathbf{4}$'' } (j=3/2): \quad
  \raisebox{-1mm}{\begin{ytableau} \ & \ & \ \end{ytableau}}, \qquad
  \text{``$\mathbf{5}$'' } (j=2): \quad
  \raisebox{-1mm}{\begin{ytableau} \ & \ & \ & \ \end{ytableau}}, \qquad\ldots.
\end{gather}
The numbers in bold are the dimensions of the respective representations, while
the ``quantum-number'' $j=0,1/2,1,3/2,\ldots$ is related to the dimensions as
$d=2j+1$.

Let us consider a field $\Psi$ transforming under the
$\mathbf{(2j+1)}$-dimensional irreducible representation of SU(2). Without loss
of generality, we can take the possible states of $\Psi$ to be the eigenvectors
of the generator $T_3$, labelled by $j$ and the $T_3$-eigenvalue $m$,
i.e. $\ket{j,m}$ with
\begin{equation}
  T_3 \ket{j,m} = \hbar m \ket{j,m}.
\end{equation}
Note that $T_3(T_\pm\ket{j,m}) = \hbar(m\pm1) T_\pm \ket{j,m}$. It implies
\begin{equation}
  T_\pm\ket{j,m} = \hbar\lambda_{j,m}^\pm \ket{j,m\pm1}, \qquad
  T^2\ket{j,m} = \hbar^2\lambda(\lambda+1) \ket{j,m},
\end{equation}
for some constants $\lambda_{j,m}^\pm$ and $\lambda$. Using the normalisation of
states $\langle j,m|j,m\rangle = 1$, and relations $[T_+,T_-] = 2\hbar\,T_3$ and
$\{T_+,T_-\} = 2T^2 - 2T_3^2$, we get
\begin{equation}
  \lambda_{j,m}^- \lambda_{j,m-1}^+ = \lambda(\lambda+1) - m(m-1), \qquad
  |\lambda_{j,m}^+|^2 - |\lambda_{j,m}^-|^2 =   -2m.
\end{equation}
Since the dimension of the Hilbert space is finite, there must exist a state
$\ket{j,m_h}$ such that $T_+\ket{j,m_h} = 0$, and similarly $\ket{j,m_\ell}$
such that $T_-\ket{j,m_\ell} = 0$. Correspondingly, we find
$\lambda^+_{j,m_h} = \lambda^-_{j,m_\ell} = 0$, leading to $m_h = \lambda$ and
$m_\ell = -\lambda$.  Since the dimension of the Hilbert space is $2j+1$, we
must have that $(T_+)^{2j}\ket{j,m_h} \propto \ket{j,m_\ell}$, leading to
$m_h - m_\ell = 2j$ or
\begin{equation}
  \lambda = j.
\end{equation}
Finally, we can solve the recursion relations for $\lambda^\pm_{j,m}$ to get
\begin{equation}
  \lambda^\pm_{j,m} = \sqrt{j(j+1) - m(m\pm 1)}.
\end{equation}

In summary, the states in a generic $\mathbf{(2j+1)}$-dimensional representation
of SU(2), taken to be eigenstates of $T_3$, can be labelled by two quantum
numbers $\ket{j,m}$ where $j=0,1/2,1,3/2,\ldots$ and
\begin{equation}
  m=-j,-j+1,\ldots, j-1, j.
\end{equation}
The quantum numbers $j$ and $m$ are related to the eigenvalues of $T^2$ and
$T_3$ according to $T^2\ket{j,m} = \hbar^2j(j+1)\ket{j,m}$ and
$T_3\ket{j,m} = \hbar m\ket{j,m}$. The states can be arranged into a
line-segment
\begin{equation*}
  \begin{tikzpicture}
    \draw[thick,<->] (-20mm,6mm) -- (20mm,6mm);
    \node at (-25mm,6mm) {$T_-$};
    \node at (25mm,6mm) {$T_+$};
    
    \filldraw (-30mm,0) circle (2pt);
    \filldraw (-20mm,0) circle (2pt);
    \filldraw (-10mm,0) circle (2pt);
    \draw[dashed,ultra thick,lightgray](-5mm,0) -- (5mm,0);
    \filldraw (10mm,0) circle (2pt);
    \filldraw (20mm,0) circle (2pt);
    \filldraw (30mm,0) circle (2pt);

    \node at (-30mm,-6mm) {\scriptsize{$m=-j$}};
    \node at (30mm,-6mm) {\scriptsize{$m=j$}};
  \end{tikzpicture}
\end{equation*}
There are a total of $2j+1$ states. The ladder operators $T_\pm$ take us to the
state immediately to the right/left in the arrangement above.

\paragraph*{Example:} The angular momentum operators $T_i=J_i$ in quantum
mechanics furnish an $\mathfrak{su}(2)$ algebra. Quantum particles transform in
an irreducible representation of this $\SU(2)$, labelled by the quantum number
$j$ associated with the total angular momentum operator $T^2=J^2$. Independent
states of a quantum particle are labelled by the quantum number $m$ representing
the $z$-component $T_3=J_3$ of angular-momentum.

\paragraph*{Example:} The weak force in the Standard Model is associated with an
internal SU(2) symmetry. The left-handed quarks and leptons, and the Standard
Model Higgs boson transform as weak SU(2) doublets ``${\bf 2}$'' ($j=1/2$),
while the weak force gauge bosons and the electromagnetic photon collectively
transform as a weak SU(2) triplet ``${\bf 3}$'' ($j=1$) and a singlet
``${\bf 1}$'' ($j=0$). The quantum number $m$ is associated with the electric
charge $q=m\pm 1/6$ for left-handed (anti)quarks and $q=m\mp 1/2$ for
left-handed (anti)leptons. See \cref{sec:electroweak,sec:standardmodel}.

\paragraph*{Example:} The Standard Model has an approximate SU(2) isospin
symmetry. The light quarks ($u,d$) and anti-quarks ($\bar u,\bar d$) make up
SU(2) isospin doublets ``${\bf 2}$'' ($j=1/2$). Various hadrons appear as SU(2)
isospin singlets ``${\bf 1}$'' ($j=0$; e.g. $\Lambda^0$ baryon and $\eta^0$
meson), doublets ``${\bf 2}$'' ($j=1/2$; e.g. $(p,n)$ baryons and $(K^0,K^+)$
mesons), triplets ``${\bf 3}$'' ($j=1$; e.g. $(\Sigma^-,\Sigma^0,\Sigma^+)$
baryons and $(\pi^-,\pi^0,\pi^+)$ mesons), or quadruplets ``${\bf 4}$''
($j=3/2$; e.g. $(\Delta^-,\Delta^0,\Delta^+,\Delta^{2+})$ baryons). More details
on the classification can be found in the next subsection during our discussion
of the SU(3) flavour group.

\subsection{SU(3)}
\label{sec:SU3}

SU(3) is next simplest special unitary group. It is the group of all $3\times3$
unit determinant unitary matrices. The eight generators of the associated Lie
algebra $\mathfrak{su}(3)$, i.e. the set of all $3\times3$ traceless Hermitian
matrices, are given by $T_a$ written out in \cref{eq:su3-algebra}. To classify
the spectrum of states, it is instead convenient to work with the complexified
basis
\begin{equation}
  T_\pm = T_1 \pm iT_2, \qquad
  T_3, \qquad
  V_\pm = T_4 \pm iT_5, \qquad
  U_\pm = T_6 \pm iT_7, \qquad
  Y = \frac{2}{\sqrt{3}} T_8.
\end{equation}
These have commutation relations
\begin{gather}
  [T_+,T_-] = 2\hbar T_3, \qquad
  [V_+,V_-] = \frac{3\hbar }{2}  Y + \hbar T_3, \qquad
  [U_+,U_-] = \frac{3\hbar }{2} Y - \hbar T_3, \nn\\
  [T_\pm,V_\pm] = 0, \qquad
  [T_\pm,V_\mp] = \mp \hbar U_\mp, \qquad
  [T_\pm,U_\pm] = \pm \hbar V_\pm, \qquad
  [T_\pm,U_\mp] = 0, \nn\\
  [U_\pm,V_\pm] = 0, \qquad
  [U_\pm,V_\mp,] = \pm \hbar T_\mp,
\end{gather}
along with
\begin{gather}
  [T_3,T_\pm] = \pm \hbar T_\pm, \qquad
  [T_3,V_\pm] = \pm \frac\hbar2 V_\pm, \qquad
  [T_3,U_\pm] = \mp \frac\hbar2 U_\pm, \nn\\
  [Y,T_\pm]  = 0, \qquad
  [Y,T_3]  = 0, \qquad
  [Y,V_\pm] = \pm \hbar  V_\pm \qquad
  [Y,U_\pm] = \pm\hbar  U_\pm.
\end{gather}
The Casimir operators are given as $T^2 = T_a T_a$ and
$T^3 = d_{abc} T_a T_b T_c$, where $d_{abc} = 2/\hbar^3\,\tr(\{T_a,T_b\} T_c)$
is the totally symmetric structure constant of $\mathfrak{su}(3)$. These Casimir
operators commute with all the generators of the $\mathfrak{su}(3)$ algebra
$[T^2,T_a] = [T^3,T_a] = 0$.

The generators $T_\pm$, $T_3$ span a $\mathfrak{su}(2)$ subalgebra of
$\mathfrak{su}(3)$, known as the \emph{isospin} subalgebra. However, the total
isospin operator $I^2 = T_1^2+T_2^2+T_3^2 = \half\{T_+,T_-\}+T_3^2$ is not a
Casimir of the full $\mathfrak{su}(3)$ algebra; to wit
\begin{gather}
  [I^2,T_\pm] = 0, \qquad
  [I^2,T_3] = 0, \qquad
  [I^2,Y] = 0, \nn\\
  [I^2,V_\pm]
  = \pm\frac\hbar2\{V_\pm,T_3\}
  \pm\frac\hbar2 \{U_\pm,T_{\pm}\}, \qquad
  [I^2,U_\pm]
  = \mp\frac\hbar2\{U_\pm,T_3\}
  \pm\frac\hbar2 \{V_\pm,T_{\mp}\}.
\end{gather}
It will, nonetheless, be useful in the classification of the spectrum of
states. Similar $\mathfrak{su}(2)$ subalgebras are also spanned by the
generators $V_\pm$, $3Y/4+T_3/2$ and by $U_\pm$, $3Y/4-T_3/2$.

The irreducible representations of SU(3) can be labelled by two integers $p$ and
$q$. In terms of Young tableaux, the representation $D(p,q)$ is one with $p$
number of one-box columns and $q$ number of two-box columns. The representations
can be arranged in a 2-dimensional array
\vspace{1em}

{
  \centering
  \begin{tabular}[h]{cccccc}
    null
    & \begin{ytableau} \ \\ \ \end{ytableau}
    & \begin{ytableau} \ & \ \\ \ & \ \end{ytableau}
    & \begin{ytableau} \ & \ & \ \\ \ & \ & \ \end{ytableau}
    & \begin{ytableau} \ & \ & \ & \ \\ \ & \ & \ & \ \end{ytableau}
    & $\cdots$ \\[3.5ex]

    ``\textbf{1}'' $D(0,0)$
    & ``$\bar{\mathbf 3}$'' $D(0,1)$
    & ``$\bar{\mathbf 6}$'' $D(0,2)$
    & ``$\bar{\mathbf{10}}$'' $D(0,3)$
    & ``$\bar{\mathbf{15}}$'' $D(0,4)$ \\[2ex]

    \begin{ytableau} \ \end{ytableau}
    & \begin{ytableau} \ & \ \\ \ \end{ytableau}
    & \begin{ytableau} \ & \ & \ \\ \ & \ \end{ytableau}
    & \begin{ytableau} \ & \ & \ & \ \\ \ & \ & \ \end{ytableau}
    & $\ddots$ \\[3.5ex]

    ``\textbf{3}'' $D(1,0)$
    & ``$\mathbf 8$'' $D(1,1)$
    & ``$\bar{\mathbf{15}}$'' $D(1,2)$
    & ``$\bar{\mathbf{24}}$'' $D(1,3)$ \\[2ex]

    \begin{ytableau} \ & \ \end{ytableau}
    & \begin{ytableau} \ & \ & \ \\ \ \end{ytableau}
    & \begin{ytableau} \ & \ & \ & \ \\ \ & \ \end{ytableau}
    & $\ddots$ \\[3.5ex]

    ``\textbf{6}'' $D(2,0)$
    & ``$\mathbf{15}$'' $D(2,1)$
    &  ``$\mathbf{27}$'' $D(2,2)$ \\[2ex]

    \begin{ytableau} \ & \ & \ \end{ytableau}
    & \begin{ytableau} \ & \ & \ & \ \\ \ \end{ytableau}
    & $\ddots$ \\[3.5ex]

    ``\textbf{10}'' $D(3,0)$
    & ``$\mathbf{24}$'' $D(3,1)$ \\[2ex]

    \begin{ytableau} \ & \ & \ & \ \end{ytableau}
                                  & $\ddots$
    \\[1ex]

    ``\textbf{15}'' $D(4,0)$ \\[2ex]

    $\vdots$
    
  \end{tabular}
} \\

\noindent
Diagonally opposite representations $D(p,q)$ and $D(q,p)$ are conjugates to one
another, while the diagonal representations $D(p,p)$ are self-conjugates. We
have noted the dimensions of the representations in bold. Note that, unlike
$\SU(2)$, the dimensions do not uniquely characterise a representation. It is
conventional to denote the dimensions of conjugate representations $D(p,q)$ with
$q>p$ with a bar. With the barred notation in place, the dimensions do uniquely
characterise the first $4\times 4$ block of the representations above. These are
typically all we need in the Standard Model. The $T^2$ and $T^3$ eigenvalues of
the representation $D(p,q)$ are given as $\hbar^2c_2(p,q)$ and $\hbar^3c_3(p,q)$
where
\begin{equation}
  c_2(p,q) = \frac{1}{3} \lb p^2 + q^2 + 3p + 3q + pq \rb, \qquad
  c_3(p,q) = \frac{1}{18}(p-q)(3+p+2q)(3+q+2p),
\end{equation}
whereas the dimension of the representation is given as
\begin{equation}
  \dim D(p,q) = \half(p+1)(q+1)(p+q+2).
  \label{eq:dimension-SU3}
\end{equation}

Let us consider a field $\Psi$ transforming in the $D(p,q)$ representation of
$\SU(3)$. The independent states of $\Psi$ can be labelled by the eigenvalues of
the mutually commuting set of operators $T^2$, $T^3$, $I^2$, $T_3$, and $Y$,
i.e. $\ket{p,q;j,m,y}$. Note that the eigenvalues of the Casimir operators $T^2$
and $T^3$ is already fixed in terms of $p$, $q$. To wit, we have
\begin{gather}
  T^2 \ket{p,q;j,m,y} = \hbar^2 c_2(p,q) \ket{p,q;j,m,y}, \qquad
  T^3 \ket{p,q;j,m,y} = \hbar^3 c_3(p,q) \ket{p,q;j,m,y}, \nn\\
  T_3 \ket{p,q;j,m,y} = \hbar m \ket{p,q;j,m,y}, \qquad
  Y \ket{p,q;j,m,y} = \hbar y \ket{p,q;j,m,y}, \nn\\
  I^2 \ket{p,q;j,m,y} = \hbar^2 j(j+1) \ket{p,q;j,m,y}.
\end{gather}
The action of the remaining $\mathfrak{su}(3)$ operators can be derived using
the Lie algebra commutation relations, leading to
\begin{align}
  T_\pm \ket{p,q;j,m,y}
  &= \hbar \sqrt{j(j+1)-m(m\pm 1)}\,\ket{p,q;j,m\pm1,y} \nn\\
  U_\pm \ket{p,q;j,m,y}
  &= \hbar \sqrt{\frac{(j\mp m+1) f_\pm(j\pm\frac{y}{2}+1)}{(j+1) (2 j+1)}}\,
    \ket{p,q;j+{\textstyle\frac12},m\mp{\textstyle\frac12},y\pm 1} \nn\\
  &\qquad
    + \hbar \sqrt{\frac{(j\pm m) f_\mp(j\mp\frac{y}{2})}{j (2 j+1)}}\,
    \ket{p,q;j-{\textstyle\frac12},m\mp{\textstyle\frac12},y\pm 1} \nn\\
  V_\pm \ket{p,q;j,m,y}
  &= \pm \hbar\sqrt{\frac{(j\pm m+1) f_\pm(j\pm\frac{y}{2}+1)}{(j+1) (2 j+1)}}\,
    \ket{p,q;j+{\textstyle\frac12},m\pm{\textstyle\frac12},y\pm 1} \nn\\
  &\qquad
    \mp \hbar\sqrt{\frac{(j\mp m) f_\mp(j\mp\frac{y}{2})}{j (2 j+1)}}\,
    \ket{p,q;j-{\textstyle\frac12},m\pm{\textstyle\frac12},y\pm 1},
    \label{eq:SU3_statestep}
\end{align}
where $f_\pm(x) = (c_2+1-x^2)x/2\pm c_3/3$.  These expressions are considerably
more complicated than the SU(2) case. Nonetheless, to write down the spectrum of
states, we can start with the ``highest-weight'' state that is annihilated by
all the ``creation'' operators $T_+$, $U_+$, $V_+$, i.e.
\begin{equation}
  T_+\ket{p,q;j_h,m_h,y_h}
  = U_+\ket{p,q;j_h,m_h,y_h}
  = V_+\ket{p,q;j_h,m_h,y_h} = 0.
\end{equation}
Using \cref{eq:SU3_statestep}, it is straight-forward to see that this state has
$j_h=m_h=p/2$ and $y_h=p/3+2q/3$. All the remaining states in the representation
can be obtained by repeatedly applying the ``annihilation'' operators $T_-$,
$U_-$, $V_-$ on the highest weight state. It is convenient to define the quantum
number $\ell= (y-y_h)/2-j+j_h$. In terms of this, we have the complete spectrum
of states
\begin{gather}
  \ell = 0, 1, \ldots, q, \nn\\
  2j = \ell,\ell+1,\ldots,\ell+p, \nn\\
  m = -j,-j+1,\ldots, j-1,j.
\end{gather}
It can be checked that the total number of states add up to $\dim D(p,q)$ given
in \cref{eq:dimension-SU3}. These states can be neatly arranged in the $m-y$
plane into hexagonal or triangular patterns, known as the eightfold-way diagrams
proposed by Murray Gell-Mann in 1961~\cite{Gell-Mann:1961omu}. For example, the
eightfold-way diagram representing the states of the representation $D(p,q)$,
assuming $p>q$, is given as


\def\dmf{8}
\def\dhf{\fpeval{0.866*\dmf}}

\def\dashline#1#2{
\begin{scope}[shift={#1}, rotate=#2]
    \draw[dashed,ultra thick,lightgray](-3mm,0) -- (3mm,0);
\end{scope}
}

\def\state#1#2{
\begin{scope}
  \filldraw (\fpeval{#1*\dmf}mm,\fpeval{#2*\dhf}mm) circle (2pt);
\end{scope}
}
\def\duostate#1#2{
\begin{scope}
  \filldraw (\fpeval{#1*\dmf}mm,\fpeval{#2*\dhf}mm) circle (2pt);
  \draw (\fpeval{#1*\dmf}mm,\fpeval{#2*\dhf}mm) circle (4pt);
\end{scope}
}
\def\triostate#1#2{
\begin{scope}
  \filldraw (\fpeval{#1*\dmf}mm,\fpeval{#2*\dhf}mm) circle (2pt);
  \draw (\fpeval{#1*\dmf}mm,\fpeval{#2*\dhf}mm) circle (4pt);
  \draw (\fpeval{#1*\dmf}mm,\fpeval{#2*\dhf}mm) circle (6pt);
\end{scope}
}

\def\xtick#1#2{
\begin{scope}
  \draw[](#1,\fpeval{-10.5*\dhf-.5}mm) -- (#1,\fpeval{-10.5*\dhf+.5}mm);
  \node at (#1,\fpeval{-10.5*\dhf-3}mm) {\tiny{#2}};
\end{scope}
}
\def\xxtick#1#2{
\begin{scope}
  \draw[](#1,\fpeval{2*\dhf-.5}mm) -- (#1,\fpeval{2*\dhf+.5}mm);
  \node at (#1,\fpeval{2*\dhf+3}mm) {\tiny{#2}};
\end{scope}
}
\def\ytick#1#2{
\begin{scope}
  \draw[](\fpeval{-9*\dmf-.5}mm,#1) -- (\fpeval{-9*\dmf+.5}mm,#1);
  \node at (\fpeval{-9*\dmf-7}mm,#1) {\tiny{#2}};
\end{scope}
}
\def\yytick#1#2{
\begin{scope}
  \draw[](\fpeval{3*\dmf-.5}mm,#1) -- (\fpeval{3*\dmf+.5}mm,#1);
  \node at (\fpeval{3*\dmf-7}mm,#1) {\tiny{#2}};
\end{scope}
}

\begin{equation*}
  \begin{tikzpicture}
    \draw[thick,<->] (\fpeval{(-5.5)*\dmf}mm,\fpeval{1*\dhf}mm) --
    (\fpeval{-.5*\dmf}mm,\fpeval{1*\dhf}mm);
    \node at (\fpeval{-6*\dmf}mm,\fpeval{1*\dhf}mm) {$T_-$};
    \node at (\fpeval{0*\dmf}mm,\fpeval{1*\dhf}mm) {$T_+$};

    \draw[thick,<->] (\fpeval{(-8)*\dmf}mm,\fpeval{-4*\dhf}mm) --
    (\fpeval{-5.5*\dmf}mm,\fpeval{-9*\dhf}mm);
    \node at (\fpeval{(-8.25)*\dmf}mm,\fpeval{-3.5*\dhf}mm) {$U_+$};
    \node at (\fpeval{-5.1*\dmf}mm,\fpeval{-9.5*\dhf}mm) {$U_-$};
    
    \draw[thick,<->] (\fpeval{2*\dmf}mm,\fpeval{-4*\dhf}mm) --
    (\fpeval{-.5*\dmf}mm,\fpeval{-9*\dhf}mm);
    \node at (\fpeval{2.25*\dmf}mm,\fpeval{-3.5*\dhf}mm) {$V_+$};
    \node at (\fpeval{-.4*\dmf}mm,\fpeval{-9.5*\dhf}mm) {$V_-$};
    
    \node at (\fpeval{3.5*\dmf}mm,\fpeval{-11*\dhf}mm) {$m$};
    \node at (\fpeval{-9.5*\dmf}mm,\fpeval{2.25*\dhf}mm) {$y$};
    
    \draw[<-] (\fpeval{-9*\dmf}mm,\fpeval{2.5*\dhf}mm) --
    (\fpeval{-9*\dmf}mm,\fpeval{-1.5*\dhf}mm);
    \ytick{\fpeval{0*\dhf}mm}{$\frac{p+2q}{3}$};
    \ytick{\fpeval{-1*\dhf}mm}{$\frac{p+2q}{3}{-}1$};
    \draw[dashed] (\fpeval{-9*\dmf}mm,\fpeval{-1.5*\dhf}mm) --
    (\fpeval{-9*\dmf}mm,\fpeval{-2.5*\dhf}mm);
    \draw[] (\fpeval{-9*\dmf}mm,\fpeval{-2.5*\dhf}mm) --
    (\fpeval{-9*\dmf}mm,\fpeval{-5.5*\dhf}mm);
    \ytick{\fpeval{-3*\dhf}mm}{$\frac{p-q}{3}$};
    \ytick{\fpeval{-4*\dhf}mm}{$\frac{p-q}{3}{-}1$};
    \ytick{\fpeval{-5*\dhf}mm}{$\frac{p-q}{3}{-}2$};
    \draw[dashed] (\fpeval{-9*\dmf}mm,\fpeval{-5.5*\dhf}mm) --
    (\fpeval{-9*\dmf}mm,\fpeval{-6.5*\dhf}mm);
    \draw[] (\fpeval{-9*\dmf}mm,\fpeval{-6.5*\dhf}mm) --
    (\fpeval{-9*\dmf}mm,\fpeval{-10.5*\dhf}mm);
    \ytick{\fpeval{-7*\dhf}mm}{$\frac{-2p-q}{3}{+}2$};
    \ytick{\fpeval{-8*\dhf}mm}{$\frac{-2p-q}{3}{+}1$};
    \ytick{\fpeval{-9*\dhf}mm}{$\frac{-2p-q}{3}$};

    \draw[<-] (\fpeval{3*\dmf}mm,\fpeval{2.5*\dhf}mm) --
    (\fpeval{3*\dmf}mm,\fpeval{-1.5*\dhf}mm);
    \yytick{\fpeval{0*\dhf}mm}{};
    \yytick{\fpeval{-1*\dhf}mm}{};
    \draw[dashed] (\fpeval{3*\dmf}mm,\fpeval{-1.5*\dhf}mm) --
    (\fpeval{3*\dmf}mm,\fpeval{-2.5*\dhf}mm);
    \draw[] (\fpeval{3*\dmf}mm,\fpeval{-2.5*\dhf}mm) --
    (\fpeval{3*\dmf}mm,\fpeval{-5.5*\dhf}mm);
    \yytick{\fpeval{-3*\dhf}mm}{};
    \yytick{\fpeval{-4*\dhf}mm}{};
    \yytick{\fpeval{-5*\dhf}mm}{};
    \draw[dashed] (\fpeval{3*\dmf}mm,\fpeval{-5.5*\dhf}mm) --
    (\fpeval{3*\dmf}mm,\fpeval{-6.5*\dhf}mm);
    \draw[] (\fpeval{3*\dmf}mm,\fpeval{-6.5*\dhf}mm) --
    (\fpeval{3*\dmf}mm,\fpeval{-10.5*\dhf}mm);
    \yytick{\fpeval{-7*\dhf}mm}{};
    \yytick{\fpeval{-8*\dhf}mm}{};
    \yytick{\fpeval{-9*\dhf}mm}{};

    \draw[] (\fpeval{-9*\dmf}mm,\fpeval{-10.5*\dhf}mm) --
    (\fpeval{-6.5*\dmf}mm,\fpeval{-10.5*\dhf}mm);
    \draw[dashed] (\fpeval{(-6.5)*\dmf}mm,\fpeval{-10.5*\dhf}mm) --
    (\fpeval{-5.5*\dmf}mm,\fpeval{-10.5*\dhf}mm);
    \draw[] (\fpeval{(-5.5)*\dmf}mm,\fpeval{-10.5*\dhf}mm) --
    (\fpeval{-4*\dmf}mm,\fpeval{-10.5*\dhf}mm);
    \xtick{\fpeval{-7.5*\dmf}mm}{$\frac{-p-q}{2}$};
    \xtick{\fpeval{-4.5*\dmf}mm}{$\frac{-q}{2}$};
    \draw[dashed] (\fpeval{(-4)*\dmf}mm,\fpeval{-10.5*\dhf}mm) --
    (\fpeval{-3*\dmf}mm,\fpeval{-10.5*\dhf}mm);
    \draw[] (\fpeval{(-3)*\dmf}mm,\fpeval{-10.5*\dhf}mm) --
    (\fpeval{-.5*\dmf}mm,\fpeval{-10.5*\dhf}mm);
    \xtick{\fpeval{-2.5*\dmf}mm}{$\frac{q-1}{2}$};
    \xtick{\fpeval{-1.5*\dmf}mm}{$\frac{q}{2}$};
    \draw[dashed] (\fpeval{(-.5)*\dmf}mm,\fpeval{-10.5*\dhf}mm) --
    (\fpeval{.5*\dmf}mm,\fpeval{-10.5*\dhf}mm);
    \draw[->] (\fpeval{(.5)*\dmf}mm,\fpeval{-10.5*\dhf}mm) --
    (\fpeval{3.5*\dmf}mm,\fpeval{-10.5*\dhf}mm);
    \xtick{\fpeval{1.5*\dmf}mm}{$\frac{p+q}{2}$};

    \draw[] (\fpeval{-9*\dmf}mm,\fpeval{2*\dhf}mm) --
    (\fpeval{-7.25*\dmf}mm,\fpeval{2*\dhf}mm);
        \xxtick{\fpeval{-7.5*\dmf}mm}{$\frac{-p-q}{2}$};
    \draw[dashed] (\fpeval{(-7.25)*\dmf}mm,\fpeval{2*\dhf}mm) --
    (\fpeval{-6.25*\dmf}mm,\fpeval{2*\dhf}mm);
    \draw[] (\fpeval{(-6.25)*\dmf}mm,\fpeval{2*\dhf}mm) --
    (\fpeval{-3.5*\dmf}mm,\fpeval{2*\dhf}mm);
    \xxtick{\fpeval{-6*\dmf}mm}{$\frac{-p}{2}$};
    \xxtick{\fpeval{-5*\dmf}mm}{$\frac{-p+1}{2}$};
    \xxtick{\fpeval{-4*\dmf}mm}{$\frac{-p+2}{2}$};
    \draw[dashed] (\fpeval{(-3.5)*\dmf}mm,\fpeval{2*\dhf}mm) --
    (\fpeval{-2.5*\dmf}mm,\fpeval{2*\dhf}mm);
    \draw[] (\fpeval{(-2.5)*\dmf}mm,\fpeval{2*\dhf}mm) --
    (\fpeval{.25*\dmf}mm,\fpeval{2*\dhf}mm);
    \xxtick{\fpeval{-2*\dmf}mm}{$\frac{p-2}{2}$};
    \xxtick{\fpeval{-1*\dmf}mm}{$\frac{p-1}{2}$};
    \xxtick{\fpeval{-0*\dmf}mm}{$\frac{p}{2}$};
    \draw[dashed] (\fpeval{(.25)*\dmf}mm,\fpeval{2*\dhf}mm) --
    (\fpeval{1.25*\dmf}mm,\fpeval{2*\dhf}mm);
    \draw[->] (\fpeval{(1.25)*\dmf}mm,\fpeval{2*\dhf}mm) --
    (\fpeval{3.5*\dmf}mm,\fpeval{2*\dhf}mm);
    \xxtick{\fpeval{1.5*\dmf}mm}{$\frac{p+q}{2}$};
    
    \filldraw (\fpeval{-6*\dmf}mm,0) circle (2pt);
    \filldraw (\fpeval{-5*\dmf}mm,0) circle (2pt);
    \filldraw (\fpeval{-4*\dmf}mm,0) circle (2pt);
    \dashline{(\fpeval{-3*\dmf}mm,0)}{0};
    \filldraw (\fpeval{-2*\dmf}mm,0) circle (2pt);
    \filldraw (\fpeval{-1*\dmf}mm,0) circle (2pt);
    \filldraw[red] (\fpeval{0*\dmf}mm,0) circle (2pt);
    \filldraw (\fpeval{(-7+.5)*\dmf}mm,\fpeval{-\dhf}mm) circle (2pt);
    \filldraw (\fpeval{(-6+.5)*\dmf}mm,\fpeval{-\dhf}mm) circle (2pt);
    \filldraw (\fpeval{(-5+.5)*\dmf}mm,\fpeval{-\dhf}mm) circle (2pt);
    \filldraw (\fpeval{(-4+.5)*\dmf}mm,\fpeval{-\dhf}mm) circle (2pt);
    \dashline{(\fpeval{(-3+.5)*\dmf}mm,\fpeval{-\dhf}mm)}{0};
    \filldraw (\fpeval{(-2+.5)*\dmf}mm,\fpeval{-\dhf}mm) circle (2pt);
    \filldraw (\fpeval{(-1+.5)*\dmf}mm,\fpeval{-\dhf}mm) circle (2pt);
    \filldraw (\fpeval{(-0+.5)*\dmf}mm,\fpeval{-\dhf}mm) circle (2pt);

    \draw (\fpeval{(-6+.5)*\dmf}mm,\fpeval{-\dhf}mm) circle (4pt);
    \draw (\fpeval{(-5+.5)*\dmf}mm,\fpeval{-\dhf}mm) circle (4pt);
    \draw (\fpeval{(-4+.5)*\dmf}mm,\fpeval{-\dhf}mm) circle (4pt);
    \draw (\fpeval{(-2+.5)*\dmf}mm,\fpeval{-\dhf}mm) circle (4pt);
    \draw (\fpeval{(-1+.5)*\dmf}mm,\fpeval{-\dhf}mm) circle (4pt);
    \dashline{(\fpeval{-7*\dmf}mm,\fpeval{-2*\dhf}mm)}{60}
    \dashline{(\fpeval{-6*\dmf}mm,\fpeval{-2*\dhf}mm)}{60}
    \dashline{(\fpeval{-5*\dmf}mm,\fpeval{-2*\dhf}mm)}{120}
    \dashline{(\fpeval{-4*\dmf}mm,\fpeval{-2*\dhf}mm)}{120}
    \dashline{(\fpeval{-3*\dmf}mm,\fpeval{-2*\dhf}mm)}{120}
    \dashline{(\fpeval{-2*\dmf}mm,\fpeval{-2*\dhf}mm)}{60}
    \dashline{(\fpeval{-1*\dmf}mm,\fpeval{-2*\dhf}mm)}{60}
    \dashline{(\fpeval{0*\dmf}mm,\fpeval{-2*\dhf}mm)}{120}
    \dashline{(\fpeval{1*\dmf}mm,\fpeval{-2*\dhf}mm)}{120}
    \filldraw (\fpeval{(-8+.5)*\dmf}mm,\fpeval{-3*\dhf}mm) circle (2pt);
    \filldraw (\fpeval{(-7+.5)*\dmf}mm,\fpeval{-3*\dhf}mm) circle (2pt);
    \dashline{(\fpeval{(-6+.5)*\dmf}mm,\fpeval{-3*\dhf}mm)}{0};
    \filldraw (\fpeval{(-5+.5)*\dmf}mm,\fpeval{-3*\dhf}mm) circle (2pt);
    \filldraw (\fpeval{(-4+.5)*\dmf}mm,\fpeval{-3*\dhf}mm) circle (2pt);
    \dashline{(\fpeval{(-3+.5)*\dmf}mm,\fpeval{-3*\dhf}mm)}{0};
    \filldraw (\fpeval{(-2+.5)*\dmf}mm,\fpeval{-3*\dhf}mm) circle (2pt);
    \dashline{(\fpeval{(-1+.5)*\dmf}mm,\fpeval{-3*\dhf}mm)}{0};
    \filldraw (\fpeval{(-0+.5)*\dmf}mm,\fpeval{-3*\dhf}mm) circle (2pt);
    \filldraw (\fpeval{(1+.5)*\dmf}mm,\fpeval{-3*\dhf}mm) circle (2pt);

    \draw (\fpeval{(-7+.5)*\dmf}mm,\fpeval{-3*\dhf}mm) circle (4pt);
    \draw (\fpeval{(-5+.5)*\dmf}mm,\fpeval{-3*\dhf}mm) circle (4pt);
    \draw (\fpeval{(-5+.5)*\dmf}mm,\fpeval{-3*\dhf}mm) circle (6pt);
    \draw (\fpeval{(-5+.5)*\dmf}mm,\fpeval{-3*\dhf}mm) circle (8pt);
    \draw (\fpeval{(-4+.5)*\dmf}mm,\fpeval{-3*\dhf}mm) circle (4pt);
    \draw (\fpeval{(-4+.5)*\dmf}mm,\fpeval{-3*\dhf}mm) circle (6pt);
    \draw (\fpeval{(-4+.5)*\dmf}mm,\fpeval{-3*\dhf}mm) circle (8pt);
    \draw (\fpeval{(-2+.5)*\dmf}mm,\fpeval{-3*\dhf}mm) circle (4pt);
    \draw (\fpeval{(-2+.5)*\dmf}mm,\fpeval{-3*\dhf}mm) circle (6pt);
    \draw (\fpeval{(-2+.5)*\dmf}mm,\fpeval{-3*\dhf}mm) circle (8pt);
    \draw (\fpeval{(-0+.5)*\dmf}mm,\fpeval{-3*\dhf}mm) circle (4pt);
    \filldraw (\fpeval{-7*\dmf}mm,\fpeval{-4*\dhf}mm) circle (2pt);
    \filldraw (\fpeval{-6*\dmf}mm,\fpeval{-4*\dhf}mm) circle (2pt);
    \dashline{(\fpeval{-5*\dmf}mm,\fpeval{-4*\dhf}mm)}{0};
    \dashline{(\fpeval{-4*\dmf}mm,\fpeval{-4*\dhf}mm)}{120};
    \filldraw (\fpeval{-3*\dmf}mm,\fpeval{-4*\dhf}mm) circle (2pt);
    \dashline{(\fpeval{-2*\dmf}mm,\fpeval{-4*\dhf}mm)}{60};
    \dashline{(\fpeval{-1*\dmf}mm,\fpeval{-4*\dhf}mm)}{0};
    \filldraw (\fpeval{-0*\dmf},\fpeval{-4*\dhf}mm) circle (2pt);
    \filldraw (\fpeval{1*\dmf}mm,\fpeval{-4*\dhf}mm) circle (2pt);

    \draw (\fpeval{-6*\dmf}mm,\fpeval{-4*\dhf}mm) circle (4pt);
    \draw (\fpeval{-3*\dmf}mm,\fpeval{-4*\dhf}mm) circle (4pt);
    \draw (\fpeval{-3*\dmf}mm,\fpeval{-4*\dhf}mm) circle (6pt);
    \draw (\fpeval{-3*\dmf}mm,\fpeval{-4*\dhf}mm) circle (8pt);
    \draw (\fpeval{-0*\dmf},\fpeval{-4*\dhf}mm) circle (4pt);
    \filldraw (\fpeval{(-7+.5)*\dmf}mm,\fpeval{-5*\dhf}mm) circle (2pt);
    \dashline{(\fpeval{(-6+.5)*\dmf}mm,\fpeval{-5*\dhf}mm)}{120};
    \dashline{(\fpeval{(-5+.5)*\dmf}mm,\fpeval{-5*\dhf}mm)}{60};
    \filldraw (\fpeval{(-4+.5)*\dmf}mm,\fpeval{-5*\dhf}mm) circle (2pt);
    \filldraw (\fpeval{(-3+.5)*\dmf}mm,\fpeval{-5*\dhf}mm) circle (2pt);
    \dashline{(\fpeval{(-2+.5)*\dmf}mm,\fpeval{-5*\dhf}mm)}{120};
    \dashline{(\fpeval{(-1+.5)*\dmf}mm,\fpeval{-5*\dhf}mm)}{60};
    \filldraw (\fpeval{(0+.5)*\dmf}mm,\fpeval{-5*\dhf}mm) circle (2pt);

    \draw (\fpeval{(-4+.5)*\dmf}mm,\fpeval{-5*\dhf}mm) circle (4pt);
    \draw (\fpeval{(-4+.5)*\dmf}mm,\fpeval{-5*\dhf}mm) circle (6pt);
    \draw (\fpeval{(-4+.5)*\dmf}mm,\fpeval{-5*\dhf}mm) circle (8pt);
    \draw (\fpeval{(-3+.5)*\dmf}mm,\fpeval{-5*\dhf}mm) circle (4pt);
    \draw (\fpeval{(-3+.5)*\dmf}mm,\fpeval{-5*\dhf}mm) circle (6pt);
    \draw (\fpeval{(-3+.5)*\dmf}mm,\fpeval{-5*\dhf}mm) circle (8pt);
    \dashline{(\fpeval{-6*\dmf}mm,\fpeval{-6*\dhf}mm)}{120};
    \filldraw (\fpeval{-5*\dmf}mm,\fpeval{-6*\dhf}mm) circle (2pt);
    \dashline{(\fpeval{-4*\dmf}mm,\fpeval{-6*\dhf}mm)}{60};
    \filldraw (\fpeval{-3*\dmf}mm,\fpeval{-6*\dhf}mm) circle (2pt);
    \dashline{(\fpeval{-2*\dmf}mm,\fpeval{-6*\dhf}mm)}{120};
    \filldraw (\fpeval{-1*\dmf}mm,\fpeval{-6*\dhf}mm) circle (2pt);
    \dashline{(\fpeval{-0*\dmf},\fpeval{-6*\dhf}mm)}{60};
    \draw (\fpeval{-5*\dmf}mm,\fpeval{-6*\dhf}mm) circle (4pt);
    \draw (\fpeval{-3*\dmf}mm,\fpeval{-6*\dhf}mm) circle (4pt);
    \draw (\fpeval{-3*\dmf}mm,\fpeval{-6*\dhf}mm) circle (6pt);
    \draw (\fpeval{-3*\dmf}mm,\fpeval{-6*\dhf}mm) circle (8pt);
    \draw (\fpeval{-1*\dmf}mm,\fpeval{-6*\dhf}mm) circle (4pt);
    \filldraw (\fpeval{(-6+.5)*\dmf}mm,\fpeval{-7*\dhf}mm) circle (2pt);
    \filldraw (\fpeval{(-5+.5)*\dmf}mm,\fpeval{-7*\dhf}mm) circle (2pt);
    \dashline{(\fpeval{(-4+.5)*\dmf}mm,\fpeval{-7*\dhf}mm)}{60};
    \dashline{(\fpeval{(-3+.5)*\dmf}mm,\fpeval{-7*\dhf}mm)}{120};
    \filldraw (\fpeval{(-2+.5)*\dmf}mm,\fpeval{-7*\dhf}mm) circle (2pt);
    \filldraw (\fpeval{(-1+.5)*\dmf}mm,\fpeval{-7*\dhf}mm) circle (2pt);

    \draw (\fpeval{(-5+.5)*\dmf}mm,\fpeval{-7*\dhf}mm) circle (4pt);
    \draw (\fpeval{(-2+.5)*\dmf}mm,\fpeval{-7*\dhf}mm) circle (4pt);
    \filldraw (\fpeval{-5*\dmf}mm,\fpeval{-8*\dhf}mm) circle (2pt);
    \filldraw (\fpeval{-4*\dmf}mm,\fpeval{-8*\dhf}mm) circle (2pt);
    \dashline{(\fpeval{-3*\dmf}mm,\fpeval{-8*\dhf}mm)}{0};
    \filldraw (\fpeval{-2*\dmf}mm,\fpeval{-8*\dhf}mm) circle (2pt);
    \filldraw (\fpeval{-1*\dmf}mm,\fpeval{-8*\dhf}mm) circle (2pt);

    \draw (\fpeval{-4*\dmf}mm,\fpeval{-8*\dhf}mm) circle (4pt);
    \draw (\fpeval{-2*\dmf}mm,\fpeval{-8*\dhf}mm) circle (4pt);
    \filldraw (\fpeval{(-5+.5)*\dmf}mm,\fpeval{-9*\dhf}mm) circle (2pt);
    \dashline{(\fpeval{(-4+.5)*\dmf}mm,\fpeval{-9*\dhf}mm)}{0};
    \filldraw (\fpeval{(-3+.5)*\dmf}mm,\fpeval{-9*\dhf}mm) circle (2pt);
    \filldraw (\fpeval{(-2+.5)*\dmf}mm,\fpeval{-9*\dhf}mm) circle (2pt);
  \end{tikzpicture}
\end{equation*}
The diagram is a non-regular hexagon with $p+1$ and $q+1$ number of states on
its alternate sides starting from the top. The overlapping circles denote the
multiplicity of states in the total isospin quantum number $j$. The states lying
on the outermost hexagon are non-degenerate, those lying on the hexagon
immediately inside are doubly degenerate, and so on until we hit a triangle in
the center. All the states on and inside this triangle have degeneracy
$\min(p,q)+1$. The highest weight state is denoted in red. The ladder operators
$T_\pm$ take us horizontally to a state lying immediately to the right/left,
$V_\pm$ diagonally to a linear combination of the states lying immediately to
the top-right/bottom-left, while $U_\pm$ diagonally to a linear combination of
the states lying immediately to the top-left/bottom-right. The eightfold-way
diagram for $D(q,p)$ has the same form, but reflected in the $y$-direction
$y\to-y$.

For instance, the states in the ``trivial/singlet'' ``{\bf 1}'' $D(0,0)$,
``fundamental/triplet'' ``{\bf 3}'' $D(1,0)$, ``sextet'' ``{\bf 6}'' $D(2,0)$,
``decuplet'' ``{\bf 10}'' $D(3,0)$, and ``{\bf 15}'' $D(4,0)$ representations,
and so on, are respectively arranged as
\begin{equation*}
  \begin{tikzpicture}
    \filldraw (\fpeval{0*\dmf}mm,0) circle (2pt);
    \node at (0*\fpeval{\dmf}mm,\fpeval{-1*\dhf}mm)
    {$D(0,0)$};
  \end{tikzpicture} \qquad
  \begin{tikzpicture}
    \filldraw (\fpeval{-1*\dmf}mm,0) circle (2pt);
    \filldraw (\fpeval{0*\dmf}mm,0) circle (2pt);
    \filldraw (\fpeval{-.5*\dmf}mm,\fpeval{-1*\dhf}mm) circle (2pt);
    \node at (-.5*\fpeval{\dmf}mm,\fpeval{-2*\dhf}mm) {$D(1,0)$};
  \end{tikzpicture} \qquad\qquad
  \begin{tikzpicture}
    \filldraw (\fpeval{-2*\dmf}mm,0) circle (2pt);
    \filldraw (\fpeval{-1*\dmf}mm,0) circle (2pt);
    \filldraw (\fpeval{0*\dmf}mm,0) circle (2pt);
    \filldraw (\fpeval{-1.5*\dmf}mm,\fpeval{-1*\dhf}mm) circle (2pt);
    \filldraw (\fpeval{-.5*\dmf}mm,\fpeval{-1*\dhf}mm) circle (2pt);
    \filldraw (\fpeval{-1*\dmf}mm,\fpeval{-2*\dhf}mm) circle (2pt);
    \node at (-1*\fpeval{\dmf}mm,\fpeval{-3*\dhf}mm) {$D(2,0)$};
  \end{tikzpicture} \qquad\qquad
  \begin{tikzpicture}
    \filldraw (\fpeval{-3*\dmf}mm,0) circle (2pt);
    \filldraw (\fpeval{-2*\dmf}mm,0) circle (2pt);
    \filldraw (\fpeval{-1*\dmf}mm,0) circle (2pt);
    \filldraw (\fpeval{0*\dmf}mm,0) circle (2pt);
    \filldraw (\fpeval{-2.5*\dmf}mm,\fpeval{-1*\dhf}mm) circle (2pt);
    \filldraw (\fpeval{-1.5*\dmf}mm,\fpeval{-1*\dhf}mm) circle (2pt);
    \filldraw (\fpeval{-.5*\dmf}mm,\fpeval{-1*\dhf}mm) circle (2pt);
    \filldraw (\fpeval{-2*\dmf}mm,\fpeval{-2*\dhf}mm) circle (2pt);
    \filldraw (\fpeval{-1*\dmf}mm,\fpeval{-2*\dhf}mm) circle (2pt);
    \filldraw (\fpeval{-1.5*\dmf}mm,\fpeval{-3*\dhf}mm) circle (2pt);
    \node at (-1.5*\fpeval{\dmf}mm,\fpeval{-4*\dhf}mm) {$D(3,0)$};
  \end{tikzpicture} \qquad\qquad
  \begin{tikzpicture}
    \filldraw (\fpeval{-4*\dmf}mm,0) circle (2pt);
    \filldraw (\fpeval{-3*\dmf}mm,0) circle (2pt);
    \filldraw (\fpeval{-2*\dmf}mm,0) circle (2pt);
    \filldraw (\fpeval{-1*\dmf}mm,0) circle (2pt);
    \filldraw (\fpeval{0*\dmf}mm,0) circle (2pt);
    \filldraw (\fpeval{-3.5*\dmf}mm,\fpeval{-1*\dhf}mm) circle (2pt);
    \filldraw (\fpeval{-2.5*\dmf}mm,\fpeval{-1*\dhf}mm) circle (2pt);
    \filldraw (\fpeval{-1.5*\dmf}mm,\fpeval{-1*\dhf}mm) circle (2pt);
    \filldraw (\fpeval{-.5*\dmf}mm,\fpeval{-1*\dhf}mm) circle (2pt);
    \filldraw (\fpeval{-3*\dmf}mm,\fpeval{-2*\dhf}mm) circle (2pt);
    \filldraw (\fpeval{-2*\dmf}mm,\fpeval{-2*\dhf}mm) circle (2pt);
    \filldraw (\fpeval{-1*\dmf}mm,\fpeval{-2*\dhf}mm) circle (2pt);
    \filldraw (\fpeval{-2.5*\dmf}mm,\fpeval{-3*\dhf}mm) circle (2pt);
    \filldraw (\fpeval{-1.5*\dmf}mm,\fpeval{-3*\dhf}mm) circle (2pt);
    \filldraw (\fpeval{-2*\dmf}mm,\fpeval{-4*\dhf}mm) circle (2pt);
    \node at (-2*\fpeval{\dmf}mm,\fpeval{-5*\dhf}mm) {$D(4,0)$};
  \end{tikzpicture}
\end{equation*}
while the states in the conjugate ``anti-fundamental/anti-triplet''
``$\bar{\bf 3}$'' $D(0,1)$, ``anti-sextet'' ``$\bar{\bf 6}$'' $D(0,2)$,
``anti-decuplet'' ``$\bar{\bf 10}$'' $D(0,3)$, and ``$\bar{\bf 15}$'' $D(0,4)$
representations, and so on, are arranged as
\begin{equation*}
  \begin{tikzpicture}
    \filldraw (\fpeval{-1*\dmf}mm,0) circle (2pt);
    \filldraw (\fpeval{0*\dmf}mm,0) circle (2pt);
    \filldraw (\fpeval{-.5*\dmf}mm,\fpeval{1*\dhf}mm) circle (2pt);
    \node at (-.5*\fpeval{\dmf}mm,\fpeval{-1*\dhf}mm) {$D(0,1)$};
  \end{tikzpicture} \qquad\qquad
  \begin{tikzpicture}
    \filldraw (\fpeval{-2*\dmf}mm,0) circle (2pt);
    \filldraw (\fpeval{-1*\dmf}mm,0) circle (2pt);
    \filldraw (\fpeval{0*\dmf}mm,0) circle (2pt);
    \filldraw (\fpeval{-1.5*\dmf}mm,\fpeval{1*\dhf}mm) circle (2pt);
    \filldraw (\fpeval{-.5*\dmf}mm,\fpeval{1*\dhf}mm) circle (2pt);
    \filldraw (\fpeval{-1*\dmf}mm,\fpeval{2*\dhf}mm) circle (2pt);
    \node at (-1*\fpeval{\dmf}mm,\fpeval{-1*\dhf}mm) {$D(0,2)$};
  \end{tikzpicture} \qquad\qquad
  \begin{tikzpicture}
    \filldraw (\fpeval{-3*\dmf}mm,0) circle (2pt);
    \filldraw (\fpeval{-2*\dmf}mm,0) circle (2pt);
    \filldraw (\fpeval{-1*\dmf}mm,0) circle (2pt);
    \filldraw (\fpeval{0*\dmf}mm,0) circle (2pt);
    \filldraw (\fpeval{-2.5*\dmf}mm,\fpeval{1*\dhf}mm) circle (2pt);
    \filldraw (\fpeval{-1.5*\dmf}mm,\fpeval{1*\dhf}mm) circle (2pt);
    \filldraw (\fpeval{-.5*\dmf}mm,\fpeval{1*\dhf}mm) circle (2pt);
    \filldraw (\fpeval{-2*\dmf}mm,\fpeval{2*\dhf}mm) circle (2pt);
    \filldraw (\fpeval{-1*\dmf}mm,\fpeval{2*\dhf}mm) circle (2pt);
    \filldraw (\fpeval{-1.5*\dmf}mm,\fpeval{3*\dhf}mm) circle (2pt);
    \node at (-1.5*\fpeval{\dmf}mm,\fpeval{-1*\dhf}mm) {$D(0,3)$};
  \end{tikzpicture} \qquad\qquad
  \begin{tikzpicture}
    \filldraw (\fpeval{-4*\dmf}mm,0) circle (2pt);
    \filldraw (\fpeval{-3*\dmf}mm,0) circle (2pt);
    \filldraw (\fpeval{-2*\dmf}mm,0) circle (2pt);
    \filldraw (\fpeval{-1*\dmf}mm,0) circle (2pt);
    \filldraw (\fpeval{0*\dmf}mm,0) circle (2pt);
    \filldraw (\fpeval{-3.5*\dmf}mm,\fpeval{1*\dhf}mm) circle (2pt);
    \filldraw (\fpeval{-2.5*\dmf}mm,\fpeval{1*\dhf}mm) circle (2pt);
    \filldraw (\fpeval{-1.5*\dmf}mm,\fpeval{1*\dhf}mm) circle (2pt);
    \filldraw (\fpeval{-.5*\dmf}mm,\fpeval{1*\dhf}mm) circle (2pt);
    \filldraw (\fpeval{-3*\dmf}mm,\fpeval{2*\dhf}mm) circle (2pt);
    \filldraw (\fpeval{-2*\dmf}mm,\fpeval{2*\dhf}mm) circle (2pt);
    \filldraw (\fpeval{-1*\dmf}mm,\fpeval{2*\dhf}mm) circle (2pt);
    \filldraw (\fpeval{-2.5*\dmf}mm,\fpeval{3*\dhf}mm) circle (2pt);
    \filldraw (\fpeval{-1.5*\dmf}mm,\fpeval{3*\dhf}mm) circle (2pt);
    \filldraw (\fpeval{-2*\dmf}mm,\fpeval{4*\dhf}mm) circle (2pt);
    \node at (-2*\fpeval{\dmf}mm,\fpeval{-1*\dhf}mm) {$D(0,4)$};
  \end{tikzpicture}
\end{equation*}
The eightfold-way diagrams for the more non-trivial ``adjoint/octet''
``${\bf 8}$'' $D(1,1)$, ``${\bf 15}$'' $D(2,1)$, ``${\bf 24}$'' $D(3,1)$,
``$\bar{\bf 15}$'' $D(1,2)$, ``$\bar{\bf 24}$'' $D(1,3)$, and ``${\bf 27}$''
$D(2,2)$ representations are given as
\begin{gather*}
   \begin{tikzpicture}
    \filldraw (\fpeval{-1*\dmf}mm,0) circle (2pt);
    \filldraw (\fpeval{0*\dmf}mm,0) circle (2pt);
    \filldraw (\fpeval{-1.5*\dmf}mm,\fpeval{-1*\dhf}mm) circle (2pt);
    \filldraw (\fpeval{-.5*\dmf}mm,\fpeval{-1*\dhf}mm) circle (2pt);
    \draw (\fpeval{-.5*\dmf}mm,\fpeval{-1*\dhf}mm) circle (4pt);
    \filldraw (\fpeval{.5*\dmf}mm,\fpeval{-1*\dhf}mm) circle (2pt);
    \filldraw (\fpeval{-1*\dmf}mm,\fpeval{-2*\dhf}mm) circle (2pt);
    \filldraw (\fpeval{0*\dmf}mm,\fpeval{-2*\dhf}mm) circle (2pt);
    \node at (-.5*\fpeval{\dmf}mm,\fpeval{-3*\dhf}mm) {$D(1,1)$};
  \end{tikzpicture} \qquad\qquad
  \begin{tikzpicture}
    \state{-2}{0};
    \state{-1}{0};
    \state{0}{0};
    \state{-2.5}{-1};
    \duostate{-1.5}{-1};
    \duostate{-.5}{-1};
    \state{.5}{-1};
    \state{-2}{-2};
    \duostate{-1}{-2};
    \state{0}{-2};
    \state{-1.5}{-3};
    \state{-.5}{-3};
    \node at (-1*\fpeval{\dmf}mm,\fpeval{-4*\dhf}mm) {$D(2,1)$};
  \end{tikzpicture} \qquad\qquad
  \begin{tikzpicture}
    \state{-3}{0};
    \state{-2}{0};
    \state{-1}{0};
    \state{0}{0};
    \state{-3.5}{-1};
    \duostate{-2.5}{-1};
    \duostate{-1.5}{-1};
    \duostate{-.5}{-1};
    \state{.5}{-1};
    \state{-3}{-2};
    \duostate{-2}{-2};
    \duostate{-1}{-2};
    \state{0}{-2};
    \state{-2.5}{-3};
    \duostate{-1.5}{-3};
    \state{-.5}{-3};
    \state{-2}{-4};
    \state{-1}{-4};
    \node at (-1.5*\fpeval{\dmf}mm,\fpeval{-5*\dhf}mm) {$D(3,1)$};
  \end{tikzpicture} \nn\\[1ex] 
  \begin{tikzpicture}
    \state{-2}{0};
    \state{-1}{0};
    \state{0}{0};
    \state{-2.5}{1};
    \duostate{-1.5}{1};
    \duostate{-.5}{1};
    \state{.5}{1};
    \state{-2}{2};
    \duostate{-1}{2};
    \state{0}{2};
    \state{-1.5}{3};
    \state{-.5}{3};
    \node at (-1*\fpeval{\dmf}mm,\fpeval{-1*\dhf}mm) {$D(1,2)$};
  \end{tikzpicture} \qquad\qquad
  \begin{tikzpicture}
    \state{-3}{0};
    \state{-2}{0};
    \state{-1}{0};
    \state{0}{0};
    \state{-3.5}{1};
    \duostate{-2.5}{1};
    \duostate{-1.5}{1};
    \duostate{-.5}{1};
    \state{.5}{1};
    \state{-3}{2};
    \duostate{-2}{2};
    \duostate{-1}{2};
    \state{0}{2};
    \state{-2.5}{3};
    \duostate{-1.5}{3};
    \state{-.5}{3};
    \state{-2}{4};
    \state{-1}{4};
    \node at (-1.5*\fpeval{\dmf}mm,\fpeval{-1*\dhf}mm) {$D(1,3)$};
  \end{tikzpicture} \qquad\qquad
  \begin{tikzpicture}
    \state{-2}{0};
    \state{-1}{0};
    \state{0}{0};
    \state{-2.5}{1};
    \duostate{-1.5}{1};
    \duostate{-.5}{1};
    \state{.5}{1};
    \state{-3}{2};
    \duostate{-2}{2};
    \triostate{-1}{2};
    \duostate{0}{2};
    \state{1}{2};
    \state{-2.5}{3};
    \duostate{-1.5}{3};
    \duostate{-.5}{3};
    \state{.5}{3};
    \state{-2}{4};
    \state{-1}{4};
    \state{0}{4};
    \node at (-1*\fpeval{\dmf}mm,\fpeval{-1*\dhf}mm) {$D(2,2)$};
  \end{tikzpicture}
\end{gather*}

\paragraph*{Example:} The QCD colour force in the Standard Model is associated
with an internal SU(3) symmetry.  The quarks and anti-quarks transform in the
triplet ``${\bf 3}$'' $D(1,0)$ and anti-triplet ``$\bar{\bf 3}$'' $D(0,1)$
representations respectively of the colour SU(3) group, whereas the gluons
transform in the adjoint ``${\bf 8}$'' $D(1,1)$ representation. The remaining
Standard Model fundamental particles, as well as the naturally appearing
hadrons, are colour SU(3) singlets. See \cref{sec:QED-QCD,sec:standardmodel}.

\paragraph*{Example:} The Standard Model has an approximate SU(3) flavour
symmetry, which has the isospin SU(2) symmetry as a subgroup. The light quarks
($u,d,s$) and antiquarks ($\bar u, \bar d, \bar s$) make up the triplet
``${\bf 3}$'' $D(1,0)$ and anti-triplet ``$\bar{\bf 3}$'' $D(0,1)$
representations of the flavour SU(3) group: {\def\dmf{10}%
\begin{equation*}
  \begin{tikzpicture}
    \state{-1}{0}
    \node at (\fpeval{-.75*\dmf}mm,0) {$d$};
    \state{0}{0};
    \node at (\fpeval{0.25*\dmf}mm,0) {$u$};
    \state{-.5}{-1};
    \node at (\fpeval{-.25*\dmf}mm,\fpeval{-1*\dhf}mm) {$s$};
    \node at (\fpeval{-.5*\dmf}mm,\fpeval{-1.7*\dhf}mm) {quark triplet};
  \end{tikzpicture} \qquad\qquad
  \begin{tikzpicture}
    \state{-1}{0}
    \node at (\fpeval{-.75*\dmf}mm,0) {$\bar u$};
    \state{0}{0};
    \node at (\fpeval{0.25*\dmf}mm,0) {$\bar d$};
    \state{-.5}{1};
    \node at (\fpeval{-.25*\dmf}mm,\fpeval{1*\dhf}mm) {$\bar s$};
    \node at (\fpeval{-.5*\dmf}mm,\fpeval{-0.7*\dhf}mm) {antiquark anti-triplet};
  \end{tikzpicture}
\end{equation*}}%
\noindent
The baryons are made of 3 quarks and transform in the octet ``${\bf 8}$''
$D(1,1)$ or decuplet ``${\bf 10}$'' $D(3,0)$ representations of the flavour
SU(3) group, while the anti-baryons are made of 3 anti-quarks and transform in
the respective octet ``${\bf 8}$'' $D(1,1)$ or anti-decuplet ``$\bar{\bf 10}$''
$D(0,3)$ representations: {\def\dmf{10}%
\begin{gather*}
  \begin{tikzpicture}
    \state{-2}{0}
    \node at (\fpeval{-1.65*\dmf}mm,0) {$n^0$};
    \state{-1}{0}
    \node at (\fpeval{-0.65*\dmf}mm,0) {$p^{+}$};
    \state{-2.5}{-1};
    \node at (\fpeval{-2.1*\dmf}mm,\fpeval{-1*\dhf}mm) {$\Sigma^{-}$};
    \duostate{-1.5}{-1};
    \node at (\fpeval{-1.45*\dmf}mm,\fpeval{-0.5*\dhf}mm) {$\Lambda^0$};
    \node at (\fpeval{-1.1*\dmf}mm,\fpeval{-1*\dhf}mm) {$\Sigma^{0}$};
    \state{-.5}{-1};
    \node at (\fpeval{-.1*\dmf}mm,\fpeval{-1*\dhf}mm) {$\Sigma^{+}$};
    \state{-2}{-2};
    \node at (\fpeval{-1.6*\dmf}mm,\fpeval{-2*\dhf}mm) {$\Xi^{-}$};
    \state{-1}{-2};
    \node at (\fpeval{-0.6*\dmf}mm,\fpeval{-2*\dhf}mm) {$\Xi^{0}$};
    \node at (\fpeval{-1.5*\dmf}mm,\fpeval{-2.7*\dhf}mm) {baryon octet};
  \end{tikzpicture} \quad
  \begin{tikzpicture}
    \state{-3}{0}
    \node at (\fpeval{-2.6*\dmf}mm,0) {$\Delta^{-}$};
    \state{-2}{0}
    \node at (\fpeval{-1.6*\dmf}mm,0) {$\Delta^{0}$};
    \state{-1}{0}
    \node at (\fpeval{-0.6*\dmf}mm,0) {$\Delta^{+}$};
    \state{0}{0};
    \node at (\fpeval{0.45*\dmf}mm,0) {$\Delta^{2+}$};
    \state{-2.5}{-1};
    \node at (\fpeval{-2.05*\dmf}mm,\fpeval{-1*\dhf}mm) {$\Sigma^{*-}$};
    \state{-1.5}{-1};
    \node at (\fpeval{-1.05*\dmf}mm,\fpeval{-1*\dhf}mm) {$\Sigma^{*0}$};
    \state{-.5}{-1};
    \node at (\fpeval{-.05*\dmf}mm,\fpeval{-1*\dhf}mm) {$\Sigma^{*+}$};
    \state{-2}{-2};
    \node at (\fpeval{-1.55*\dmf}mm,\fpeval{-2*\dhf}mm) {$\Xi^{*-}$};
    \state{-1}{-2};
    \node at (\fpeval{-0.55*\dmf}mm,\fpeval{-2*\dhf}mm) {$\Xi^{*0}$};
    \state{-1.5}{-3};
    \node at (\fpeval{-1.1*\dmf}mm,\fpeval{-3*\dhf}mm) {$\Omega^{-}$};
    \node at (\fpeval{-1.45*\dmf}mm,\fpeval{-3.7*\dhf}mm) {baryon decuplet};
  \end{tikzpicture} \qquad
  \begin{tikzpicture}
    \state{-2}{0}
    \node at (\fpeval{-1.65*\dmf}mm,0) {$\bar p^-$};
    \state{-1}{0}
    \node at (\fpeval{-0.6*\dmf}mm,0) {$\bar n^0$};
    \state{-2.5}{1};
    \node at (\fpeval{-2.1*\dmf}mm,\fpeval{1*\dhf}mm) {$\bar\Sigma^{-}$};
    \duostate{-1.5}{1};
    \node at (\fpeval{-1.4*\dmf}mm,\fpeval{1.45*\dhf}mm) {$\bar\Lambda^0$};
    \node at (\fpeval{-1.1*\dmf}mm,\fpeval{1*\dhf}mm) {$\bar\Sigma^{0}$};
    \state{-.5}{1};
    \node at (\fpeval{-.1*\dmf}mm,\fpeval{1*\dhf}mm) {$\bar\Sigma^{+}$};
    \state{-2}{2};
    \node at (\fpeval{-1.6*\dmf}mm,\fpeval{2*\dhf}mm) {$\bar\Xi^{0}$};
    \state{-1}{2};
    \node at (\fpeval{-0.6*\dmf}mm,\fpeval{2*\dhf}mm) {$\bar\Xi^{+}$};
    \node at (\fpeval{-1.45*\dmf}mm,\fpeval{-.7*\dhf}mm) {anti-baryon octet};
  \end{tikzpicture} \quad
  \begin{tikzpicture}
    \state{-3}{0}
    \node at (\fpeval{-2.55*\dmf}mm,0) {$\bar\Delta^{2-}$};
    \state{-2}{0}
    \node at (\fpeval{-1.6*\dmf}mm,0) {$\bar\Delta^{-}$};
    \state{-1}{0}
    \node at (\fpeval{-0.6*\dmf}mm,0) {$\bar\Delta^{0}$};
    \state{0}{0};
    \node at (\fpeval{0.4*\dmf}mm,0) {$\bar\Delta^{+}$};
    \state{-2.5}{1};
    \node at (\fpeval{-2.05*\dmf}mm,\fpeval{1*\dhf}mm) {$\bar\Sigma^{*-}$};
    \state{-1.5}{1};
    \node at (\fpeval{-1.05*\dmf}mm,\fpeval{1*\dhf}mm) {$\bar\Sigma^{*0}$};
    \state{-.5}{1};
    \node at (\fpeval{-.05*\dmf}mm,\fpeval{1*\dhf}mm) {$\bar\Sigma^{*+}$};
    \state{-2}{2};
    \node at (\fpeval{-1.55*\dmf}mm,\fpeval{2*\dhf}mm) {$\bar\Xi^{*0}$};
    \state{-1}{2};
    \node at (\fpeval{-0.55*\dmf}mm,\fpeval{2*\dhf}mm) {$\bar\Xi^{*+}$};
    \state{-1.5}{3};
    \node at (\fpeval{-1.1*\dmf}mm,\fpeval{3*\dhf}mm) {$\bar\Omega^{+}$};
    \node at (\fpeval{-1.45*\dmf}mm,\fpeval{-.7*\dhf}mm) {anti-baryon anti-decuplet};
  \end{tikzpicture} 
\end{gather*}}%
This follows because the tensor decomposition of 3-quark states is
${\bf 3}\otimes{\bf 3}\otimes{\bf 3} = {\bf 10}\oplus{\bf 8}\oplus{\bf
  8}\oplus{\bf 1}$ and that of 3-antiquark states is
$\bar{\bf 3}\otimes\bar{\bf 3}\otimes\bar{\bf 3} = \bar{\bf 10}\oplus{\bf
  8}\oplus{\bf 8}\oplus{\bf 1}$. Similarly, mesons are made of a quark and an
antiquark and transform in the singlet ``${\bf 1}$'' $D(0,0)$ or octet
``${\bf 8}$'' $D(1,1)$ representations of the SU(3) flavour group:
{\def\dmf{10}%
\begin{gather*}
  \begin{tikzpicture}
    \state{0}{0};
    \node at (\fpeval{0.35*\dmf}mm,0) {$\eta'^{0}$};
    \node at (0,\fpeval{-.7*\dhf}mm) {meson singlet};
  \end{tikzpicture}\qquad\qquad
  \begin{tikzpicture}
    \state{-2}{0}
    \node at (\fpeval{-1.65*\dmf}mm,0) {$K^0$};
    \state{-1}{0}
    \node at (\fpeval{-0.6*\dmf}mm,0) {$K^{+}$};
    \state{-2.5}{-1};
    \node at (\fpeval{-2.1*\dmf}mm,\fpeval{-1*\dhf}mm) {$\pi^{-}$};
    \duostate{-1.5}{-1};
    \node at (\fpeval{-1.35*\dmf}mm,\fpeval{-0.55*\dhf}mm) {$\eta^0$};
    \node at (\fpeval{-1.1*\dmf}mm,\fpeval{-1*\dhf}mm) {$\pi^{0}$};
    \state{-.5}{-1};
    \node at (\fpeval{-.1*\dmf}mm,\fpeval{-1*\dhf}mm) {$\pi^{+}$};
    \state{-2}{-2};
    \node at (\fpeval{-1.6*\dmf}mm,\fpeval{-2*\dhf}mm) {$\bar K^{-}$};
    \state{-1}{-2};
    \node at (\fpeval{-0.6*\dmf}mm,\fpeval{-2*\dhf}mm) {$\bar K^{0}$};
    \node at (\fpeval{-1.45*\dmf}mm,\fpeval{-2.7*\dhf}mm) {meson octet};
  \end{tikzpicture}
\end{gather*}}%
This is due to the the tensor decomposition of quark-antiquark states
${\bf 3}\otimes\bar{\bf 3} = {\bf 8}\oplus{\bf 1}$. The states appearing in the
same horizontal line in the diagrams above make up isospin multiplets. Note that
the electric charge of a state is given as $q=m+y/2$. Note also that these
diagrams only cover light baryons and mesons, made up of the 3 light quarks
$(u,d,s)$ and the respective anti-quarks. The complete set of baryons and
mesons, made up of the 6 quarks $(u,d,c,s,t,b)$ and the respective anti-quarks,
can be similarly classified using the representations of SU(6).

\newpage

\section{Lorentz and Poincar\'e groups}
\label{sec:LorentzPoincare}

Physical theories are required to be invariant under a variety of spacetime
symmetries corresponding to the change of inertial reference frame observing the
system under consideration. These include spacetime translations, rotations, and
boosts, formally making up the Poincar\'e group. The subgroup of all rotations
and boosts is known as the Lorentz group. In this section, we dig into a
detailed discussion of the representation theory of Lorentz and Poincar\'e
groups. This also allows us to opportunity to introduce spinor fields as a
representation of the Lorentz group, along with other crucial concepts such as
chirality, helicity, and spin. Since most of the matter fields occurring in
nature are spinorial, these concepts will prove quintessential in setting up the
Standard Model of particle physics. The discussion in this section heavily
relies on Hugh Osborn's notes on group theory~\cite{HughOsborn}.

\subsection{Lorentz group}

Consider a physical theory defined on our $(3+1)$-dimensional spacetime manifold
with coordinates $x^\mu = (ct,x,y,z)$. The Lorentz transformations form a matrix
group given by the coordinate transformations
\begin{equation}
  x^\mu \to \Lambda^\mu{}_\nu x^\nu,
\end{equation}
such that the proper distance between a spacetime point and origin remains
invariant\footnote{We are using the mostly positive convention for the
  Minkowskian metric, as opposed to the mostly negative convention where
  $\eta_{\mu\nu} = \diag(1,-1,-1,-1)$.}
\begin{equation}
  x^\mu x^\nu \eta_{\mu\nu} \to x^\mu x^\nu \eta_{\mu\nu}, \qquad
  \eta_{\mu\nu} = \diag(-1,1,1,1).
\end{equation}
Suppressing the indices, we get the Lorentz group O(3,1) as all $4\times4$
matrices $\Lambda$ satisfying 
\begin{equation}
  \Lambda^\rmT \eta \Lambda = \eta.
\end{equation}
This is a generalisation of the orthogonal group O(4) defined as all $4\times4$
matrices $M$ satisfying the relation $M^\rmT \mathbb{1} M = \mathbb{1}$.

Similar to $\SO(4)$, the \emph{proper Lorentz group} $\SO(3,1)$ contains all the
orientation preserving Lorentz transformations, defined as all $4\times4$
matrices $\Lambda$ satisfying $\Lambda^\rmT \eta \Lambda = \eta$ and
$\det\Lambda = 1$. This excludes the parity operation $\rmP:x^i \to -x^i$ and
the time-reversal operation $\rmT: x^0\to -x^0$, but includes the spacetime
parity operation $\mathrm{PT}: x^\mu \to - x^\mu$. The simply connected piece of
O(3,1) or SO(3,1) further requires
\begin{equation}
  \Lambda^\rmT \eta \Lambda = \eta, \qquad
  \det\Lambda = 1, \qquad
  \Lambda^0{}_0 > 0,
\end{equation}
known as $\SO^+(3,1)$ or \emph{proper orthochronous Lorentz group}, which
excludes all P, T, and PT.

\paragraph*{Lorentz generators:} Let us focus on the connected piece
$\SO^+(3,1)$. Writing in the exponential parametrisation, we can express an
arbitrary element of $\Lambda\in\SO^+(3,1)$ in terms of arbitrary parameters
$\omega^{\mu\nu} = -\omega^{\nu\mu}$ and generators $M_{\mu\nu} = -M_{\nu\mu}$
spanning the associated Lie algebra $\mathfrak{so}(3,1)$, leading to
\begin{equation}
  \Lambda = \exp\lb \frac{i}{2\hbar} \omega^{\mu\nu}M_{\mu\nu} \rb.
  \label{eq:expParamLambda}
\end{equation}
Here $M_{\mu\nu}$ are six linearly independent $4\times4$ matrices satisfying
\begin{equation}
  (M_{\mu\nu})^\rmT = - \eta M_{\mu\nu} \eta.
\end{equation}
We can take the generators to be: the rotation generators
\begin{gather}
  J_1 \equiv M_{23} = -M_{32} = \hbar
  \begin{pmatrix}
    0 & 0 & 0 & 0 \\
    0 & 0 & 0 & 0 \\
    0 & 0 & 0 & -i \\
    0 & 0 & i & 0
  \end{pmatrix}, \qquad
  J_2 \equiv M_{31} = -M_{13} = \hbar
  \begin{pmatrix}
    0 & 0 & 0 & 0 \\
    0 & 0 & 0 & i \\
    0 & 0 & 0 & 0 \\
    0 & -i & 0 & 0
  \end{pmatrix}, \nn\\
  J_3 \equiv M_{12} = -M_{21} = \hbar
  \begin{pmatrix}
    0 & 0 & 0 & 0 \\
    0 & 0 & -i & 0 \\
    0 & i & 0 & 0 \\
    0 & 0 & 0 & 0
  \end{pmatrix},
\end{gather}
and the boost generators
\begin{gather}
  K_1 \equiv M_{10} = - M_{01} = \hbar
  \begin{pmatrix}
    0 & i & 0 & 0 \\
    i & 0 & 0 & 0 \\
    0 & 0 & 0 & 0 \\
    0 & 0 & 0 & 0
  \end{pmatrix}, \qquad
  K_2 \equiv M_{20} = - M_{02} = \hbar
  \begin{pmatrix}
    0 & 0 & i & 0 \\
    0 & 0 & 0 & 0 \\
    i & 0 & 0 & 0 \\
    0 & 0 & 0 & 0
  \end{pmatrix}, \nn\\
  K_3 \equiv M_{30} = - M_{03} = \hbar
  \begin{pmatrix}
    0 & 0 & 0 & i \\
    0 & 0 & 0 & 0 \\
    0 & 0 & 0 & 0 \\
    i & 0 & 0 & 0
  \end{pmatrix},
\end{gather}
satisfying the commutation relations
\begin{equation}
  [J_i,J_j] = i\hbar\epsilon_{ijk}J_k, \qquad
  [K_i,K_j] = -i\hbar\epsilon_{ijk}J_k, \qquad
  [J_i,K_j] = i\hbar\epsilon_{ijk}K_k.
  \label{eq:JKAlgebra}
\end{equation}
In terms of $M_{\mu\nu}$, these can be covariantly represented as
\begin{equation}
  [M_{\mu\nu},M_{\rho\sigma}]
  = i\hbar\lb \eta_{\mu\rho}M_{\nu\sigma}
  - \eta_{\nu\rho}M_{\mu\sigma}
  - \eta_{\mu\sigma}M_{\nu\rho}
  + \eta_{\nu\sigma}M_{\mu\rho} \rb.
  \label{eq:LorentzAlgebra}
\end{equation}
Note that
\begin{equation}
  (M_{\mu\nu})^\rho{}_{\sigma}
  = -i\hbar\lb\delta^\rho_{\mu} \eta_{\nu\sigma}
  - \delta^\rho_{\nu} \eta_{\mu\sigma}\rb
  \quad\implies\quad
  \Lambda^\mu{}_\nu
  = \exp(\omega)^\mu{}_\nu.
\end{equation}

\paragraph*{Double cover:} The connected piece of the Lorentz group $\SO^+(3,1)$
can be mapped to the group $\mathrm{SL}(2,\bbC)$ of all $2\times 2$ unit
determinant complex matrices. Consider rewriting the coordinates $x^\mu$ into a
$2\times 2$ matrix
\begin{equation}
  X = \sigma_\mu x^\mu =
  \begin{pmatrix}
    x^0 + x^3 & x^1 -i x^2 \\
    x^1 + ix^2 & x^0 - x^3
  \end{pmatrix}, \qquad
  X^\dagger = X, \qquad
  \det X = - x^\mu x^\nu \eta_{\mu\nu},
\end{equation}
where $\sigma_\mu = (\mathbb{1},\sigma_i)$,
with $\sigma_i$ being the Pauli matrices. Raising/lowering of $\mu,\nu,\ldots$
indices is done by $\eta^{\mu\nu}$ and $\eta_{\mu\nu}$ respectively.  A generic
$\SO^+(3,1)$ transformation preserves the Hermiticity and determinant of $X$,
leading to an $A\in\SL(2,\bbC)$ transformation\footnote{The explicit mapping is
  given by
  \begin{equation}
    \Lambda^\mu{}_\nu \to A = \E{i\alpha} \frac{\sigma_\mu \Lambda^\mu{}_\nu \bar\sigma^\nu
    }{2\sqrt{\Lambda^\mu{}_\mu}},
\end{equation}
where the phase $\E{i\alpha}$ is determined up to $\pm 1$ by the condition
$\det A = 1$.}
\begin{equation}
  x^\mu \to \Lambda^\mu{}_\nu x^\nu
  \implies
  X \to AXA^\dagger, \qquad
  \det A = 1.
\end{equation}
Note that an arbitrary $\SL(2,\bbC)$ matrix also has 6 independent real
components, similar to Lorentz transformations. Note also that two distinct
elements $\pm A\in\SL(2,\bbC)$ correspond to the same Lorentz transformation
$\Lambda\in\SO^+(3,1)$. Hence, the group $\SL(2,\bbC)$ is often referred to as
the double-cover of the proper orthochronous Lorentz group $\SO^+(3,1)$. This
should be contrasted with the double cover $\SU(2)$ of the rotation group
$\SO(3)$ we discussed around \cref{eq:so3-double-cover}.

The Lie algebra $\mathfrak{sl}(2,\bbC)$ of $\SL(2,\bbC)$ is the set of all
$2\times 2$ complex traceless matrices. The respective generators can be taken
to be the traceless Hermitian matrices $\tilde J_i = \hbar/2\,\sigma_i$ and the
traceless anti-Hermitian matrices $\tilde K_i = \pm i\hbar/2\,\sigma_i$ with the
same commutation relations as \cref{eq:JKAlgebra} for either of the signs in
$\tilde K_i$. Defining
\begin{equation}
  X^\pm_i = \half (\tilde J_i \pm i\tilde K_i),
\end{equation}
the commutation relations reduce to
\begin{equation}
  [X^+_i,X^+_j] = i\hbar\epsilon_{ijk}X^+_k, \qquad
  [X^-_i,X^-_j] = i\hbar\epsilon_{ijk}X^-_k, \qquad
  [X^+_i,X^-_j] = 0.
\end{equation}
Note that the generators $X^+_i$ and $X^-_i$ independently span two copies of
$\mathfrak{su}(2)$ algebras. It follows that
$\mathfrak{sl}(2,\bbC)\cong\mathfrak{su}(2)\oplus\mathfrak{su}(2)$ or
correspondingly $\SL(2,\bbC)\cong\SU(2)\times\SU(2)$. Taking the sign in
$\tilde K_i$ to be $+$ or $-$, one sees that either $X^+_i$ or $X^-_i$ is
identically zero. This is because $\tilde K_i = \pm i\tilde J_i$. However, since
this property does not need to be satisfied by an arbitrary matrix
representation $D$ of $\mathfrak{sl}(2,\bbC)$, both $D(X^\pm_i)$ will be
generically nonzero. The exponential parametrisation of $\Lambda\in\SO^+(3,1)$
in \cref{eq:expParamLambda}, in this language, leads to\footnote{Note that
  $M_{ij} = \epsilon_{ijk} J_k$.}
\begin{equation}
  \Lambda = \exp\lb \frac{i}{\hbar}\alpha^i J_i
  + \frac{i}{\hbar} \beta^{i} K_{i} \rb
  \to  \exp\lb \frac{i}{\hbar}\omega_+^i X^+_i
  + \frac{i}{\hbar} \omega_-^i X^-_i \rb
  = \exp\lb \frac{i}{\hbar}\omega_+^i X^+_i\rb
  \exp\lb \frac{i}{\hbar} \omega_-^i X^-_i \rb,
\end{equation}
where $\alpha^i = 1/2\,\epsilon^{ijk}\omega_{jk}$ are the rotation parameters
and $\beta^i = \omega^{i0}$ are the boost parameters, while
$\omega_\pm^i = \alpha^i \mp i\beta^i$ are the parameters of the two copies of
$\SU(2)$ respectively. Note that $\omega_\mp^i = (\omega_\pm^i)^*$. The final
step above is justified because $[X^+_i,X^-_j] = 0$.

\subsection{Lorentz representations}
\label{sec:LorentzReps}

A representation $D$ of $\SO^+(3,1)$ is a mapping onto the set of matrices such
that
\begin{equation}
  D(\mathbb{1}) = \mathbb{1}, \qquad
  D(\Lambda)D(\Lambda') = D(\Lambda\Lambda'),
\end{equation}
for all $\Lambda,\Lambda'\in\SO^+(3,1)$. A field $\Phi(x)$ is said to transform
in the representation $D$ of $\SO^+(3,1)$, if $\Lambda\in\SO^+(3,1)$ acts as
\begin{equation}
  \Phi(x) \to D(\Lambda) \Phi(\Lambda^{-1}x).
\end{equation}
Note that the symmetry also acts on the coordinate arguments of the field,
justifying it to be a ``local'' spacetime symmetry. Using the exponential
parametrisation of $\Lambda\in\SO^+(3,1)$ given in \cref{eq:expParamLambda}, the
representation $D(\Lambda)$ can be written in terms of the representations of
the Lie algebra generators $D(M_{\mu\nu})$, i.e.
\begin{equation}
  D(\Lambda) = \exp\lb \frac{i}{2\hbar}\omega^{\mu\nu} D(M_{\mu\nu}) \rb.
\end{equation}

A representation $D$ of $\SO^+(3,1)$ can also be mapped onto a direct product
representation $D_+\times D_-$ of $\SL(2,\bbC)\cong\SU(2)\times\SU(2)$, with
\begin{equation}
  D^\pm(\Lambda) = \exp\lb \frac{i}{\hbar} \omega_\pm^i D(X^\pm_i) \rb, \qquad
  D(X^\pm_i) = \half\lb D(J_i) \pm iD(K_i) \rb.
\end{equation}
A field $\Phi(x)$, in this representation, transforms as
\begin{equation}
  \Phi(x) \to (D^+(\Lambda)\otimes D^-(\Lambda)) \Phi(\Lambda^{-1}x).
\end{equation}
The irreducible representations of (the double cover of) $\SO^+(3,1)$ can hence
be labelled by the ``highest weights'' $(j_-,j_+)$ with
$j_\pm=0,1/2,1,3/2,\ldots$ corresponding to the two copies of SU(2); see
\cref{sec:SU2}. The states in an irreducible representation can respectively be
labelled by the eigenvalues of $(\vec X^\pm)^2$ and $X^\pm_3$ operators, i.e.
\begin{align}
  (\vec X^\pm)^2\ket{j_-,j_+;m_-,m_+}
  &= \hbar^2 j_\pm(j_\pm+1)\ket{j_-,j_+;m_-,m_+}, \nn\\
  X^\pm_3\ket{j_-,j_+;m_-,m_+}
  &= \hbar m_\pm\ket{j_-,j_+;m_-,m_+}.
\end{align}
Few interesting cases of fields transforming under such representations are
\begin{itemize}
\item $(0,0)$: scalar fields.
\item $(1/2,0)$, $(0,1/2)$: left-handed and right-handed Weyl spinor fields.
\item $(1/2,1/2)$: vector fields.
\end{itemize}
We will look at these in some detail below. It is also convenient to define the
\emph{chirality operator}
\begin{equation}
  \Gamma
  = (\vec X^+)^2 - (\vec X^+)^2 
  = i\vec K\cdot \vec J
  = \frac{i}{8} \epsilon^{\mu\nu\rho\sigma} M_{\mu\nu}M_{\rho\sigma}.
\end{equation}
By definition, $\Gamma$ commutes with all the generators of the Lorentz algebra,
and has eigenvalue given by $\hbar^2j_+(j_++1)-\hbar^2j_-(j_-+1)$. For the three
examples above, scalars and vectors have chirality $0$, while left-handed and
right-handed Weyl spinors have chirality $-3\hbar/4$ and $+3\hbar/4$
respectively. In general, the fields with positive chirality are said to be
right-handed (or right-chiral), while the fields with negative chirality are
said to be left-handed (or left-chiral).

Given a Lorentz representation $(j_-,j_+)$,
the conjugate representation is given by $(j_+,j_-)$. The states in the
representation respectively map to
$\ket{j_-,j_+;m_-,m_+} \to \ket{j_+,j_-;-m_+,-m_-}$, up to a phase factor.

\subsubsection{Scalars}

A complex field $\phi(x)$ is said to be a scalar if it transforms in the trivial
$(0,0)$ representation of the double cover of the connected Lorentz group
$\SL(2,\bbC)$, i.e.
\begin{equation}
  D_{(0,0)}(X_{i}^\pm) = 0, \qquad
  D_{(0,0)}(M_{\mu\nu}) = 0,
\end{equation}
or equivalently
\begin{equation}
  D^+_{(0,0)}(\Lambda) = D^-_{(0,0)}(\Lambda) = \mathbb 1, \qquad
  D_{(0,0)}(\Lambda) = \mathbb 1.
\end{equation}
Correspondingly, the transformation rule of a scalar field $\phi(x)$ is given as
\begin{equation}
  \phi(x) \to \phi(\Lambda^{-1}x).
\end{equation}

\subsubsection{Spinors}
\label{sec:spinors}

\paragraph*{Right-handed Weyl spinors:} A right-handed Weyl spinor
$\psi_\alpha(x)$, with $\alpha=1,2$, transforms in the $(0,1/2)$ irreducible
representation of $\SL(2,\bbC)$. By definition, this representation is
fundamental in the $X^+_i$ generators, while trivial in the $X^-_i$ generators,
i.e.
\begin{gather}
  D_{(0,1/2)}(X^+_i) = \frac\hbar2 \sigma_i, \qquad
  D_{(0,1/2)}(X^-_i) = 0 \nn\\
  \implies\qquad
  D_{(0,1/2)}(M_{\mu\nu})
  = s_{\mu\nu}
  = \frac{i\hbar}{2} \sigma_{[\mu} \bar\sigma_{\nu]}
  = \frac{i\hbar}{4} \lb \sigma_{\mu} \bar\sigma_{\nu} - \sigma_\nu \bar\sigma_\mu
  \rb
  \in \mathfrak{sl}(2,\bbC),
\end{gather}
where $\sigma_\mu = (-\mathbb{1},\sigma_i)$ and
$\bar\sigma_\mu = (-\mathbb{1},-\sigma_i)$. Raising the Lorentz indices, we get
$\sigma^\mu = (\mathbb{1},\sigma_i)$ and
$\bar\sigma^\mu = (\mathbb{1},-\sigma_i)$. This corresponds to
$D_{(0,1/2)}(J_i) = \hbar/2\,\sigma_i$ and
$D_{(0,1/2)}(K_i) = -i\hbar/2\,\sigma_i$. Note that each component of
$\sigma_\mu$ and $\bar\sigma_\mu$ is a $2\times2$ complex matrix. It is
customary to denote the components in terms of ``undotted'' and ``dotted''
indices
\begin{equation}
  (\sigma_\mu)_{\alpha\dot\alpha}, \qquad
  (\bar\sigma_\mu)^{\dot\alpha\alpha},
\end{equation}
where both the kinds of indices run over
$\alpha,\beta,\ldots,\dot\alpha,\dot\beta,\ldots = 1,2$. In terms of these
components, we can denote
\begin{equation}
  (s_{\mu\nu})_\alpha{}^\beta
  = \frac{i\hbar}{4} \lb
  (\sigma_{\mu})_{\alpha\dot\alpha} (\bar\sigma_{\nu})^{\dot\alpha\beta}
  - (\sigma_{\nu})_{\alpha\dot\alpha} (\bar\sigma_{\mu})^{\dot\alpha\beta} \rb.
\end{equation}
We can check that $s_{\mu\nu}$ satisfies the Lorentz algebra
\eqref{eq:LorentzAlgebra}. Correspondingly, in the exponential parametrisation,
we have
\begin{equation}
  D_{(0,1/2)}(\Lambda)
  = A(\Lambda)
  = \exp\lb \frac{i}{2\hbar} \omega^{\mu\nu} s_{\mu\nu}\rb
  \in \SL(2,\bbC),
\end{equation}
and the right-handed Weyl spinor $\psi_\alpha(x)$ transforms as
\begin{equation}
  \psi_\alpha(x) \to A(\Lambda)_\alpha{}^\beta \psi_\beta(\Lambda^{-1}x).
\end{equation}
We can also define the ``anti-fundamental'' right-handed Weyl spinor by raising
the index with the Levi-Civita tensor $\varepsilon^{\alpha\beta}$, with
$\varepsilon^{12}=1$, i.e.
\begin{equation}
  \psi^\alpha(x) = \varepsilon^{\alpha\beta} \psi_\beta(x), \qquad
  \psi^\alpha(x) \to
  \varepsilon^{\alpha\beta} A(\Lambda)_\beta{}^\gamma \varepsilon_{\gamma\delta}
  \psi^\delta(\Lambda^{-1}x)
  = \psi^\beta(\Lambda^{-1}x) (A(\Lambda)^{-1})_\beta{}^\alpha.
\end{equation}
The lowered $\varepsilon_{\alpha\beta}$ is defined such that
$\varepsilon_{12}=-1$.  Note that for a $2\times2$ unit determinant matrix,
$(A^{-1})_\beta{}^\alpha = \varepsilon^{\alpha\gamma}A_\gamma{}^\delta
\varepsilon_{\delta\beta}$.

\paragraph*{Left-handed Weyl spinors:} The discussion for a left-handed Weyl
spinor $\bar\psi^{\dot\alpha}(x)$ follows in an analogous manner, transforming
in the $(1/2,0)$ irreducible representation of $\SL(2,\bbC)$, given by
\begin{gather}
  D_{(1/2,0)}(X^+_i) = 0, \qquad
  D_{(1/2,0)}(X^-_i) = \frac\hbar2 \sigma_i \nn\\
  \implies\qquad
  D_{(1/2,0)}(M_{\mu\nu})
  = \bar s_{\mu\nu}
  = \frac{i\hbar}{2} \bar\sigma_{[\mu} \sigma_{\nu]}
  = \frac{i\hbar}{4} \lb \bar\sigma_{\mu} \sigma_{\nu} - \bar\sigma_\nu \sigma_\mu
  \rb
  \in \mathfrak{sl}(2,\bbC).
\end{gather}
This still corresponds to $D_{(1/2,0)}(J_i) = \hbar/2\,\sigma_i$, but
$D_{(1/2,0)}(K_i) = i\hbar/2\,\sigma_i$. Therefore, the left-handed and
right-handed Weyl spinors transform in the same way under rotations, but
differently under boosts. In components
\begin{equation}
  (\bar s_{\mu\nu})^{\dot\alpha}{}_{\dot\beta}
  = \frac{i\hbar}{4} \lb (\bar\sigma_{\mu})^{\dot\alpha\alpha} (\sigma_{\nu})_{\alpha\dot\beta}
  - (\bar\sigma_{\nu})^{\dot\alpha\alpha} (\sigma_{\mu})_{\alpha\dot\beta} \rb.
\end{equation}
It can again be checked that $\bar s_{\mu\nu}$ satisfies the Lorentz algebra
commutation relations \eqref{eq:LorentzAlgebra}. Denoting
\begin{equation}
  D_{(1/2,0)}(\Lambda) = \bar A(\Lambda)
  = \exp\lb \frac{i}{2\hbar} \omega^{\mu\nu} \bar s_{\mu\nu}\rb
  \in \SL(2,\bbC),
\end{equation}
the left-handed Weyl spinor $\psi^{\dot\alpha}(x)$ transforms as
\begin{equation}
  \bar\psi^{\dot\alpha}(x) \to \bar A(\Lambda)^{\dot\alpha}{}_{\dot\beta}
  \psi^{\dot\beta}(\Lambda^{-1}x).
\end{equation}
In a similar manner, one can define the ``anti-fundamental'' Weyl spinor as 
\begin{equation}
  \bar\psi_{\dot\alpha}(x) = \varepsilon_{\dot\alpha\dot\beta}
  \psi^{\dot\beta}(x), 
  \qquad
  \bar\psi_{\dot\alpha}(x) \to
  \varepsilon_{\dot\alpha\dot\beta}
  \bar A(\Lambda)^{\dot\beta}{}_{\dot\gamma} \varepsilon^{\dot\gamma\dot\delta}
  \bar\psi_{\dot\delta}(\Lambda^{-1}x)
  = \bar\psi_{\dot\beta}(\Lambda^{-1}x)
  (\bar A(\Lambda)^{-1})^{\dot\beta}{}_{\dot\alpha}.
\end{equation}
Here $\varepsilon_{\dot\alpha\dot\beta}$, $\varepsilon^{\dot\alpha\dot\beta}$
are again defined so that $\varepsilon^{12}=1$, $\varepsilon_{12}=-1$.

Note that since $\sigma_\mu^\dagger = \sigma_\mu$ and
$\bar\sigma_\mu^\dagger = \bar\sigma_\mu$, it follows that
$\bar s_{\mu\nu} = s_{\mu\nu}^\dagger$, and as a consequence,
$\bar A(\Lambda)^{-1} = A(\Lambda)^{\dagger}$. Therefore, if $\psi_\alpha(x)$ is
a right-handed Weyl spinor, its complex conjugate $(\psi_\alpha)^*$ is a
left-handed (anti-fundamental) Weyl spinor, and vice-versa. This is a
characteristic feature of $\SO^+(3,1)$ (or $\SL(2,\bbC)$ or
$\SU(2)\times\SU(2)$) representations, as opposed to $\SU(2)$ representations
discussed in \cref{sec:SU2}, where complex conjugation yields equivalent
representations.

\paragraph*{Lorentz invariants:} We would like to observe that the Kronecker
delta symbols $\delta^\alpha_\beta$, $\delta^{\dot\alpha}_{\dot\beta}$, as well
as the Levi-Civita symbols $\varepsilon_{\alpha\beta}$,
$\varepsilon^{\alpha\beta}$, $\varepsilon_{\dot\alpha\dot\beta}$,
$\varepsilon^{\dot\alpha\dot\beta}$ are all Lorentz invariants. Furthermore,
$(\sigma_\mu)_{\alpha\dot\alpha}$ and $(\bar\sigma_\mu)^{\dot\alpha\alpha}$ are
also Lorentz invariants, i.e.
\begin{equation}
  \sigma_\mu \to (\Lambda^{-1})^\nu{}_\mu
  A(\Lambda)\sigma_\nu A(\Lambda)^\dagger = \sigma_\mu,
  \qquad
  \bar\sigma_\mu \to (\Lambda^{-1})^\nu{}_\mu
  \bar A(\Lambda)\bar\sigma_\nu \bar A(\Lambda)^\dagger = \bar\sigma_\mu,
  \label{eq:sigmaTransformation}
\end{equation}
or in components
\begin{gather}
  (\Lambda^{-1})^\nu{}_\mu\,
  A(\Lambda)_\alpha{}^\beta\, (\sigma_\nu)_{\beta\dot\beta}\,
  (\bar A(\Lambda)^{-1})^{\dot\beta}{}_{\dot\alpha}
  = (\sigma_\mu)_{\alpha\dot\alpha}, \nn\\
  (\Lambda^{-1})^\nu{}_\mu\,
  \bar A(\Lambda)^{\dot\alpha}{}_{\dot\beta}\, (\bar\sigma_\nu)^{\dot\beta\beta}\,
  (A(\Lambda)^{-1})_{\beta}{}^{\alpha}
  = (\bar\sigma_\mu)^{\dot\alpha\alpha}.
\end{gather}
This can be proved using the identities
\begin{equation}
  s_{\mu\nu}\sigma_\lambda - \sigma_\lambda \bar s_{\mu\nu}
  = i\hbar(\eta_{\mu\lambda}\sigma_\nu - \eta_{\nu\lambda}\sigma_\mu), \qquad
  \bar s_{\mu\nu}\bar\sigma_\lambda - \bar\sigma_\lambda s_{\mu\nu}
  = i\hbar(\eta_{\mu\lambda}\bar\sigma_\nu - \eta_{\nu\lambda}\bar\sigma_\mu).
\end{equation}
Note that
$\omega^{\mu\nu}s_{\mu\nu}\sigma_\lambda = \sigma_\lambda \omega^{\mu\nu}\bar
s_{\mu\nu} - 2i\hbar\sigma_\nu \omega^{\nu}{}_\lambda$ and
$\omega^{\mu\nu}\bar s_{\mu\nu}\bar\sigma_\lambda = \bar\sigma_\lambda
\omega^{\mu\nu} s_{\mu\nu} - 2i\hbar\bar\sigma_\nu \omega^{\nu}{}_\lambda$. This
leads to, for instance
\begin{align}
  A(\Lambda)\sigma_\lambda
  &= \sum_{n=0}^\infty \frac{1}{n!} \lb\frac{i}{2\hbar}\omega^{\mu\nu}s_{\mu\nu}\rb^n
    \sigma_\lambda \nn\\
  &= \sum_{n=0}^\infty \frac{1}{n!}
    \sum_{r=0}^n \frac{n!}{r!(n-r)!} (\sigma\omega^{n-r})_\lambda
    \lb\frac{i}{2\hbar}\omega^{\mu\nu}\bar s_{\mu\nu}\rb^r \nn\\
  &= \sum_{r=0}^\infty\frac{1}{r!}\lb \sum_{n=r}^\infty \frac{1}{(n-r)!}
    (\sigma\omega^{n-r})_\lambda
    \rb \lb\frac{i}{2\hbar}\omega^{\mu\nu}\bar s_{\mu\nu}\rb^r \nn\\
  &= \sigma_\mu \exp(i\omega^\mu{}_\lambda)
    \exp\lb\frac{i}{2\hbar}\omega^{\mu\nu}\bar s_{\mu\nu}\rb \nn\\
  &= \sigma_\mu \Lambda^\mu{}_\lambda \bar A(\Lambda),
\end{align}
where
$(\sigma\omega^{m})_\lambda =
\sigma_{\mu_1}\omega^{\mu_1}{}_{\mu_2}\omega^{\mu_2}{}_{\mu_3}\ldots
\omega^{\mu_m}{}_{\lambda}$. In the third step, we have exchanged the
summations. The other identity follows in a similar manner.

\paragraph*{Dirac spinors:} A Dirac spinor $\Psi^\sA(x)$, with $\sA=1,2,3,4$, is
a 4-component spinor field that transforms in the $(1/2,0)\oplus(0,1/2)$
\emph{reducible} representation of $\SL(2,\bbC)$, given by
\begin{equation}
  D_{\text{dirac}}(\Lambda) =
  \begin{pmatrix}
    D_{(0,1/2)}(\Lambda) & 0 \\ 0 & D_{(1/2,0)}(\Lambda)
  \end{pmatrix} =
  \begin{pmatrix}
    A(\Lambda) & 0 \\ 0 & \bar A(\Lambda)
  \end{pmatrix},
\end{equation}
or equivalently
\begin{equation}
  D_{\text{dirac}}(M_{\mu\nu}) =
  \begin{pmatrix}
    s_{\mu\nu} & 0 \\ 0 & \bar s_{\mu\nu}
  \end{pmatrix}.
\end{equation}
A Dirac spinor $\Psi^\sA(x)$ can itself be understood as a direct sum of a
left-handed and a right-handed Weyl spinor
\begin{equation}
  \Psi^\sA(x) =
  \begin{pmatrix}
    \psi_\alpha(x) \\ \bar\psi^{\dot\alpha}(x)
  \end{pmatrix},
\end{equation}
so that
\begin{equation}
  \Psi^\sA(x) \to D_{\text{dirac}}(\Lambda)^\sA{}_{\sB} \Psi^\sB(\Lambda^{-1}x).
\end{equation}

Dirac representation can also be written in terms of the gamma-matrices (in Weyl
basis)
\begin{equation}
  \gamma_\mu =
  \begin{pmatrix}
    0 & \sigma_\mu \\
    \bar\sigma_\mu & 0
  \end{pmatrix}, \qquad
  \{\gamma_\mu,\gamma_\nu\} = - 2\eta_{\mu\nu}, \qquad
  D_{\text{Dirac}}(M_{\mu\nu}) = \frac{i\hbar}{4}[\gamma_\mu,\gamma_\nu].
\end{equation}
Let us define the chiral projection operators
\begin{equation}
  P_\pm = \half \lb 1 \pm \gamma_5 \rb, \qquad
  \gamma_5
  = \frac43 D_{\text{Dirac}}(\Gamma)
  = i\gamma_0\gamma_1\gamma_2\gamma_3
  =
  \begin{pmatrix}
    \mathbb{1} & 0 \\ 0 & -\mathbb{1}
  \end{pmatrix}.
\end{equation}
Recall that
$\Gamma = i/(8\hbar^2) \epsilon^{\mu\nu\rho\sigma}M_{\mu\nu}M_{\rho\sigma}$.
Since $\sigma_\mu$ and $\bar\sigma_\mu$ are left invariant by Lorentz
transformations, so are the gamma-matrices $\gamma_\mu$, and by extension
$\gamma_5$ and the projectors $P_\pm$. This allows us to project out irreducible
Weyl components of the Dirac spinor as
\begin{equation}
  P_+ \Psi(x) =
  \begin{pmatrix}
    \psi_\alpha(x) \\ 0
  \end{pmatrix}, \qquad
  P_- \Psi(x) =
  \begin{pmatrix}
    0 \\ \bar\psi^{\dot\alpha}(x)
  \end{pmatrix}.
\end{equation}

\subsubsection{Vectors (1/2,1/2)}

A vector field $V^\mu(x)$ transforms in the fundamental representation of
$\SO^+(3,1)$, given by
\begin{equation}
  V^\mu(x) \to \Lambda^\mu{}_\nu V^\nu(\Lambda^{-1}x).
\end{equation}
In terms of the $\SL(2,\bbC)$ classification, vectors can be understood as
transforming under the $(1/2,1/2)$ irreducible representation; to wit
\begin{equation}
  \sigma_{\mu}V^\mu(x)
  \to \sigma_\mu \Lambda^\mu{}_{\nu} V^\nu(\Lambda^{-1}x)
  = A(\Lambda) (\sigma_\mu V^\mu(\Lambda^{-1}x)) \bar A(\Lambda)^{-1},
\end{equation}
or equivalently
\begin{equation}
  \bar\sigma_{\mu}V^\mu(x)
  \to \sigma_\mu \Lambda^\mu{}_{\nu} V^\nu(\Lambda^{-1}x)
  = \bar A(\Lambda) (\sigma_\mu V^\mu(\Lambda^{-1}x)) A(\Lambda)^{-1},
\end{equation}
Here we have used \cref{eq:sigmaTransformation}.

\subsection{Poincar\'e group}

Poincar\'e transformations are an extension of Lorentz transformations that
leave the proper distance between arbitrary spacetime points $x^\mu$, $x'^\mu$
invariant, i.e.
\begin{equation}
  (x^\mu - x'^\mu)(x^\nu - x'^\nu)\eta_{\mu\nu}
  \to (x^\mu - x'^\mu)(x^\nu - x'^\nu)\eta_{\mu\nu}.
\end{equation}
These are given by
\begin{equation}
  x^\mu \to \Lambda^\mu{}_\nu x^\nu + a^\mu,
\end{equation}
where $(\Lambda^\mu{}_\nu)\in\rmO(3,1)$ are Lorentz transformations, and $a^\mu$
are constant spacetime translations. Let us denote an element of the Poincar\'e
group as $(\Lambda,a)$. We can check that these satisfy the group axioms:
\begin{itemize}
\item $(\Lambda_1,a_1)(\Lambda_2,a_2)
  = (\Lambda_1\Lambda_2,\Lambda_1 a_2+a_1)$.
\item $ \Big( (\Lambda_1,a_1)(\Lambda_2,a_2) \Big) (\Lambda_3,a_3)
  = (\Lambda_1\Lambda_2\Lambda_3,\Lambda_1\Lambda_2a_3 + \Lambda_1 a_2+a_1)
  = (\Lambda_1,a_1)\Big( (\Lambda_2,a_2)(\Lambda_3,a_3)\Big)$.
\item $(\mathbb 1, 0)(\Lambda,a) = (\Lambda,a)(\mathbb 1,0) = (\Lambda,a)$.
\item $(\Lambda,a)^{-1} = (\Lambda^{-1},-\Lambda^{-1}a)$.
\end{itemize}
The Poincar\'e group is mathematically
denoted as $\bbR^{3,1}\rtimes\rmO(3,1)$.\footnote{The ``semi-direct product''
  symbol $\rtimes$ essentially means that $\bbR^{3,1}$ is a normal subgroup of
  the Poincar\'e group, but $\rmO(3,1)$ is not.} The simply connected piece of
the Poincar\'e group is accordingly $\bbR^{3,1}\rtimes\SO^+(3,1)$, and its
double cover is $\bbR^{3,1}\rtimes\SL(2,\bbC)$.

Let us consider a representation $D$ of the simply connected Poincar\'e group
$\bbR^{3,1}\rtimes\SO^+(3,1)$. The exponential parametrisation of an element
$D(\Lambda,a)$ can be written as
\begin{equation}
  D(\Lambda,a) = \exp\lb\frac{i}{2\hbar}\omega^{\mu\nu}M_{\mu\nu}
  + \frac{i}{\hbar}a^\mu P_\mu\rb,
\end{equation}
where $M_{\mu\nu}$ and $P_\mu$ are the Lie algebra generators of Lorentz
transformations and spacetime translations respectively, while $\omega^{\mu\nu}$
and $a^\mu$ are the respective parameters. These satisfy the commutation
relations
\begin{gather}
  [P_\mu,P_\nu] = 0, \nn\\
  [M_{\mu\nu},P_\rho] =
  i\hbar\lb \eta_{\mu\rho}P_\nu - \eta_{\nu\rho} P_\mu \rb, \nn\\
  [M_{\mu\nu},M_{\rho\sigma}]
  = i\hbar\lb \eta_{\mu\rho}M_{\nu\sigma}
  - \eta_{\nu\rho}M_{\mu\sigma}
  - \eta_{\mu\sigma}M_{\nu\rho}
  + \eta_{\nu\sigma}M_{\mu\rho} \rb.
  \label{eq:PoinAlgebra}
\end{gather}
In terms of the rotation generators (angular momenta)
$J_i = 1/2\,\epsilon_{ijk}M_{jk}$, boost generators $K_i = M_{i0}$, translation
generators (momenta) $P_i$, and time-translation generator (Hamiltonian)
$H = -cP_0$, where $c$ is the speed of light, the Lie algebra is given as
\begin{gather}
  [H,P_i] = 0, \qquad
  [P_i,P_j] = 0, \nn\\
  [J_i,H] = 0, \qquad
  [J_i,P_j] = i\hbar \epsilon_{ijk} P_k, \qquad
  [K_i,H] = -i\hbar c P_i, \qquad
  [K_i,P_j] = -\frac{i\hbar}{c} \delta_{ij}H, \nn\\
  [J_i,J_j] = i\hbar \epsilon_{ijk}J_k, \qquad
  [K_i,K_j] = -i\hbar \epsilon_{ijk}J_k, \qquad
  [J_i,K_j] = i\hbar\epsilon_{ijk}K_k.
\end{gather}
At least some of these might be recognisable from quantum mechanics. Note that
the subalgebra spanned by $P_i,H$ is an invariant subalgebra of the Poincar\'e
algebra.

\subsection{Poincar\'e representations}
\label{sec:poincare-reps}

The irreducible representations of the Poincar\'e group can be labelled using
the Casimir operators $P^2 = \eta_{\mu\nu}P^\mu P^\nu$ and
$W^2 = \eta_{\mu\nu}W^\mu W^\nu$, where $W^\mu$ is the Pauli-Lubanski vector
\begin{equation}
  W^\mu = \half \epsilon^{\mu\nu\rho\sigma} M_{\nu\rho} P_\sigma.
\end{equation}
It satisfies
\begin{equation}
  [P_\mu,W^\nu]=0, \qquad
  [M_{\mu\nu},W^\rho] = i\hbar \lb \delta^\rho_\mu W_\nu - \delta^\rho_\nu W_\mu \rb,
  \qquad
  [W^\mu,W^\nu] = -i\hbar\epsilon^{\mu\nu\rho\sigma}P_\rho W_\sigma,
\end{equation}
along with $W^\mu P_\mu = 0$. It can be checked that
\begin{equation}
  [P^2,M_{\mu\nu}] = [P^2,P_\mu]
  = [W^2,M_{\mu\nu}] = [W^2,P_\mu] = 0.
\end{equation}
We can take the states in the Hilbert space to be the eigenstates of the
mutually commuting set of operators: four-momenta $P_\mu$ and helicity operator
$W_0 = J_i P^i$, in addition to $P^2$ and $W^2$. We can label a state
$\ket{m,s;\vec p,h}$ in a given representation by its mass $m$, spin $s$,
momenta $p_i$, and helicity $h$, defined as
\begin{align}
  W^2 \ket{m,s;\vec p,h} &= m^2c^2\hbar^2s(s+1)\ket{m,s;\vec p,h}, \nn\\
  P_\mu \ket{m,s;\vec p,h} &= p_\mu \ket{m,s;\vec p,h}, \nn\\
  W_0 \ket{m,s;\vec p,h} &= \hbar|\vec p| h \ket{m,s;\vec p,h},
\end{align}
where $p_\mu = (-\sqrt{m^2c^2+\vec p^2},\vec p)$ is the four-momentum of the
state. Note that these states are automatically eigenstates of the $P^2$
operator as
\begin{equation}
  P^2 \ket{m,s;\vec p,h}
  = \eta^{\mu\nu} P_\mu P_\nu \ket{m,s;\vec p,h}
  = - m^2c^2\ket{m,s;\vec p,h}.
\end{equation}

To appreciate the physical meaning of the spin quantum number $s$ and helicity
$h$, let us first focus on massive representations with $m> 0$. We can always
Lorentz boost arbitrary states in the representation to go to the local rest
frame and set $p_\mu = (-mc,\vec 0)$ or $\vec p = 0$. For such states
\begin{equation}
  W^0 |m,s;\vec 0,h\rangle = 0, \qquad
  W^i |m,s;\vec 0,h\rangle = -\frac1c J^i H |m,s;\vec 0,h\rangle
  = -mcJ^i|m,s;\vec 0,h\rangle.
\end{equation}
It implies that
\begin{equation}
  W^2|m,s;\vec 0,h\rangle
  = m^2c^2 \vec J^2|m,s;\vec 0,h\rangle
  = m^2c^2 \hbar^2 s(s+1)|m,s;\vec 0,h\rangle.
\end{equation}
Hence, the spin $s$ is the highest-weight quantum number associated with the
total angular-momentum $\vec J^2$ in the local rest frame of the particle. To
interpret $h$, on the other hand, we need to spatially rotate an arbitrary state
to set $\vec p = p\hat z = (0,0,p)$. We note that
\begin{equation}
  W_0 |m,s;p\hat z,h\rangle
  = J_i P^i |m,s;p\hat z,h\rangle
  = p J_3 |m,s;p\hat z,h\rangle
  = \hbar p h |m,s;p\hat z,h\rangle.
\end{equation}
Hence $h$ is the $J_3$ eigenvalue of a state where the momenta is aligned along
the $z$-direction. It also follows that $s$ and $h$ take half integer values
\begin{equation}
  s=0,\half,1,\frac32,\ldots, \qquad
  h=-s,-s+1,\ldots, s-1, s.
\end{equation}
For instance, electron (massive Dirac spinor) has $s=1/2$ and $h=\pm 1/2$, while
Z-boson (massive vector) has $s=1$ and $h=-1,0,+1$. Massive irreducible Poincare
representations (particles) can be labelled by their mass and spin $(m,s)$.

For massless representations $m=0$, on the other hand, we can imagine going to a
Lorentz frame moving arbitrarily close to the speed of light, such that
$p_\mu \to 0$. From the above analysis, this implies that both $P^2$ and $W^2$
eigenvalues are 0. Since both these operators are Poincar\'e invariant, the same
must be true in any Lorentz frame. Hence, the spin quantum number $s$ loses all
meaning and we can label the states as simply $\ket{h;\vec p}$. Now, if we were
to perform a rotation to set $p_\mu = (-E/c,0,0,E/c)$, where $E$ is the energy
of the state, we see that
\begin{equation}
  W_0 |h;E/c\hat z\rangle
  = \frac{E}{c} J_3|h;E/c\hat z\rangle
  = \frac{\hbar E}{c} h|h;E/c\hat z\rangle.
\end{equation}
Therefore $h=0,\pm 1/2,\pm1,\pm3/2,\ldots$ is still the $J_3$ eigenvalue in a
state where the momenta is aligned along the $z$-direction. An important
distinction between massless and massive particles is that helicity for a
massless particle is Poincar\'e invariant.\footnote{To see this, note that
  $W_0\ket{\Psi} = \hbar Eh/c\ket\psi$ and
  $W_3\ket\psi = -\hbar E h/c\ket{\Psi}$, where
  $\ket{\Psi} = |h;E/c\hat z\rangle$. Since $W^2\ket\psi = 0$, it follows that
  $W_2\ket\psi = W_3\ket\psi = 0$. Furthermore, assuming the Hamiltonian to not
  admit any zero-energy states, let us define the inverse Hamiltonian operator
  $H^{-1}$ so that $H^{-1}\ket\psi = E^{-1}\ket\psi$ and
  $H^{-1}W_0\ket\psi = \hbar h/c\ket\psi$. It can be checked that the
  commutators $[H,H^{-1}W_0]=[P_i,H^{-1}W_0]=[J_i,H^{-1}W_0]=0$ are identically
  zero, while
  \begin{align}
    [K_i,H^{-1}W_0]\ket\psi
    &= [K_i,H^{-1}]W_0\ket\psi
    + H^{-1}[K_i,W_0]\ket\psi
    = i\hbar c H^{-1}P_i H^{-1}W_0\ket\psi
      + i\hbar H^{-1}W_i\ket\psi \nn\\
    &= i\hbar H^{-1}\lb P_i \hbar h + W_i \rb \ket\psi
      = 0.
  \end{align}
  Hence, $H^{-1}W_0$ is Poincar\'e invariant for massless representations.
}  Hence, massless irreducible representations (particles) can be labelled by
just their helicity $(h)$. For instance, photons come in either left-polarised
$h=-1$ or right-polarised $h=+1$ varieties. Similarly gravitons come in $h=-2$
and $h=+2$ varieties.

The concept of helicity should be contrasted with the concept of chirality
mentioned before. The helicity operator $W_0 = \vec J\cdot\vec P$ is conserved,
i.e. it commutes with the Hamiltonian $[H,W_0] = 0$, but is not generically
Lorentz invariant because $[K_i,W_0]=i\hbar W_i$. On the other hand, the
chirality operator $\Gamma = i\vec J\cdot\vec K$ is Lorentz invariant, but is
not generically conserved, i.e. $[H,\Gamma] = \hbar c W_0$. The two concepts are
equivalent for massless particles.

\newpage

\section{Symmetries in field theory}
\label{field-theory-symm}

In previous sections, we have provided an abstract discussion of the symmetry
groups that enter the physical theories of particle physics: internal U(1) and
$\SU(N)$ symmetries and spacetime Poincar\'e symmetries. We now proceed to see
how these symmetries are implemented in the framework of quantum field
theories. In the following, we study various field theoretic topics and
techniques such as global and local (gauge) symmetries, their spontaneous
breaking, and Yukawa couplings. In the process, we also write down field
theoretic models for quantum electrodynamics (QED), quantum chromodynamics
(QCD), and the electroweak theory. The discussion here will be quite brief. An
elaborate discussion can be found in any of the standard texts on quantum field
theory, like~\cite{Peskin:1995ev, Weinberg:1995mt, Weinberg:1996kr}, or those
geared towards particle physics, like~\cite{Aitchison:1989bs, Aitchison:2004cs,
  Cheng:1984vwu}.

\subsection{Global symmetries}

Let us consider a field theory characterised by an action $S[\Phi]$, written as
a functional of the spacetime field(s) $\Phi(x)$. Assuming the field theory to
be local, we can write the action in terms of a local Lagrangian density
$\mathcal{L}(\Phi(x),\dow\Phi(x),\ldots)$, written as a function of $\Phi(x)$
and its derivatives $\dow_\mu\Phi(x)$, $\dow_\mu\dow_\nu\Phi(x)$ and so on, i.e.
\begin{equation}
  S[\Phi] = \int \df t\df^3 x\,\mathcal{L}(\Phi(x),\dow\Phi(x),\ldots).
\end{equation}
The equations of motion for $\Phi(x)$ can be obtained by varying $S[\Phi]$ with
respect to $\Phi$, leading to
\begin{equation}
  \frac{\delta S[\Phi]}{\delta \Phi(x)}
  = \frac{\dow\mathcal{L}(\Phi(x),\dow\Phi(x),\ldots)}{\dow \Phi(x)}
  - \dow_\mu \lb \frac{\dow\mathcal{L}(\Phi(x),\dow\Phi(x),\ldots)}{\dow
    (\dow_\mu\Phi(x))} \rb + \ldots
  = 0.
\end{equation}
The middle expression above is known as the Euler-Lagrange derivative of the
Lagrangian density $\mathcal{L}$, denoted by $\delta\mathcal{L}/\delta\Phi$. The
generalisation to multiple fields is straight-forward.

A field theory is said to be admit a symmetry group $G$ if the action $S[\Phi]$
is invariant under the action of $g\in G$ on the fields
$\Phi(x) \overset{g}{\to} \Phi'(x)$, i.e. $S[\Phi(x)]=S[\Phi'(x)]$. A continuous
(Lie group) symmetry $G$ is said to be \emph{global}, if the transformation
group parameters $g\in G$ are not dependent on spacetime points. Note that
spacetime Poincar\'e symmetries are global, because both the Lorentz matrix
$\Lambda^\mu{}_\nu$ and translation parameters $a^\mu$ are constants.

\subsubsection{Noether's theorem}
\label{sec:Noether-thm}

Given a field theory invariant under spacetime Poincar\'e symmetries and global
internal symmetries $G_{\text{int}}$, the fields $\Phi(x)$ transform as
\begin{equation}
  \Phi(x) \to \Phi'(x) = (D(g)\otimes D(\Lambda)) \Phi(\Lambda^{-1}x-a),
\end{equation}
for $g\in G_{\text{int}}$, $\Lambda^\mu{}_\nu\in\SO^+(3,1)$, and
$a^\mu\in\bbR^{3,1}$. Due to this transformation being a symmetry of the action,
the Lagrangian density must return to itself up to a total derivative term
\begin{equation}
  \mathcal{L}(\Phi',\dow\Phi',\ldots)
  = \mathcal{L}(\Phi,\dow\Phi,\ldots)
  - \dow_\mu K^\mu,
\end{equation}
for some $K^\mu$. Let us consider an infinitesimal such symmetry transformation,
given by
\begin{equation}
  D(g) = \mathbb 1 + \frac{i}{\hbar}\theta^a D(T_a) + \ldots, \qquad
  D(\Lambda) = \mathbb 1 + \frac{i}{2\hbar} \omega^{\mu\nu} D(M_{\mu\nu})
  + \ldots.
\end{equation}
We note that
\begin{align}
  \Phi'(x)
  &= \Phi(x) + \frac{i}{\hbar}\theta^a D(T_a) \Phi(x)
  + \frac{i}{2\hbar}\omega^{\mu\nu} D(M_{\mu\nu})\Phi(x)
  - \lb \omega^\mu{}_\nu x^\nu + a^\mu \rb
  \dow_\mu \Phi(x)
  + \ldots \nn\\
  &= \Phi(x) + \delta\Phi(x) + \ldots.
\end{align}
It follows that
\begin{align}
  0
  &= \mathcal{L}(\Phi',\dow\Phi',\ldots)
    - \mathcal{L}(\Phi,\dow\Phi,\ldots)
    + \dow_\mu K^\mu \nn\\
  &= \frac{\dow\mathcal{L}}{\dow\Phi} \delta\Phi
    + \frac{\dow\mathcal{L}}{\dow(\dow_\mu\Phi)} \dow_\mu\delta\Phi
    + \frac{\dow\mathcal{L}}{\dow(\dow_\mu\dow_\nu\Phi)}
    \dow_\mu\dow_\nu\delta\Phi
    + \ldots
    + \dow_\mu K^\mu \nn\\
  &= \dow_\mu\lb
    \frac{\dow\mathcal{L}}{\dow(\dow_\mu\Phi)} \delta\Phi
    + \frac{\dow\mathcal{L}}{\dow(\dow_\mu\dow_\nu\Phi)}
    \dow_\nu\delta\Phi
    - \dow_\nu \bfrac{\dow\mathcal{L}}{\dow(\dow_\nu\dow_\mu\Phi)}
    \delta\Phi
    + \ldots
    + K^\mu \rb \nn\\
  &= \dow_\mu\lb
    \frac{\delta\mathcal{L}}{\delta(\dow_\mu\Phi)} \delta\Phi
    + \frac{\delta\mathcal{L}}{\delta(\dow_\mu\dow_\nu\Phi)}
    \dow_\nu\delta\Phi
    + \ldots
    + K^\mu \rb.
    \label{eq:fullJ0} 
\end{align}
In the third line, we have used the Euler-Lagrange equations of motion for
$\Phi$, while in the forth line, we have used the Euler Lagrange derivatives of
the Lagrangian with respect to the derivatives of $\Phi$. Assuming the
Lagrangian density $\mathcal{L}(x) = \mathcal{L}(\Phi(x),\dow\Phi(x),\ldots)$ to
be a scalar, and defining
$\mathcal{L}'(x) = \mathcal{L}(\Phi'(x),\dow\Phi'(x),\ldots)$, it must be true
that
\begin{gather}
  \mathcal{L}'(x) = \mathcal{L}(\Lambda^{-1}x-a) = \mathcal{L}(x) -
  (\omega^{\mu}{}_\nu x^\nu + a^\mu) \dow_\mu \mathcal{L}(x) + \ldots =
  \mathcal{L}(x)
  - \dow_\mu \Big((\omega^{\mu}{}_\nu x^\nu + a^\mu) \mathcal{L}(x) + \ldots \Big), \nn\\
  \implies K^\mu = (\omega^{\mu}{}_\nu x^\nu + a^\mu) \mathcal{L}(x) + \ldots.
\end{gather}
With this in place, we note that \cref{eq:fullJ0} must be satisfied for all
values of the infinitesimal transformation parameters $\theta^a$,
$\omega^{\mu\nu}$, and $a^\mu$. This results in a set of conserved currents
\begin{align}
  J^\mu_a
  &= -\frac{\delta\mathcal{L}}{\delta(\dow_\mu\Phi)} \frac{i}{\hbar}D(T_a)\Phi
    + \ldots, \nn\\
  L^{\mu}{}_{\rho\sigma}
  &= - \frac{\delta\mathcal{L}}{\delta(\dow_\mu\Phi)}
    \lb \frac{i}{\hbar} D(M_{\rho\sigma}) \Phi
    - 2x_{[\sigma}\dow_{\rho]}\Phi
 \rb
    + \ldots
    - 2\delta^\mu_{[\rho} x_{\sigma]}\mathcal{L} \nn\\
  &= - \frac{\delta\mathcal{L}}{\delta(\dow_\mu\Phi)}
    \frac{i}{\hbar} D(M_{\rho\sigma}) \Phi
    + \ldots
    - 2T_{\text{can}}{}^{\mu}{}_{[\rho} x_{\sigma]}, \nn\\
  T_{\text{can}}{}^{\mu}{}_\nu
  &= - \frac{\delta\mathcal{L}}{\delta(\dow_\mu\Phi)} \dow_\nu \Phi
    + \ldots
    + \delta^\mu_\nu\mathcal{L}.
    \label{eq:NoetherCurrents}
\end{align}
For a canonical field theory, where the Lagrangian only depends on
$\dow_\mu\Phi$ and not on the higher derivatives, the ellipsis in the
expressions above go away. Generalisation to multiple fields is
straight-forward. These are the internal conserved currents $J^\mu_a$, angular
momentum current $L^{\mu}{}_{\rho\sigma}$, and canonical energy-momentum tensor
$T_{\text{can}}{}^{\mu}{}_\nu$ respectively, satisfying
\begin{equation}
  \dow_\mu J^\mu_a = \dow_\mu L^{\mu}{}_{\rho\sigma}
  = \dow_\mu  T_{\text{can}}{}^{\mu}{}_\nu = 0.
\end{equation}
This is the statement of the Noether's theorem: \emph{corresponding to every
  continuous global symmetry, there exists a locally conserved current}.

\paragraph*{Spin current and Belinfante energy-momentum tensor:} The
angular-momentum current can be split into orbital angular momentum
$(L_{\text{orbital}})^\mu{}_{\rho\sigma}$ and spin angular momentum
$S^\mu{}_{\rho\sigma}$ as
\begin{equation}
  (L_{\text{orbital}})^\mu{}_{\rho\sigma}
  = -2T_{\text{can}}{}^{\mu}{}_{[\rho} x_{\sigma]}, \qquad
  S^\mu{}_{\rho\sigma}
  = -\frac{\delta\mathcal{L}}{\delta(\dow_\mu\Phi)} \frac{i}{\hbar}
  D(M_{\rho\sigma}) \Phi(x).
\end{equation}
Only the sum of these components is conserved. In fact,
$\dow_\mu S^\mu{}_{\rho\sigma} = - \dow_\mu
(L_{\text{orbital}})^\mu{}_{\rho\sigma} = - T^{\text{can}}_{[\rho\sigma]}$. We
can also define the symmetric Belinfante energy-momentum tensor
\begin{equation}
  T^{\mu\nu} = T_{\text{can}}^{(\mu\nu)} + 2\dow_\lambda S^{(\mu\nu)\lambda},
\end{equation}
which is conserved $\dow_\mu T^{\mu\nu} = 0$.

\paragraph*{Conserved charges:} Conserved currents imply conserved charges,
i.e. quantities that remain constant in time. These are given by\footnote{The
  conservation trivially follows; for instance for the Hamiltonian we have
  \begin{equation}
    \frac{\df \hat H}{\df t}
    = c\int \df^3 x\, \dow_0 T_{\text{can}}^{00}
    = - c\int \df^3 x\, \dow_i T_{\text{can}}^{i0}
    = - c\oint \df^2 x\, n_iT_{\text{can}}^{i0},
  \end{equation}
  where $n_i$ is an outward-pointing unit vector transverse to the boundary of
  the region. A similar argument follows for all the remaining conserved charges
  as well. Physically, this implies that the rate of change of a conserved
  charge in a region is equal to the flux flowing into the region from its
  boundary.}
\begin{align}
  \hat T_a = \frac1c \int \df^3 x\, J^0_a
  &= - \int \df^3 x\, \frac{\delta\mathcal{L}}{\delta(\dow_t\Phi)}
    \frac{i}{\hbar} D(T_a)\Phi
    + \ldots, \nn\\
  \hat K_i = 
  \frac1c \int \df^3 x\, L^{0}{}_{i0}
  &= - \int \df^3 x\, \frac{\delta\mathcal{L}}{\delta(\dow_t\Phi)}
    \lb \frac{i}{\hbar} D(K_i) \Phi
    + ct \dow_i \Phi
    + \frac{x_i}{c} \dow_t \Phi
    \rb
    + \ldots
    + \frac1c \int \df^3 x\, x_i\mathcal{L}, \nn\\
  \hat J_i = 
  \frac1c \int \df^3 x\, \half\epsilon_{ijk} L^{0}{}_{jk}
  &= - \int \df^3 x\, \frac{\delta\mathcal{L}}{\delta(\dow_t\Phi)}
    \lb \frac{i}{\hbar} D(J_i) \Phi + \epsilon_{ijk}x_j \dow_k \Phi \rb
    + \ldots, \nn\\
  \hat H = \int \df^3 x\, T_{\text{can}}^{00}
  &= \int \df^3 x \lB \frac{\delta\mathcal{L}}{\delta(\dow_t\Phi)} \dow_t \Phi
    + \ldots
    - \mathcal{L} \rB, \nn\\
  \hat P^i = \frac1c \int \df^3 x\, T_{\text{can}}^{0i}
  &= - \int \df^3 x\, \frac{\delta\mathcal{L}}{\delta(\dow_t\Phi)} \dow^i \Phi
    + \ldots.
    \label{eq:NoetherCharges}
\end{align}
Again, for a canonical field theory, the ellipsis in these expressions go away
and $\delta\mathcal{L}/\delta(\dow_t\Phi)$ just becomes the conjugate momentum
field $\Pi = \dow\mathcal{L}/\dow(\dow_t\Phi)$. In this case, the definition of
the Hamiltonian $\hat H = \int\df^3 x\lb \Pi\dow_t\Phi - \mathcal{L}\rb$ must be
familiar from classical mechanics.

Upon quantisation of fields, the conserved charges get promoted to operators on
the Hilbert space of states, and themselves furnish a Hermitian representation
of the symmetry algebra. Given as such, the states in the Hilbert space can be
parametrised by the eigenvalues of a set of mutually commuting
operators. Defining $\hat P_0 = -\hat H/c$, $\hat M_{i0} = \hat K_i$, and
$\hat M_{ij} = \epsilon_{ijk}\hat J_k$, in the Poincar\'e sector, these
operators can be taken to be $\hat P_\mu$, $\hat W^\mu \hat W_\mu$, and
$\hat W_0$, where
$\hat W^\mu =1/2\,\epsilon^{\mu\nu\rho\sigma}\hat M_{\nu\rho}\hat P_\sigma$; see
\cref{sec:poincare-reps} for more details. In the internal symmetry sector, the
choice of mutually commuting operators depends on the symmetry group
$G_{\text{int}}$: for U(1) it is just
$\hat T = - \int \df^3 x\,i\Pi\Phi$, for SU(2) these can be taken to be
$\hat T^2$ and $\hat T_3$ (see \cref{sec:SU2}), while for SU(3) we have
$\hat T^2$, $\hat T^3$, $\hat T_3$, $\hat Y = 2/\sqrt{3}\, \hat T_8$, and
$\hat I^2$ (see \cref{sec:SU3}).

\subsubsection{Scalar field theory}
\label{sec:GlobalExamples-Scalar}

Let us consider the example of a complex scalar field $\varphi(x)$ with
Lagrangian
\begin{equation}
  \mathcal{L}_{\text{scalar}}
  = - \hbar c\,\dow_\mu\varphi^* \dow^\mu\varphi
  - \frac{m^2c^3}{\hbar} \varphi^*\varphi,
\end{equation}
where the constant $m$ represents the mass of the field. We have chosen the
coefficients in the Lagrangian such that $\varphi$ has units of $L^{-1}$. This
Lagrangian has spacetime Poincar\'e and an internal U(1) symmetry given by
\begin{equation}
  \varphi(x) \to \E{iq\theta}\varphi(\Lambda^{-1}x-a), \qquad
  \varphi^*(x) \to \E{-iq\theta}\varphi^*(\Lambda^{-1}x-a),
\end{equation}
where $\E{i\theta}\in\rmU(1)$ and $q$ is a dimensionless constant. Note that the
scalar fields $\varphi(x)$, $\varphi^*(x)$ transform in the trivial $(0,0)$
representation of the Lorentz group and the $D_q(\alpha)=\alpha^q$,
$D_{-q}(\alpha)=\alpha^{-q}$ representations of $\alpha\in \rmU(1)$
respectively. Taking the Lie algebra generator of $\mathfrak{u}(1)$ to be
``$\hbar$'', the respective representations are given as
$D(1\in\mathfrak{u}(1))= \hbar q$ and $D(1\in\mathfrak{u}(1))= -\hbar q$
respectively. Substituting $\Phi=(\varphi,\varphi^*)$ in the general discussion
above, we can find the conserved currents
\begin{align}
  J^\mu
  &= iq\hbar c\lb (\dow^\mu\varphi^*) \varphi - \varphi^* \dow^\mu\varphi \rb, \nn\\
  S^\mu{}_{\rho\sigma}
  &= 0, \nn\\
  T_{\text{can}}^{\mu\nu}
  = T^{\mu\nu}
  &= 2\hbar c\, \dow^{(\mu} \varphi^* \dow^{\nu)}\varphi
    - \eta^{\mu\nu}
    \lb \hbar c\,\dow_\lambda\varphi^* \dow^\lambda\varphi
    + \frac{m^2c^3}{\hbar} \varphi^*\varphi \rb.
\end{align}
Note that the spin current is zero for a scalar field and the canonical
energy-momentum tensor is the same as the symmetric Belinfante energy-momentum
tensor. We can also add interactions in the theory by adding terms like
$-\lambda(\varphi^*\varphi)^2$ to the Lagrangian.

For a non-Abelian example, let us consider a theory of $\SU(N)$ fundamental
scalars
\begin{align}
  \mathcal{L}_{\text{scalar}}
  &= - \hbar c\, \dow_\mu\varphi^\dagger \dow^\mu\varphi
  - \frac{m^2c^3}{\hbar} \varphi^\dagger\varphi, 
\end{align}
where $\varphi = (\varphi_1,\ldots,\varphi_N)$ is an $N$-component complex
scalar field. The theory is invariant under spacetime Poincar\'e and an internal
$\SU(N)$ symmetry\footnote{This theory also has a $\rmU(1)^N$ symmetry,
  corresponding to an independent phase change of each of the scalars. We ignore
  this for now.}
\begin{equation}
  \varphi(x)
  \to U\varphi(\Lambda^{-1}x-a), \qquad
  \varphi^\dagger
  \to \varphi^\dagger(\Lambda^{-1}x-a) U^\dagger,
\end{equation}
where $U = \exp(i/\hbar\,\lambda^a T_a) \in \SU(N)$. We can find the conserved
currents
\begin{align}
  J^\mu_a
    &= ic\lb \dow^\mu\varphi^\dagger T_a \varphi
      - \varphi^\dagger T_a \dow^\mu\varphi\rb, \nn\\
  S^\mu{}_{\rho\sigma}
    &= 0, \nn\\
  T_{\text{can}}^{\mu\nu}
  = T^{\mu\nu}
    &= 2\hbar c\,\dow^{(\mu} \varphi^\dagger \dow^{\nu)}\varphi
      - \eta^{\mu\nu}
      \lb \hbar c\,\dow_\lambda\varphi^\dagger \dow^\lambda\varphi
      + \frac{m^2c^3}{\hbar} \varphi^\dagger\varphi \rb.
\end{align}
Note that $T_a$'s are $N\times N$ matrices. So, for instance, the non-Abelian
charge current is really
$J^\mu_a = c(\dow^\mu(\varphi^\dagger)^i (T_a)_i{}^j \varphi_j -
(\varphi^\dagger)^i (T_a)_i{}^j \dow^\mu\varphi_j)$, where
$(\varphi^\dagger)^i = \varphi_i^*$.
  
\subsubsection{Spinor field theory}
\label{sec:GlobalExamples-Spinor}

Let us consider the example of right- and left-handed Weyl spinors $\psi(x)$ and
$\chi(x)$, and Dirac spinor $\Psi(x)$. The respective Lagrangians are given
as\footnote{Note that, up to a total derivative term, the Dirac Lagrangian can
  also be expressed into a more popular form
    \begin{equation}
      \mathcal{L} = i\hbar c\bar\Psi \gamma^\mu\dow_\mu \Psi - mc^2 \bar\Psi \Psi.
    \end{equation}
  }
\begin{align}
  \mathcal{L}_{\text{weyl-right}}
  &= \frac{i\hbar c}{2}\lb \psi^\dagger \bar\sigma^\mu\dow_\mu \psi
    - \dow_\mu \psi^\dagger \bar\sigma^\mu \psi\rb, \nn\\
  \mathcal{L}_{\text{weyl-left}}
  &= \frac{i\hbar c}{2}\lb \chi^\dagger \sigma^\mu\dow_\mu \chi
    - \dow_\mu \chi^\dagger \sigma^\mu \chi\rb, \nn\\
  \mathcal{L}_{\text{dirac}}
  &=
  \frac{i\hbar c}{2}\lb
    \bar\Psi \gamma^\mu\dow_\mu \Psi
    - \dow_\mu \bar\Psi \gamma^\mu \Psi\rb
    - mc^2 \bar\Psi \Psi,
\end{align}
where $\bar\Psi = \Psi^\dagger\gamma^0$. The spinor fields have dimensions
$L^{-3/2}$. The Lagrangians have spacetime Poincar\'e and internal U(1) symmetry
given by
\begin{gather}
  \psi(x) \to \E{iq\theta} A(\Lambda)\psi(\Lambda^{-1}x-a), \qquad
  \psi^\dagger(x) \to \E{-iq\theta} \psi^\dagger(\Lambda^{-1}x-a)
  A^\dagger(\Lambda), \nn\\
  \chi(x) \to \E{iq\theta} \bar A(\Lambda)\chi(\Lambda^{-1}x-a), \qquad
  \chi^\dagger(x) \to \E{-iq\theta} \chi^\dagger(\Lambda^{-1}x-a)
  \bar A^\dagger(\Lambda), \nn\\
  \Psi(x) \to \E{iq\theta} D_{\text{dirac}}(\Lambda)\Psi(\Lambda^{-1}x-a), \qquad
  \bar\Psi(x) \to \E{-iq\theta} \bar\Psi(\Lambda^{-1}x-a)
  D_{\text{dirac}}(\Lambda^{-1}),
\end{gather}
where $\E{i\theta}\in\rmU(1)$ and $q$ is a constant representing the charge of
the spinor fields. See the definition of the matrices $A(\Lambda)$,
$\bar A(\Lambda)$, $D_{\text{dirac}}(\Lambda)$ and $\sigma^\mu$,
$\bar\sigma^\mu$, $\gamma^\mu$ in \cref{sec:spinors}. The identities in
\cref{eq:sigmaTransformation} shall be useful in showing the Poincar\'e
invariance of the Lagrangians. A curious feature of Weyl spinors as opposed to
Dirac spinors is that Poincar\'e invariance does not allow a mass term; this
will turn out to be crucial in the Standard Model. Substituting
$\Phi=(\psi,\psi^\dagger)$ in the general discussion above, we can find the
conserved currents for the right-handed Weyl spinor
\begin{align}
  J^\mu
  &= q\hbar c\,\psi^\dagger \bar\sigma^\mu \psi, \nn\\
  S^\mu{}_{\rho\sigma}
  &= \frac{c}{2} \psi^\dagger
    \lb \bar\sigma^\mu s_{\rho\sigma}
    + \bar s_{\rho\sigma} \bar\sigma^\mu \rb \psi, \nn\\
  T_{\text{can}}^{\mu\nu}
  &= -\frac{i\hbar c}{2}\lb
    \psi^\dagger \bar\sigma^\mu\dow^\nu \psi
    - \dow^\nu \psi^\dagger \bar\sigma^\mu \psi\rb, \qquad
    T^{\mu\nu}
    = -\frac{i\hbar c}{2}\lb
    \psi^\dagger \bar\sigma^{(\mu}\dow^{\nu)} \psi
    - \dow^{(\nu} \psi^\dagger \bar\sigma^{\mu)} \psi\rb.
\end{align}
We have used the equations of motion to simplify the energy-momentum tensor. It
can be checked that $\dow_\lambda S^{(\mu\nu)\lambda} = 0$, therefore the
Belinfante energy-momentum is merely the symmetric part of the canonical
one. Similarly, for left-handed Weyl spinor, we take $\Phi=(\chi,\chi^\dagger)$
and find
\begin{align}
  J^\mu
  &= q\hbar c\,\chi^\dagger \sigma^\mu \chi, \nn\\
  S^\mu{}_{\rho\sigma}
  &= \frac{c}{2} \chi^\dagger
    \lb \sigma^\mu \bar s_{\rho\sigma}
    + s_{\rho\sigma} \sigma^\mu \rb \chi, \nn\\
  T_{\text{can}}^{\mu\nu}
  &= -\frac{i\hbar c}{2}\lb
    \chi^\dagger \sigma^\mu\dow^\nu \chi
    - \dow^\nu \chi^\dagger \sigma^\mu \chi\rb, \qquad
    T^{\mu\nu}
    = -\frac{i\hbar c}{2}\lb
    \chi^\dagger \sigma^{(\mu}\dow^{\nu)} \chi
    - \dow^{(\nu} \chi^\dagger \sigma^{\mu)} \chi\rb.
\end{align}
For Dirac spinor, we take $\Phi=(\Psi,\bar\Psi)$ and find 
\begin{align}
  J^\mu
  &= q\hbar c\,\bar\Psi \gamma^\mu \Psi, \nn\\
  S^\mu{}_{\rho\sigma}
  &= \frac{i\hbar c}{8}\bar\Psi \{\gamma^\mu, [\gamma_\rho,\gamma_\sigma]\} \Psi, \nn\\
  T_{\text{can}}^{\mu\nu}
  &= -\frac{i\hbar c}{2}\lb
    \bar\Psi \gamma^\mu\dow^\nu \Psi
    - \dow^\nu \bar\Psi \gamma^\mu \Psi\rb, \qquad
    T^{\mu\nu}
    = -\frac{i\hbar c}{2}\lb
    \bar\Psi \gamma^{(\mu}\dow^{\nu)} \Psi
    - \dow^{(\nu} \bar\Psi \gamma^{\mu)} \Psi\rb.
\end{align}

The discussion can be repeated when the spinors are $\SU(N)$-fundamental. The
respective Lagrangians stay notationally the same
\begin{align}
  \mathcal{L}_{\text{weyl-right}}
  &= \frac{i\hbar c}{2}\lb \psi^\dagger \bar\sigma^\mu\dow_\mu \psi
    - \dow_\mu \psi^\dagger \bar\sigma^\mu \psi \rb, \nn\\
  \mathcal{L}_{\text{weyl-left}}
  &= \frac{i\hbar c}{2}\lb \chi^\dagger \sigma^\mu\dow_\mu \chi
    - \dow_\mu \chi^\dagger \sigma^\mu \chi \rb, \nn\\
  \mathcal{L}_{\text{dirac}}
  &=
  \frac{i\hbar c}{2}\lb
    \bar\Psi \gamma^\mu\dow_\mu \Psi
    - \dow_\mu \bar\Psi \gamma^\mu \Psi \rb
    - mc^2 \bar\Psi \Psi,
\end{align}
but now have spacetime Poincar\'e and internal $\SU(N)$ symmetry given by
\begin{gather}
  \psi(x) \to U A(\Lambda)\psi(\Lambda^{-1}x-a), \qquad
  \psi^\dagger(x) \to \psi^\dagger(\Lambda^{-1}x-a)
  A^\dagger(\Lambda) U^\dagger, \nn\\
  \chi(x) \to U \bar A(\Lambda)\chi(\Lambda^{-1}x-a), \qquad
  \chi^\dagger(x) \to \chi^\dagger(\Lambda^{-1}x-a)
  \bar A^\dagger(\Lambda) U^\dagger, \nn\\
  \Psi(x) \to U D_{\text{dirac}}(\Lambda)\Psi(\Lambda^{-1}x-a), \qquad
  \bar\Psi(x) \to \bar\Psi^j(\Lambda^{-1}x-a)
  D_{\text{dirac}}(\Lambda^{-1}) U^\dagger,
\end{gather}
where $U=\exp(i/\hbar\,\lambda^a T_a)\in\SU(N)$. Note that the $\SU(N)$ and
Poincar\'e symmetries act independently. For example, for the right-handed Weyl
spinor, in components we have the transformation rule
$\psi_{\alpha i}(x) \to U_i{}^j A(\Lambda)_\alpha{}^\beta\psi_{\beta
  j}(\Lambda^{-1}x-a)$. The conserved currents for the right-handed Weyl spinor
are
\begin{align}
  J^\mu_a
  &= c\psi^\dagger \bar\sigma^\mu T_a\psi, \nn\\
  S^\mu{}_{\rho\sigma}
  &= \frac{c}{2} \psi^\dagger
    \lb \bar\sigma^\mu s_{\rho\sigma}
    + \bar s_{\rho\sigma} \bar\sigma^\mu \rb \psi, \nn\\
  T_{\text{can}}^{\mu\nu}
  &= -\frac{i\hbar c}{2}\lb
    \psi^\dagger \bar\sigma^\mu\dow^\nu \psi
    - \dow^\nu \psi^\dagger \bar\sigma^\mu \psi \rb, \qquad
    T^{\mu\nu}
    = -\frac{i\hbar c}{2}\lb
    \psi^\dagger \bar\sigma^{(\mu}\dow^{\nu)} \psi
    - \dow^{(\nu} \psi^\dagger \bar\sigma^{\mu)} \psi \rb.
\end{align}
To avoid confusion with indices, we further note that the non-Abelian charge
current is expanded as
$c(\psi^\dagger)_{\dot\alpha i} (\bar\sigma^\mu)^{\dot\alpha\alpha}
(T_a)_i{}^j\psi_{\alpha j}$, but it is neater to drop the indices. For the
left-handed Weyl spinor we similarly have
\begin{align}
  J_a^\mu
  &= c \chi^\dagger \sigma^\mu T_a \chi, \nn\\
  S^\mu{}_{\rho\sigma}
  &= \frac{c}{2} \chi^\dagger
    \lb \sigma^\mu \bar s_{\rho\sigma}
    + s_{\rho\sigma} \sigma^\mu \rb \chi, \nn\\
  T_{\text{can}}^{\mu\nu}
  &= -\frac{i\hbar c}{2}\lb
    \chi^\dagger \sigma^\mu\dow^\nu \chi
    - \dow^\nu \chi^\dagger \sigma^\mu \chi \rb, \qquad
    T^{\mu\nu}
    = -\frac{i\hbar c}{2}\lb
    \chi^\dagger \sigma^{(\mu}\dow^{\nu)} \chi
    - \dow^{(\nu} \chi^\dagger \sigma^{\mu)} \chi \rb.
\end{align}
And finally for Dirac spinors we get
\begin{align}
  J^\mu_a
  &= c \bar\Psi \gamma^\mu T_a \Psi, \nn\\
  S^\mu{}_{\rho\sigma}
  &= \frac{i\hbar c}{8}\bar\Psi \{\gamma^\mu, [\gamma_\rho,\gamma_\sigma]\}
    \Psi, \nn\\
  T_{\text{can}}^{\mu\nu}
  &= -\frac{i\hbar c}{2}\lb
    \bar\Psi \gamma^\mu\dow^\nu \Psi
    - \dow^\nu \bar\Psi \gamma^\mu \Psi \rb, \qquad
  T^{\mu\nu}
  = -\frac{i\hbar c}{2}\lb
    \bar\Psi \gamma^{(\mu}\dow^{\nu)} \Psi
    - \dow^{(\nu} \bar\Psi \gamma^{\mu)} \Psi \rb.
\end{align}




\subsection{Local (gauge) symmetries}

Let us consider a field theory that is invariant under spacetime Poincar\'e
transformations and global internal symmetry group $G_{\text{int}}$, i.e.
$U\in G_{\text{int}}$ is not dependent on spacetime. If the field theory happens
to be further invariant under arbitrary spacetime-dependent group
transformations, i.e. $U(x)\in G_{\text{int}}$ is a smooth function of
spacetime, the internal symmetry group $G_{\text{int}}$ is said to be a local
symmetry.\footnote{Spacetime Poincar\'e symmetry,
  $x^\mu \to \Lambda^\mu{}_\mu x^\nu + a^\mu$, can also be promoted to a local
  symmetry, i.e. invariance under arbitrary coordinate transformations
  $x^\mu \to \Lambda^\mu{}_\mu(x) x^\nu + a^\mu(x) = x'^\mu(x)$, known as
  diffeomorphism symmetry. This symmetry is present, for instance, in the
  general theory of relativity describing Einstein's gravity. Since the Standard
  Model of particle physics does not describe gravity, we will not focus on
  these in this course.}

\subsubsection{Gauging}

A field theory described by an action $S[\Phi]$, given as
\begin{equation}
  S[\Phi] = \int \df t\df^3 x\,\mathcal{L}(\Phi(x),\dow\Phi(x),\ldots),
\end{equation}
that is invariant under a global internal symmetry $G_{\text{int}}$, is
typically not invariant under the local version of the symmetry
automatically. This is because under a local internal transformation
$U(x)\in G_{\text{int}}$, the field $\Phi(x) \to D(U(x))\Phi(x)$ transforms
``covariantly'', but its derivative does not
\begin{equation}
  \dow_\mu \Phi(x) \to D(U(x))\, \dow_\mu \Phi(x) + \dow_\mu D(U(x))\,\Phi(x).
\end{equation}
The second ``inhomogeneous'' piece in this transformation results from the
spacetime dependence of the transformation parameter $U(x)$.

The situation can be remedied by introducing a gauge field
$A_\mu \in \fg_{\text{int}}$ valued in the Lie algebra, that transforms as
\begin{equation}
  A_\mu(x) \to U(x)\lb A_\mu(x) + \frac{i}{g}\dow_\mu \rb U^{-1}(x).
\end{equation}
Here $g$ is a dimensionless constant called the interaction strength of the
gauge field.\footnote{With this convention, the dimension of the gauge field is
  $[A_\mu]=L^{-1}$.} Assuming the exponential parametrisation
$U(x) = \exp(i/\hbar\,\theta^a(x) T_a)$ and decomposing the gauge field into
components $A_\mu(x) = 1/\hbar\,A_\mu^a(x) T_a$, the infinitesimal
transformation is given as
\begin{equation}
  A_\mu^a(x) \to A_\mu^a(x) - \theta^b(x) A^c_\mu(x) f_{bca}
  + \frac{1}{g} \dow_\mu \theta^a(x)
  + \mathcal{O}(\theta^2).
\end{equation}
The gauge field can be used to define a gauge covariant derivative
$\Df_\mu\Phi(x)$ of the fields, such that
$\Df_\mu\Phi(x) \to D(U(x))\,\Df_\mu\Phi(x)$ under the action of the group. For
instance, taking $\Phi(x)$ to transform in the fundamental representation,
i.e. $\Phi(x) \to U(x)\Phi(x)$, the covariant derivative is defined as
\begin{equation}
  \Df_\mu\Phi(x) = \dow_\mu\Phi(x) - ig A_\mu(x) \Phi(x).
\end{equation}
Similarly, for an anti-fundamental field $\Phi(x) \to \Phi(x)U(x)^{-1}$, we have
\begin{equation}
  \Df_\mu\Phi(x) = \dow_\mu\Phi(x) + ig\, \Phi(x) A_\mu(x).
\end{equation}
And, for an adjoint field $\Phi(x) \to U(x)\Phi(x)U(x)^{-1}$, we have
\begin{equation}
  \Df_\mu\Phi(x) = \dow_\mu \Phi(x) - ig[A_\mu(x),\Phi(x)].
\end{equation}
A field theory with global $G_{\text{int}}$ symmetry can typically be turned
into a theory with local $G_{\text{int}}$ symmetry by simply converting all
partial derivatives $\dow_\mu$ to covariant derivatives $\Df_\mu$, i.e.
\begin{equation}
  S[\Phi] = \int \df t\df^3 x\,\mathcal{L}(\Phi(x),\dow\Phi(x),\ldots)
  \to
  S[\Phi;A]
  = \int \df t\df^3 x\,\mathcal{L}(\Phi(x),\Df\Phi(x),\ldots).
\end{equation}

\paragraph*{Conserved currents via variation:} A curious byproduct of the
gauging procedure is that the conserved current $J^\mu_a$, associated with
global $G_{\text{int}}$ symmetry, can be obtained by varying $S[\Phi]$ with
respect to the gauge field components $A^a_\mu$, i.e.
\begin{equation}
  J^\mu_a = \frac{1}{g}\frac{\delta S[\Phi,A]}{\delta A_\mu^a}.
\end{equation}
In the limit where $A_\mu^a=0$, this formula results in the same expression as
derived in \cref{eq:NoetherCurrents}. The ``conservation equation'' follows from
requiring the action to be invariant under infinitesimal symmetry
transformations $\delta\Phi = i/\hbar\,\theta^a D(T_a)\Phi$ and
$\delta A^a_\mu = - \theta^b A^c_\mu f_{bca} + 1/g\, \dow_\mu\theta^a$,
leading to
\begin{align}
  0
  &= \int \df^4 x \lB
    \frac{\delta S[\Phi,A]}{\delta\Phi} \frac{i}{\hbar}\theta^a D(T_a)\Phi
    + \frac{\delta S[\Phi,A]}{\delta A_\mu^a} \lb - \theta^b A^c_\mu f_{bca} +
    \frac{1}{g} \dow_\mu\theta^a \rb \rB \nn\\
  &= - \int \df^4 x \lb
    g f_{abc} A^b_\mu J^\mu_c
    + \dow_\mu J^\mu_a \rb \theta^a,
\end{align}
implying
\begin{equation}
  \dow_\mu J^\mu_a + g f_{abc} A^b_\mu J^\mu_c = 0.
\end{equation}
In the second line above, we have used the $\Phi$ equations of motion
$\delta S/\delta\Phi = 0$. We see that $J^\mu_a$ is conserved in the absence of
gauge field. Defining $J^\mu = 1/\hbar\,J^\mu_a T_a$ and using total
anti-symmetry of $f_{abc}$, this can be more compactly denoted as
\begin{equation}
  \Df_\mu J^\mu = \dow_\mu J^\mu - ig [A_\mu,J^\mu] = 0.
\end{equation}
Hence, the covariant derivative of $J^\mu$ is zero.

\paragraph*{Local U(1) symmetry:} Let us consider complex scalar and spinor
field theories from
\cref{sec:GlobalExamples-Scalar,sec:GlobalExamples-Spinor}. The respective
Lagrangians get promoted to
\begin{align}
  \mathcal{L}_{\text{scalar}}
  &= -\hbar c\,\Df_\mu\varphi^* \Df^\mu\varphi - \frac{m^2c^3}{\hbar}
    \varphi^*\varphi, \nn\\
  \mathcal{L}_{\text{weyl-right}}
  &= \frac{i\hbar c}{2}\lb \psi^\dagger \bar\sigma^\mu\Df_\mu \psi
    - \Df_\mu \psi^\dagger \bar\sigma^\mu \psi\rb, \nn\\
  \mathcal{L}_{\text{weyl-left}}
  &= \frac{i\hbar c}{2}\lb \chi^\dagger \sigma^\mu\Df_\mu \chi
    - \Df_\mu \chi^\dagger \sigma^\mu \chi\rb, \nn\\
  \mathcal{L}_{\text{dirac}}
  &= \frac{i\hbar c}{2}\lb
    \bar\Psi \gamma^\mu\Df_\mu \Psi
    - \Df_\mu \bar\Psi \gamma^\mu \Psi\rb
    - mc^2 \bar\Psi \Psi,
\end{align}
where the covariant derivatives are defined as
\begin{gather}
  \Df_\mu\varphi = \dow_\mu\varphi - iqe A_\mu \varphi, \qquad
  \Df_\mu\varphi^* = \dow_\mu\varphi^* + iqe \varphi^* A_\mu, \nn\\
  \Df_\mu\psi = \dow_\mu\psi - iqe A_\mu\psi, \qquad
  \Df_\mu\psi^\dagger = \dow_\mu\psi^\dagger + iqe \psi^\dagger A_\mu, \nn\\
  \Df_\mu\chi = \dow_\mu\chi - iqe A_\mu\chi, \qquad
  \Df_\mu\chi^\dagger = \dow_\mu\chi^\dagger + iqe \chi^\dagger A_\mu, \nn\\
  \Df_\mu\Psi = \dow_\mu\Psi - iqe A_\mu\Psi, \qquad
  \Df_\mu\bar\Psi = \dow_\mu\bar\Psi + iqe \bar\Psi A_\mu,
\end{gather}
where $e = \sqrt{4\pi\alpha}$ is the fundamental dimensionless electronic charge
serving as the U(1) interaction strength, and $\alpha$ is the fine structure
constant. These theories are now invariant under local U(1) symmetry in addition
to Poincar\'e symmetries. The transformation of the U(1) gauge field is given as
$A_\mu \to A_\mu + 1/e\,\dow_\mu\theta$.
Varying the respective Lagrangians with respect to the gauge field $A_\mu$
results in the U(1) conserved currents. For spinors, we get the exact same
results as in \cref{sec:GlobalExamples-Spinor}, while for scalar field we get
\begin{align}
  J^\mu_{\text{scalar}}
  &= iq\hbar c\lb (\Df^\mu\varphi^*)\varphi - \varphi^*\Df^\mu\varphi \rb,
\end{align}
which reduces to the expression in \cref{sec:GlobalExamples-Scalar} upon setting
the gauge field to zero.
  
\paragraph*{Local SU(N) symmetry:}
Similarly, the $\SU(N)$ scalar and spinors field theories become
\begin{align}
  \mathcal{L}_{\text{scalar}}
  &= -\hbar c\,\Df_\mu\varphi^\dagger\Df^\mu\varphi
    - \frac{m^2c^3}{\hbar} \varphi^\dagger\varphi, \nn\\
  \mathcal{L}_{\text{weyl-right}}
  &= \frac{i\hbar c}{2}\lb \psi^\dagger \bar\sigma^\mu\Df_\mu \psi
    - \Df_\mu \psi^\dagger \bar\sigma^\mu \psi \rb, \nn\\
  \mathcal{L}_{\text{weyl-left}}
  &= \frac{i\hbar c}{2}\lb \chi^\dagger \sigma^\mu\Df_\mu \chi
    - \Df_\mu \chi^\dagger \sigma^\mu \chi\rb, \nn\\
    \mathcal{L}_{\text{dirac}}
  &= \frac{i\hbar c}{2}\lb
    \bar\Psi \gamma^\mu\Df_\mu \Psi
    - \Df_\mu \bar\Psi \gamma^\mu \Psi\rb
    - mc^2 \bar\Psi\Psi,
\end{align}
where the covariant derivatives are now defined as
\begin{gather}
  \Df_\mu\varphi = \dow_\mu\varphi
  - \frac{ig}{\hbar} A_\mu^a T_a \varphi,
  \qquad \Df_\mu\varphi^\dagger
  = \dow_\mu\varphi^\dagger
  + \frac{ig}{\hbar} \varphi^\dagger T_a A_\mu^a,
  \nn\\
  \Df_\mu\psi = \dow_\mu\psi
  - \frac{ig}{\hbar} A_\mu^a T_a \psi,
  \qquad \Df_\mu\psi^\dagger
  = \dow_\mu\psi^\dagger
  + \frac{ig}{\hbar} \psi^\dagger T_a A_\mu^a,
  \nn\\
  \Df_\mu\chi = \dow_\mu\chi
  - \frac{ig}{\hbar} A_\mu^a T_a \chi, \qquad
  \Df_\mu\chi^\dagger = \dow_\mu\chi^\dagger
  + \frac{ig}{\hbar} \chi^\dagger T_a A^a_\mu,
  \nn\\
  \Df_\mu\Psi = \dow_\mu\Psi
  - \frac{ig}{\hbar} A_\mu^a T_a \Psi, \qquad
  \Df_\mu\bar\Psi
  = \dow_\mu\bar\Psi + \frac{ig}{\hbar} \bar\Psi T_a A^a_\mu.
\end{gather}
The resultant theories are now invariant under $\SU(N)$ gauge transformations in
addition to Poincar\'e symmetries. Varying with respect to the gauge field
components $A^a_\mu$ results in the conserved currents. Again, the results for
spinor fields is the same as in \cref{sec:GlobalExamples-Spinor}, while for the
scalar fields we get
\begin{align}
  J^\mu_{a,\text{scalar}}
  &= ic\lb \Df^\mu\varphi^\dagger T_a \varphi
    - \varphi^\dagger T_a \Df^\mu\varphi\rb, 
\end{align}
which reduces to the expression in \cref{sec:GlobalExamples-Scalar} upon setting
the gauge field to zero.

\subsubsection{Dynamical gauge fields}

In the gauging procedure above, we introduced gauge fields to promote a global
internal symmetry into a local one. However, the gauge fields themselves do not
have any dynamics. In other words, gauging procedure only introduces coupling
terms of the kind $\Phi^* A_\mu \dow^\mu \Phi$ or $\Phi^* A^\mu A_\mu \Phi$ in
the action, but no kinetic term for $A_\mu$. These are known as ``background
gauge fields''. To make gauge the fields dynamical, we need to introduce a
kinetic term by hand.

\paragraph*{Field strength:} The kinetic term we seek must include a
$1/2\,\dow_t A^a_\mu \dow_t A_a^\mu$ piece, and must respect all the symmetries
of the theory -- Poincare symmetry and local internal gauge symmetry. To
motivate such a term, let us consider a fundamental field $\Phi$ and compute
\begin{align}
  [D_\mu,D_\nu]\Phi
  &= D_\mu D_\nu \Phi - D_\nu D_\mu \Phi \nn\\
  &= -ig \lb \dow_\mu A_\nu - \dow_\nu A_\mu
    - ig [A_\mu, A_\nu] \rb \Phi \nn\\
  &= -ig F_{\mu\nu} \Phi,
\end{align}
where
\begin{equation}
  F_{\mu\nu} = \dow_\mu A_\nu - \dow_\nu A_\mu - ig [A_\mu, A_\nu],
\end{equation}
is the \emph{gauge field strength}. Since $\Phi \to U\Phi$ under a gauge
transformation and, by definition, $[D_\mu,D_\nu]\Phi \to U\,[D_\mu,D_\nu]\Phi$,
it must be true that
\begin{equation}
  F_{\mu\nu} \to U F_{\mu\nu} U^{-1}.
\end{equation}
Hence $F_{\mu\nu}$ transforms in the adjoint representation of the group under
local symmetry transformations. In components
$F_{\mu\nu} = 1/\hbar\,F^a_{\mu\nu} T_a$, the field strength is given as
\begin{equation}
  F^a_{\mu\nu} = \dow_\mu A^a_\nu - \dow_\nu A^a_\mu + g f_{bca} A^b_\mu A^c_\nu.
\end{equation}

The following ``Bianchi identities'' satisfied by the field strength can be
verified
\begin{equation}
  \Df_\mu \Df_\nu F^{\mu\nu} = 0, \qquad
  \Df_{[\lambda} F_{\mu\nu]} = 0.
  \label{eq:Bianchi}
\end{equation}
Note that
$\Df_\lambda F^{\mu\nu} = \dow_\lambda F_{\mu\nu} - ig[A_\lambda,F_{\mu\nu}]$.

\paragraph*{Yang-Mills action:} The field strength $F_{\mu\nu}$ has a term
linear in $\dow_t A_\mu$, so a natural contender for the kinetic term would be
$F^{\mu\nu} F_{\mu\nu}$. However, this is not gauge invariant, because
$F^{\mu\nu}F_{\mu\nu} \to UF^{\mu\nu}U^{-1}UF_{\mu\nu}U^{-1} =
UF^{\mu\nu}F_{\mu\nu}U^{-1}$. This situation can be remedied by taking a trace,
leading to the Yang-Mills action
\begin{equation}
  S_{\text{YM}}[A]
  = -\frac{\hbar c}{4C} \int \df t\df^3 x\, \tr(F^{\mu\nu}F_{\mu\nu}).
\end{equation}
The factor of $\hbar$ is included in the action due to dimensional reasons.
Using $\tr(T_a,T_b) = C\hbar^2\delta_{ab}$, in components we have
\begin{align}
  S_{\text{YM}}[A]
  &= -\frac{\hbar c}{4} \int \df t\df^3 x\, F_a^{\mu\nu}F^a_{\mu\nu} \nn\\
  &= -\frac{\hbar c}{2} \int \df t\df^3 x \bigg(
    \dow^\mu A_{a}^\nu \dow_\mu A^a_\nu
    - \dow^\nu A_{a}^\mu \dow_\mu A^a_\nu \nn\\
  &\qquad\qquad
    + 2g f_{bca}A^b_\mu A^c_\nu \dow^\mu A_a^\nu
    + \half g^2 f_{efa} f_{bca} A^\mu_e A^\nu_f  A^b_\mu A^c_\nu
    \bigg).
\end{align}
We get a kinetic term, quadratic in $\dow_\mu A_\nu$, plus two interaction
terms: a 3-point self-interaction with coupling constant $g$ and a 4-point
self-interaction with coupling constant $g^2$. The ensuing equations of motion
can be written as 
\begin{equation}
  \Df_\mu F^{\mu\nu} = 0.
\end{equation}
The Bianchi identities \eqref{eq:Bianchi} further results in
\begin{equation}
  \Df_\mu {\star F}^{\mu\nu} = 0,
\end{equation}
where $\star F^{\mu\nu} =
1/2\,\epsilon^{\mu\nu\rho\sigma}F_{\rho\sigma}$. Yang-Mills theories describe
(non-Abelian) gauge fields in vacuum.

We can bring the Yang-Mills action $S_{\text{YM}}[A]$ and the gauge-invariant
matter action $S_{m}[\Phi;A]$ together to obtain the full theory of dynamical
gauge fields coupled to matter
\begin{align}
  S[\Phi,A]
  &= S_{\text{YM}}[A] + S_{m}[\Phi;A] \nn\\
  &= \int \df t\df^3 x \lb  -\frac{\hbar c}{4C} \tr(F^{\mu\nu}F_{\mu\nu})
    + \mathcal{L}_m(\Phi,\Df\Phi,\ldots)
    \rb.
\end{align}
The equations of motion now read
\begin{equation}
  \Df_\mu F^{\mu\nu} = \frac{g}{\hbar c}J^\mu.
\end{equation}
The conservation equation $\Df_\mu J^\mu = 0$ follows using the Bianchi
identities \eqref{eq:Bianchi}.

\paragraph*{Maxwell's action:} A special case of Yang-Mills action is when the
gauge symmetry group is U(1). In this case the gauge field strength is simply
the Maxwell field strength tensor
$F_{\mu\nu} = \dow_\mu A_\nu - \dow_\nu A_\mu$. This results in the Maxwell's
action
\begin{equation}
  \mathcal{L}_{\text{Maxwell}}
  = -\frac{\hbar c}{4} \int \df t\df^3 x\, F^{\mu\nu}F_{\mu\nu},
\end{equation}
with associated equations
$\dow_\mu F^{\mu\nu} = \dow_\mu{\star F}^{\mu\nu} = 0$. This theory describes
free electromagnetic fields in vacuum. A qualitative difference from the generic
non-Abelian case is that this theory is free, i.e. it does not have any
interactions. This signals to the physical fact that U(1) electromagnetic
photons do not interact among themselves, while, for example, $\SU(3)$ gluons
do. When coupled to matter, the Lagrangian reads
\begin{align}
  S[\Phi,A]
  &= S_{\text{Maxwell}}[A] + S_{m}[\Phi;A] \nn\\
  &= \int \df t\df^3 x \lb  -\frac{\hbar c}{4} F^{\mu\nu}F_{\mu\nu}
    + \mathcal{L}_m(\Phi,\Df\Phi,\ldots)
    \rb,
\end{align}
resulting in the Maxwell's equations\footnote{Few comments are in order on the
  dimensions. We have take the gauge field $A_\mu$ to have dimensions
  $L^{-1}$. However, the usual electromagnetic gauge field has dimensions
  $MLT^{-1}Q^{-1}$, which can be obtained by rescaling
  $A_\mu \to A_\mu/\sqrt{\hbar c\mu_0}$ where $\mu_0$ is the permeability of
  free space. This results in the familiar Maxwell's action
  $-\frac{1}{4\mu_0} \int \df t\df^3 x\, F^{\mu\nu}F_{\mu\nu}$. Similarly, the
  charge current should be rescaled as $J^\mu \to J^\mu \sqrt{\hbar
    c\mu_0}/e$. The Maxwell's equations then become
  $\dow_\mu F^{\mu\nu} = \mu_0 J^\nu$.}
\begin{equation}
  \dow_\mu F^{\mu\nu} = \frac{e}{\hbar c} J^\mu.
\end{equation}
The remaining Maxwell's equations are obtained by the Bianchi identity
$\dow_\mu{\star F}^{\mu\nu} = 0$.

\subsection{QED and QCD}
\label{sec:QED-QCD}

Quantum electrodynamics (QED) is a theory with U(1) gauge symmetry. This can be
obtained by coupling U(1) matter fields (scalars or spinors) to Maxwell's
action. For instance, we have scalar- and Dirac-QED Lagrangians
\begin{align}
  \mathcal{L}_{\text{scalar-QED}}
  &= - \hbar c\,\Df_\mu\varphi^* \Df^\mu\varphi
    - \frac{m^2c^3}{\hbar}\varphi^*\varphi
    - \frac{\hbar c}{4} F^{\mu\nu}F_{\mu\nu}, \nn\\
    \mathcal{L}_{\text{dirac-QED}}
  &= \frac{i\hbar c}{2}\lb
    \bar\Psi \gamma^\mu\Df_\mu \Psi
    - \Df_\mu \bar\Psi \gamma^\mu \Psi\rb
    - mc^2 \bar\Psi \Psi
    - \frac{\hbar c}{4} F^{\mu\nu}F_{\mu\nu},
\end{align}
The latter ``Dirac-QED'' or simply ``QED'' can describe, for instance, electrons
coupled to electromagnetic fields. These are interacting theories. For instance,
denoting matter fields with solid and photons by wavy lines, scalar-QED has two
interaction vertices
\begin{equation*}
  \begin{tikzpicture}
    \draw[thick] (-1,1) -- (0,0);
    \draw[thick] (-1,-1) -- (0,0);
    \draw[thick,wave] (0,0) -- (1.2,0);
    \node at (0.2,-0.3) {$e$};
  \end{tikzpicture} \qquad\qquad
  \begin{tikzpicture}
    \draw[thick] (-1,1) -- (0,0);
    \draw[thick] (-1,-1) -- (0,0);
    \draw[thick,wave] (0,0) -- (1,1);
    \draw[thick,wave] (0,0) -- (1,-1);
    \node at (0,-0.4) {$e^2$};
  \end{tikzpicture}
\end{equation*}
that scale as $e$ and $e^2$ respectively. On the other hand, Dirac-QED only has
the first 3-point interaction vertex.

Quantum chromodynamics (QCD), on the other hand, is a theory with SU(3) gauge
symmetry. This can be obtained by coupling SU(3) matter fields (scalars or
spinors) to Yang-Mills action. The Lagrangians for scalar- and Dirac-QCD are
given as
\begin{align}
  \mathcal{L}_{\text{scalar-QCD}}
  &= - \hbar c\,\Df_\mu\varphi^\dagger \Df^\mu\varphi
    - \frac{m^2c^3}{\hbar} \varphi^\dagger\varphi
    - \frac{\hbar c}{2} \tr(F^{\mu\nu} F_{\mu\nu}), \nn\\
  \mathcal{L}_{\text{dirac-QCD}}
  &= \frac{i\hbar c}{2}\lb
    \bar\Psi \gamma^\mu\Df_\mu \Psi
    - \Df_\mu \bar\Psi \gamma^\mu \Psi\rb
    - mc^2 \bar\Psi\Psi
    - \frac{\hbar c}{2} \tr(F^{\mu\nu} F_{\mu\nu}).
\end{align}
``Dirac-QCD'' or simply ``QCD'' can describe quarks coupled to gluons. The
qualitative difference in QCD compared to QED is that we still have the matter
interaction vertices given above
\begin{equation*}
  \begin{tikzpicture}
    \draw[thick] (-1,1) -- (0,0);
    \draw[thick] (-1,-1) -- (0,0);
    \draw[thick,coil] (0,0) -- (1.2,0);
    \node at (0.2,-0.3) {$g$};
  \end{tikzpicture} \qquad\qquad
  \begin{tikzpicture}
    \draw[thick] (-1,1) -- (0,0);
    \draw[thick] (-1,-1) -- (0,0);
    \draw[thick,coil] (0,0) -- (1,1);
    \draw[thick,coil] (0,0) -- (1,-1);
    \node at (0,-0.4) {$g^2$};
  \end{tikzpicture}
\end{equation*}
but, in addition, we also have gluons self interaction vertices
\begin{equation*}
  \begin{tikzpicture}
    \draw[thick,coil] (-1,1) -- (0,0);
    \draw[thick,coil] (-1,-1) -- (0,0);
    \draw[thick,coil] (0,0) -- (1.2,0);
    \node at (0.2,-0.3) {$g$};
  \end{tikzpicture} \qquad\qquad
  \begin{tikzpicture}
    \draw[thick,coil] (-1,1) -- (0,0);
    \draw[thick,coil] (-1,-1) -- (0,0);
    \draw[thick,coil] (0,0) -- (1,1);
    \draw[thick,coil] (0,0) -- (1,-1);
    \node at (0,-0.4) {$g^2$};
  \end{tikzpicture}
\end{equation*}
This is the fundamental reason why we see free photons in nature, but we do not
see free gluons.

\subsection{Spontaneous symmetry breaking}

It is often the case in field theory that the full action respects a certain
symmetry group, but the ground state is not invariant under the
symmetry. Therefore, the said symmetry is not apparent in the low-energy
spectrum of the theory. This phenomenon is formally known as \emph{spontaneous
  symmetry breaking}.

\subsubsection{Global symmetry breaking and Goldstone theorem}

\begin{figure}[t]
  \centering
  \begin{subfigure}[b]{0.5\textwidth}
    \includegraphics[trim=0 150 0 100,clip,width=\textwidth]{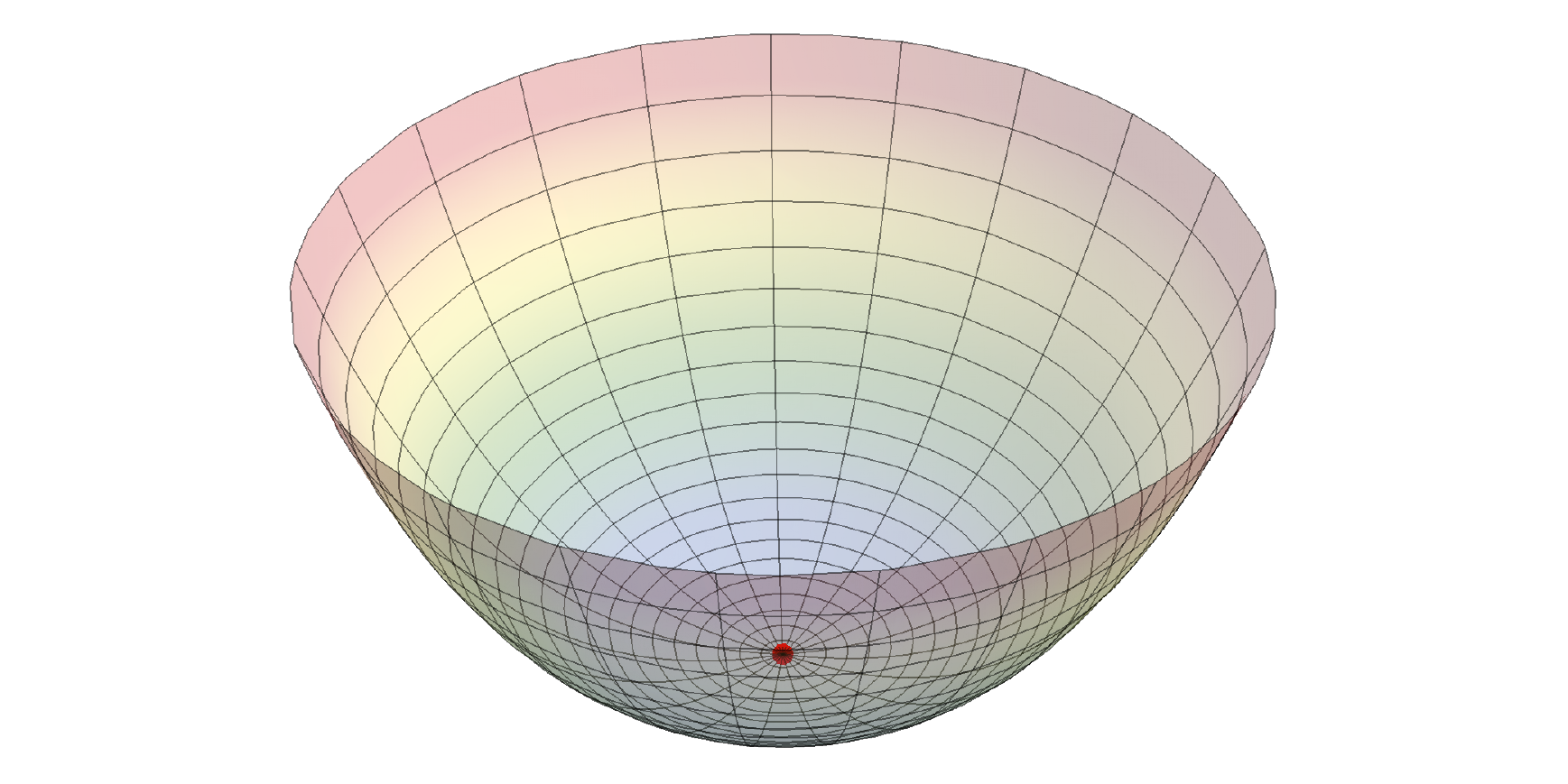}
    \caption{}
  \end{subfigure}%
  \begin{subfigure}[b]{0.5\textwidth}
    \includegraphics[trim=0 105 0 145,clip,width=\textwidth]{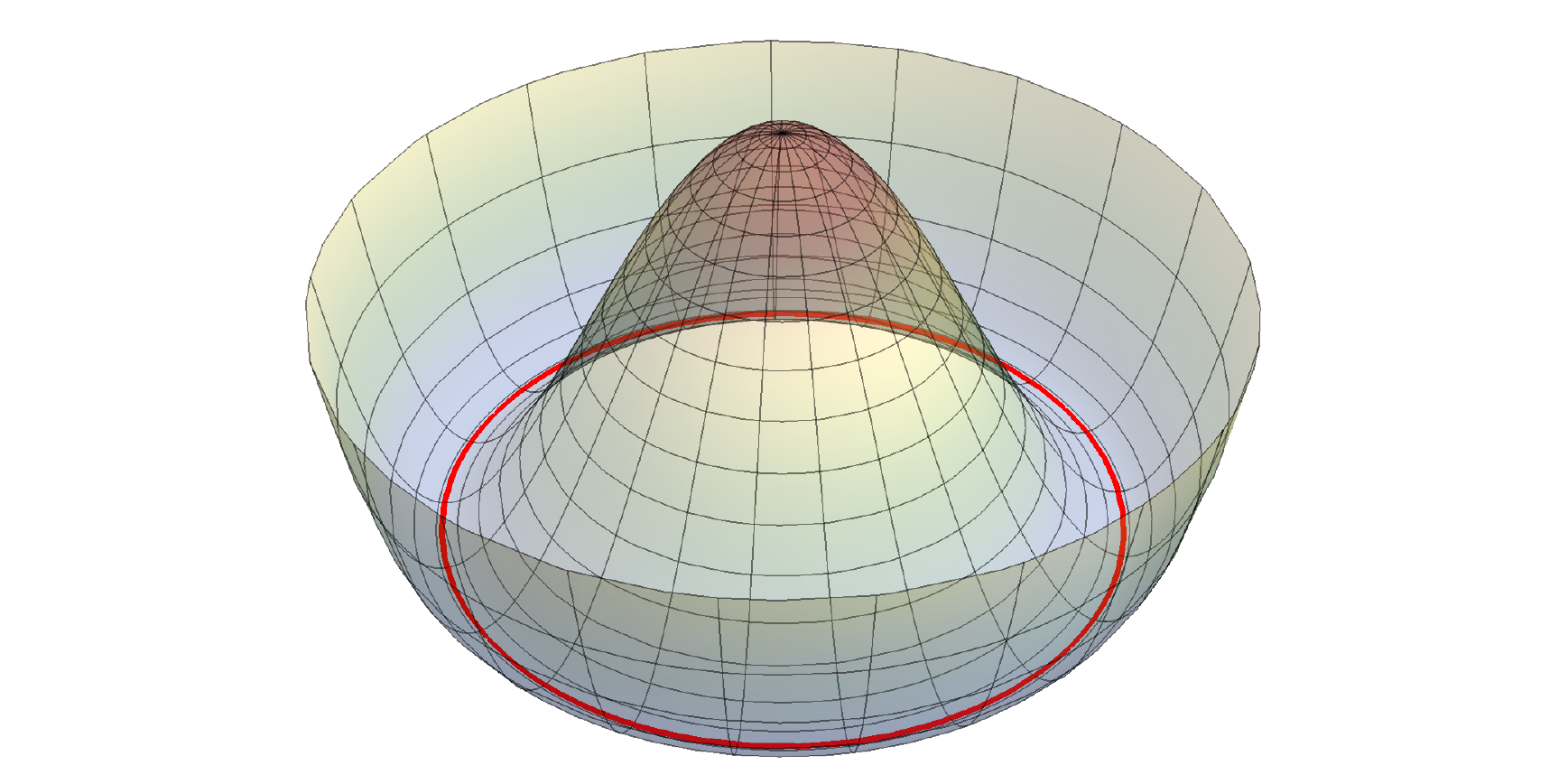}
    \caption{}
  \end{subfigure} 
  \caption{Potential $V(\varphi^*\varphi)$ plotted against complex $\varphi$ for
    \textbf{(a)} unbroken symmetry and \textbf{(b)} spontaneously broken
    symmetry. In the former case, the potential has a unique minima at
    $\varphi_0=0$, denoted by a red dot, which is invariant under the action of
    the U(1) symmetry. In the latter case, the potential admits an entire
    one-parameter family of minima $\varphi_0=v/\sqrt{2}\,\exp(i\vartheta)$,
    denoted by a red ring. Under the action of the U(1) symmetry, the possible
    minima transform into each other.\label{fig:ssb}}
\end{figure}

Consider a complex scalar field theory living in an arbitrary potential
$V(\varphi^*\varphi)$, described by the Lagrangian
\begin{equation}
  \mathcal{L} = - \hbar c\,\dow_\mu\varphi^* \dow^\mu\varphi
  - V(\varphi^*\varphi).
  \label{eq:scalarL-ssb}
\end{equation}
This Lagrangian admits a global U(1) symmetry given by
\begin{equation}
  \varphi \to \exp(iq\theta) \varphi, \qquad
  \varphi^* \to \varphi^*\exp(-iq\theta),
\end{equation}
for a constant $q$ representing the U(1) charge of $\varphi$, and a constant
symmetry parameter $\theta$. Normally, we take the potential to be the simple
mass term, with possible interactions, e.g.
$V(\varphi^*\varphi) = m^2c^3/\hbar\,\varphi^*\varphi +
\lambda(\varphi^*\varphi)^2$. This potential has a minima at
$\varphi=0$. Notably, the $\varphi=0$ state is invariant under the action of the
U(1) symmetry, and hence the symmetry is \emph{not} spontaneously
broken. Consider now the potential
\begin{equation}
  V(\varphi^*\varphi) = -\hbar c\mu^2\varphi^*\varphi
  + \hbar c\lambda(\varphi^*\varphi)^2,
  \label{eq:scalar-pot-ssb}
\end{equation}
colloquially known as the ``Mexican-hat potential''.  This potential has a
maxima at $\varphi=0$, whereas it admits an infinite number of minima
parametrised by $\varphi_0 = v/\sqrt{2}\,\exp(iq\vartheta)$ for arbitrary
$\vartheta$, where $v=\mu/\sqrt{\lambda}$. These states are no longer invariant
under the U(1) symmetry, but rather transform into each other according to
$\vartheta\to\vartheta+\theta$. Since all these minima states are equivalent and
have exactly the same potential energy, the system arbitrarily picks one as its
ground state, spontaneously breaking the U(1) symmetry. Without loss of
generality, we can take this state to be $\vartheta=0$ or
$\varphi_0 = v/\sqrt{2}$. Excitations about this ground state no longer respect
the original U(1) symmetry of the field theory. See \cref{fig:ssb}.

To obtain an effective theory of aforementioned excitations, let us expand
$\varphi$ about the ground state $\varphi_0 = v/\sqrt{2}$ and parametrise
\begin{equation}
  \varphi(x) = \frac{1}{\sqrt{2}}\exp(iq\pi(x)) (v+\eta(x)).
  \label{eq:ab-higgs-decomposition}
\end{equation}
The real scalar field $\eta$ denotes excitations in the magnitude of $\varphi$,
going up and down the valley of the Mexican-hat potential. On the other hand,
the real scalar field $\pi$ denotes excitations in the phase of $\varphi$, going
around the ring in the potential. As a consequence, $\eta$ is a massive field,
while $\pi$ is massless. Generically, we get one massless field, known as a
\emph{Goldstone boson}, for every spontaneously broken symmetry generator. This
is known as the Goldstone theorem. To be more precise, let us write the
Lagrangian in terms of $\pi$ and $\eta$. We get
\begin{align}
  \mathcal{L}
  &= - \frac{\hbar c}{2} \dow_\mu \eta\dow^\mu\eta
  - \frac{q^2\hbar c}{2} (v+\eta)^2 \dow_\mu\pi\dow^\mu\pi
    + \frac{\hbar c\mu^2}{2}(v+\eta)^2
    - \frac{\hbar c\lambda}{4}(v+\eta)^4 \nn\\
  &= - \frac{\hbar c}{2} \dow_\mu \eta\dow^\mu\eta
    - \hbar c\mu^2\eta^2
    - \frac{q^2v^2\hbar c}{2} \dow_\mu\pi\dow^\mu\pi
    + \text{interactions}.
\end{align}
In the second step, we have ignored all cubic and higher-order interactions in
$\eta$ and $\pi$. We have also ignored a constant piece in the action
$\hbar c\mu^4/(4\lambda)$ that only shifts the zero point of the energy and does
not affect the equations of motion. We get a massive real scalar field $\eta$,
with mass $\sqrt{2}\mu\hbar/c$, and a massless Goldstone field $\pi$.

For a non-Abelian realisation of this idea, consider a field $\varphi$
transforming in some $n$-dimensional unitary representation $D(G)$ of the
internal symmetry group $G$,\footnote{A unitary representation $D(G)$ of a group
  $G$ is a matrix representation such that $D(U^{-1})=D(U)^\dagger$ for all
  $U\in G$.}  i.e. an element $U\in G$ acts as
\begin{equation}
  \varphi \to D(U)\varphi, \qquad
  \varphi^\dagger \to \varphi^\dagger D(U^{-1}).
\end{equation}
Consider the $G$-invariant Lagrangian
\begin{equation}
  \mathcal{L}
  = - \hbar c\,\dow_\mu \varphi^\dagger\dow^\mu \varphi
  - V(\varphi^\dagger\varphi).
  \label{eq:scalar-nonab-L-ssb}
\end{equation}
where we take the potential to have a similar spontaneous symmetry breaking form
\begin{equation}
  V(\varphi^\dagger\varphi)
  = -\hbar c\mu^2\varphi^\dagger\varphi
  + \hbar c\lambda(\varphi^\dagger\varphi)^2.
\end{equation}
The potential has a family of minima at $\varphi^\dagger\varphi = v^2/2$ for the
system to choose from for the ground state. Without loss of generality, we can
take this ground state to be
\begin{equation}
  \varphi_0 = \frac{v}{\sqrt{2}} \hat\varphi_0, \qquad
  \hat\varphi_0 =
  \begin{pmatrix}
    0 \\ \vdots \\ 0 \\ 1
  \end{pmatrix}.
\end{equation}
We can parametrise the fluctuations around the ground state as
\begin{equation}
  \varphi(x) = \frac{v+\eta(x)}{\sqrt{2}}
  \exp\lb \frac{i}{\hbar}\pi^a(x)D(T_a) \rb \hat\varphi_0.
  \label{eq:non-ab-higgs-decomposition}
\end{equation}
Note that $n$-dimensional complex field $\varphi$ has $2n$ real components. On
the other hand, the fields $\eta$ and $\pi^a$ together have $\dim(\fg)+1$
components. Hence, generically, some components of $\pi^a$ must leave
$\hat\varphi_0$ invariant, and hence the respective $T_a$ must be unbroken. For
example, with $\varphi$ furnishing a $2$-dimensional fundamental representation
of $G=\SU(2)$, none of the generators are unbroken. On the other hand, with
$\varphi$ furnishing a $3$-dimensional fundamental representation of $G=\SU(3)$,
three generators are unbroken; these are $J_{1,2,3}=1/2\,\lambda_{1,2.3}$, where
$\lambda_{i}$'s are the Gell-Mann matrices. Putting in the decomposition
\eqref{eq:non-ab-higgs-decomposition} into the Lagrangian, we find
\begin{align}
  \mathcal{L}
  &= - \frac{\hbar c}{2} \dow_\mu\eta \dow^\mu\eta
    - \frac{c}{2\hbar} (v+\eta)^2
    (\hat\varphi_0 D(T_a)D(T_b) \hat\varphi_0)
    \dow_\mu \pi^a \dow^\mu \pi^b
    + \frac{\hbar c\mu^2}{2}(v+\eta)^2 - \frac{\hbar c\lambda}{4}(v+\eta)^4 \nn\\
  &= - \frac{\hbar c}{2} \dow_\mu \eta\dow^\mu\eta
    - \hbar c\mu^2\eta^2
    - \frac{v^2c}{4\hbar} \hat\varphi_0\{D(T_a),D(T_b)\} \hat\varphi_0
    \dow_\mu\pi^a \dow^\mu\pi^b
    + \text{interactions}.
\end{align}
For the generators that leave the ground state $\hat\varphi_0$ invariant, we
have $D(T_a)\hat\varphi_0 = 0$, and the associated $\pi^a$ fields do not show up
in the Lagrangian. For instance, with $\varphi$ being a $\SU(2)$ fundamental
field, all $\pi^{1,2,3}$ are present in the Lagrangian and make up the Goldstone
bosons. However, with $\varphi$ being a $\SU(3)$ fundamental field, we only have
$\pi^{4,5,6,7,8}$ Goldstone bosons in the Lagrangian, while $\pi^{1,2,3}$ drop
out.

\subsubsection{Local symmetry breaking and Higgs mechanism}
\label{sec:localSSB}

In the previous subsection, we considered spontaneous breaking of global
symmetries. The discussion is qualitatively different when the broken symmetries
are local. Consider, for instance, the complex scalar theory from
\cref{eq:scalarL-ssb}, but coupled to U(1) gauge field $A_\mu$, i.e.
\begin{align}
  \mathcal{L}
  &= - \hbar c\,\Df_\mu\varphi^* \Df^\mu\varphi
    - V(\varphi^*\varphi)
    - \frac{\hbar c}{4} F^{\mu\nu}F_{\mu\nu} \nn\\
  &= - \hbar c\,\dow_\mu\varphi^* \dow^\mu\varphi
    - ieq\hbar c \lb \varphi^* A^\mu \dow_\mu\varphi
    - \dow_\mu\varphi^* A^\mu \varphi \rb
    - e^2q^2 \hbar c\,\varphi^*\varphi A_\mu A^\mu \nn\\
  &\qquad
    - V(\varphi^*\varphi)
    - \frac{\hbar c}{4} F^{\mu\nu}F_{\mu\nu},
\end{align}
with the same potential \eqref{eq:scalar-pot-ssb}. Putting in the decomposition
\eqref{eq:ab-higgs-decomposition} for $\varphi$, we get
\begin{align}
  \mathcal{L}
  &= - \frac{\hbar c}{2} \dow_\mu \eta\dow^\mu\eta
    - \hbar c\mu^2\eta^2
    - \half q^2 v^2\hbar c\, \dow_\mu\pi \dow^\mu\pi
    \nn\\
  &\qquad
    + eq^2 v^2\hbar c\, A^\mu \dow_\mu\pi
    - \half e^2q^2 v^2\hbar c\, A_\mu A^\mu
    - \frac{\hbar c}{4} F^{\mu\nu}F_{\mu\nu}
    + \text{interactions}.
\end{align}
As it turns out, in this theory, we can get rid of the Goldstone boson entirely
by a redefinition of the gauge field. Let us redefine
\begin{equation}
  A_\mu \to A_\mu + \frac{1}{e}\dow_\mu\pi,
  \label{eq:ab-A-higgs-trans}
\end{equation}
leading to
\begin{align}
  \mathcal{L}
  &= - \frac{\hbar c}{2} \dow_\mu \eta\dow^\mu\eta
    - \hbar c\mu^2\eta^2
    - \frac{\hbar c}{4} F^{\mu\nu}F_{\mu\nu}
    - \half \frac{m_A^2c^3}{\hbar} A_\mu A^\mu
    + \text{interactions}.
\end{align}
where $m_A = eqv\hbar/c$. Importantly, we note that the gauge field $A_\mu$
obtains a mass term that was previously disallowed by the gauge invariance of
the action. Colloquially, we say that ``the gauge field eats the Goldstone
bosons and becomes massive''. Spontaneous symmetry breaking can, hence, be
thought of as a mechanism to generate masses for gauge fields. This is called
the \emph{Higgs mechanism}, and is vital in particle physics to generate masses
for $W^\pm_\mu$ and $Z_\mu$ gauge bosons mediating the weak forces.

The aforementioned idea of Higgs mechanism is more general than mere U(1)
symmetries. To appreciate this, let us note that
\cref{eq:ab-higgs-decomposition,eq:ab-A-higgs-trans} can be together thought of
as a local U(1) symmetry transformation with $\theta(x)=\pi(x)$. This is the
reason why the ``parameter'' of this symmetry transformation $\pi$ drops out of
the final Lagrangian. The final symmetry-broken Lagrangian is simply given by
substituting $\varphi\to(v+\eta)/\sqrt{2}$. More generally, let us consider a
field $\varphi$ transforming in some $n$-dimensional unitary representation
$D(G)$ of the internal symmetry group $G$. We can couple the Lagrangian
\eqref{eq:scalar-nonab-L-ssb} to a non-Abelian gauge field $A_\mu$ and make it
locally $G$-invariant
\begin{equation}
  \mathcal{L}
  = - \hbar c\, \Df_\mu\varphi^\dagger \Df^\mu\varphi
  - V(\varphi^\dagger\varphi)
  - \frac{\hbar c}{4C} \tr(F^{\mu\nu}F_{\mu\nu}),
\end{equation}
where $\tr(T_aT_b)=C\delta_{ab}$. The gauge covariant derivative is defined as
$\Df_\mu\varphi = \dow_\mu\varphi + ig A_\mu\varphi = \dow_\mu\varphi -
ig/\hbar\,A^a_\mu D(T_a)\varphi$, and similarly for the Hermitian
conjugate. Substituting the scalar field decomposition
\eqref{eq:non-ab-higgs-decomposition} into the Lagrangian, along with a symmetry
transformation of the gauge field
\begin{equation}
  A_\mu \to
  \exp\lb \frac{i}{\hbar}\pi^a T_a \rb
  \lb A_\mu + \frac{i}{g} \dow_\mu \rb \exp\lb -\frac{i}{\hbar}\pi^a T_a \rb,
\end{equation}
we note that the Lagrangian remains invariant except $\varphi\to
(v+\eta)/\sqrt{2}\,\hat\varphi_0$. We find
\begin{align}
  \mathcal{L}
  &= - \frac{\hbar c}{2} \dow_\mu\eta\dow^\mu\eta
    - \frac{g^2\hbar c}{2}(v+\eta)^2\, \hat\varphi_0 A_\mu A^\mu \hat\varphi_0 
    + \frac{\hbar c\mu^2}{2} (v+\eta)^2
    - \frac{\hbar c\lambda}{4} (v+\eta)^4
    - \frac{\hbar c}{4C} \tr(F^{\mu\nu}F_{\mu\nu}) \nn\\
  &= - \frac{\hbar c}{2}\dow_\mu\eta\dow^\mu\eta
    - \frac{c^3}{2\hbar}\frac{(v+\eta)^2}{v^2} m_{ab} A^a_\mu A^{b\mu}
    + \frac{\hbar c\mu^2}{2} (v+\eta)^2
    - \frac{\hbar c\lambda}{4} (v+\eta)^4
    - \frac{\hbar c}{4} F^a_{\mu\nu}F^{\mu\nu}_a \nn\\
  &= - \frac{\hbar c}{2}\dow_\mu\eta\dow^\mu\eta
    - \hbar c\mu^2\eta^2
    - \frac{c^3}{2\hbar} m_{ab} A^a_\mu A^{b\mu}
    - \frac{\hbar c}{4} F^a_{\mu\nu}F^{\mu\nu}_a
    + \text{interactions},
\end{align}
where
\begin{equation}
  m_{ab} = \frac{g^2v^2}{2c^2} \hat\varphi_0\{D(T_a),D(T_b)\} \hat\varphi_0.
\end{equation}
is the gauge field mass squared matrix. Note that not all gauge field components
get massive. For the generators that leave the ground state $\hat\varphi_0$
invariant, we have $D(T_a)\hat\varphi_0 = 0$, and the associated $m_{ab}$
components are zero. For instance, with $\varphi$ being a $\SU(2)$ fundamental,
the entire $m_{ab}\neq 0$ and all the three gauge field components obtain a
mass. However, with $\varphi$ being a $\SU(3)$ fundamental, we have
$m_{11} = m_{12} = m_{13} = m_{22} = m_{23} = m_{33} = 0$, and only the five
components $A_\mu^{4,5,6,7,8}$ obtain a mass. We note that there is a one-to-one
correspondence between the ``would be'' Goldstone bosons, if the symmetry was
global, and the massive gauge fields. This is the general Higgs mechanism.

\subsection{Electroweak theory}
\label{sec:electroweak}

The electroweak theory is a field theory describing electromagnetic and weak
forces in a unified framework. The theory contains $\SU(2)_L\times\rmU(1)_Y$
gauge bosons $W_\mu = 1/\hbar\, W_\mu^A T_A$ and $B_\mu$ (with
$T_A = \hbar/2\,\sigma_A$, where $\sigma_{A=1,2,3}$ are Pauli matrices
generating the $\mathfrak{su}(2)$ algebra), coupled to a left-handed Weyl spinor
$\chi^{\dot\alpha}_{I}$ transforming in the fundamental representation of
$\SU(2)_L$, with the indices $I,J,\ldots = 1,2$ and a right-handed Weyl spinor
$\psi_\alpha$ transforming in the trivial representation of
$\SU(2)_L$. Generalisation to multiple spinor fields is straightforward. We also
introduce a complex scalar ``Higgs'' field $\varphi_{I}$ transforming in the
fundamental representation of $\SU(2)_L$ with a ``Mexican-hat potential''
$V(\varphi^\dagger\varphi)$. The theory utilises spontaneous symmetry breaking
mechanism outlined above to generate masses for the weak force gauge fields and
the spinors. The subscript ``$L$'' on $\SU(2)_L$ is to remind ourselves that the
symmetry only acts on left-handed particles, while the subscript ``$Y$'' on
$\rmU(1)_Y$ is to distinguish it from the electromagnetic $\rmU(1)$ obtained as
the residual symmetry after spontaneous symmetry breaking.

\subsubsection{Electroweak Lagrangian}

The Lagrangian for the electroweak theory is given as (suppressing group and
spinor indices)
\begin{align}
  \frac{1}{\hbar c}\mathcal{L}
  &= \frac{i}{2} \lb \psi^\dagger \bar\sigma^{\mu}\Df_\mu \psi
    - \Df_\mu \psi^\dagger\bar\sigma^{\mu} \psi \rb
    + \frac{i}{2} \lb \chi^\dagger \sigma^{\mu}\Df_\mu\chi
    - \Df_\mu\chi^\dagger \sigma^{\mu}\chi \rb
    - \Df_\mu\varphi^\dagger \Df^\mu\varphi
    - \frac{1}{\hbar c}V(\varphi^\dagger\varphi) \nn\\
  &\qquad
    - \frac{1}{2} \tr(W^{\mu\nu}W_{\mu\nu})
    - \frac{1}{4} B^{\mu\nu} B_{\mu\nu},
\end{align}
where $B_{\mu\nu} = \dow_\mu B_\nu - \dow_\nu B_\mu$ and
$W_{\mu\nu} = \dow_\mu W_\nu - \dow_\nu W_\mu - ig_w[W_\mu,W_\nu]$. The
covariant derivatives are defined as
\begin{align}
  \Df_\mu\psi
  &= \dow_\mu\psi - iq_\psi g_y B_\mu\psi, \nn\\
  \Df_\mu\chi
  &= \dow_\mu\chi - iq_\chi g_y B_\mu\chi
    - ig_w W_\mu\chi, \nn\\
  \Df_\mu\varphi
  &= \dow_\mu\varphi - iq_\varphi g_y B_\mu\varphi
    - ig_w W_\mu \varphi,
\end{align}
and similarly for the Hermitian conjugates. Here $g_w$ and $g_y$ are $\SU(2)_L$
and $\rmU(1)_Y$ coupling constants, while $q_\psi$, $q_\chi$, $q_\varphi$ are
the fundamental $\rmU(1)_Y$ ``hypercharges'' of the respective fields. The
theory is invariant under the action of $\E{i\theta}\in\rmU(1)_Y$ and
$U\in\SU(2)_L$ acting as
\begin{gather}
  \psi \to \E{iq_\psi\theta}\,\psi, \qquad
  \psi^\dagger \to \psi^\dagger \E{-iq_\psi\theta}, \nn\\
  \chi \to \E{iq_\chi\theta}\,U\chi, \qquad
  \chi^\dagger \to \chi^\dagger\, U^{-1} \E{-iq_\chi\theta}, \nn\\
  \varphi \to \E{i q_\varphi\theta}\,U\varphi, \qquad
  \varphi^\dagger \to \varphi^\dagger\,U^{-1}\E{-i q_\varphi\theta}, \nn\\
  %
  %
  B_\mu \to B_\mu + \frac{1}{g_y}\dow_\mu\theta, \qquad
  W_\mu \to UW_\mu U^{-1} + \frac{i}{g_w} U\dow_\mu U^{-1}.
\end{gather}

Finally, the Higgs potential is taken to be
\begin{equation}
  V(\varphi^\dagger\varphi)
  = -\hbar c\mu^2 \varphi^\dagger\varphi
  + \hbar c\lambda\,(\varphi^\dagger\varphi)^2.
\end{equation}
The Higgs potential has a minima at $\varphi^\dagger\varphi = v^2/2$, where
$v=\mu/\sqrt{\lambda}$, and spontaneously breaks $\SU(2)_L\times\rmU(1)_Y$ down
to a subgroup U(1). The system spontaneously picks from one of the minima for
its ground state, which we take to be $\varphi_0=v/\sqrt{2}\,\hat\varphi_0$,
where $\hat\varphi_0 = (0,1)$.

\subsubsection{Electromagnetic U(1)}

To inspect the residual U(1) symmetry after spontaneous symmetry breaking, we
use the exponential parametrisation of $\SU(2)_L$ and identify the subgroup of
$\SU(2)_L\times\rmU(1)_Y$ transformations that leave the ground state invariant
\begin{gather}
  \exp(iq_\varphi\theta)\exp\lb\frac{i}{\hbar}\theta^AT_A\rb\varphi_0 = \varphi_0
  \quad\implies\quad
  \lb q_\varphi \theta\, \mathbb{1} + \frac{1}{\hbar}\theta^AT_A \rb \varphi_0 = 0 \nn\\
  \implies \theta^1 = \theta^2 = 0, \qquad
  \theta^3 = 2q_\varphi\theta.
\end{gather}
Setting these parameters as such, this transformation acts on the remaining
fields as
\begin{gather}
  \psi \to \exp(iq_\psi\theta)\,\psi, \qquad
  \psi^\dagger \to \psi^\dagger \exp(-iq_\psi\theta), \nn\\
  \chi\to\exp\lb i\theta(q_\chi\mathbb{1} + q_\varphi\sigma_3)\rb\chi, \qquad
  \chi^\dagger\to\chi^\dagger \exp\lb -i\theta(q_\chi\mathbb{1}
  + q_\varphi\sigma_3)\rb, \nn\\
  \eta \to \eta, \nn\\
  B_\mu \to B_\mu + \frac{1}{g_y}\dow_\mu\theta, \qquad
  W_\mu \to \exp(iq_\varphi\theta\sigma_3)W_\mu\exp(-iq_\varphi\theta\sigma_3)
  + \frac{q_\varphi}{g_w} \dow_\mu \theta\,\sigma_3.
\end{gather}
In components, it results in
\begin{gather}
  \psi \to \exp(iq_\psi\theta)\,\psi, \qquad
  \psi^\dagger \to \psi^\dagger \exp(-iq_\psi\theta), \nn\\
  \chi_1 \to \exp\lb i\theta(q_\chi + q_\varphi)\rb\chi_1, \qquad
  \chi^\dagger_1 \to \chi^\dagger_1
  \exp\lb -i\theta(q_\chi + q_\varphi)\rb, \nn\\
  \chi_2 \to \exp\lb i\theta(q_\chi - q_\varphi)\rb\chi_2, \qquad
  \chi^\dagger_2 \to \chi^\dagger_2
  \exp\lb -i\theta(q_\chi - q_\varphi)\rb, \nn\\
  \eta \to \eta, \nn\\
  W^\pm_\mu \to \exp\lb \pm 2iq_\varphi\theta \rb W^\pm_\mu, \qquad
  Z_\mu \to Z_\mu, \qquad
  A_\mu \to A_\mu
  + \frac{1}{e} \dow_\mu\theta.
\end{gather}
where we have defined
\begin{equation}
  W^{\pm}_\mu = \frac{1}{\sqrt{2}} \lb W^{1}_\mu\mp iW^{2}_\mu\rb, \qquad
  \begin{pmatrix}
    A_\mu \\ Z_\mu
  \end{pmatrix}
  =
  \begin{pmatrix}
    \cos\theta_w & \sin\theta_w \\
    -\sin\theta_w & \cos\theta_w
  \end{pmatrix}
  \begin{pmatrix}
    B_\mu \\ W^3_\mu
  \end{pmatrix},
  \label{eq:WZA}
\end{equation}
with $\theta_w = \arctan(2q_\varphi g_y/g_w)$ being the weak mixing angle and
$e=g_y\cos\theta_w$ being the fundamental dimensionless electronic charge
(electromagnetic coupling constant). We can read out the electronic charges of
various fields in terms of the hypercharges
\begin{gather}
  q^e_\psi = q_\psi, \qquad
  q^e_{\chi_1} = q_\chi + q_\varphi, \qquad
  q^e_{\chi_2} = q_\chi - q_\varphi, \qquad
  q^e_\eta = 0, \nn\\
  q^e_{W^\pm} = \pm 2q_\varphi, \qquad
  q^e_{Z} = q^e_A = 0,
\end{gather}
while the Hermitian conjugate fields have opposite charges. We note that the
residual Higgs field $\eta$, as well as gauge fields $Z_\mu$ and $A_\mu$, are
electrically neutral. On the other hand, the $W^\pm_\mu$ gauge fields are
electrically charged. The electric charges of the remaining fields follow the
constraint
\begin{equation}
  q^e_{\chi_1} - q^e_{\chi_2} = q^e_{W^+} = - q^e_{W^-}.
\end{equation}

We shall require the electromagnetic U(1) symmetry to preserve chirality. This
means that right- and ($\SU(2)_L$ components of) left-handed spinors must either
come in pairs of same electric charge, or their respective electric charge must
be zero. This implies three possibilities: either there are two right-handed
spinors $\psi=\psi_1,\psi_2$ corresponding to the left-handed spinor
$\chi_{I=1,2}$ such that
\begin{align}
  q^e_{\chi_1} = q^e_{\psi_1}
  &\quad\implies\quad
    q_{\psi_1} = q_\chi + q_\varphi, \nn\\
  q^e_{\chi_2} = q^e_{\psi_2}
  &\quad\implies\quad
    q_{\psi_2} = q_\chi - q_\varphi.
\end{align}
These fields model quarks; see \cref{sec:standardmodel}. Or, the field $\chi_1$
has electric charge zero, while the field $\psi$ couples to $\chi_2$ such that
\begin{align}
  q^e_{\chi_1} = 0
  &\quad\implies\quad
    q_\chi = - q_\varphi, \nn\\
  q^e_{\chi_2} = q^e_{\psi}
  &\quad\implies\quad
    q_{\psi} = q_\chi - q_\varphi = -2q_\varphi.
\end{align}
This is the case of leptons; see \cref{sec:standardmodel}. Or, finally, the
field $\chi_2$ has electric charge zero, while the field $\psi$ couples to
$\chi_1$ such that
\begin{align}
  q^e_{\chi_1} = q^e_{\psi}
  &\quad\implies\quad
    q_{\psi} = q_\chi + q_\varphi = 2q_\varphi, \nn\\
  q^e_{\chi_2} = 0
  &\quad\implies\quad
    q_\chi = q_\varphi.
\end{align}
The second and third cases are equivalent up to $q_\varphi\to-q_\varphi$.


\subsubsection{Higgs mechanism and massive gauge fields}
\label{sec:electroweak-Higgs}

To see the repercussions of spontaneous symmetry breaking on the electroweak
Lagrangian, we can proceed in the manner similar to the one employed in
\cref{sec:localSSB} and decompose the Higgs field as
\begin{equation}
  \varphi(x) = \frac{1}{\sqrt{2}} (v+\eta(x))\,
  \exp(iq_\varphi\pi(x))
  \exp\lb \frac{i}{\hbar} \pi^A(x)T_A\rb \hat\varphi_0, \qquad
  \hat\varphi_0 =
  \begin{pmatrix}
    0 \\ 1
  \end{pmatrix},
\end{equation}
and perform a field redefinition
\begin{align}
  \psi(x)
  &\to \exp\lb iq_\psi \pi(x) \rb \psi(x), \nn\\
  \chi(x)
  &\to \exp\lb iq_\chi\pi(x) \rb
  \exp\lb \frac{i}{\hbar} \pi^A(x)T_A\rb \chi(x), \nn\\
  B_\mu(x)
  &\to B_\mu(x)
    + \frac{1}{g_y} \dow_\mu \pi(x), \nn\\
  W_\mu(x)
  &\to \exp\lb \frac{i}{\hbar} \pi^A(x)T_A\rb
    \lb W_\mu(x) + \frac{i}{g_w} \dow_\mu \rb
    \exp\lb -\frac{i}{\hbar} \pi^A(x)T_A\rb.
\end{align}
The amounts to a symmetry transformation of the Lagrangian with $\theta=\pi$ and
$U = \exp(i/\hbar\,\pi^A T_A)$, and only has the effect that
$\varphi\to(v+\eta)/\sqrt{2}$. We are left with
\begin{align}
  \frac{1}{\hbar c}\mathcal{L}
  &= \frac{i}{2} \lb \psi^\dagger \bar\sigma^{\mu}\Df_\mu \psi
    - \Df_\mu \psi^\dagger \bar\sigma^{\mu} \psi \rb
    + \frac{i}{2} \lb \chi^\dagger \sigma^{\mu}\Df_\mu\chi
    - \Df_\mu\chi^\dagger \sigma^{\mu}\chi \rb \nn\\
  &\qquad
    - \half \dow_\mu\eta\dow^\mu\eta
    - \frac{1}{2} (v+\eta)^2 \hat\varphi_0
    \lb q_\varphi g_y B_\mu \mathbb{1} + g_w W_\mu \rb
    \lb q_\varphi g_y  B^\mu \mathbb{1} + g_w W^\mu \rb \hat\varphi_0
    - \frac{1}{\hbar c}V((v+\eta)^2/2) \nn\\
  &\qquad
    - \frac{1}{2} \tr(W^{\mu\nu}W_{\mu\nu})
    - \frac{1}{4} B^{\mu\nu} B_{\mu\nu},
\end{align}
which expands to give
\begin{align}
  \frac{1}{\hbar c}\mathcal{L}
  &= \frac{i}{2} \lb \psi^\dagger \bar\sigma^{\mu}\tilde\Df_\mu \psi
    - \tilde\Df_\mu \psi^\dagger \bar\sigma^{\mu} \psi \rb
    + \frac{i}{2} \lb \chi^\dagger_1 \sigma^{\mu}\tilde\Df_\mu\chi_1
    - \tilde\Df_\mu\chi^\dagger_1 \sigma^{\mu}\chi_1 \rb
    + \frac{i}{2} \lb \chi^\dagger_2 \sigma^{\mu}\tilde\Df_\mu\chi_2
    - \tilde\Df_\mu\chi^\dagger_2 \sigma^{\mu}\chi_2 \rb \nn\\
  &\qquad
    - \lb\lambda_\psi \psi^\dagger \bar\sigma^{\mu}\psi
    + \lambda_{\chi_1} \chi^\dagger_1 \sigma^{\mu}\chi_1
    + \lambda_{\chi_2} \chi^\dagger_2 \sigma^{\mu}\chi_2 \rb Z_\mu
    + \frac{g_w}{\sqrt2} \lb
    W^+_\mu \chi^\dagger_1 \sigma^{\mu}\chi_2
    + W^-_\mu \chi^\dagger_2 \sigma^{\mu}\chi_1 \rb \nn\\
  &\qquad
    - \half \dow_\mu\eta\dow^\mu\eta
    + \frac{\mu^2}{2}(v+\eta)^2
    - \frac{\lambda}{4}(v+\eta)^4
    - \frac14 g_w^2 (v+\eta)^2 \lb W^-_\mu W^{+\mu}
    + \half \sec^2\theta_w\, Z^\mu Z_\mu \rb
    \nn\\
  &\qquad
    - \frac{1}{2} W^+_{\mu\nu} W^{-\mu\nu}
    - \frac{1}{4} Z_{\mu\nu} Z^{\mu\nu}
    - \frac{1}{4} F_{\mu\nu} F^{\mu\nu} \nn\\
  &\qquad
    - ig_w\cos\theta_w \lb W^{-\mu\nu} W^+_{\mu} Z_{\nu}
    - W^{+\mu\nu}  W^-_{\mu} Z_{\nu}
    - Z^{\mu\nu} W^+_{\mu} W^-_{\nu} \rb
    + ig_w \sin\theta_w F^{\mu\nu} W^+_{\mu}W^-_{\nu} \nn\\
  &\qquad
    - g_w^2\cos^2\theta_w \lb W^+_{\mu} W^{-\mu}Z_{\nu}Z^{\nu}
    - W^+_{\nu}Z^{\nu} W^-_\mu Z^{\mu} \rb \nn\\
  &\qquad
    + \half g^2_w \lb W^+_{\mu}W^{+\mu} W^-_{\nu} W^{-\nu}
    - (W^+_{\mu}W^{-\mu})^2 \rb,
    \label{eq:electroweak_SSB_L}
\end{align}
where
\begin{gather}
  \lambda_\psi = q_\psi g_y\sin\theta_w, \qquad \lambda_{\chi_1} = q_\chi g_y
  \sin\theta_w - \half g_w \cos\theta_w, \qquad \lambda_{\chi_2} = q_\chi g_y
  \sin\theta_w + \half g_w \cos\theta_w.
\end{gather}
In \cref{eq:electroweak_SSB_L}, we have utilised the electromagnetic covariant
derivatives of all charged fields
\begin{align}
  \tilde\Df_\mu \psi
  &= \dow_\mu\psi - iq^e_\psi e A_\mu \psi, \nn\\
  \tilde\Df_\mu\chi_1
  &= \dow_\mu\chi_1 - iq^e_{\chi_1} e A_\mu\chi_1, \nn\\
  \tilde\Df_\mu\chi_2
  &= \dow_\mu\chi_2 - iq^e_{\chi_2} e A_\mu\chi_2,  \nn\\
  \tilde\Df_\mu W^\pm_\nu
  &= \dow_\mu W^\pm_\nu \mp 2iq_\varphi e A_\mu W^\pm_\nu,
\end{align}
along with field strengths
$W^\pm_{\mu\nu} = \tilde\Df_\mu W^\pm_\nu - \tilde\Df_\nu W^\pm_\mu$,
$Z_{\mu\nu} = \dow_\mu Z_\nu - \dow_\nu Z_\mu$, and
$F_{\mu\nu} = \dow_\mu A_\nu - \dow_\nu A_\mu$. We have used the following
identities
\begin{align}
  \Df_\mu\psi
  &= \tilde\Df_\mu\psi + iq_\psi g_y \sin\theta_w Z_\mu \psi, \nn\\
  (\Df_\mu\chi)_1
  &= \tilde\Df_\mu\chi_1
    + \frac{i}{2}\lb 2q_\chi g_y \sin\theta_w 
    - g_w \cos\theta_w \rb Z_\mu \chi_1
    - \frac{ig_w}{\sqrt2} W^+_\mu \chi_2, \nn\\
  (\Df_\mu\chi)_2
  &= \tilde\Df_\mu\chi_2
    + \frac{i}{2}\lb 2q_\chi g_y \sin\theta_w 
    + g_w \cos\theta_w \rb Z_\mu \chi_2
    - \frac{ig_w}{\sqrt2} W^-_\mu \chi_1, \nn\\
  W^1_{\mu\nu}
  &= \frac{1}{\sqrt2}\lb W^+_{\mu\nu} + W^-_{\mu\nu}
    + 2ig_w \lb  W^+_{[\mu} Z_{\nu]} -  W^-_{[\mu} Z_{\nu]} \rb \cos\theta_w \rb,
    \nn\\
  W^2_{\mu\nu}
  &= \frac{i}{\sqrt2}\lb  W^+_{\mu\nu} - W^-_{\mu\nu}
    + 2i g_w \lb W^+_{[\mu} Z_{\nu]} + W^-_{[\mu} Z_{\nu]} \rb \cos\theta_w \rb,
    \nn\\
  W^3_{\mu\nu}
  &= Z_{\mu\nu}\cos\theta_w + F_{\mu\nu}\sin\theta_w - 2i g_w W^+_{[\mu}
    W^-_{\nu]}, \nn\\
  B_{\mu\nu}
  &= F_{\mu\nu}\cos\theta_w - Z_{\mu\nu}\sin\theta_w.
\end{align}

The full form of the Lagrangian is not terribly useful. To get some physical
intuition, let us ignore all cubic and quartic interactions, which leads to
\begin{align}
  \frac{1}{\hbar c}\mathcal{L}
  &= \frac{i}{2} \lb \psi^\dagger \bar\sigma^{\mu}\dow_\mu \psi
    - \dow_\mu \psi^\dagger \bar\sigma^{\mu} \psi \rb
    + \frac{i}{2} \lb \chi^\dagger_1 \sigma^{\mu}\dow_\mu\chi_1
    - \dow_\mu\chi^\dagger_1 \sigma^{\mu}\chi_1 \rb
    + \frac{i}{2} \lb \chi^\dagger_2 \sigma^{\mu}\dow_\mu\chi_2
    - \dow_\mu\chi^\dagger_2 \sigma^{\mu}\chi_2 \rb \nn\\
  &\qquad
    - \frac{1}{2} W^-_{\mu\nu} W^{+\mu\nu}
    - \frac{m_W^2c^2}{\hbar^2} W^-_\mu W^{+\mu}
    - \frac{1}{4} Z_{\mu\nu} Z^{\mu\nu}
    - \frac{m_Z^2c^2}{2\hbar^2} Z^\mu Z_\mu
    - \frac{1}{4} F_{\mu\nu} F^{\mu\nu} \nn\\
  &\qquad
    - \half \dow_\mu\eta\dow^\mu\eta
    - \mu^2\eta^2
    + \text{interactions}.
\end{align}
We observe that the gauge field components $W^\pm_\mu$ and $Z_\mu$ acquire
masses
\begin{equation}
  m_W = \frac{v\hbar}{2c} g_w, \qquad
  m_Z = \frac{v\hbar}{2c} \sqrt{g_w^2 + 4q_\varphi^2 g_y^2},
\end{equation}
while the gauge field $A_\mu$ is massless. The gauge field combinations
$W^\pm_\mu$, $Z_\mu$ can be thought of as the carriers of weak force, while
$A_\mu$ can be identified with the electromagnetic photon. We are also left with
a residual massive real scalar Higgs field $\eta$, with mass
$\sqrt{2}\mu\hbar/c$.

\subsubsection{Yukawa couplings}

So far, our model of electroweak forces does not give masses to various spinor
fields. We already know that Weyl spinors cannot have their own mass terms. The
regular Dirac mass term
\begin{equation}
  - mc^2 \bar\Psi \Psi
  = - mc^2 \lb \chi^\dagger\psi + \psi^\dagger\chi \rb,
  \qquad\text{where}\quad
  \Psi =
  \begin{pmatrix}
    \psi \\ \chi
  \end{pmatrix},
\end{equation}
is also disallowed by $\SU(2)_L$-invariance, which only acts on $\chi$ and not
on $\psi$. As it turns out, such a term can indeed be generated via spontaneous
symmetry breaking. To wit, we can add the so-called Yukawa couplings to the
Lagrangian
\begin{align}
  \frac{1}{\hbar c}\mathcal{L}
  &= \frac{i}{2} \lb \psi^\dagger \bar\sigma^{\mu}\Df_\mu \psi
    - \Df_\mu \psi^\dagger \bar\sigma^{\mu} \psi \rb
    + \frac{i}{2} \lb \chi^\dagger \sigma^{\mu}\Df_\mu\chi
    - \Df_\mu\chi^\dagger \sigma^{\mu}\chi \rb
    - \Df_\mu\varphi^\dagger \Df^\mu\varphi
    - \frac{1}{\hbar c}V(\varphi^\dagger\varphi) \nn\\
  &\qquad
    - \frac{1}{2} \tr(W^{\mu\nu}W_{\mu\nu})
    - \frac{1}{4} B^{\mu\nu} B_{\mu\nu}
    - y\lb \chi^\dagger \varphi \psi
    + \psi^\dagger\varphi^\dagger\chi \rb
    - y'\lb \chi^\dagger \tilde\varphi \psi
    + \psi^\dagger\tilde\varphi^\dagger\chi \rb,
\end{align}
with arbitrary Yukawa coupling constants $y$ and $y'$. Here
$\tilde\varphi_I = \epsilon_{IJ}(\varphi^\dagger)^J$ is the conjugate Dirac
field. It transforms under the $\SU(2)_L\times\rmU(1)_Y$ transformation as
\begin{equation}
  \tilde\varphi \to \exp(-iq_\varphi\theta)\,U\tilde\varphi, \qquad
  \tilde\varphi^\dagger \to \tilde\varphi\,U^{-1}\exp(i q_\varphi\theta).
\end{equation}
It immediately follows that the two Yukawa coupling terms are
$\SU(2)_L$-invariant. However, for these to be $\rmU(1)_Y$-invariant, the
respective hypercharges must be constrained
\begin{equation}
  y\neq 0\quad\text{if}\quad q_\psi = q_\chi - q_\varphi, \qquad\qquad
  y'\neq 0\quad\text{if}\quad q_\psi = q_\chi + q_\varphi.
\end{equation}
Or in the electromagnetic terms
\begin{equation}
  y\neq 0\quad\text{if}\quad q^e_\psi = q^e_{\chi_2}, \qquad\qquad
  y'\neq 0\quad\text{if}\quad q^e_\psi = q^e_{\chi_1}.
\end{equation}
Note that, as long as $q_\varphi \neq 0$, both these Yukawa couplings cannot be
turned on at the same time for a given right-handed spinor $\psi$. Note also
that the Yukawa couplings force us to set the same electric charges for the
left- and right-handed spinors. After spontaneous symmetry breaking, Yukawa
couplings generate masses for the spinor fields
\begin{align}
  \frac{1}{\hbar c}\mathcal{L}
  &= \frac{i}{2} \lb \psi^\dagger \bar\sigma^{\mu}\dow_\mu \psi
    - \dow_\mu \psi^\dagger \bar\sigma^{\mu} \psi \rb
    + \frac{i}{2} \lb \chi^\dagger_1 \sigma^{\mu}\dow_\mu \chi_1
    - \dow_\mu \chi_1^\dagger \sigma^{\mu} \chi_1 \rb
    + \frac{i}{2} \lb \chi_2^\dagger \sigma^{\mu}\dow_\mu \chi_2
    - \dow_\mu \chi_2^\dagger \sigma^{\mu} \chi_2 \rb \nn\\
  &\qquad
    - \frac{1}{2} W^-_{\mu\nu} W^{+\mu\nu}
    - \frac{m_W^2c^2}{\hbar^2} W^-_\mu W^{+\mu}
    - \frac{1}{4} Z_{\mu\nu} Z^{\mu\nu}
    - \frac{m_Z^2c^2}{2\hbar^2} Z^\mu Z_\mu
    - \frac{1}{4} F_{\mu\nu} F^{\mu\nu} \nn\\
  &\qquad
    - \half \dow_\mu\eta\dow^\mu\eta
    - \mu^2\eta^2
    - \frac{m_{\chi_2}c}{\hbar}
    \lb \chi^\dagger_2 \psi + \psi^\dagger\chi_2 \rb
    - \frac{m_{\chi_1}c}{\hbar}
    \lb \chi^\dagger_1 \psi + \psi^\dagger\chi_1 \rb
    + \text{int.},
\end{align}
where
\begin{equation}
  m_{\chi_1} = \frac{vy'\hbar}{\sqrt{2}\,c}, \qquad
  m_{\chi_2} = \frac{vy\hbar}{\sqrt{2}\,c}.
\end{equation}
If $y\neq0$, the Yukawa coupling gives the same mass to $\chi_2$ and $\psi$
spinors, while leaving $\chi_1$ massless. On the other hand, if $y'\neq 0$, the
spinors $\chi_1$ and $\psi$ are massive, while the $\chi_2$ is massless. If we
intend to give masses to both $\chi_1$ and $\chi_2$, we will need to introduce
two right-handed spinors $\psi_1$ and $\psi_2$ with the same respective electric
charges.

\subsubsection{Interactions}
\label{sec:electroweak-interactions}

The final observation we want to make are the allowed electroweak
interactions. Due to spontaneous symmetry breaking, the interaction structure of
electroweak forces is far richer than QED or QCD. Looking at the spontaneously
broken Lagrangian in \cref{eq:electroweak_SSB_L}, we first have the
electromagnetic and weak interactions of the spinor fields coming from the first
and second lines
\begin{gather*}
  \begin{tikzpicture}
    \draw[thick] (-1,1) -- (0,0);
    \node at (-1.8,0.9) {$\psi$, $\chi_1$, $\chi_2$};
    \draw[thick] (-1,-1) -- (0,0);
    \node at (-1.9,-0.9) {$\psi^\dagger$, $\chi_1^\dagger$, $\chi_2^\dagger$};
    \draw[thick,wave] (0,0) -- (1.2,0);
    \node at (1.4,0) {$A$};
  \end{tikzpicture} \qquad\qquad
  \begin{tikzpicture}
    \draw[thick] (-1,1) -- (0,0);
    \node at (-1.8,0.9) {$\psi$, $\chi_1$, $\chi_2$};
    \draw[thick] (-1,-1) -- (0,0);
    \node at (-1.9,-0.9) {$\psi^\dagger$, $\chi_1^\dagger$, $\chi_2^\dagger$};
    \draw[thick,weak] (0,0) -- (1.2,0);
    \node at (1.4,0) {$Z$};
  \end{tikzpicture} \nn\\
  \begin{tikzpicture}
    \draw[thick] (-1,1) -- (0,0); 
    \node at (-1.3,0.9) {$\chi_2$};
    \draw[thick] (-1,-1) -- (0,0);
    \node at (-1.3,-0.9) {$\chi_1^\dagger$};
    \draw[thick,weak] (0,0) -- (1.2,0);
    \node at (1.6,0) {$W^+$};
  \end{tikzpicture} \qquad\qquad
  \begin{tikzpicture}
    \draw[thick] (-1,1) -- (0,0); 
    \node at (-1.3,0.9) {$\chi_1$};
    \draw[thick] (-1,-1) -- (0,0);
    \node at (-1.3,-0.9) {$\chi_2^\dagger$};
    \draw[thick,weak] (0,0) -- (1.2,0);
    \node at (1.6,0) {$W^-$};
  \end{tikzpicture}
\end{gather*}
In particular, we note that $\chi_1$ spinors (left-handed neutrino/up-quark) can
turn into $\chi_2$ spinors (left-handed electron/down-quark), and vice-versa, by
absorbing/emitting charged weak bosons $W^\pm_\mu$. Then, we have the self and
weak interactions of the residual Higgs field coming from the third line
\begin{gather*}
  \begin{tikzpicture}
    \draw[thick,dashed] (-1,1) -- (0,0);
    \node at (-1.2,0.9) {$\eta$};
    \draw[thick,dashed] (-1,-1) -- (0,0);
    \node at (-1.2,-0.9) {$\eta$};
    \draw[thick,dashed] (0,0) -- (1.2,0);
    \node at (1.4,0) {$\eta$};
  \end{tikzpicture} \qquad\qquad
  \begin{tikzpicture}
    \draw[thick,dashed] (-1,1) -- (0,0);
    \node at (-1.2,0.9) {$\eta$};
    \draw[thick,dashed] (-1,-1) -- (0,0);
    \node at (-1.2,-0.9) {$\eta$};
    \draw[thick,dashed] (0,0) -- (1,1);
    \node at (1.2,0.9) {$\eta$};
    \draw[thick,dashed] (0,0) -- (1,-1);
    \node at (1.2,-0.9) {$\eta$};
  \end{tikzpicture} \nn\\
  \begin{tikzpicture}
    \draw[thick,dashed] (-1.2,0) -- (0,0);
    \node at (-1.4,0) {$\eta$};
    \draw[thick,weak] (0,0) -- (1,1);
    \node at (1.4,0.9) {$W^+$};
    \draw[thick,weak] (0,0) -- (1,-1);
    \node at (1.4,-0.9) {$W^-$};
  \end{tikzpicture} \qquad
  \begin{tikzpicture}
    \draw[thick,dashed] (-1.2,0) -- (0,0);
    \node at (-1.4,0) {$\eta$};
    \draw[thick,weak] (0,0) -- (1,1);
    \node at (1.3,0.9) {$Z$};
    \draw[thick,weak] (0,0) -- (1,-1);
    \node at (1.3,-0.9) {$Z$};
  \end{tikzpicture} \qquad
  \begin{tikzpicture}
    \draw[thick,dashed] (-1,1) -- (0,0);
    \node at (-1.2,0.9) {$\eta$};
    \draw[thick,dashed] (-1,-1) -- (0,0);
    \node at (-1.2,-0.9) {$\eta$};
    \draw[thick,weak] (0,0) -- (1,1);
    \node at (1.4,0.9) {$W^+$};
    \draw[thick,weak] (0,0) -- (1,-1);
    \node at (1.4,-0.9) {$W^-$};
  \end{tikzpicture} \qquad
  \begin{tikzpicture}
    \draw[thick,dashed] (-1,1) -- (0,0);
    \node at (-1.2,0.9) {$\eta$};
    \draw[thick,dashed] (-1,-1) -- (0,0);
    \node at (-1.2,-0.9) {$\eta$};
    \draw[thick,weak] (0,0) -- (1,1);
    \node at (1.3,0.9) {$Z$};
    \draw[thick,weak] (0,0) -- (1,-1);
    \node at (1.3,-0.9) {$Z$};
  \end{tikzpicture}
\end{gather*}
In addition, we have mutual interactions of various gauge fields. We have mixed
interactions in the electromagnetic and weak sector coming from the U(1)
covariant derivatives in the forth line, and a few terms in the fifth line
\begin{equation*}
   \qquad\qquad
  \begin{tikzpicture}
    \draw[thick,weak] (-1,1) -- (0,0); 
    \node at (-1.4,0.9) {$W^+$};
    \draw[thick,weak] (-1,-1) -- (0,0);
    \node at (-1.4,-0.9) {$W^-$};
    \draw[thick,wave] (0,0) -- (1.2,0);
    \node at (1.4,0) {$A$};
  \end{tikzpicture} \qquad
  \begin{tikzpicture} 
    \draw[thick,weak] (-1,1) -- (0,0);
    \node at (-1.4,0.9) {$W^+$};
    \draw[thick,weak] (-1,-1) -- (0,0);
    \node at (-1.4,-0.9) {$W^-$};
    \draw[thick,wave] (0,0) -- (1,1);
    \node at (1.2,0.9) {$A$};
    \draw[thick,wave] (0,0) -- (1,-1);
    \node at (1.2,-0.9) {$A$};
  \end{tikzpicture} \qquad\qquad
  \begin{tikzpicture} 
    \draw[thick,weak] (-1,1) -- (0,0);
    \node at (-1.4,0.9) {$W^+$};
    \draw[thick,weak] (-1,-1) -- (0,0);
    \node at (-1.4,-0.9) {$W^-$};
    \draw[thick,weak] (0,0) -- (1,1);
    \node at (1.2,0.9) {$Z$};
    \draw[thick,wave] (0,0) -- (1,-1);
    \node at (1.2,-0.9) {$A$};
  \end{tikzpicture}
\end{equation*}
Then, we have weak force self-interactions coming from the last three lines
\begin{equation*}
  \begin{tikzpicture}
    \draw[thick,weak] (-1,1) -- (0,0); 
    \node at (-1.4,0.9) {$W^+$};
    \draw[thick,weak] (-1,-1) -- (0,0);
    \node at (-1.4,-0.9) {$W^-$};
    \draw[thick,weak] (0,0) -- (1.2,0);
    \node at (1.4,0) {$Z$};
  \end{tikzpicture} \qquad\qquad
  \begin{tikzpicture} 
    \draw[thick,weak] (-1,1) -- (0,0);
    \node at (-1.4,0.9) {$W^+$};
    \draw[thick,weak] (-1,-1) -- (0,0);
    \node at (-1.4,-0.9) {$W^-$};
    \draw[thick,weak] (0,0) -- (1,1);
    \node at (1.2,0.9) {$Z$};
    \draw[thick,weak] (0,0) -- (1,-1);
    \node at (1.2,-0.9) {$Z$};
  \end{tikzpicture} \qquad\qquad
  \begin{tikzpicture} 
    \draw[thick,weak] (-1,1) -- (0,0);
    \node at (-1.4,0.9) {$W^+$};
    \draw[thick,weak] (-1,-1) -- (0,0);
    \node at (-1.4,-0.9) {$W^-$};
    \draw[thick,weak] (0,0) -- (1,1);
    \node at (1.4,0.9) {$W^+$};
    \draw[thick,weak] (0,0) -- (1,-1);
    \node at (1.4,-0.9) {$W^-$};
  \end{tikzpicture}
\end{equation*}
Finally, we have Yukawa interactions between spinors and Higgs field
\begin{gather*}
  \begin{tikzpicture}
    \draw[thick] (-1,1) -- (0,0);
    \node at (-1.6,0.9) {$\psi$, $\chi_2$};
    \draw[thick] (-1,-1) -- (0,0);
    \node at (-1.6,-0.9) {$\chi_2^\dagger$, $\psi^\dagger$};
    \draw[thick,dashed] (0,0) -- (1.2,0);
    \node at (1.4,0) {$\eta$};
  \end{tikzpicture}
  \qquad\qquad\raisebox{1.2cm}{\text{or}}\qquad\qquad
  \begin{tikzpicture}
    \draw[thick] (-1,1) -- (0,0);
    \node at (-1.6,0.9) {$\psi$, $\chi_1$};
    \draw[thick] (-1,-1) -- (0,0);
    \node at (-1.6,-0.9) {$\chi_1^\dagger$, $\psi^\dagger$};
    \draw[thick,dashed] (0,0) -- (1.2,0);
    \node at (1.4,0) {$\eta$};
  \end{tikzpicture}
\end{gather*}
For a given spinor $\psi$, only one of the above interactions can be present in
a model, depending on the hypercharge. These Yukawa interactions mix the left-
and right-handed spinor sectors and are only present if the respective spinors
admit a Yukawa mass term.


\newpage

\section{Standard Model of particle physics}
\label{sec:standardmodel}

We now have all the requisite tools to write down the Standard Model of particle
physics. At the time Standard Model was proposed, neutrinos were known to be
massless. For the present discussion, we will assume as such. Discussion of
neutrino masses and neutrino flavor oscillations goes somewhat beyond the
minimal Standard Model and will not be covered in this course. The discussion in
this section follows the standard texts on the Standard Model of particle
physics such as~\cite{Burgess:2006hbd}.

The local internal symmetry group of the Standard Model is a direct product of
quantum chromodynamics symmetry and electroweak symmetry,
i.e. $\SU(3)\times\SU(2)_L\times\rmU(1)_Y$. On the other hand, the spacetime
symmetry group is still Poincar\'e.
The model employs the Higgs mechanism in the electroweak sector to spontaneously
break the symmetry down to
$\SU(3)\times\SU(2)_L\times\rmU(1)_Y \to \SU(3)\times\rmU(1)$, where $\rmU(1)$
corresponds to the residual $\rmU(1)$ electromagnetic transformation. In the
process, we generate masses for $W^\pm_\mu$ and $Z_\mu$ gauge bosons mediating
weak forces. Prior to symmetry breaking, the QCD symmetry $\SU(3)$ preserves
chirality, i.e. acts uniformly on left- and right-handed spinors, however the
electroweak symmetry $\SU(2)_L\times\rmU(1)_Y$ violates it. Interestingly, post
symmetry breaking, the residual symmetry group $\SU(3)\times\rmU(1)$ preserves
chirality.

\subsection{Field content and symmetries}

\begin{table}[p]
  \centering
  \begin{tabular}{c|ccc|c}
    \toprule
    Particle & $\SU(3)$ & $\SU(2)_L$ & $\rmU(1)_Y$
    & $\SO^+(3,1)$ \\
    \toprule
    \multicolumn{5}{c}{\textbf{Gauge sector}} \\ 
    \toprule

    $G^a_\mu$ & $\mathbf{8}$ & $\mathbf{1}$ & $0$ & $(\half,\half)$ \\[0.5em]

    $W^A_\mu$ & $\mathbf{1}$ & $\mathbf{3}$ & $0$ & $(\half,\half)$ \\[0.5em]

    $B_\mu$ & $\mathbf{1}$ & $\mathbf{1}$ & $0$ & $(\half,\half)$ \\[0.5em]

    \toprule
    \multicolumn{5}{c}{\textbf{Quark sector}} \\ 
    \toprule
    
    $(Q^L_m)^{\dot\alpha}_{Ii} =
    \begin{pmatrix}
      (u^L_m)^{\dot\alpha}_{i} \\
      (d^L_m)^{\dot\alpha}_{i}
    \end{pmatrix}$
             & $\mathbf{3}$
                    & $\mathbf{2}$
                           & $1/6$ & $(\half,0)$ \\[1.5em]

    $(u^R_m)_{\alpha i}$
                                  & $\mathbf{3}$
                    & $\mathbf{1}$
 & $2/3$ & $(0,\half)$ \\[0.5em]
    
    $(d^R_m)_{\alpha i}$
                                  & $\mathbf{3}$
                    & $\mathbf{1}$
 & $-1/3$ & $(0,\half)$ \\[0.5em]

    \midrule

    $(Q^{L\dagger}_m)^{\alpha Ii} =
    \begin{pmatrix} 
      (u^{L\dagger}_m)^{\alpha i} &
      (d^{L\dagger}_m)^{\alpha i}
    \end{pmatrix}$
             & $\bar{\mathbf{3}}$
                    & $\bar{\mathbf{2}}$
                           & $-1/6$ & $(0,\half)$ \\[1.5em]

    $(u^{R\dagger}_m)_{\dot\alpha}^i$
                                  & $\bar{\mathbf{3}}$
                    & $\mathbf{1}$
 & $-2/3$ & $(\half,0)$ \\[0.5em]
    
    $(d^{R\dagger}_m)_{\dot\alpha}^{i}$
                                  & $\bar{\mathbf{3}}$
                    & $\mathbf{1}$
 & $1/3$ & $(\half,0)$ \\[0.5em]

    \toprule
    \multicolumn{5}{c}{\textbf{Lepton sector}} \\ 
    \toprule

    $(L^L_m)^{\dot\alpha}_{I} =
    \begin{pmatrix}
      (\nu^L_m)^{\dot\alpha} \\
      (e^L_m)^{\dot\alpha}
    \end{pmatrix}$
             & $\mathbf{1}$
                    & $\mathbf{2}$
                           & $-1/2$ & $(\half,0)$ \\[1.5em]

    $(e^R_m)_{\alpha}$
                                  & $\mathbf{1}$
                    & $\mathbf{1}$
                                                          & $-1$ & $(0,\half)$ \\[0.5em]

    \midrule

    $(L^{L\dagger}_m)^{\alpha I} =
    \begin{pmatrix}
      (\nu^{L\dagger}_m)^{\alpha} &
      (e^{L\dagger}_m)^{\alpha}
    \end{pmatrix}$
             & $\mathbf{1}$
                    & $\bar{\mathbf{2}}$
                           & $1/2$ & $(0,\half)$ \\[1.5em]

    $(e^{R\dagger}_m)_{\dot\alpha}$ 
                                  & $\mathbf{1}$
                    & $\mathbf{1}$
 & $1$ & $(\half,0)$ \\[0.5em]

    \toprule
    \multicolumn{5}{c}{\textbf{Higgs sector}} \\ 
    \toprule

    $\varphi_I =
    \begin{pmatrix} 
      \varphi^+ \\
      \varphi^0
    \end{pmatrix}$        & $\mathbf{1}$
                    & $\mathbf{2}$
                                             & $1/2$  & $(0,0)$ \\[0.5em]

    \midrule

    $(\varphi^\dagger)^I  =
    \begin{pmatrix} 
      \varphi^{+*} &
      \varphi^{0*}
    \end{pmatrix}$   & $\mathbf{1}$
                    & $\bar{\mathbf{2}}$
                                             & $-1/2$  & $(0,0)$ \\[0.5em]

    \midrule

    $\tilde\varphi_I = \epsilon_{IJ}(\varphi^\dagger)^I =
    \begin{pmatrix} 
      \varphi^{0*} \\
      - \varphi^{+*}
    \end{pmatrix}$  & $\mathbf{1}$
                    & $\mathbf{2}$
                                             & $-1/2$  & $(0,0)$ \\[0.5em]

    \midrule

    $(\tilde\varphi^\dagger)^I = \epsilon^{IJ}\varphi_J =
    \begin{pmatrix} 
      \varphi^{0} &
      - \varphi^{+}
    \end{pmatrix}$   & $\mathbf{1}$
                    & $\bar{\mathbf{2}}$
                                             & $1/2$  & $(0,0)$ \\[0.5em]

    \bottomrule
                                                            
  \end{tabular}
  \caption{Field content of the Standard Model of particle physics. The $\SU(3)$
    and $\SU(2)_L$ columns contain the dimensions of the representations
    respectively, while the $\rmU(1)_Y$ column contains the hypercharges. The
    $\SO^+(3,1)$ column contains the highest-weights $(j_-,j_+)$ of the
    respective fields. \label{tab:particle-content}}
\end{table}

The field content of Standard Model, pre-symmetry breaking, consists of gauge
bosons associated with the local symmetry group
$\SU(3)\times\SU(2)_L\times\rmU(1)_Y$, i.e.
$(G_\mu)_i{}^j = 1/\hbar\,G^a_\mu (T_a)_i{}^j$,
$(W_\mu)_I{}^J = 1/\hbar\,W^A_\mu (T_A)_I{}^J$, and $B_\mu$ respectively,
coupled to matter fields. Here $\mu,\nu,\ldots=0,1,2,3$ are Lorentz indices,
$a,b,\ldots=1,\ldots,8$ are $\mathfrak{su}(3)$ indices, $i,j,\ldots=1,2,3$ are
$\SU(3)$ indices, $A,B,\ldots=1,2,3$ are $\mathfrak{su}(2)_L$ indices, while
$I,J,\ldots=1,2$ are $\SU(2)_L$ indices. Here $T_a = \hbar/2\,\lambda_a$, with
$\lambda_a$ being the Gell-Mann matrices, and $T_A = \hbar/2\,\sigma_A$, with
$\sigma_A$ being the Pauli matrices.

The matter fields can be broadly classified into three sectors: quarks, leptons,
and Higgs. The quark sector contains a left-handed Weyl spinor
$(Q^L_m)^{\dot\alpha}_{Ii}$, transforming in the fundamental (triplet)
representation of $\SU(3)$, fundamental (doublet) representation of $\SU(2)_L$,
and carrying hypercharge $1/6$ under $\rmU(1)_Y$. We denote the respective
$\SU(2)_L$ components as $(Q^L_m)^{\dot\alpha}_{1i}=(u^L_m)^{\dot\alpha}_{i}$
and $(Q^L_m)^{\dot\alpha}_{2i}=(d^L_m)^{\dot\alpha}_{i}$. In addition, we have
two right-handed Weyl spinors $u^R_{m\alpha i}$ and $d^R_{m\alpha i}$
transforming in the fundamental (triplet) representation of $\SU(3)$, trivial
(singlet) representation of $\SU(2)_L$, and carrying $\rmU(1)_Y$ hypercharges
$2/3$ and $-1/3$ respectively.  Here $\alpha,\beta,\ldots=1,2$ and
$\dot\alpha,\dot\beta,\ldots=1,2$ are right- and left-handed Weyl spinor indices
respectively. All the quarks come in 3 \emph{generations}, labelled by the index
$m=1,2,3$. Dropping all spinor and group indices, these are identified with
$u=u_1$ (up), $d=d_1$ (down), $c=u_2$ (charm), $s=d_2$ (strange), $t=u_3$ (top),
and $b=d_3$ (bottom) quarks observed in nature. Anti-quarks are given by the
respective Hermitian conjugate fields.

In the lepton sector, we have a a left-handed Weyl spinor
$(L^L_m)^{\dot\alpha}_{I}$, transforming in the trivial (singlet) representation
of $\SU(3)$, fundamental (doublet) representation of $\SU(2)_L$, and carrying
hypercharge $-1/2$ under $\rmU(1)_Y$. We denote the respective $\SU(2)_L$
components as $(L^L_m)^{\dot\alpha}_{1}=(\nu^L_m)^{\dot\alpha}$ and
$(L^L_m)^{\dot\alpha}_{2}=(e^L_m)^{\dot\alpha}$. In addition, we have a
right-handed Weyl spinor $e^R_{m\alpha}$ transforming in the singlet (trivial)
representations of both $\SU(3)$ and $\SU(2)_L$, and carrying $\rmU(1)_Y$
hypercharge $-1$. Leptons also come in 3 \emph{generations}. Dropping all spinor
and group indices, these are identified with $e=e_1$ (electron), $\nu_e=\nu_1$
(electron neutrino), $\mu=e_2$ (muon), $\nu_\mu=\nu_2$ (muon neutrino),
$\tau=e_3$ (tau), $\nu_\tau=\nu_3$ (tau neutrino). Importantly, there are no
right-handed neutrinos in the Standard Model, which is a vital assumption that
renders the Standard Model neutrinos massless. Anti-leptons are given by the
respective Hermitian conjugate fields.

Finally, in the Higgs sector, we have a single scalar Higgs field $\varphi_I$,
transforming in the trivial (singlet) representation of $\SU(3)$, fundamental
(doublet) representation of $\SU(2)_L$, and carrying hypercharge $1/2$ under
$\rmU(1)_Y$. The Higgs field is assumed to have a spontaneous symmetry breaking
potential. It is also convenient to define a conjugate Higgs field
$\tilde\varphi_I = \epsilon_{IJ}(\varphi^{\dagger})^I$, which also transforms in
the trivial (singlet) representation of $\SU(3)$ and fundamental (doublet)
representation of $\SU(2)_L$, but carries hypercharge $-1/2$ under
$\rmU(1)_Y$. Anti-Higgs and conjugate anti-Higgs fields are given by the
respective Hermitian conjugates. We have summarised the particle content in
\cref{tab:particle-content}.

The symmetry transformations act on various fields according to the
representations stated above. The global Poincar\'e transformations
$(\Lambda^\mu{}_\nu,a^\mu)$ act as
\begin{gather}
  G^a_\mu \to (\Lambda^{-1})^\nu{}_\mu\, G^a_\nu, \qquad
  W^A_\mu \to (\Lambda^{-1})^\nu{}_\mu\, W^A_\nu, \qquad
  B_\mu \to (\Lambda^{-1})^\nu{}_\mu\, B_\nu, \nn\\
  (Q^L_m)^{\dot\alpha}_{Ii}
  \to \bar A^{\dot\alpha}{}_{\dot\beta}\, (Q^L_m)^{\dot\beta}_{Ii}, \qquad
  (Q^{L\dagger}_m)^{\alpha Ii}
  \to (Q^{L\dagger}_m)^{\beta Ii}\, (A^{-1})_\beta{}^\alpha,\nn\\
  (u^R_m)_{\alpha i} \to  A_{\alpha}{}^{\beta}\, (u^R_m)_{\beta i}, \qquad
  (u^{R\dagger}_m)_{\dot\alpha}^i \to  (u^{R\dagger}_m)^i_{\dot\beta}\,
  (\bar A^{-1})^{\dot\beta}{}_{\dot\alpha}, \nn\\
  (d^R_m)_{\alpha i} \to A_{\alpha}{}^{\beta}\, (d^R_m)_{\beta i}, \qquad
  (d^{R\dagger}_m)_{\dot\alpha}^i \to (d^{R\dagger}_m)^i_{\dot\beta}\,
  (\bar A^{-1})^{\dot\beta}{}_{\dot\alpha}, \nn\\
  (L^L_m)^{\dot\alpha}_{I}
  \to  \bar A^{\dot\alpha}{}_{\dot\beta}\,(L^L_m)^{\dot\beta}_{I}, \qquad
  (L^{L\dagger}_m)^{\alpha I}
  \to (L^{L\dagger}_m)^{\beta I}\, (A^{-1})_\beta{}^\alpha, \nn\\
  (e^R_m)_{\alpha} \to A_{\alpha}{}^{\beta}\, (e^R_m)_{\beta}, \qquad
  (e^{R\dagger}_m)_{\dot\alpha} \to (e^{R\dagger}_m)_{\dot\beta}\,
  (\bar A^{-1})^{\dot\beta}{}_{\dot\alpha}, \nn\\
  \varphi_I \to \varphi_I, \qquad
  (\varphi^\dagger)^I \to (\varphi^\dagger)^I, \qquad
  \tilde\varphi_I \to \tilde\varphi_I, \qquad
  (\tilde\varphi^\dagger)^I \to (\tilde\varphi^\dagger)^I.
\end{gather}
Here $A = D_{(1/2,0)}(\Lambda)$ and $\bar A = D_{(0,1/2)}(\Lambda)$. The
spacetime arguments of all the fields also transform under Poincar\'e
transformations as $x^\mu \to (\Lambda^{-1})^\mu{}_\nu x^\nu - a^\mu$, which we
have suppressed in the expressions above for clarity. Note that Hermitian
conjugates of (left-) right-handed spinors transform as (right-) left-handed
anti-fundamental fermions. We have kept all the indices explicit to avoid any
confusion. On the other hand, local $U\in\SU(3)$ transformations act on various
fields as
\begin{gather}
  G^a_\mu (T_a)_i{}^j \to
  U_i{}^k\lb G^a_\mu (T_a)_k{}^l
    + \frac{i\hbar}{g_s} \delta_k^l \dow_\mu \rb (U^{-1})_l{}^j, \nn\\
    (Q^L_m)^{\dot\alpha}_{Ii} \to  U_i{}^j\, (Q^L_m)^{\dot\alpha}_{Ij}, \qquad
    (Q^{L\dagger}_m)^{\alpha Ii}
  \to (Q^{L\dagger}_m)^{\alpha Ij}\, (U^{-1})_j{}^i, \nn\\
  (u^R_m)_{\alpha i} \to  U_i{}^j\, (u^R_m)_{\alpha j}, \qquad
  (u^{R\dagger}_m)_{\dot\alpha}^i \to  (u^{R\dagger}_m)^j_{\dot\alpha}\,
  (U^{-1})_j{}^i, \nn\\
  (d^R_m)_{\alpha i} \to  U_i{}^j\, (d^R_m)_{\alpha j},  \qquad
  (d^{R\dagger}_m)_{\dot\alpha}^i \to  (d^{R\dagger}_m)^j_{\dot\alpha}\,
  (U^{-1})_j{}^i,
\end{gather}
while all other fields are $\SU(3)$-invariant. Here $g_s$ is the strong force
coupling constant. Similarly, local $V\in\SU(2)_L$
transformations act as
\begin{gather}
  W^A_\mu(T_A)_I{}^J \to V_I{}^K\lb
  W^A_\mu (T_A)_{K}{}^L
    + \frac{i\hbar}{g_w} \delta_K^L \dow_\mu \rb (V^{-1})_L{}^J, \nn\\
    (Q^L_m)^{\dot\alpha}_{Ii} \to  V_I{}^J\, (Q^L_m)^{\dot\alpha}_{Ji}, \qquad
    (Q^L_m)^{\alpha Ii} \to (Q^L_m)^{\alpha Ji}\, (V^{-1})_J{}^I, \nn\\
    (L^L_m)^{\dot\alpha}_{I}(x) \to  V_I{}^J\, (L^L_m)^{\dot\alpha}_{J}, \qquad
    (L^L_m)^{\alpha I} \to (L^L_m)^{\alpha J}\, (V^{-1})_J{}^I,\nn\\
  \varphi_I \to V_I{}^J\, \varphi_J,\qquad
  (\varphi^\dagger)^I \to (\varphi^\dagger)^J\, (V^{-1})_J{}^I, \nn\\
  \tilde\varphi_I \to V_I{}^J\, \tilde\varphi_J, \qquad
  (\tilde\varphi^\dagger)^I \to (\tilde\varphi^\dagger)^J\, (V^{-1})_J{}^I,
\end{gather}
where $g_w$ is the weak force coupling constant. Finally, local
$\E{i\theta}\in\rmU(1)_Y$ transformations act as
\begin{gather}
  B_\mu \to B_\mu
  + \frac{i}{g_y} \E{i\theta}\dow_\mu\E{-i\theta}
  = B_\mu + \frac{1}{g_y}\dow_\mu\theta, \nn\\
  (Q^L_m)^{\dot\alpha}_{Ii} \to \E{i\theta/6}\,(Q^L_m)^{\dot\alpha}_{Ii}, \qquad
  (Q^{L\dagger}_m)^{\alpha Ii} \to \E{-i\theta/6}\,(Q^{L\dagger}_m)^{\alpha Ii}, \nn\\
  (u^R_m)_{\alpha i} \to \E{2i\theta/3}\,(u^R_m)_{\alpha i}, \qquad
  (u^{R\dagger}_m)_{\dot\alpha}^i \to
  \E{-2i\theta/3}\,(u^{R\dagger}_m)_{\dot\alpha}^i, \nn\\
  (d^R_m)_{\alpha i} \to  \E{-i\theta/3}\,(d^R_m)_{\alpha i}, \qquad
  (d^{R\dagger}_m)_{\dot\alpha}^i \to \E{i\theta/3}\, (d^{R\dagger}_m)_{\dot\alpha}^i, \nn\\
  (L^L_m)^{\dot\alpha}_{I} \to \E{-i\theta/2}\,(L^L_m)^{\dot\alpha}_{I}, \qquad
  (L^{L\dagger}_m)^{\alpha I} \to \E{i\theta/2}\, (L^{L\dagger}_m)^{\alpha I},
  \nn\\
  (e^R_m)_{\alpha} \to  \E{-i\theta}\,(e^R_m)_\alpha, \qquad
  (e^{R\dagger}_m)_{\dot\alpha} \to  \E{i\theta}\,(e^{R\dagger}_m)_{\dot\alpha}, \nn\\
  \varphi_I\to \E{i\theta/2}\,\varphi_I,\qquad
  (\varphi^\dagger)^I \to \E{-i\theta/2}\,(\varphi^\dagger)^I, \nn\\
  \tilde\varphi_I \to \E{-i\theta/2}\,\tilde\varphi_I, \qquad
  (\tilde\varphi^\dagger)^I \to \E{i\theta/2}\,(\tilde\varphi^\dagger)^I,
\end{gather}
where $g_y$ is the hypercharge coupling constant. The Standard Model of
particle physics is required to be invariant under all of these transformations.

\subsection{Standard Model Lagrangian}

We will now write down the Lagrangian for the Standard Model based on our
discussion in the course. The Lagrangian will be required to be invariant under
all the spacetime and internal symmetries mentioned above. The construction,
pretty much, follows from our discussion in \cref{sec:QED-QCD,sec:electroweak}.

\subsubsection{Gauge sector}

The most straight-forward contribution to the Lagrangian comes from the gauge
fields. We simply get the respective Yang-Mills terms
\begin{align}
  \frac{1}{\hbar c}\mathcal{L}_{\text{gauge}}
  &= - \frac{1}{2} \tr(G_{\mu\nu} G^{\mu\nu})
    - \frac{1}{2} \tr(W_{\mu\nu} W^{\mu\nu})
    - \frac{1}{4} B_{\mu\nu} B^{\mu\nu} \nn\\
  &= - \frac{1}{4} G^a_{\mu\nu} G^{\mu\nu}_a
    - \frac{1}{4} W^A_{\mu\nu} W^{\mu\nu}_A
    - \frac{1}{4} B_{\mu\nu} B^{\mu\nu}.
\end{align}
Here, the field strengths are defined as
\begin{align}
  G_{\mu\nu}
  &= \dow_\mu G_\nu - \dow_\nu G_\mu - ig_s[G_\mu,G_\nu], \nn\\
  W_{\mu\nu}
  &= \dow_\mu W_\nu - \dow_\nu W_\mu - ig_w[W_\mu,W_\nu], \nn\\
  B_{\mu\nu}
  &= \dow_\mu B_\nu - \dow_\nu B_\mu.
\end{align}
To avoid confusion, we also note these in components
\begin{align}
  G^a_{\mu\nu}
  &= \dow_\mu G^a_\nu - \dow_\nu G^a_\mu + g_sG_\mu^b G_\nu^c f_{bca}, \nn\\
  W^A_{\mu\nu}
  &= \dow_\mu W^A_\nu - \dow_\nu W^A_\mu + g_w W^B_\mu W^C_\nu \epsilon_{BCA}, \nn\\
  B_{\mu\nu}
  &= \dow_\mu B_\nu - \dow_\nu B_\mu,
\end{align}
where $f_{abc}$ are the $\SU(3)$ structure constants, while the Levi-Civita
$\epsilon_{ABC}$ functions as the structure constants of $\SU(2)_L$.

\subsubsection{Higgs sector}

The Lagrangian for the Higgs field comprises of the kinetic term with a SSB
potential
\begin{align}
  \frac{1}{\hbar c}\mathcal{L}_{\text{higgs}}
  &= - \Df_\mu\varphi^\dagger \Df^\mu\varphi + \mu^2\varphi^\dagger\varphi
  - \lambda (\varphi^\dagger\varphi)^2 \nn\\
  &= - (\Df_\mu\varphi^\dagger)^I (\Df^\mu\varphi)_I
  + \mu^2(\varphi^\dagger)^I\varphi_I
  - \lambda \lb (\varphi^\dagger)^I\varphi_I \rb^2,
\end{align}
where $\mu$ and $\lambda$ are constants. The covariant derivatives are defined
as
\begin{align}
  \Df_\mu \varphi
  &\to \dow_\mu \varphi - ig_w W_\mu \varphi
    - \frac{i}{2} g_y B_\mu \varphi, \nn\\
  \Df_\mu \varphi^\dagger
  &\to \dow_\mu \varphi^\dagger + ig_w \varphi^\dagger W_\mu
    + \frac{i}{2} g_y \varphi^\dagger B_\mu,
\end{align}
or in components
\begin{align}
  (\Df_\mu\varphi)_{I}
  &\to \dow_\mu \varphi_{I}
    - \frac{ig_w}{\hbar} W^A_\mu (T_A)_I{}^J \varphi_J
    - \frac{i}{2} g_y B_\mu \varphi_I, \nn\\
  (\Df_\mu \varphi^\dagger)^{I}
  &\to \dow_\mu (\varphi^\dagger)^{I}
    + \frac{ig_w}{\hbar} (\varphi^\dagger)^{J} (T_A)_J{}^I W^A_\mu
    + \frac{i}{2} g_y (\varphi^\dagger)^{I} B_\mu.
\end{align}

\subsubsection{Quark sector} 

Next, we have the quark sector. The Lagrangian is given as
\begin{align}
  \frac{1}{\hbar c}\mathcal{L}_{\text{quark}}
  &= \frac{i}{2} \lb Q^{L\dagger}_m \sigma^\mu\Df_\mu Q^L_m
    - \Df_\mu Q^{L\dagger}_m \sigma^\mu Q^L_m
    \rb \nn\\
  &\qquad
    + \frac{i}{2} \lb u^{R\dagger}_m \bar\sigma^\mu\Df_\mu u^R_m
    - \Df_\mu u^{R\dagger}_m \bar\sigma^\mu u^R_m
    \rb
    + \frac{i}{2} \lb d^{R\dagger}_m \bar\sigma^\mu\Df_\mu d^R_m
    - \Df_\mu d^{R\dagger}_m \bar\sigma^\mu d^R_m
    \rb \nn\\
  &\qquad
    - Y^d_{mn}\lb Q^{L\dagger}_m\varphi\,d^R_n + d^{R\dagger}_n\varphi^\dagger
    Q^L_m \rb
    - Y^u_{mn}\lb Q^{L\dagger}_m\tilde\varphi\,u^R_n
    + u^{R\dagger}_n\tilde\varphi^\dagger Q^L_m \rb \nn\\
  &= \frac{i}{2} \lb (Q^{L\dagger}_m)^{\alpha Ii}
    \sigma^\mu_{\alpha\dot\alpha}
    (\Df_\mu Q^L_m)^{\dot\alpha}_{Ii}
    - (\Df_\mu Q^{L\dagger}_m)^{\alpha Ii} \sigma^\mu_{\alpha\dot\alpha}
    (Q^L_m)^{\dot\alpha}_{Ii}
    \rb \nn\\
  &\qquad
    + \frac{i}{2} \lb (u^{R\dagger}_m)_{\dot\alpha}^i
    \bar\sigma^{\mu\dot\alpha \alpha} (\Df_\mu u^R_m)_{\alpha i}
    - (\Df_\mu u^{R\dagger}_m)_{\dot\alpha}^i
    \bar\sigma^{\mu\dot\alpha \alpha} (u^R_m)_{\alpha i}
    \rb \nn\\
  &\qquad
    + \frac{i}{2} \lb (d^{R\dagger}_m)_{\dot\alpha}^i
    \bar\sigma^{\mu\dot\alpha \alpha} (\Df_\mu d^R_m)_{\alpha i}
    - (\Df_\mu d^{R\dagger}_m)_{\dot\alpha}^i
    \bar\sigma^{\mu\dot\alpha \alpha} (d^R_m)_{\alpha i}
    \rb \nn\\
  &\qquad - \lb Y^d_{mn} (Q^{L\dagger}_m)^{\alpha Ii}\varphi_I (d^R_n)_{\alpha i}
    + Y^{d*}_{mn} (d^{R\dagger}_n)_{\dot\alpha}^i (\varphi^\dagger)^I
    (Q^L_m)_{Ii}^{\dot\alpha} \rb \nn\\
  &\qquad
    - \lb Y^u_{mn} (Q^{L\dagger}_m)^{\alpha Ii}\tilde\varphi_I (u^R_n)_{\alpha i}
    + Y^{u*}_{mn} (u^{R\dagger}_n)_{\dot\alpha}^i (\tilde\varphi^\dagger)^I
    (Q^L_m)_{Ii}^{\dot\alpha} \rb.
\end{align}
Sum over the generation index $m$ is understood. Here $Y^d_{mn}$ and $Y^u_{mn}$
are arbitrary $3\times 3$ Yukawa coupling complex matrices that will determine
the quark masses. Note that the regular Dirac mass terms for quarks,
e.g. $u_m^{L\dagger}u_n^R+u_n^{R\dagger}u_m^L$, are not permitted by $\SU(2)_L$
invariance, which only rotates the left-handed quarks and leaves the
right-handed quarks invariant.

The covariant derivatives of quarks are defined according to their
transformation properties
\begin{align}
  \Df_\mu Q^L_m
  &\to \dow_\mu Q^L_m - ig_s G_\mu Q^L_m - ig_w W_\mu Q^L_m
    - \frac{i}{6} g_y B_\mu Q^L_m, \nn\\
  \Df_\mu Q^{L\dagger}_m
  &\to \dow_\mu Q^{L\dagger}_m
    + ig_s Q^{L\dagger}_m G_\mu + ig_w Q^{L\dagger}_m W_\mu
    + \frac{i}{6} g_y Q^{L\dagger}_m B_\mu, \nn\\
  \Df_\mu u^R_m
  &\to \dow_\mu u^R_m - ig_s G_\mu u^R_m 
    - \frac{2i}{3} g_y B_\mu u^R_m, \nn\\
  \Df_\mu u^{R\dagger}_m
  &\to \dow_\mu u^{R\dagger}_m
    + ig_s u^{R\dagger}_m G_\mu
    + \frac{2i}{3} g_y u^{R\dagger}_m B_\mu, \nn\\
  \Df_\mu d^R_m
  &\to \dow_\mu d^R_m - ig_s G_\mu d^R_m 
    + \frac{i}{3} g_y B_\mu d^R_m, \nn\\
  \Df_\mu d^{R\dagger}_m
  &\to \dow_\mu d^{R\dagger}_m
    + ig_s d^{R\dagger}_m G_\mu
    - \frac{i}{3} g_y d^{R\dagger}_m B_\mu,
\end{align}
or in components
\begin{align}
  (\Df_\mu Q^L_m)^{\dot\alpha}_{Ii}
  &\to \dow_\mu (Q^L_m)^{\dot\alpha}_{Ii}
    - \frac{ig_s}{\hbar} G_\mu^a (T_a)_i{}^j (Q^L_m)^{\dot\alpha}_{Ij}
    - \frac{ig_w}{\hbar} W^A_\mu (T_A)_I{}^J (Q^L_m)^{\dot\alpha}_{Ji}
    - \frac{i}{6} g_y B_\mu (Q^L_m)^{\dot\alpha}_{Ii}, \nn\\
  (\Df_\mu Q^{L\dagger}_m)^{\alpha Ii}
  &\to \dow_\mu (Q^{L\dagger}_m)^{\alpha Ii}
    + \frac{ig_s}{\hbar} (Q^{L\dagger}_m)^{\alpha Ij} (T_a)_j{}^i G^a_\mu
    + \frac{ig_w}{\hbar} (Q^{L\dagger}_m)^{\alpha Ji} (T_A)_J{}^I W^A_\mu
    + \frac{i}{6} g_y (Q^{L\dagger}_m)^{\alpha Ii} B_\mu, \nn\\
  (\Df_\mu u^R_m)_{\alpha i}
  &\to \dow_\mu (u^R_m)_{\alpha i}
    - \frac{ig_s}{\hbar} G^a_\mu (T_a)_i{}^j (u^R_m)_{\alpha j}
    - \frac{2i}{3} g_y B_\mu (u^R_m)_{\alpha i}, \nn\\
  (\Df_\mu u^{R\dagger}_m)^i_{\dot\alpha}
  &\to \dow_\mu (u^{R\dagger}_m)^i_{\dot\alpha}
    + \frac{ig_s}{\hbar} (u^{R\dagger}_m)^j_{\dot\alpha} (T_a)_j{}^i G_\mu^a
    + \frac{2i}{3} g_y (u^{R\dagger}_m)^i_{\dot\alpha} B_\mu, \nn\\
  (\Df_\mu d^R_m)_{\alpha i}
  &\to \dow_\mu (d^R_m)_{\alpha i}
    - \frac{ig_s}{\hbar} G^a_\mu (T_a)_i{}^j (d^R_m)_{\alpha j}
    + \frac{i}{3} g_y B_\mu (d^R_m)_{\alpha i}, \nn\\
  (\Df_\mu d^{R\dagger}_m)^i_{\dot\alpha}
  &\to \dow_\mu (u^{R\dagger}_m)^i_{\dot\alpha}
    + \frac{ig_s}{\hbar} (d^{R\dagger}_m)^j_{\dot\alpha} (T_a)_j{}^i G_\mu^a
    - \frac{i}{3} g_y (d^{R\dagger}_m)^i_{\dot\alpha} B_\mu.
\end{align}
It can explicitly be checked that the Lagrangian is invariant under all the
symmetries. 

\subsubsection{Lepton sector}

Next, we have the lepton sector. The Lagrangian is given as
\begin{align}
  \frac{1}{\hbar c} \mathcal{L}_{\text{lepton}}
  &= \frac{i}{2} \lb L^{L\dagger}_m \sigma^\mu\Df_\mu L^L_m
    - \Df_\mu L^{L\dagger}_m \sigma^\mu L^L_m
    \rb
    + \frac{i}{2} \lb e^{R\dagger}_m \bar\sigma^\mu\Df_\mu e^R_m
    - \Df_\mu e^{R\dagger}_m \bar\sigma^\mu e^R_m
    \rb \nn\\
  &\qquad
    - Y^e_{mn}\lb L^{L\dagger}_m \varphi\,e^R_n
    + e^{R\dagger}_n \varphi^\dagger L^L_m \rb \nn\\
  &= \frac{i}{2} \lb (L^{L\dagger}_m)^{\alpha I}
    \sigma^\mu_{\alpha\dot\alpha}
    (\Df_\mu L^L_m)^{\dot\alpha}_{I}
    - (\Df_\mu L^{L\dagger}_m)^{\alpha I} \sigma^\mu_{\alpha\dot\alpha}
    (L^L_m)^{\dot\alpha}_{I}
    \rb \nn\\
  &\qquad
    + \frac{i}{2} \lb (e^{R\dagger}_m)_{\dot\alpha}
    \bar\sigma^{\mu\dot\alpha \alpha} (\Df_\mu e^R_m)_{\alpha}
    - (\Df_\mu e^{R\dagger}_m)_{\dot\alpha}
    \bar\sigma^{\mu\dot\alpha \alpha} (e^R_m)_{\alpha}
    \rb \nn\\
  &\qquad
    - \lb Y^e_{mn} (L^{L\dagger}_m)^{\alpha I}\varphi_I(e^R_n)_\alpha
    +  Y^{e*}_{mn}
    (e^{R\dagger}_n)_{\dot\alpha}(\varphi^\dagger)^I (L^L_m)^{\dot\alpha}_I \rb.
\end{align}
Sum over the generation index $m$ is understood. Here $Y^e_{mn}$ is an arbitrary
$3\times3$ Yukawa coupling complex matrix that determines the mass of electron,
muon, and taon. A similar Yukawa mass term for neutrinos is not admitted by the
Standard Model due the absence of right-handed neutrinos.

The covariant derivatives appearing above are defined as
\begin{align}
  \Df_\mu L^L_m
  &\to \dow_\mu L^L_m - ig_w W_\mu L^L_m
    + \frac{i}{2} g_y B_\mu L^L_m, \nn\\
  \Df_\mu L^{L\dagger}_m
  &\to \dow_\mu L^{L\dagger}_m + ig_w L^{L\dagger}_m W_\mu
    - \frac{i}{2} g_y L^{L\dagger}_m B_\mu, \nn\\
  \Df_\mu e^R_m
  &\to \dow_\mu e^R_m + ig_y B_\mu e^R_m, \nn\\
  \Df_\mu e^{R\dagger}_m
  &\to \dow_\mu e^{R\dagger}_m - ig_y e^{R\dagger}_m B_\mu,
\end{align}
or in components
\begin{align}
  (\Df_\mu L^L_m)^{\dot\alpha}_{I}
  &\to \dow_\mu (L^L_m)^{\dot\alpha}_{I}
    - \frac{ig_w}{\hbar} W^A_\mu (T_A)_I{}^J (L^L_m)^{\dot\alpha}_{J}
    + \frac{i}{2} g_y B_\mu (L^L_m)^{\dot\alpha}_{I}, \nn\\
  (\Df_\mu L^{L\dagger}_m)^{\alpha I}
  &\to \dow_\mu (L^{L\dagger}_m)^{\alpha I}
    + \frac{ig_w}{\hbar} (L^{L\dagger}_m)^{\alpha J} (T_A)_J{}^I W^A_\mu
    - \frac{i}{2} g_y (L^{L\dagger}_m)^{\alpha I} B_\mu, \nn\\
  (\Df_\mu e^R_m)_{\alpha}
  &\to \dow_\mu (e^R_m)_{\alpha}
    + ig_y B_\mu (e^R_m)_{\alpha}, \nn\\
  (\Df_\mu e^{R\dagger}_m)_{\dot\alpha}
  &\to \dow_\mu (e^{R\dagger}_m)_{\dot\alpha}
    - ig_y (e^{R\dagger}_m)_{\dot\alpha} B_\mu.
\end{align}
It can explicitly be checked that the Lagrangian is invariant under all the
symmetries.

\subsection{Higgs mechanism and mass generation}

Following our discussion from the electroweak theory, the Higgs potential has a
minima at $\varphi^\dagger\varphi = v^2/2$, where $v=\mu/\sqrt{\lambda}$. Let us
spontaneously choose the minima $\varphi_{0I}=v/\sqrt{2}\,\hat\varphi_{0I}$,
where $\hat\varphi_{0I} = (0,1)$. We can expand the Standard Model Lagrangian
around this state according to
\begin{equation}
  \varphi_I = \frac{1}{\sqrt 2}(v+\eta)\,
  \E{i/2\,\pi}(V_\pi)_I{}^J\hat\varphi_{0J},
\end{equation}
where $V_\pi = \exp(i/\hbar\,\pi^AT_A)$, with $\pi^A$ and $\pi$ being the
plausible Goldstone fields capturing the fluctuations in
$\SU(2)_L\times\rmU(1)_Y$ phase of $\varphi$, while $\eta$ being the fluctuation
in the magnitude. We follow this by a redefinition of the remaining fields
\begin{gather}
  G^a_\mu \to G^a_\mu, \qquad W^A_\mu(T_A)_I{}^J \to (V_\pi)_I{}^K\lb W^A_\mu
  (T_A)_{K}{}^L + \frac{i\hbar}{g_w} \delta_K^L \dow_\mu \rb (V_\pi^{-1})_L{}^J,
  \qquad
  B_\mu \to B_\mu + \frac{1}{g_y}\dow_\mu\pi, \nn\\
  (Q^L_m)^{\dot\alpha}_{Ii} \to \E{i\pi/6}\,(V_\pi)_I{}^J\,
  (Q^L_m)^{\dot\alpha}_{Ji}, \qquad
  (u^R_m)_{\alpha i} \to \E{2i\pi/3}\,(u^R_m)_{\alpha i}, \qquad
  (d^R_m)_{\alpha i} \to  \E{-i\pi/3}\,(d^R_m)_{\alpha i}, \nn\\
  (L^L_m)^{\dot\alpha}_{I} \to \E{-i\pi/2}\,(V_\pi)_I{}^J\,
  (L^L_m)^{\dot\alpha}_{J}, \qquad
  (e^R_m)_{\alpha} \to  \E{-i\pi}\,(e^R_m)_\alpha,
\end{gather}
that, being a symmetry transformation, renders the Lagrangian independent of the
$\pi^A$ and $\pi$ fields. The only effect of the procedure above on the
Lagrangian is to the Higgs field
$\varphi_I \to 1/\sqrt{2}(v+\eta)\hat\varphi_{0I}$. Note that this sets the
conjugate Higgs to
$\tilde\varphi_I \to 1/\sqrt{2}(v+\eta)\hat{\tilde\varphi}_{0I}$, where
$\hat{\tilde\varphi}_{0I} = (0,1)$. In the following, we shall explore the
repercussions of this on the Standard Model Lagrangian.

\subsubsection{Higgs and gauge sector}

The Higgs and gauge sector Lagrangian after spontaneous symmetry breaking
follows directly from our electroweak discussion. We get
\begin{align}
  \frac{1}{\hbar c}\Big( \mathcal{L}^{\text{SSB}}_{\text{higgs}} 
  &+ \mathcal{L}_{\text{gauge}}^{\text{SSB}} \Big) \nn\\
  &=
    - \half \dow_\mu\eta\dow^\mu\eta
    + \half \mu^2 (v+\eta)^2
    - \frac14 \lambda (v+\eta)^4
    - \frac14 (v+\eta)^2 g_w^2 \lb W^-_\mu W^{+\mu}
    + \half \sec^2\theta_w\, Z^\mu Z_\mu \rb \nn\\
  &\qquad
    - \frac{1}{4} G^a_{\mu\nu} G^{\mu\nu}_a
    - \frac{1}{2} W^+_{\mu\nu} W^{-\mu\nu}
    - \frac{1}{4} Z_{\mu\nu} Z^{\mu\nu}
    - \frac{1}{4} F_{\mu\nu} F^{\mu\nu} \nn\\
  &\qquad
    - ig_w \cos\theta_w \lb W^{-\mu\nu} W^+_{\mu} Z_{\nu}
    - W^{+\mu\nu}  W^-_{\mu} Z_{\nu}
    - Z^{\mu\nu} W^+_{\mu} W^-_{\nu} \rb
    + ig_w \sin\theta_w F^{\mu\nu} W^+_{\mu}W^-_{\nu} \nn\\
  &\qquad
    - g_w^2 \cos^2\theta_w \lb W^+_{\mu} W^{-\mu}Z_{\nu}Z^{\nu}
    - W^+_{\nu}Z^{\nu} W^-_\mu Z^{\mu} \rb \nn\\
  &\qquad
    + \half g^2_w \lb W^+_{\mu}W^{+\mu} W^-_{\nu} W^{-\nu}
    - (W^+_{\mu}W^{-\mu})^2 \rb.
\end{align}
Focusing on the quadratic sector and ignoring interactions, we get
\begin{align}
  \frac{1}{\hbar c}\Big( \mathcal{L}^{\text{SSB}}_{\text{higgs}} 
  &+ \mathcal{L}_{\text{gauge}}^{\text{SSB}} \Big) \nn\\
  &= - \frac{1}{4} G^a_{\mu\nu} G^{\mu\nu}_a
    - \frac{1}{2} W^-_{\mu\nu} W^{+\mu\nu}
    - \frac{m_W^2c^2}{\hbar^2} W^-_\mu W^{+\mu}
    - \frac{1}{4} Z_{\mu\nu} Z^{\mu\nu}
    - \frac{m_Z^2c^2}{2\hbar^2} Z^\mu Z_\mu
    - \frac{1}{4} F_{\mu\nu} F^{\mu\nu} \nn\\
  &\qquad
    - \half \dow_\mu\eta\dow^\mu\eta
    - \mu^2\eta^2
    + \text{interactions}.
\end{align}
We get a massive residual Higgs field $\eta$ with mass $\sqrt2\mu\hbar/c$, three
massive weak force gauge fields $W^\pm_\mu$, $Z_\mu$ with respective masses
$m_W = v/2\, g_w \hbar/c $, $m_Z = m_W\sec\theta_w$, one massless
electromagnetic gauge field (photon) $A_\mu$, and eight massless strong force
gauge fields (gluons) $G^a_\mu$.

The interactions in this sector are also the same as the electroweak
interactions in \cref{sec:electroweak-interactions}, but with additional gluon
self interactions. We have the self and weak interactions of the Higgs field:
\begin{gather*}
  \begin{tikzpicture}
    \draw[thick,dashed] (-1,1) -- (0,0);
    \node at (-1.2,0.9) {$\eta$};
    \draw[thick,dashed] (-1,-1) -- (0,0);
    \node at (-1.2,-0.9) {$\eta$};
    \draw[thick,dashed] (0,0) -- (1.2,0);
    \node at (1.4,0) {$\eta$};
  \end{tikzpicture} \qquad\qquad
  \begin{tikzpicture}
    \draw[thick,dashed] (-1,1) -- (0,0);
    \node at (-1.2,0.9) {$\eta$};
    \draw[thick,dashed] (-1,-1) -- (0,0);
    \node at (-1.2,-0.9) {$\eta$};
    \draw[thick,dashed] (0,0) -- (1,1);
    \node at (1.2,0.9) {$\eta$};
    \draw[thick,dashed] (0,0) -- (1,-1);
    \node at (1.2,-0.9) {$\eta$};
  \end{tikzpicture} \nn\\
  \begin{tikzpicture}
    \draw[thick,dashed] (-1.2,0) -- (0,0);
    \node at (-1.4,0) {$\eta$};
    \draw[thick,weak] (0,0) -- (1,1);
    \node at (1.4,0.9) {$W^+$};
    \draw[thick,weak] (0,0) -- (1,-1);
    \node at (1.4,-0.9) {$W^-$};
  \end{tikzpicture} \qquad
  \begin{tikzpicture}
    \draw[thick,dashed] (-1.2,0) -- (0,0);
    \node at (-1.4,0) {$\eta$};
    \draw[thick,weak] (0,0) -- (1,1);
    \node at (1.3,0.9) {$Z$};
    \draw[thick,weak] (0,0) -- (1,-1);
    \node at (1.3,-0.9) {$Z$};
  \end{tikzpicture} \qquad
  \begin{tikzpicture}
    \draw[thick,dashed] (-1,1) -- (0,0);
    \node at (-1.2,0.9) {$\eta$};
    \draw[thick,dashed] (-1,-1) -- (0,0);
    \node at (-1.2,-0.9) {$\eta$};
    \draw[thick,weak] (0,0) -- (1,1);
    \node at (1.4,0.9) {$W^+$};
    \draw[thick,weak] (0,0) -- (1,-1);
    \node at (1.4,-0.9) {$W^-$};
  \end{tikzpicture} \qquad
  \begin{tikzpicture}
    \draw[thick,dashed] (-1,1) -- (0,0);
    \node at (-1.2,0.9) {$\eta$};
    \draw[thick,dashed] (-1,-1) -- (0,0);
    \node at (-1.2,-0.9) {$\eta$};
    \draw[thick,weak] (0,0) -- (1,1);
    \node at (1.3,0.9) {$Z$};
    \draw[thick,weak] (0,0) -- (1,-1);
    \node at (1.3,-0.9) {$Z$};
  \end{tikzpicture}
\end{gather*}
mixed interactions between photon and weak bosons:
\begin{equation*}
   \qquad\qquad
  \begin{tikzpicture}
    \draw[thick,weak] (-1,1) -- (0,0); 
    \node at (-1.4,0.9) {$W^+$};
    \draw[thick,weak] (-1,-1) -- (0,0);
    \node at (-1.4,-0.9) {$W^-$};
    \draw[thick,wave] (0,0) -- (1.2,0);
    \node at (1.4,0) {$A$};
  \end{tikzpicture} \qquad
  \begin{tikzpicture} 
    \draw[thick,weak] (-1,1) -- (0,0);
    \node at (-1.4,0.9) {$W^+$};
    \draw[thick,weak] (-1,-1) -- (0,0);
    \node at (-1.4,-0.9) {$W^-$};
    \draw[thick,wave] (0,0) -- (1,1);
    \node at (1.2,0.9) {$A$};
    \draw[thick,wave] (0,0) -- (1,-1);
    \node at (1.2,-0.9) {$A$};
  \end{tikzpicture} \qquad\qquad
  \begin{tikzpicture} 
    \draw[thick,weak] (-1,1) -- (0,0);
    \node at (-1.4,0.9) {$W^+$};
    \draw[thick,weak] (-1,-1) -- (0,0);
    \node at (-1.4,-0.9) {$W^-$};
    \draw[thick,weak] (0,0) -- (1,1);
    \node at (1.2,0.9) {$Z$};
    \draw[thick,wave] (0,0) -- (1,-1);
    \node at (1.2,-0.9) {$A$};
  \end{tikzpicture}
\end{equation*}
self interactions of weak bosons:
\begin{equation*}
  \begin{tikzpicture}
    \draw[thick,weak] (-1,1) -- (0,0); 
    \node at (-1.4,0.9) {$W^+$};
    \draw[thick,weak] (-1,-1) -- (0,0);
    \node at (-1.4,-0.9) {$W^-$};
    \draw[thick,weak] (0,0) -- (1.2,0);
    \node at (1.4,0) {$Z$};
  \end{tikzpicture} \qquad\qquad
  \begin{tikzpicture} 
    \draw[thick,weak] (-1,1) -- (0,0);
    \node at (-1.4,0.9) {$W^+$};
    \draw[thick,weak] (-1,-1) -- (0,0);
    \node at (-1.4,-0.9) {$W^-$};
    \draw[thick,weak] (0,0) -- (1,1);
    \node at (1.2,0.9) {$Z$};
    \draw[thick,weak] (0,0) -- (1,-1);
    \node at (1.2,-0.9) {$Z$};
  \end{tikzpicture} \qquad\qquad
  \begin{tikzpicture} 
    \draw[thick,weak] (-1,1) -- (0,0);
    \node at (-1.4,0.9) {$W^+$};
    \draw[thick,weak] (-1,-1) -- (0,0);
    \node at (-1.4,-0.9) {$W^-$};
    \draw[thick,weak] (0,0) -- (1,1);
    \node at (1.4,0.9) {$W^+$};
    \draw[thick,weak] (0,0) -- (1,-1);
    \node at (1.4,-0.9) {$W^-$};
  \end{tikzpicture}
\end{equation*}
and self interactions of gluons
\begin{equation*}
  \begin{tikzpicture}
    \draw[thick,coil] (-1,1) -- (0,0);
    \node at (-1.2,0.9) {$G$};
    \draw[thick,coil] (-1,-1) -- (0,0);
    \node at (-1.2,-0.9) {$G$};
    \draw[thick,coil] (0,0) -- (1.2,0);
    \node at (1.4,0) {$G$};
  \end{tikzpicture} \qquad\qquad
  \begin{tikzpicture}
    \draw[thick,coil] (-1,1) -- (0,0);
    \node at (-1.2,0.9) {$G$};
    \draw[thick,coil] (-1,-1) -- (0,0);
    \node at (-1.2,-0.9) {$G$};
    \draw[thick,coil] (0,0) -- (1,1);
    \node at (1.2,0.9) {$G$};
    \draw[thick,coil] (0,0) -- (1,-1);
    \node at (1.2,-0.9) {$G$};
  \end{tikzpicture}
\end{equation*}
Note that the residual Higgs field is uncharged under the residual
$\SU(3)\times\rmU(1)$ symmetry, hence it does not interact with photons or
gluons.

\subsubsection{Quark sector}

Spontaneous symmetry breaking of the quark sector Lagrangian follows from the
generic discussion in \cref{sec:electroweak-Higgs}, with $\chi=Q^L_m$ and
$\psi=u^R_m$, $d^R_m$
\begin{align}
  \frac{1}{\hbar c}\mathcal{L}_{\text{quark}}^{\text{SSB}}
  &= \frac{i}{2} \lb u^{R\dagger}_m \bar\sigma^{\mu}\tilde\Df_\mu u^R_m
    - \tilde\Df_\mu u^{R\dagger}_m \bar\sigma^{\mu} u^R_m \rb
    + \frac{i}{2} \lb d^{R\dagger}_m \bar\sigma^{\mu}\tilde\Df_\mu d^R_m
    - \tilde\Df_\mu d^{R\dagger}_m \bar\sigma^{\mu} d^R_m \rb \nn\\
  &\qquad
    + \frac{i}{2} \lb u^{L\dagger}_m \sigma^{\mu}\tilde\Df_\mu u^L_m
    - \tilde\Df_\mu u^{L\dagger}_m \sigma^{\mu} u^L_m \rb
    + \frac{i}{2} \lb d^{L\dagger}_m \sigma^{\mu}\tilde\Df_\mu d^L_m
    - \tilde\Df_\mu d^{L\dagger}_m \sigma^{\mu} d^L_m \rb \nn\\
  &\qquad
    - \lb \lambda_u^R\, u^{R\dagger}_m \bar\sigma^{\mu} u^{R}_m
    + \lambda_d^R\, d^{R\dagger}_m \bar\sigma^{\mu} d^{R}_m
    + \lambda_u^L\, u^{L\dagger}_m \sigma^{\mu} u^{L}_m
    + \lambda_d^L\, d^{L\dagger}_m \sigma^{\mu} d^{L}_m \rb Z_\mu \nn\\
  &\qquad
    + \frac{g_w}{\sqrt2} \lb
    W^+_\mu u^{L\dagger}_m \sigma^{\mu} d^{L}_m
    + W^-_\mu d^{L\dagger}_m \sigma^{\mu} u^{L}_m \rb \nn\\
  &\qquad
    - \frac{v+\eta}{\sqrt2} 
    \lb Y^u_{mn} u^{L\dagger}_mu^{R}_n + Y^{u*}_{mn} u^{R\dagger}_n u^{L}_m \rb
    - \frac{v+\eta}{\sqrt2} 
    \lb Y^d_{mn} d^{L\dagger}_md^{R}_n + Y^{d*}_{mn} d^{R\dagger}_n d^{L}_m \rb,
\end{align}
where
\begin{gather}
  \lambda_u^R = \frac{2}{3} g_y\sin\theta_w, \qquad
  \lambda_d^R = -\frac{1}{3} g_y\sin\theta_w, \nn\\
  \lambda_{u}^L = \frac{1}{6} g_y
  \sin\theta_w - \half g_w \cos\theta_w, \qquad
  \lambda_{d}^L = \frac{1}{6} g_y
  \sin\theta_w + \half g_w \cos\theta_w,
\end{gather}
and the residual $\SU(3)\times\rmU(1)$ covariant derivatives are defined as
\begin{align}
  \tilde\Df_\mu u^{L,R}_m
  &= \dow_\mu u^{L,R}_m - ig_s G_\mu u^{L,R}_m 
    - \frac{2i}{3} e A_\mu u^{L,R}_m, \nn\\
  \tilde\Df_\mu d^{L,R}_m
  &= \dow_\mu d^{L,R}_m - ig_s G_\mu d^{L,R}_m 
    + \frac{i}{3} e A_\mu d^{L,R}_m.
\end{align}

Focusing on the quadratic sector of the Lagrangian, we get
\begin{align}
  \frac{1}{\hbar c}\mathcal{L}_{\text{quark}}^{\text{SSB}}
  &= \frac{i}{2} \lb u^{R\dagger}_m \bar\sigma^{\mu}\dow_\mu u^R_m
    - \dow_\mu u^{R\dagger}_m \bar\sigma^{\mu} u^R_m \rb
    + \frac{i}{2} \lb d^{R\dagger}_m \bar\sigma^{\mu}\dow_\mu d^R_m
    - \dow_\mu d^{R\dagger}_m \bar\sigma^{\mu} d^R_m \rb \nn\\
  &\qquad
    + \frac{i}{2} \lb u^{L\dagger}_m \sigma^{\mu}\dow_\mu u^L_m
    - \dow_\mu u^{L\dagger}_m \sigma^{\mu} u^L_m \rb
    + \frac{i}{2} \lb d^{L\dagger}_m \sigma^{\mu}\dow_\mu d^L_m
    - \dow_\mu d^{L\dagger}_m \sigma^{\mu} d^L_m \rb \nn\\
  &\qquad
    - \frac{v}{\sqrt2} 
    \lb Y^u_{mn} u^{L\dagger}_mu^{R}_n + Y^{u*}_{mn} u^{R\dagger}_n u^{L}_m \rb
    - \frac{v}{\sqrt2} 
    \lb Y^d_{mn} d^{L\dagger}_md^{R}_n + Y^{d*}_{mn} d^{R\dagger}_n d^{L}_m \rb
    + \text{int.},
\end{align}
To convert this into a standard form, we can perform a unitary transformation in
the flavour (generation) space
\begin{equation}
  u^{L}_{m}\to \alpha^{uL}_{mn}u^{L}_n, \qquad
  u^{R}_{m}\to \alpha^{uR}_{mn}u^{R}_n, \qquad
  d^{L}_{m}\to \alpha^{dL}_{mn}d^{L}_n, \qquad
  d^{R}_{m}\to \alpha^{dR}_{mn}d^{R}_n,
\end{equation}
for arbitrary unitary matrices $\alpha^{uL}_{mn}$, $\alpha^{uR}_{mn}$,
$\alpha^{dL}_{mn}$, and $\alpha^{dR}_{mn}$, such that the transformed Yukawa
coupling matrix is diagonal with positive eigenvalues\footnote{This can be done
  generically for any non-singular complex matrix $M$. Note that the Hermitian
  matrix $M^\dagger M$ can always be diagonalised by some unitary matrix $U$ as
  $U^\dagger M^\dagger M U = D^2$, where $D^2$ is a matrix with positive
  eigenvalues and $D$ is a matrix with its positive square-roots as diagonal
  entries. Then, we can check that $V^\dagger M U = D$ where $V = M U D^{-1}$ is
  also a unitary matrix.}
\begin{equation}
  \frac{v\hbar}{\sqrt{2} c}\alpha^{uL\dagger}_{mr}\, Y_{rs}^{u}\, \alpha^{uR}_{sn}
  = m^u_{mn}, \qquad
  \frac{v\hbar}{\sqrt{2} c}\alpha^{dL\dagger}_{mr}\, Y_{rs}^{d}\, \alpha^{dR}_{sn}
  = m^d_{mn},
\end{equation}
where $m^{u,d}_{mn}$ are diagonal matrices with positive entries. The
transformed free Lagrangian can be represented in the Dirac representation by
defining 4-component Dirac spinors
\begin{equation}
  u_m =
  \begin{pmatrix}
    u_m^R \\ u_m^L
  \end{pmatrix}, \qquad
  d_m =
  \begin{pmatrix}
    d_m^R \\ d_m^L
  \end{pmatrix},
\end{equation}
leading to
\begin{align}
  \mathcal{L}_{\text{quark}}^{\text{SSB}}
  &= \frac{i\hbar c}{2} \lb \bar u_m \gamma^\mu\dow_\mu u_m
    - \dow_\mu \bar u_m \gamma^\mu u_m \rb
    - m^u_{mn}c^2 \bar u_m u_n \nn\\
  &\qquad
    + \frac{i\hbar c}{2} \lb \bar d_m \gamma^\mu\dow_\mu d_m
    - \dow_\mu \bar d_m \gamma^\mu d_m \rb
    - m^d_{mn}c^2 \bar d_m d_n
    + \text{interactions},
\end{align}
In this basis, we get the mass $m^{u,d}_{11}$ for up and down quarks,
$m^{u,d}_{22}$ for charm and strange quarks, and $m^{u,d}_{33}$ for top and
bottom quarks.

Shifting to the ``mass basis'', however, has non-trivial implications for
interactions. Since the weak interactions in the forth line mix $u^L_m$ and
$d^L_m$ quarks, which have been redefined independently, these interactions are
not invariant under the said redefinition. To wit, the full quark sector
Lagrangian becomes
\begin{align}
  \frac{1}{\hbar c}\mathcal{L}_{\text{quark}}^{\text{SSB}}
  &= \frac{i}{2} \lb u^{R\dagger}_m \bar\sigma^{\mu}\tilde\Df_\mu u^R_m
    - \tilde\Df_\mu u^{R\dagger}_m \bar\sigma^{\mu} u^R_m \rb
    + \frac{i}{2} \lb d^{R\dagger}_m \bar\sigma^{\mu}\tilde\Df_\mu d^R_m
    - \tilde\Df_\mu d^{R\dagger}_m \bar\sigma^{\mu} d^R_m \rb \nn\\
  &\qquad
    + \frac{i}{2} \lb u^{L\dagger}_m \sigma^{\mu}\tilde\Df_\mu u^L_m
    - \tilde\Df_\mu u^{L\dagger}_m \sigma^{\mu} u^L_m \rb
    + \frac{i}{2} \lb d^{L\dagger}_m \sigma^{\mu}\tilde\Df_\mu d^L_m
    - \tilde\Df_\mu d^{L\dagger}_m \sigma^{\mu} d^L_m \rb \nn\\
  &\qquad
    - \lb \lambda_u^R\, u^{R\dagger}_m \bar\sigma^{\mu} u^{R}_m
    + \lambda_d^R\, d^{R\dagger}_m \bar\sigma^{\mu} d^{R}_m
    + \lambda_u^L\, u^{L\dagger}_m \sigma^{\mu} u^{L}_m
    + \lambda_d^L\, d^{L\dagger}_m \sigma^{\mu} d^{L}_m \rb Z_\mu \nn\\
  &\qquad
    + \frac{g_w}{\sqrt2} \lb
    V_{mn}\,W^+_\mu u^{L\dagger}_m \sigma^{\mu} d^{L}_n
    + V^\dagger_{mn}W^-_\mu d^{L\dagger}_m \sigma^{\mu} u^{L}_n \rb \nn\\
  &\qquad
    - \frac{v+\eta}{v\hbar/c} m^u_{mn}
    \lb u^{L\dagger}_mu^{R}_n + u^{R\dagger}_n u^{L}_m \rb
    - \frac{v+\eta}{v\hbar/c} \tilde m^d_{mn}
    \lb d^{L\dagger}_md^{R}_n + d^{R\dagger}_n d^{L}_m \rb,
    \label{eq:final-quark-sector}
\end{align}
where the CKM (Cabibbo-Kobayashi-Maskawa) matrix $V_{mn}$ is defined as the
unitary matrix
\begin{equation}
  V_{mn} = \alpha^{uL\dagger}_{mr}\alpha^{dL}_{rn}.
\end{equation}
Since the CKM matrix is generally non-diagonal, the weak interactions mediated
by $W^\pm_\mu$ bosons can switch the generation of quarks.\footnote{A unitary
  matrix can generically be parametrised by 9 independent real
  components. However, we can absorb 5 of these in the CKM matrix into a
  redefinition of the left-handed quarks $u^{L}_n$, $d^{L}_n$ with relative
  phases. This leaves us with 4 independent parameters: three Euler angles and a
  complex ``CP-violating'' phase. See \cref{sec:discrete_symmetries} for a
  discussion on CP symmetry and its violation.}  This is to say that, during
interactions, up quarks can turn into charm or top quarks, while down quarks can
turn into strange or bottom quarks, and vice-versa. These are known as quark
flavor oscillations. Experimentally, the CKM matrix is found to be extremely
close to diagonal, meaning that such quark flavor changing interactions are
extremely rare. All in all, quarks can interact via electromagnetic forces
\begin{equation*}
  \begin{tikzpicture}
    \draw[thick] (-1,1) -- (0,0);
    \node at (-2.4,0.9) {$u^L_m$, $d^L_m$, $u^R_m$, $d^R_m$};
    \draw[thick] (-1,-1) -- (0,0);
    \node at (-2.5,-0.9) {$u^{L\dagger}_m$, $d^{L\dagger}_m$,
      $u^{R\dagger}_m$, $d^{R\dagger}_m$};
    \draw[thick,wave] (0,0) -- (1.2,0);
    \node at (1.4,0) {$A$};
  \end{tikzpicture}
\end{equation*}
weak forces
\begin{gather*}
  \begin{tikzpicture}
    \draw[thick] (-1,1) -- (0,0);
    \node at (-2.4,0.9) {$u^L_m$, $d^L_m$, $u^R_m$, $d^R_m$};
    \draw[thick] (-1,-1) -- (0,0);
    \node at (-2.5,-0.9) {$u^{L\dagger}_m$, $d^{L\dagger}_m$,
      $u^{R\dagger}_m$, $d^{R\dagger}_m$};
    \draw[thick,weak] (0,0) -- (1.2,0);
    \node at (1.4,0) {$Z$};
  \end{tikzpicture} \nn\\
  \begin{tikzpicture}
    \draw[thick] (-1,1) -- (0,0); 
    \node at (-1.3,0.9) {$d^L_m$};
    \draw[thick] (-1,-1) -- (0,0);
    \node at (-1.3,-0.9) {$u^{L\dagger}_n$};
    \draw[thick,weak] (0,0) -- (1.2,0);
    \node at (1.6,0) {$W^+$};
  \end{tikzpicture} \qquad\qquad
  \begin{tikzpicture}
    \draw[thick] (-1,1) -- (0,0); 
    \node at (-1.3,0.9) {$u^L_m$};
    \draw[thick] (-1,-1) -- (0,0);
    \node at (-1.3,-0.9) {$d^{L\dagger}_n$};
    \draw[thick,weak] (0,0) -- (1.2,0);
    \node at (1.6,0) {$W^-$};
  \end{tikzpicture}
\end{gather*}
strong forces
\begin{equation*}
  \begin{tikzpicture}
    \draw[thick] (-1,1) -- (0,0);
    \node at (-2.4,0.9) {$u^L_m$, $d^L_m$, $u^R_m$, $d^R_m$};
    \draw[thick] (-1,-1) -- (0,0);
    \node at (-2.5,-0.9) {$u^{L\dagger}_m$, $d^{L\dagger}_m$,
      $u^{R\dagger}_m$, $d^{R\dagger}_m$};
    \draw[thick,coil] (0,0) -- (1.2,0);
    \node at (1.4,0) {$G$};
  \end{tikzpicture}
\end{equation*}
and the residual Higgs field
\begin{gather*}
  \begin{tikzpicture}
    \draw[thick] (-1,1) -- (0,0);
    \node at (-1.6,0.9) {$u^R_m$, $u^L_m$};
    \draw[thick] (-1,-1) -- (0,0);
    \node at (-1.6,-0.9) {$u^{L\dagger}_m$, $u^{R\dagger}_m$};
    \draw[thick,dashed] (0,0) -- (1.2,0);
    \node at (1.4,0) {$\eta$};
  \end{tikzpicture}
  \qquad\qquad
  \begin{tikzpicture}
    \draw[thick] (-1,1) -- (0,0);
    \node at (-1.6,0.9) {$d^R_m$, $d^L_m$};
    \draw[thick] (-1,-1) -- (0,0);
    \node at (-1.6,-0.9) {$d^{L\dagger}_m$, $d^{R\dagger}_m$};
    \draw[thick,dashed] (0,0) -- (1.2,0);
    \node at (1.4,0) {$\eta$};
  \end{tikzpicture}
\end{gather*}
Importantly, the $W^\pm_\mu$ weak interactions only act on the left-handed
quarks, while the strong, electromagnetic, and Higgs interactions act
democratically on the left- and right-handed quarks. Note also that the Higgs
interactions can change the handedness, while $W^\pm_\mu$ interactions can
change the generation of quarks.

\subsubsection{Lepton sector}

Spontaneous symmetry breaking of the lepton sector Lagrangian follows from the
generic discussion in \cref{sec:electroweak-Higgs}, with $\chi=L^L_m$ and
$\psi=e^R_m$
\begin{align}
  \frac{1}{\hbar c}\mathcal{L}_{\text{lepton}}^{\text{SSB}}
  &= \frac{i}{2} \lb e^{R\dagger}_m \bar\sigma^{\mu}\tilde\Df_\mu e^R_m
    - \tilde\Df_\mu e^{R\dagger}_m \bar\sigma^{\mu} e^R_m \rb \nn\\
  &\qquad
    + \frac{i}{2} \lb \nu^{L\dagger}_m \sigma^{\mu}\dow_\mu \nu^L_m
    - \dow_\mu \nu^{L\dagger}_m \sigma^{\mu} \nu^L_m \rb
    + \frac{i}{2} \lb e^{L\dagger}_m \sigma^{\mu}\tilde\Df_\mu e^L_m
    - \tilde\Df_\mu e^{L\dagger}_m \sigma^{\mu} e^L_m \rb \nn\\
  &\qquad
    - \lb \lambda_e^R\, e^{R\dagger}_m \bar\sigma^{\mu} e^{R}_m
    + \lambda_\nu^L\, \nu^{L\dagger}_m \sigma^{\mu} \nu^{L}_m
    + \lambda_e^L\, e^{L\dagger}_m \sigma^{\mu} e^{L}_m \rb Z_\mu \nn\\
  &\qquad
    + \frac{g_w}{\sqrt2} \lb
    W^+_\mu \nu^{L\dagger}_m \sigma^{\mu} e^{L}_m
    + W^-_\mu e^{L\dagger}_m \sigma^{\mu} \nu^{L}_m \rb
    - \frac{v+\eta}{\sqrt2} 
    \lb Y^e_{mn} e^{L\dagger}_me^{R}_n + Y^{e*}_{mn} e^{R\dagger}_n e^{L}_m \rb, 
\end{align}
where
\begin{gather}
  \lambda_e^R = - g_y\sin\theta_w, \nn\\
  \lambda_{\nu}^L = - \half g_y
  \sin\theta_w - \half g_w \cos\theta_w, \qquad
  \lambda_{e}^L = - \half g_y
  \sin\theta_w + \half g_w \cos\theta_w.
\end{gather}
and the residual $\SU(3)\times\rmU(1)$ transformations act as
\begin{align}
  \tilde\Df_\mu e^{L,R}_m
  &\to \dow_\mu e^R_m + ie A_\mu e^R_m.
\end{align}
Note that neutrinos are uncharged under strong and electromagnetic forces
$\SU(3)\times\rmU(1)$.

Focusing on the quadratic sector of the Lagrangian, we obtain
\begin{align}
  \frac{1}{\hbar c}\mathcal{L}_{\text{lepton}}^{\text{SSB}}
  &= \frac{i}{2} \lb e^{R\dagger}_m \bar\sigma^{\mu}\dow_\mu e^R_m
    - \dow_\mu e^{R\dagger}_m \bar\sigma^{\mu} e^R_m \rb
    + \frac{i}{2} \lb e^{L\dagger}_m \sigma^{\mu}\dow_\mu e^L_m
    - \dow_\mu e^{L\dagger}_m \sigma^{\mu} e^L_m \rb \nn\\
  &\qquad
    + \frac{i}{2} \lb \nu^{L\dagger}_m \sigma^{\mu}\dow_\mu \nu^L_m
    - \dow_\mu \nu^{L\dagger}_m \sigma^{\mu} \nu^L_m \rb 
    - \frac{v}{\sqrt2} 
    \lb Y^e_{mn} e^{L\dagger}_me^{R}_n + Y^{e*}_{mn} e^{R\dagger}_n e^{L}_m \rb, 
\end{align}
Similar to the quark sector, we can perform a unitary transformation on the
leptons in the flavour (generation) space
\begin{equation}
  e^{L}_{m}\to \alpha^{eL}_{mn}e^{L}_n, \qquad
  e^{R}_{m}\to \alpha^{eR}_{mn}e^{R}_n, \qquad
  \nu^{L}_{m}\to \alpha^{eL}_{mn}\nu^{L}_n, \qquad
  \nu^{R}_{m}\to \alpha^{eR}_{mn}\nu^{R}_n,
\end{equation}
for arbitrary unitary matrices $\alpha^{eL}_{mn}$ and $\alpha^{eR}_{mn}$, so
that the Yukawa coupling matrix diagonalises to 
\begin{equation}
  \frac{v\hbar}{\sqrt{2}c}\alpha^{eL\dagger}_{mr}\, Y_{rs}^{e}\, \alpha^{eR}_{sn}
  = m^e_{mn},
\end{equation}
where $m^{e}_{mn}$ is a diagonal matrix with positive entries. Note that we
transformed both $e^{L,R}_m$ and $\nu^{L,R}_m$ with the same matrix
$\alpha^{eL,R}_{mn}$, which is fine because there is no ``neutrino mass matrix''
to diagonalise. The transformed free Lagrangian can be represented in the Dirac
representation by defining 4-component Dirac spinors
\begin{equation}
  \nu_m =
  \begin{pmatrix}
    0 \\ \nu_m^L
  \end{pmatrix}, \qquad
  e_m =
  \begin{pmatrix}
    e_m^R \\ e_m^L
  \end{pmatrix},
\end{equation}
leading to
\begin{align}
  \mathcal{L}_{\text{lepton}}
  &= \frac{i\hbar c}{2} \lb \bar\nu_m \gamma^\mu\dow_\mu \nu_m
    - \dow_\mu \bar\nu_m \gamma^\mu\nu_m \rb \nn\\
  &\qquad
    + \frac{i\hbar c}{2} \lb \bar e_m \gamma^\mu\dow_\mu e_m
    - \dow_\mu \bar e_m \gamma^\mu e_m \rb
    - m^e_{mn}c^2 \bar e_m e_n
    + \text{interactions}.
\end{align}
In this basis, we get the mass $m^{e}_{11}$ for electrons, $m^{e}_{22}$ for
muons, and $m^{e}_{33}$ for tau leptons.

Since we have transformed both $u_m$ and $\nu_m$ with the same matrix, we do not
get an analogue of the CKM matrix for leptons. To wit, the full Lagrangian takes
the form
\begin{align}
  \frac{1}{\hbar c}\mathcal{L}_{\text{lepton}}^{\text{SSB}}
  &= - \frac{i}{2} \lb e^{R\dagger}_m \bar\sigma^{\mu}\tilde\Df_\mu e^R_m
    - \tilde\Df_\mu e^{R\dagger}_m \bar\sigma^{\mu} e^R_m \rb \nn\\
  &\qquad
    - \frac{i}{2} \lb \nu^{L\dagger}_m \sigma^{\mu}\dow_\mu \nu^L_m
    - \dow_\mu \nu^{L\dagger}_m \sigma^{\mu} \nu^L_m \rb
    - \frac{i}{2} \lb e^{L\dagger}_m \sigma^{\mu}\tilde\Df_\mu e^L_m
    - \tilde\Df_\mu e^{L\dagger}_m \sigma^{\mu} e^L_m \rb \nn\\
  &\qquad
    + \lb \lambda_e^R\, e^{R\dagger}_m \bar\sigma^{\mu} e^{R}_m
    + \lambda_\nu^L\, \nu^{L\dagger}_m \sigma^{\mu} \nu^{L}_m
    + \lambda_e^L\, e^{L\dagger}_m \sigma^{\mu} e^{L}_m \rb Z_\mu \nn\\
  &\qquad
    - \frac{g_w}{\sqrt2} \lb
    W^+_\mu \nu^{L\dagger}_m \sigma^{\mu} e^{L}_m
    + W^-_\mu e^{L\dagger}_m \sigma^{\mu} \nu^{L}_m \rb
    - \frac{v+\eta}{v\hbar/c} m^e_{mn}
    \lb e^{L\dagger}_me^{R}_n + e^{R\dagger}_n e^{L}_m \rb.
\end{align}
The (generations of) electrons can interact via electromagnetic forces
\begin{equation*}
  \begin{tikzpicture}
    \draw[thick] (-1,1) -- (0,0);
    \node at (-1.6,0.9) {$e^L_m$, $e^R_m$};
    \draw[thick] (-1,-1) -- (0,0);
    \node at (-1.7,-0.9) {$e^{L\dagger}_m$, $e^{R\dagger}_m$};
    \draw[thick,wave] (0,0) -- (1.2,0);
    \node at (1.4,0) {$A$};
  \end{tikzpicture}
\end{equation*}
while both electrons and neutrons can interact via weak forces
\begin{gather*}
  \begin{tikzpicture}
    \draw[thick] (-1,1) -- (0,0);
    \node at (-2,0.9) {$\nu^L_m$, $e^L_m$, $e^R_m$};
    \draw[thick] (-1,-1) -- (0,0);
    \node at (-2.1,-0.9) {$\nu^{L\dagger}_m$, $e^{L\dagger}_m$, $e^{R\dagger}_m$};
    \draw[thick,weak] (0,0) -- (1.2,0);
    \node at (1.4,0) {$Z$};
  \end{tikzpicture} \nn\\
  \begin{tikzpicture}
    \draw[thick] (-1,1) -- (0,0); 
    \node at (-1.3,0.9) {$e^L_m$};
    \draw[thick] (-1,-1) -- (0,0);
    \node at (-1.3,-0.9) {$\nu^{L\dagger}_m$};
    \draw[thick,weak] (0,0) -- (1.2,0);
    \node at (1.6,0) {$W^+$};
  \end{tikzpicture} \qquad\qquad
  \begin{tikzpicture}
    \draw[thick] (-1,1) -- (0,0); 
    \node at (-1.3,0.9) {$\nu^L_m$};
    \draw[thick] (-1,-1) -- (0,0);
    \node at (-1.3,-0.9) {$e^{L\dagger}_m$};
    \draw[thick,weak] (0,0) -- (1.2,0);
    \node at (1.6,0) {$W^-$};
  \end{tikzpicture}
\end{gather*}
Only the electrons interact with the residual Higgs field
\begin{gather*}
  \begin{tikzpicture}
    \draw[thick] (-1,1) -- (0,0);
    \node at (-1.6,0.9) {$e^R_m$, $e^L_m$};
    \draw[thick] (-1,-1) -- (0,0);
    \node at (-1.6,-0.9) {$e^{L\dagger}_m$, $e^{R\dagger}_m$};
    \draw[thick,dashed] (0,0) -- (1.2,0);
    \node at (1.4,0) {$\eta$};
  \end{tikzpicture}
\end{gather*}
Leptons do not talk to the strong force gluons. Notably, there are no flavor
(generation) changing interactions in this model.

We could extend the Standard Model to include right-handed neutrinos. However,
at the time the Standard Model was proposed, neutrinos were known to be
massless, i.e. they do not admit a Yukawa interaction. As such, we know that
neutrinos are uncharged under electromagnetic and strong forces, while the
right-handed neutrinos will also be uncharged under weak forces. In the absence
of a Yukawa coupling, the right handed neutrinos will not talk to the residual
Higgs field either. Consequently, right-handed neutrinos completely decouple
from the rest of the Standard Model and leave no physically distinguishable
signatures. However, now we have experimental evidence that neutrinos do have a
nonzero mass. It implies that we can indeed add right-handed neutrinos with
non-trivial Yukawa interactions. These right-handed neutrinos talk to the rest
of the Standard Model via Higgs interactions. In the process of diagonalising
the neutrino mass matrix, we will encounter a non-diagonal PMNS
(Pontecorvo-Maki-Nakagawa-Sakata) matrix, similar to the CKM matrix for quarks,
leading to neutrino and electron flavor oscillations. Unlike the CKM matrix, the
PMNS matrix is experimentally observed to be quite far from diagonal. Hence,
neutrino oscillations are more readily observed in nature.

\newpage

\section{Discrete symmetries}
\label{sec:discrete_symmetries}

Let us rewind back to the spacetime Poincar\'e symmetries discussed in
\cref{sec:LorentzPoincare}. We noted that the Poincar\'e group
$\bbR^{3,1}\rtimes\rmO(3,1)$ (or its Lorentz subgroup $\rmO(3,1)$) is not
connected. The connected piece is the proper orthochronous Poincar\'e group
$\bbR^{3,1}\rtimes\SO^+(3,1)$, which excludes the discrete transformations:
parity P and time-reversal T, as well as their combination PT. Then we quickly
specialised to just the connected piece of the Poincar\'e group and ignored the
discrete transformations altogether. The reason is that physical theories,
including the Standard Model of particle physics, are not generically invariant
under P, T, or PT. As it turns out, physical theories are invariant under CPT
symmetry, including a charge conjugation transformation C that complex
conjugates all the fields and converts all particles to anti-particles and
vice-versa. There is, in fact, a \emph{CPT theorem} that states that all
Lorentz-invariant local field theories with a Hermitian Lagrangian (Hamiltonian)
is CPT-invariant.

\paragraph*{Symmetry operators and algebra:} The discrete transformations C, P,
and T individually form cyclic $\bbZ_2$ groups, i.e. operating twice with any of
these operations returns to the identity operation:
$\rmC^2,\rmP^2,\rmT^2\propto\mathbb1$. All the three discrete transformations
mutually commute. Their action on the Poincar\'e generators is given as
\begin{gather}
  \rmC H\rmC^{-1} = H, \qquad
  \rmC P_i\rmC^{-1} = P_i, \qquad
  \rmC J_i\rmC^{-1} = J_i, \qquad
  \rmC K_i\rmC^{-1} = K_i, \nn\\
  \rmP H\rmP^{-1} = H, \qquad
  \rmP P_i\rmP^{-1} = -P_i, \qquad
  \rmP J_i\rmP^{-1} = J_i, \qquad
  \rmP K_i\rmP^{-1} = -K_i, \nn\\
  \rmT H\rmT^{-1} = H, \qquad
  \rmT P_i\rmT^{-1} = -P_i, \qquad
  \rmT J_i\rmT^{-1} = -J_i, \qquad
  \rmT K_i\rmT^{-1} = K_i.
\end{gather}
Note that all these transformations leave the Hamiltonian invariant, so that the
energy of a state is not changed. As a consequence, the time-reversal operator T
is anti-unitary and implements a complex conjugation
$\rmT i\rmT^{-1} = -i$.\footnote{This curious feature follows from the
  observation that
  \begin{equation}
    \rmT i\hbar c P_i\rmT^{-1}
    = \rmT [H,K_i]\rmT^{-1}
    = [\rmT H\rmT^{-1},\rmT K_i\rmT^{-1}]
    = [H,K_i] = i\hbar c P_i.
  \end{equation}
  However, since $\rmT P_i\rmT^{-1} = -P_i$, it must be true that
  $\rmT i\rmT^{-1} = -i$. This is to say that $\rmT$ is an ``anti-unitary
  operator'' that also implements a complex conjugation.} Note that the charge
conjugation operator C leaves all the Poincar\'e generators invariant. However,
given an internal symmetry Lie group $G$ (such as the QCD group $\SU(3)$ or the
electroweak group $\SU(2)\times\rmU(1)$), both P and T commute with the
respective generators $T_a$, while C flips them, i.e.
\begin{equation}
  \rmC T_a\rmC^{-1} = -T_a, \qquad
  \rmP T_a\rmP^{-1} = T_a, \qquad
  \rmT T_a\rmT^{-1} = T_a.
\end{equation}
The combined CPT symmetry acts on the generators as
\begin{gather}
  (\mathrm{CPT}) H (\mathrm{CPT})^{-1} = H, \qquad
  (\mathrm{CPT}) P_i (\mathrm{CPT})^{-1} = P_i, \nn\\
  (\mathrm{CPT}) J_i (\mathrm{CPT})^{-1} = -J_i, \qquad
  (\mathrm{CPT}) K_i (\mathrm{CPT})^{-1} = -K_i, \nn\\
  (\mathrm{CPT}) T_a (\mathrm{CPT})^{-1} = -T_a,
\end{gather}
or covariantly
\begin{gather}
  (\mathrm{CPT}) P_\mu (\mathrm{CPT})^{-1} = P_\mu, \qquad
  (\mathrm{CPT}) M_{\mu\nu} (\mathrm{CPT})^{-1} = -M_{\mu\nu}, \nn\\
  (\mathrm{CPT}) T_a (\mathrm{CPT})^{-1} = -T_a.
\end{gather}

\paragraph*{Action on fields:}

Let us go back to the Lorentz representations, which were classified in
\cref{sec:LorentzReps} using the eigenvalues $(\vec X^\pm)^2$ operators, where
$X_i^\pm = 1/2(J_i\pm iK_i)$. Under parity transformation,
$\rmP:X^+_i \leftrightarrow X^-_i$, while under time-reversal,
$\rmT:X^\pm_i \to -X^\pm_i$. Hence, a representation $(j_-,j_+)$ goes to
$(j_+,j_-)$ under parity, but stays invariant under time-reversal. We see that
the parity operation turns a left-chiral particle into a right-chiral one and
vice-versa. This also follows from the action of discrete symmetries on the
chirality operator $\Gamma$, i.e. $\rmP^{-1}\Gamma\rmP = -\Gamma$ and
$\rmT^{-1}\Gamma\rmT = \Gamma$.  As a consequence, parity-invariant or
\emph{achiral} theories must either include particles in the representations
$(j,j)$, like vectors, or $(j_-,j_+)\oplus(j_+,j_-)$ for $j_+\neq j_-$, like
Dirac spinors. The charge conjugation operator C, on the other hand, leaves the
Lorentz representation invariant but converts particles to anti-particles and
vice-versa. To wit, the action of C, P, T on a scalar field $\varphi(t,\vec x)$
is defined as
\begin{align}
  \rmC&: \varphi(t,\vec x) \to \eta_C\varphi^*(t,\vec x), \nn\\
  \rmP&: \varphi(t,\vec x) \to \eta_P\varphi(t,-\vec x), \nn\\
  \rmT&: \varphi(t,\vec x) \to \eta_T\varphi(-t,\vec x).
\end{align}
Here $\eta_C,\eta_P,\eta_T$ are arbitrary phases, with
$|\eta_C|^2=|\eta_P|^2=|\eta_T|^2=1$, which are intrinsic properties of the
field in question. One can check that all of these are independently symmetries
of the scalar field theory from \cref{sec:GlobalExamples-Scalar}. For a pair of
left- and right-handed Weyl spinors $\psi(t,\vec x)$, $\chi(t,\vec x)$, we
instead have
\begin{align}
  &\rmC: \psi(t,\vec x)
    \to -i\eta_C\sigma_2 \chi^*(t,\vec x), 
  &&\rmC: \chi(t,\vec x)
     \to i\eta_C\sigma_2\psi^*(t,\vec x), \nn\\
  &\rmP: \psi(t,\vec x)
  \to \eta_P \chi(t,-\vec x), 
  &&\rmP: \chi(t,\vec x)
     \to \eta_P\psi(t,-\vec x), \nn\\
  &\rmT: \psi(t,\vec x)
  \to i\eta_T\sigma_2\psi(-t,\vec x), 
  &&\rmT: \chi(t,\vec x)
  \to i\eta_T\sigma_2\chi(-t,\vec x).
\end{align}
C and P individually are \emph{not} symmetries of the Weyl spinor field theories
from \cref{sec:GlobalExamples-Spinor}, whereas T and the combined transformation
CP are symmetries.\footnote{Note that fermionic fields anti-commute. So, for
  instance, $\bar\Psi^*\Psi^* = \Psi^\rmT\bar\Psi^\rmT = -\bar\Psi \Psi$. This
  property is vital to illustrate invariance of spinor field theories under CP.}
Note that T transformation is also supposed to change all $i\to-i$ in the
Lagrangian. For a Dirac spinor $\Psi(t,x)$, the transformation rules imply
\begin{align}
  &\rmC: \Psi(t,\vec x)
    \to i\eta_C \gamma^0\gamma^2 \bar\Psi^\rmT(t,\vec x),  \nn\\
  &\rmP: \Psi(t,\vec x)
  \to \eta_P \gamma^0\Psi(t,-\vec x), \nn\\
  &\rmT: \Psi(t,\vec x)
  \to i\eta_T\gamma_5\gamma^2\gamma^0\Psi(-t,\vec x), 
\end{align}
Interestingly, all of these can be individually checked to be symmetries of the
Dirac spinor field theory from \cref{sec:GlobalExamples-Spinor} up to an overall
sign in the action. For a vector field $V^\mu(t,\vec x)$, we instead have
\begin{align}
  \rmC&: V^\mu(t,\vec x) \to \eta_CV^{*\mu}(t,\vec x), \nn\\
  \rmP&: V^\mu(t,\vec x) \to -\eta_P \bbP^\mu_\nu V^\nu(t,-\vec x), \nn\\
  \rmT&: V^\mu(t,\vec x) \to \eta_T \bbP^\mu_\nu V^\nu(-t,\vec x),
\end{align}
where $\bbP^\mu_\nu = \diag(1,-1,-1,-1)$.


Moving on to Poincar\'e representations, it is easy to see that all the discrete
symmetries leave the Casimir operators $P^2$ and $W^2$ invariant, whereas the
helicity operator $W_0$ flips sign under parity. As a consequence, massive
particles transforming in some representation $(m,s)$ are invariant under C, P,
and T; that is to say that the mass and spin of a particle are invariant. On the
other hand, massless particles transforming in some representation $(h)$ turn
into $(-h)$ under parity, but remain invariant under time-reversal and charge
conjugation. As a consequence, parity-invariant massless particles must either
transform in the representation $(0)$, like massless scalars, or in the
representation $(h)\oplus(-h)$ for $h\neq 0$, like photons and
gravitons.

\paragraph*{CPT theorem:} Let us consider the combined CPT operation. On various
kinds of fields mentioned above, the action of CPT is given as
\begin{align}
  \mathrm{CPT}&: \varphi(t,\vec x) \to
                \eta_C\eta_P\eta_T\,\varphi^*(-t,-\vec x), \nn\\
  \mathrm{CPT}&: \psi(t,\vec x) \to
                -\eta_C\eta_P\eta_T\,\psi^*(-t,-\vec x), \nn\\
  \mathrm{CPT}&: \chi(t,\vec x) \to
                \eta_C\eta_P\eta_T\,\chi^*(-t,-\vec x), \nn\\
  \mathrm{CPT}&: \Psi(t,\vec x) \to
                -\eta_C\eta_P\eta_T\,\gamma_5\Psi^*(-t,-\vec x), \nn\\
  \mathrm{CPT}&: V^\mu(t,\vec x) \to
                -\eta_C\eta_P\eta_T\,V^{*\mu}(-t,-\vec x), 
\end{align}
and so on for higher-spin fields. The CPT theorem states that for any Hermitian
local Lorentz-invariant field theory, we can individually pick the phases
$\eta_C,\eta_P,\eta_T$ of the fields so that the action of the theory is
CPT-invariant.

Note that CPT transformation is anti-unitary, i.e. it flips all $i\to-i$ in the
Lagrangian. If the Lagrangian is Hermitian, i.e.
$\mathcal{L}^\dagger = \mathcal{L}$, we can undo this flip by a subsequent
complex conjugation of the entire Lagrangian. The coordinate flip
$x^\mu\to-x^\mu$ in the arguments of the fields can also be undone by a
redefinition of the integration variables $x^\mu\to-x^\mu$ in the action
$S=\int\df^4x\,\mathcal{L}(x) \to \int\df^4x\,\mathcal{L}(-x)$. Consequently, we
get the action of the CPT transformation on the Lagrangian as
\begin{align}
  \mathrm{CPT}^*&: \varphi(t,\vec x) \to
                \eta^*_C\eta^*_P\eta^*_T\,\varphi(t,\vec x), \nn\\
  \mathrm{CPT}^*&: \psi(t,\vec x) \to
                -\eta^*_C\eta^*_P\eta^*_T\,\psi(t,\vec x), \nn\\
  \mathrm{CPT}^*&: \chi(t,\vec x) \to
                \eta^*_C\eta^*_P\eta^*_T\,\chi(t,\vec x), \nn\\
  \mathrm{CPT}^*&: \Psi(t,\vec x) \to
                -\eta^*_C\eta^*_P\eta^*_T\,\gamma_5\Psi(t,\vec x), \nn\\
  \mathrm{CPT}^*&: V^\mu(t,\vec x) \to
                -\eta^*_C\eta^*_P\eta^*_T\,V^{\mu}(t,\vec x), 
\end{align}
The bi-product of this procedure is that the spacetime derivatives in the
Lagrangian flip sign
\begin{equation}
  \mathrm{CPT}^*: \dow_\mu \to -\dow_\mu,
\end{equation}
and the order of fields is flipped, which has consequences for anti-commuting
spinor fields. Lorentz invariance requires that all Lorentz indices
$\mu,\nu,\ldots$ appear in contracted pairs in the Lagrangian. Similarly, all
the spinors appear as bilinears, such as $\bar\Psi\Psi$. Therefore, for
invariance under $\mathrm{CPT}^*$, and hence under CPT, it is sufficient to
require that the phases multiply to unity $\eta_C\eta_P\eta_T=1$ for all
fields.\footnote{The argument of spinor bilinears is slightly subtle due to
  their anti-commuting nature. For instance
  \begin{equation}
    \bar\Psi\Psi
    = \Psi^\dagger\gamma^0\Psi \overset{\text{CPT}^*}{\longrightarrow}
    \big(\mathrm{CPT}^*(\Psi)^{\dagger}\gamma^0 \mathrm{CPT}^*(\Psi) \big)^\rmT
    = \big(\Psi^{\dagger}\gamma_5\gamma^0 \gamma_5\Psi \big)^\rmT
    = -(\bar\Psi\Psi)^\rmT
    = \bar\Psi\Psi.
  \end{equation}
  Note the change of order of fields implemented by the $\mathrm{CPT}^*$
  operation via the transpose. In the last step, we have used the anti-commuting
  nature of $\Psi$ fields.} Hence the CPT theorem. Note that this is only a
sufficient condition, not necessary. There can indeed be CPT-invariant theories
where the phases of fields do not multiply to unity.

CPT theorem also implies that the following discrete transformations are
equivalent for a Hermitian local Lorentz-invariant field theory:
$\rmC\rmP\approx \rmT$, $\rmP\rmT\approx\rmC$, and $\rmC\rmT\approx\rmP$. 

\paragraph*{Discrete symmetries of the Standard Model:} The electroweak symmetry
in the Standard Model distinguishes between left- and right-handed spinors. It
immediately follows that C and P are not symmetries of the Standard Model
Lagrangian. The Lagrangian does have an ``approximate'' CP or T symmetry,
provided that we take $\eta_C\eta_P=1$ and $\eta_T=1$ for the gauge fields
$B_\mu$, $W_\mu$, $G_\mu$ (and consequently also for $W^\pm_\mu$, $Z_\mu$, and
$A_\mu$). The only violation of the CP symmetry in the Standard Model arises
from the flavour-changing weak interactions in the quark sector given in the
second-last line of \cref{eq:final-quark-sector}. To wit, CP symmetry requires
that the CKM matrix $V_{mn}$ be real, which is experimentally known not to be
the case and was awarded by the 1980 Nobel Prize in
physics~\cite{nobel_physics_1980}. Introducing right-handed neutrinos in the
Standard Model, similar CP-violation can also occur in the lepton sector via the
PMNS matrix. However, the scientific community has still not reached a consensus
on whether the flavour-changing weak interactions in the lepton sector violate
CP~\cite{Kelly:2020fkv}. Finally, due to the CPT theorem, the Standard Model
does have CPT symmetry.

\newpage

\section{Outlook}
\label{sec:outlook}

Physicists really dislike freedom when it comes to model building. Good physical
models are usually those that survive the Occam's razor, i.e. reliably explain
all the known observational evidence with the least number of free
parameters. Lesser the number of free parameters a model has, lesser fine-tuning
it requires to reconcile with the known observations and more is its predictive
power for future experiments. Symmetries serve as a crucial tool to this
end. Requiring that a model respects the symmetries of the physical system it
aims to describe, or at times postulating entirely new symmetries with the hope
that they are realised by the system in question, dramatically brings down the
number of free parameters admissible by the model. For instance, the residual
U(1) symmetry in the Standard Model does not allow for any interactions in the
Lagrangian that do not conserve the total electric charge -- a fact that has
been experimentally scrutinised without a shadow of doubt.


This was a course on the mathematical foundations of symmetries and their
implementation in quantum field theories. We discussed how the symmetries
exhibited by a physical system, their action on the underlying field content,
and their spontaneous breaking pattern can enable us to construct field
theoretic models describing these systems. We started with the foundations of
group theory -- the mathematical language for symmetries -- and provided a
detailed analysis of U(1), $\SU(N)$, and Poincar\'e groups that enter the
discussion of particle physics. We discussed various representations of these
symmetry groups, providing us with a classification of fields on the basis of
their transformation properties under the action of these symmetries. We
formalised how these symmetries can be implemented in field theories -- globally
or locally -- and inspected the consequences of the ground state not respecting
(a part of) the symmetries of the full theory. We then implemented these ideas
to write down the Lagrangian for the Standard Model of particle physics starting
from the global Poincar\'e symmetries, local internal
$\SU(3)\times\SU(2)_L\times\rmU(1)_Y$ symmetries, their action on the field
content: gauge fields, quarks, leptons, and the Higgs field, and the spontaneous
breaking of the internal symmetry down to $\SU(3)\times\rmU(1)$ by the Higgs
potential.

It is generally agreed that the Standard Model is not enough. There is the
obvious issue of neutrinos being massless in the Standard Model, however this
can be fixed relatively easily by introducing right-handed neutrinos. The
Standard Model also has a number of inconsistencies with the $\Lambda$CDM model
of cosmology, including explanations for the amount of cold dark matter and dark
energy in the universe, the matter-antimatter inhomogeneity, and the mechanism
for cosmic inflation. All of these are still active topics of research in the
physics community. Then there is the issue of gravity: the Standard Model of
particle physics only describes three out of four fundamental forces, excluding
gravity. The reconciliation of the Standard Model with our present understanding
of gravity -- Einstein's general theory of relativity -- has proven to be
incredibly challenging and is still an open question in physics.

There are also a number of ``naturalness'' issues with the Standard Model. The
weak force happens to be $10^{24}$ times stronger than the weakest fundamental
force in nature -- gravity. This is known as the hierarchy problem, for which
the Standard Model offers no viable explanation. A possible resolution is
offered by ``supersymmetry'', which extends the spacetime Poincar\'e symmetries
to include fermionic generators. However, to date, there has been no conclusive
experimental evidence for supersymmetry. A somewhat different hierarchy problem
also arises due to the vastly different masses across the generations of quarks
and leptons. Then there is the ``strong CP problem'' which asks why the weak
force violate the CP symmetry but not the strong force. The Standard Model also
does not provide any explanation for why the electric charges of fundamental
fields are quantised in multiples of $e/3$. Furthermore, there are a total of 18
free parameters in the Standard Model (6 quark masses, 3 lepton masses, Higgs
mass, Higgs vev, 3 gauge coupling constants, and 4 elements of the CKM
matrix). In fact, if we were to also include right-handed neutrinos, there would
be 7 additional free parameters (3 neutrino masses and 4 elements of the PMNS
matrix). Although not a problem as such, it would be preferable if we can reduce
this number down, i.e. find physical reasons for why the parameters in the
Standard Model are what they are.

As far as reducing the free parameters goes, it is natural to speculate that
there is a stronger physical symmetry which further constrains the Standard
Model Lagrangian. This lore goes by the name of \emph{grand unified
  theories}. Another motivation for grand unified theories is that the three
gauge couplings in the Standard Model $g_y,g_w,g_s$ seem to converge at around
$10^{15}$ GeV under renormalisation group flow. It is speculated that at this
scale, the three fundamental forces combine into a single grand unified force
described by a simple group with a single coupling constant. The simplest simple
group that seems to work at a technical level is $\SU(5)$ that admits the
Standard Model group $\SU(3)\times\SU(2)_L\times\rmU(1)_Y$ as a subgroup. As a
bonus, the SU(5) grand unified model also explains the quantisation of electric
charges. Unfortunately, it has been practically ruled out by the bounds coming
from the lifetime of proton decay. There are higher groups that circumvent this
problem like $\SO(10)$ and $\rmE_6$. However, given that the present energy
scale at the LHC is around $10^4$ GeV, testing grand unified theories at
$10^{15}$ GeV is still a fantasy. The situation is worse for the ``theories of
everything'' that aim to describe gravity as well, such as string theory, where
the unification is only expected to happen at the Planck's scale of around
$10^{19}$ GeV.

Our focus has only been on the model building part in this course; we have not
really concerned ourselves with any of the features or physical predictions of
the Standard Model. Entire textbooks have been written on the subject that go
well beyond the scope of this introductory
course~\cite{Aitchison:2004cs,Burgess:2006hbd,Cheng:1984vwu}. Certain advanced
symmetry-related concepts have also been left out of the discussion. For
instance, we have mostly ignored the quantum aspects of symmetries. In
\cref{sec:Noether-thm}, we mentioned in passing that the conserved charges
furnish a representation of the symmetry algebra on the Hilbert space. However,
the quantisation process is considerably more subtle. It is possible for a
symmetry that is respected by the classical action of a theory to get violated
by quantum fluctuations. Such symmetries are said to be anomalous. Anomalies are
particularly problematic when they occur in gauge symmetries. This is because
gauge symmetries are really just redundancies in the choice of fundamental
fields and can typically be fixed by making a suitable gauge choice. Having an
anomalous gauge symmetry, therefore, signals an inconsistency in the choice of
fields at the quantum level. As it turns out, the gauge symmetries in the
Standard Model are indeed anomaly-free; see~\cite{Weinberg:1996kr}.

In this course, we have mainly focused on the implementation of symmetries in
particle physics. However, the techniques we developed are equally vital in
other areas of physics as well. As we indicated in the introduction, phases of
condensed matter systems can often be classified based on the symmetries they
respect. Fluids (liquids or gases) are phases of matter that are invariant under
spatial translations and rotations, while solids have these symmetries
broken. There are also a number of phases with an intermediate level of symmetry
breaking in between, commonly referred to as liquid crystals. Fluids often also
exhibit a global U(1) symmetry associated with the particle number conservation,
which can also be spontaneously broken to give rise to superfluids. The ensuing
massless Goldstone bosons are responsible for the superfluidity features such as
zero-viscosity flows. Similarly solids with a spontaneously broken global U(1)
symmetry lead to supersolid phases of matter. Superconductors, on the other
hand, have a spontaneously broken local U(1) symmetry. The consequent ``massive
gauge fields'' can explain the physical phenomenon in superconductivity such as
the Meissner effect. In cosmology, the universe at large scales is assumed to be
homogeneous and isotropic, which is reflected as translational and rotational
invariance in the models of the universe. At smaller scales, this symmetry gets
broken by the clusters of galaxies.

Near second-order phase transitions, physical systems can attain scale
invariance. In group theoretic terms, this is realised via an enhancement of the
spacetime Poincar\'e symmetries to include scale transformations (dilatations)
and often also the abstract special conformal transformation, together known as
the conformal group. Field theories that realise the conformal symmetry group
are called conformal field theories. Conformal symmetry is often strong enough
to completely fix all the two- and three-point correlation functions in a field
theory and even impose strong constraints on the higher-point interactions; see
\cite{DiFrancesco:1997nk}. In string theory, the worldsheet of the fundamental
string has conformal invariance. Conformal field theories also show up in the
context of the AdS/CFT correspondence: it is believed that certain conformal
field theories are dual to certain quantum gravitational theories living in
one-higher dimension. For most physical systems, conformal symmetry is too
restrictive, but it does serve as a spherical cow approximation to gain
qualitative insights into otherwise highly intractable physical systems.

Another spacetime symmetry that we have not had the scope to discuss are
diffeomorphisms. We discussed how global internal symmetries can be made into
local spacetime-dependent symmetries by introducing gauge fields. A spiritually
similar procedure can be implemented for spacetime Poincar\'e symmetries as
well. To wit, invariance of a field theory under spacetime Poincar\'e
transformations $x^\mu \to \Lambda^\mu{}_\nu x^\nu + a^\mu$ can be promoted to
arbitrary diffeomorphisms
$x^\mu\to \Lambda^\mu{}_\nu(x) x^\nu + a^\mu(x) = x'^\mu(x)$, by introducing a
spacetime metric $g_{\mu\nu}(x)$ and an affine connection
$\Gamma^\lambda_{\mu\nu}(x)$. Invariance of physics under arbitrary
diffeomorphisms is the defining feature of Einstein's general theory of
relativity that describes classical gravity. In principle, this procedure can
also be implemented to make the Standard Model Lagrangian invariant under
diffeomorphisms and obtain a model describing all four fundamental
forces. However, one runs into technical issues with renormalisability while
trying to quantise the theory.

This is where we draw this course to a close. Our journey has undoubtedly been
high on the technical side. We covered a lot of ground, starting from the
mathematical definition of a group all the way to the Standard Model
Lagrangian. I would not blame the reader if they failed to recall most of the
intricacies of the discussion in a span of a few months. However, if there is
one key message from the course that should stay with the reader, it would be
that symmetries are powerful. In the words of the German mathematician and
theoretical physicist Hermann Weyl~\cite{Weyl1980}, ``symmetry, as wide or as
narrow as you may define its meaning, is one idea by which man through the ages
has tried to comprehend and create order, beauty and perfection.''

\subsection*{Acknowledgements}

I am immensely thankful to Adam Ritz and Pavel Kovtun for their guidance and aid
throughout the preparation of these notes and the teaching of this course. I
would like to extend my sincerest gratitude to Adam Ritz for sharing his
handwritten notes on the subject that served as the backbone for this course to
expand and elaborate upon. I am also thankful to the University of Victoria for
allowing me the opportunity to teach this course and, of course, the graduate
students who suffered through it.

\newpage

\makereferences

\end{document}